\documentclass[sn-apa]{sn-jnl}

\usepackage[utf8]{inputenc}
\usepackage{amsmath,amssymb}
\usepackage{csquotes}
\usepackage{microtype}
\usepackage{float}
\usepackage{times}
\usepackage{dsfont}
\usepackage{nowidow}
\usepackage{xcolor}
\usepackage{graphicx}
\usepackage{mathtools}
\usepackage{pdflscape}
\usepackage{bibspacing}

\DeclareMathOperator{\Var}{Var}
\DeclareMathOperator{\Cov}{Cov}
\DeclareMathOperator{\diag}{diag}

\newcommand{\bo}{$\beta_1$}

\newcommand{\bi}{$\beta_{12}$}
\newcommand{\one}{$\bf{HC}_1$}
\newcommand{\zero}{$\bf{HC}_0$}
\newcommand{\two}{$\bf{HC}_2$}
\newcommand{\three}{$\bf{HC}_3$}
\newcommand{\four}{$\bf{HC}_4$}
\newcommand{\five}{$\bf{HC}_5$}

\newcommand{\vy}{\boldsymbol{y}}
\newcommand{\vX}{\boldsymbol{X}}
\newcommand{\vx}{\boldsymbol{x}}
\newcommand{\vb}{\boldsymbol{\beta}}
\newcommand{\vu}{\boldsymbol{u}}
\newcommand{\ve}{\boldsymbol{\epsilon}}
\newcommand{\vW}{\boldsymbol{\widehat{W}}}

\newcommand{\vE}{\boldsymbol{E}}
\newcommand{\vH}{\boldsymbol{H}}
\newcommand{\vD}{\boldsymbol{D}}
\newcommand{\vI}{\boldsymbol{I}}
\newcommand{\vP}{\boldsymbol{P}}
\newcommand{\vO}{\boldsymbol{\Omega}}

\jyear{2023}%

\theoremstyle{thmstyleone}%
\theoremstyle{thmstyletwo}%

\theoremstyle{thmstylethree}%

\begin{document}

\title[Robust Confidence Intervals for Meta-Regression]{Robust Confidence Intervals for Meta-Regression with Interaction Effects}


\author*[1]{\fnm{Thilo} \sur{Welz}}\email{welz@statistik.tu-dortmund.de}
\equalcont{These authors contributed equally to this work.}

\author[1]{\fnm{Eric} \sur{Knop}}
\equalcont{These authors contributed equally to this work.}

\author[2]{\fnm{Tim} \sur{Friede}}

\author[1,3]{\fnm{Markus} \sur{Pauly}}

\affil*[1]{\orgdiv{Department of Statistics}, \orgname{TU Dortmund University}, \orgaddress{\street{Vogelpothsweg 87}, \city{Dortmund}, \postcode{44221}, \country{Germany}}}

\affil[2]{\orgdiv{Medizinische Statistik}, \orgname{Universitätsmedizin Göttingen}, \orgaddress{\street{Humboldtallee 32}, \city{Göttingen}, \postcode{37073},  \country{Germany}}}

\affil[3]{\orgdiv{Research Center Trustworthy Data Science and Security}, \orgname{UA Ruhr}, \orgaddress{\street{Otto-Hahn-Straße 14}, \city{Dortmund}, \postcode{44227}, \country{Germany}}}


\abstract{Meta-analysis is an important statistical technique for synthesizing the results of multiple studies regarding the same or closely related research question. So-called meta-regression extends meta-analysis models by accounting for study-level covariates. Mixed-effects meta-regression models provide a powerful tool for evidence synthesis, by appropriately accounting for betweem-study heterogeneity. In fact, modelling the study effect in terms of random effects and moderators not only allows to examine the impact of the moderators, but often leads to more accurate estimates of the involved parameters. Nevertheless, due to the often small number of studies on a specific research topic, interactions are often neglected in meta-regression. In this work we consider the research questions (i) how moderator interactions influence inference in mixed-effects meta-regression models and (ii) whether some inference methods are more reliable than others. Here we review robust methods for confidence intervals in meta-regression models including interaction effects. These methods are based on the application of robust sandwich estimators for estimating the variance-covariance matrix of the vector of model coefficients. Furthermore, we compare different versions of these robust estimators in an extensive simulation study. We thereby investigate coverage and length of seven different confidence intervals under varying conditions. We conclude with some practical recommendations.  }




\maketitle

\section*{Introduction}
\label{sec:intro}

Meta-analysis is a statistical technique that combines the results of multiple studies to arrive at a single, more precise estimate of the effect size of a particular intervention or treatment. It aims to provide a comprehensive and quantitative summary of the available evidence on a particular topic, taking into account the heterogeneity of the studies and the sample sizes. By pooling the data from multiple studies, meta-analysis can increase the statistical power and accuracy of the results, and provide a more robust understanding of the effects of an intervention. Such statistical techniques are routinely applied in different areas of research, such as biology, medicine or psychology. In meta-regression, study-level covariates or moderators, which may influence the observed outcome in the respective study, are accounted for. A meta-regression combines the advantages of a linear regression model and a meta-analysis. On the one hand information from different studies is taken into account. On the other hand one is able to test not only for an overall effect, which is the case for most meta analyses, but also on effects of relevant study characteristics. The characteristics are used as study-level covariates and often called moderators. In contrast to a usual regression model, the mixed-effects model assumes that the estimated treatment effect is influenced by two different types of uncertainty: First, the estimated effect of a single study is assumed to be different from the studies' true effect by a random error. Second, the analyzed studies are assumed to have different true effects caused by differences between the studies, the so called between-study heterogeneity. Therefore, the treatment effects of the studies differ from the treatment effect for the entire population. It is important to account for this additional variation when confidence intervals of moderators are calculated \citep{raudenbush2009analyzing}.

 A simulation study by \cite{viechtbauer2015} showed that the choice of heterogeneity estimator had a negligible impact on performed test results. However, \cite{viechtbauer2015} also showed that when heterogeneity is present, the choice of estimator for the covariance of the vector of model coefficients has a large impact on test results. More specifically, in a model with only one moderator large differences in Type 1 error rates and the power of $t$-type tests were determined. Amongst others, tests based on a heteroscedasticity consistent ($\bf{HC}$) estimate of the covariance matrix introduced by \cite{white1980} and a modified covariance matrix estimate ($\bf{HKSJ}$) introduced by \cite{knapp2003improved} (and also \cite{sidik2005simple}) were considered. The $\bf{HC}$ estimate is an established approach in econometrics, but not commonly applied in meta-analysis, in particular when used in medical research. Because of its structure, it is also known as a sandwich estimator and is used for robust inference. The $\bf{HKSJ}$ estimate is common in meta-analyses applied in medicine. In \citeauthor{viechtbauer2015}'s (\citeyear{viechtbauer2015}) simulation study, tests based on the $\bf{HC}$ estimator turned out to be too liberal. In contrast, the test based on the $\bf{HKSJ}$ estimate performed the best among all considered tests. Since their results are limited to special settings, \cite{viechtbauer2015} suggested additional future simulation studies that consider, e.g., non-normal random effects, multiple covariates with multicollinearity and coverage probability of coefficients' confidence intervals. \cite{welzpauly2020} extended their research by comparing tests on the significance of the moderator based on six different versions of White's covariance matrix estimator and the Hartung-Knapp-Sidik-Jonkman variance-covariance matrix estimator for several random effect distributions. The six heteroscedasticity consistent covariance estimators are known as $\bf{HC}_0, ..., \bf{HC}_5$. The main difference between these different versions is how they transform the model residuals by discounting the observations' leverages \citep{neto2007,welzpauly2020}. In regression, leverage is a measure for how far away the covariate values of an observation are from those of the other observations. The newer \textbf{HC} estimators discount the leverages more strongly than earlier version. In a simulation study \cite{welzpauly2020} also found the $\bf{HKSJ}$ based tests to perform the best compared to the $\bf{HC}$ estimators. Amongst the $\bf{HC}$ estimators the $\bf{HC}_3-\bf{HC}_5$ based tests controlled the nominal significance level well and had power close to the $\bf{HKSJ}$ based tests for larger number of studies. The distribution of the random effect turned out to have almost no effect on the results \citep{welzpauly2020}.

In a recent meta-analysis, including meta-regression analyses, \cite{kimmoun2021temporal} analyzed mortality and readmission to hospital after acute heath failure. They found a statistically significant decline of death rates over calendar time. However, the median year of recruitment is correlated with the average age of the patients. This suggests that the observed trend might be explained by a neglected interaction of those variables. In fact, \cite{knop2023impact} showed in a re-analysis of the above mentioned data that it is vitally important to account for confounding and interaction effects, when making inference based on meta-regression with multiple moderators.

Motivated by this meta-analysis, the current paper extends the research of \cite{welzpauly2020} in two directions. Firstly, two moderators and, based on the important findings in \cite{knop2023impact}, their interaction term are modelled. Modelling interactions is required in situations where not only the influence of a moderator itself is of interest but its influence in the presence of other factors. Interactions are also helpful to assess the circumstances under which the influences of certain moderators on the estimated effect size are stronger or weaker \citep{aiken1991multiple}. Although modelling interaction terms is useful in providing additional insights, they are often neglected in meta-regression. However, neglecting existing interactions may dramatically alter conclusions drawn from quantitative research synthesis, as seen in a recent data analysis from acute heart failure research \citep{knop2023impact}. Secondly, confidence intervals are considered instead of hypothesis tests. 

The methodological aim of this work is to determine the performance of confidence intervals based on the seven covariance estimators $\bf{HC}_0-\bf{HC}_5$ and $\bf{HKSJ}$ in extensive simulations. On the one hand, it is investigated whether the confidence intervals of a single moderator's coefficient perform different in presence of an interaction. On the other hand, confidence intervals for the interaction coefficient itself are considered. For these more complex models it is of interest whether the estimators have the same properties as in the univariate model. Furthermore, we check how introducing non-normal distributions for the random effects influences results, similar to \cite{welzpauly2020}. The focus was additionally set on situations where the ${\bf KH}$ estimator does not perform the best among the considered estimators.

In Section \ref{sec:methods} we introduce the relevant methods, starting with the mixed-effects meta-regression model in Section \ref{sec:model}, followed by weighted least squares (WLS) estimation in Section \ref{sec:wls} and different estimators for the variance-covariance matrix of the estimated vector of coefficients in Section \ref{sec:estimators}. In Section \ref{sec:simulation} we describe the design and results of our extensive simulation study and provide recommendations for practical applications. Finally, we close with a discussion and an outlook for future research in Section \ref{sec:discussion}.

\section*{Statistical Methods}
\label{sec:methods}

\subsection*{The Mixed-Effects Meta-Regression Model}
\label{sec:model}

The study characteristics which are used as covariates in the meta-regression model are called moderators and are denoted with $\boldsymbol{x}_j = (x_{j1},\ldots,x_{jk})'$, where $k$ is the number of studies and $j \in \{0,1,\ldots,m\},$ with $m$ as the number of moderators. Functions of other moderators such as interactions of the form $x_{3i} = x_{1i} x_{2i}$ could be moderators themselves. The true outcome of an individual study $i \in \{1,\ldots,k\}$ is denoted with $\theta_i$. The model equation for the true outcome of study $i$ is

\begin{equation}
    \theta_i = \beta_0 + \beta_1 x_{1i} + \ldots + \beta_m x_{mi} + u_i.
\end{equation}

The parameters $\beta_1,\ldots,\beta_m$ are the regression coefficients of the associated moderators. We generally assume that the number of studies is greater than the number of study-level moderators, i.e. $k>m$. The deviation of the $ith$ studies' true outcome $\theta_i$ is modelled by the random effect $u_i$. The random effect $u_i$ is usually assumed to be normally distributed with $u_i \sim \mathcal{N}(0,\tau^2)$. Furthermore, the observed outcome for study $i$ is modelled as

\begin{equation}
    y_i = \theta_i + \varepsilon_i,
\end{equation}

\noindent
with model errors $\varepsilon_i \sim \mathcal{N}(0,\sigma_i^2)$. The model errors $\varepsilon_i$ and random effects $u_i$ are assumed to be independent. Together this yields what is also known as a normal-normal hierarchical model (NNHM) \citep{friede2017meta}. It is also possible to consider a more general semiparametric setting with the moment assumptions $\mathbb{E}(u_i)=0$ and $\Var(u_i)=\tau^2$ without other distributional restrictions on the random effects, as in \cite{welzpauly2020}. In matrix notation the model can be rewritten as

\begin{equation}
\label{eq:model}
    \boldsymbol{y} = \boldsymbol{X \beta} + \boldsymbol{u} + \boldsymbol{\varepsilon},
\end{equation}

\noindent
where 

\begin{align}
    \boldsymbol{y} = \begin{pmatrix}
        y_1\\
        \vdots\\
        y_k
    \end{pmatrix} \in \mathbb{R}^k, \ \boldsymbol{X} = \begin{pmatrix}
        1 \ \ldots \ x_{1m}\\
        \vdots \hspace{0.8cm}  \vdots\\
        1 \ \ldots \ x_{km}
    \end{pmatrix} \in \mathbb{R}^{k \times (m+1)},\\
\boldsymbol{u} = \begin{pmatrix}
        u_1\\
        \vdots\\
        u_k
    \end{pmatrix} \in \mathbb{R}^k \text{ and } \boldsymbol{\varepsilon} = \begin{pmatrix}
        \varepsilon_1\\
        \vdots\\
        \varepsilon_k
    \end{pmatrix} \in \mathbb{R}^k.
\end{align}

The design matrix $\boldsymbol{X}$ is assumed to have full rank. Under the assumption that $\boldsymbol{u}$ and $\boldsymbol{\varepsilon}$ are independent, the variance-covariance matrix of $\boldsymbol{y}$ is $\Var(\boldsymbol{y}) = \boldsymbol{V} = \diag(\sigma_1^2 + \tau^2,\ldots,\sigma_k^2 + \tau^2)$.

\subsection*{Weighted-Least-Squares Estimation}
\label{sec:wls}

The weighted least squares estimate for the model coefficients $\boldsymbol{\beta}$ is given by

\begin{equation}
\label{eq:beta}
    \boldsymbol{\hat{\beta}} = \boldsymbol{(X'\widehat{W}X)^{-1}X'\widehat{W}y},
\end{equation}

\noindent
with the weight matrix $\boldsymbol{\widehat{W}}$ typically (but not always) defined as the inverse variance matrix. For Model (\ref{eq:model}) it is given by $\boldsymbol{\widehat{W}} = \diag\left((\sigma_1^2+\hat{\tau}^2)^{-1},\ldots,(\sigma_k^2+\hat{\tau}^2)^{-1})\right)$. It should be noted that the sampling variances $\sigma_i$, $i=1,\ldots,k$ are assumed as known, although they are in fact estimated from the data. This is done for mathematical convenience and is common practice in meta-analysis \citep{dersimonian1986meta}. Various estimators are available for the between-study variance $\tau^2$ \citep{veroniki2016}. The recommendation for meta-analysis is to use either the restricted maximum likelihood (REML) or the Paule-Mandel estimator, both of which are iterative \citep{veroniki2016}. We denote the variance-covariance matrix of $\hat{\boldsymbol{\beta}}$ by $\boldsymbol{\Sigma} = \Cov(\hat{\boldsymbol{\beta}})$. It was shown that, given certain regularity conditions, $\hat{\boldsymbol{\beta}} \overset{a.s.}{\longrightarrow} \boldsymbol{\beta}$ as $k \longrightarrow \infty$ and $\hat{\boldsymbol{\beta}}$ asymptotically follows a normal distribution \citep{hedges2010robust}.

Given a consistent estimator $\boldsymbol{\widehat{\Sigma}}$ for the variance-covariance matrix of $\boldsymbol{\hat{\beta}}$, an approximate $(1-\alpha), \ \alpha \in (0,1)$ confidence interval (CI) for a coefficient $\beta_j$, $j \in \{0,1,\ldots,m\}$, is given by

\begin{equation}
\label{eq:ki}
    \left[\hat{\beta}_j \pm t_{k-m-1,1-\alpha/2}\sqrt{\boldsymbol{\widehat{\Sigma}}_{jj}} \right],
\end{equation}

\noindent
where $t_{k-m-1,1-\alpha/2}$ is the $(1-\alpha/2)$ quantile of the $t$-distribution with $k-m-1$ degrees of freedom and $\boldsymbol{\widehat{\Sigma}}_{jj}$ is the $j$th diagonal element of $\boldsymbol{\widehat{\Sigma}}$ \citep{sterchi2017weighted}. In the following we discuss various possibilities for estimating $\boldsymbol{\Sigma}$.

\subsection*{Estimators for the variance-covariance matrix of $\boldsymbol{\hat{\beta}}$}
\label{sec:estimators}

There are several ways to estimate the variance-covariance matrix of $\boldsymbol{\hat{\beta}}$. Here we focus on six heteroscedasticity consistent ($\bf{HC}$) estimators denoted by $\bf{HC}_0,~ \bf{HC}_1, ...,~ \bf{HC}_5$ and the Hartung-Knapp-Sidik-Jonkman ($\bf{HKSJ}$) estimator \citep{knapp2003improved, sidik2005simple}, which performed well in a meta-analytic context in previous research (\cite{viechtbauer2015}; \cite{welzpauly2020}; \cite{welz2022fisher}). In the following section we introduce  $\bf{HC}_0,~\bf{HC}_1,~\bf{HC}_2$ according to \cite{mackinnon1985some}, $\bf{HC}_3,~\bf{HC}_4$ according to \cite{neto2004} and $\bf{HC}_5$ according to \cite{neto2007} if not stated otherwise.

The $\bf{HC}$ estimators are all based on ${\bf \bf{HC}}_0$ which was originally introduced by \cite{white1980} for an ordinary least squares (OLS) estimator. For the meta-regression model in (\ref{eq:model}) and the estimator $\boldsymbol{\hat{\beta}}$ given in (\ref{eq:beta}) the estimator $\bf{\bf{HC}}_0$ can be written as 
\begin{equation}
{\bf \bf{HC}}_0=(\vX^\top\vW\vX)^{-1}\vX^\top\vW\vE\vD_0\vD_0^\top\vE^\top\vW\vX(\vX^\top\vW\vX)^{-1},
\label{eq:hc0}
\end{equation}
where $\vD_0=\vI_k$ and $\vE=\diag(\vy-\vX\hat{\vb})$ is a matrix containing the residuals $\hat{e}_i=y_i-\vx_i\boldsymbol{\hat{\beta}}$ on its diagonal \citep{welzpauly2020}.

How the formula for $\bf{HC}_0$ in (\ref{eq:hc0}) can be derived from the representation in \cite{mackinnon1985some} is shown in Section A of the Supplement. The formulas for $\bf{HC}_1-\bf{HC}_5$ can be derived analogously. Because the usual residuals tend to be too small \citep{mackinnon2013thirty}, $\bf{HC}_0$ tends to underestimate the variance of the components of $\boldsymbol{\hat{\beta}}$. A simple adjustment of this estimator is given by $ {\bf HC_1}=k(k-m-1)^{-1}{\bf HC_0},$ which takes the models' degrees of freedom $(k-m-1)$ into account. 

Another approach to fix this problem of $\bf{HC}_0$ is to modify the residuals themselves. One possible modification is to take the leverage scores $h_{ii}$ into account. The $h_{ii}$ denotes the $ith$ diagonal element of the hat matrix $\vH=\vX(\vX^\top\vW\vX)^{-1}\vX^\top\vW$. By using $\tilde{e}_i=\hat{e}_i/\sqrt{1-h_{ii}}$ instead of $\hat{e}_i$ there is more weight on residuals with higher leverage scores.  A representation of $\bf{HC}_2$ is given by (\ref{eq:hc0}) using $\vD_2=\diag((1-h_{ii})^{-\frac{1}{2}})$ instead of $\vD_0$.
Under homoscedasticity of the $\epsilon_i$ the estimator $\bf{HC}_2$ is unbiased. 

An estimator of similar form is $\bf{HC}_3$.
It can be written by using $\vD_3=\diag((1-h_{ii})^{-1})$ in place of $\vD_0$ in (\ref{eq:hc0}).
The estimator $\bf{HC}_3$ introduced here is a close approximation of Efrons' jackknife estimator \citep{efron1982jackknife}. A property of this estimator is that it takes the leverage scores stronger into account than $\bf{HC}_2$. 

The following estimator, $\bf{HC}_4$, also differs from the former estimator in the way that it incorporates the leverage scores. The idea is to weight the residuals stronger, when the leverage score $h_{ii}$ of a residual is relatively high compared to the average leverage score $\bar{h}=k^{-1}\sum_{i=1}^k h_{ii}$. This is done by using some $\delta_i$ as exponent for $(1-h_{ii})$, where $\delta_i=\min\left\{4, h_{ii}/\bar{h}\right\}.$ In this way the exponent $h_{ii}/\bar{h}$ is truncated at $\delta_i=4$. The resulting estimator $\bf{HC}_4$ is given by (\ref{eq:hc0}) with
$\vD_4=\diag((1-h_{ii})^{-\delta_i/2})$ instead of $\vD_0$, see \cite{zimmermann2020multivariate} for a similar estimator for multivariate analysis of covariance (MANCOVA).

Finally, $\bf{HC}_5$ is defined similar to $\bf{HC}_4$ but uses the exponents\\ $\alpha_i=\min\left\{h_{ii}/\bar{h}, \max\left\{4, \eta\cdot h_{max}/\bar{h}\right\}\right\}$ instead of $\delta_i$. Here, $h_{max}=\max\{h_{11},$
$\ldots,h_{kk}\}$ and $\eta\in(0,1)$ is a predefined constant used as a tuning parameter. The simulation study of \cite{neto2007} suggests $\eta=0.7$ as a reliable choice for finite samples; we follow this recommendation here. Notably $\alpha_i$ is only different from $\delta_i$ when $(\eta\cdot h_{max})/\bar{h}>4$. In this situation $\alpha_i$ is not truncated at $\alpha_i=4$ but at $\alpha_i=(\eta\cdot h_{max})/\bar{h}$. A representation of $\bf{HC}_5$ is given by (\ref{eq:hc0}) plugging in  $\vD_5=\diag((1-h_{ii})^{-\alpha_i/2})$ for $\vD_0$.

The Hartung-Knapp-Sidik-Jonkman estimator for the mixed-effects meta-regression model was independently introduced by \cite{knapp2003improved} and \cite{sidik2005simple}. It can be derived as follows. Let $\vP=\vI-\vX(\vX^\top\vW\vX)^{-1}\vX^\top\vW$ and $s^2=(k-m-1)^{-1}(\vy^\top\vP^\top\vW\vP\vy)=(k-m-1)^{-1}(\vy^\top\vW\vP\vy).$ Then the \textbf{HKSJ} estimator for Cov($\boldsymbol{\hat{\beta}}$) is given as \[ {\bf HKSJ}=s^2(\vX^\top\vW\vX)^{-1}.\]

\section*{Simulation Study}
\label{sec:simulation}

\subsection*{Simulation Design}

The simulation was conducted using the open source software package \texttt{R}. Relevant packages that were used for the analyses are \texttt{metafor}, \texttt{MASS} and \texttt{mvtnorm}. Visualizations, such as boxplots, were created using the \texttt{ggplot2}, \texttt{reshape2}, \texttt{grid} and \texttt{gridExtra} packages. The code is provided as supplementary material. The simulation setup expands upon the one by \cite{welzpauly2020}.

We start with a description of relevant effect measures for the simulation study. We consider the standardized mean difference (SMD), estimates of which are therefore the dependent variable in our meta-regression models. In many applications, $\theta_i$ is considered as the true SMD between the means of an experimental and a control group in the $ith$ study. An unbiased estimator $y_i$ for $\theta_i$ can be derived via a modification of Hedges' $g$. We describe the effect measure in the following, according to \cite{hedges1981distribution}. An unbiased estimator for the SMD is given by \citep{lin2021evaluation}

\begin{equation}
    g := \frac{\Gamma(n/2)}{\sqrt{n/2} \Gamma((n-1)/2)}d
\end{equation}

\noindent
with $n = n_T+n_C-2$, where $n_T$ and $n_C$ refer to the treatment and control group sizes. The regular Hedges' $g$ is defined as $d = (\Bar{x}_T - \Bar{x}_C)/s$, where $s$ is the pooled standard deviation with $s = \sqrt{\frac{(n_T-1)s_T^2+(n_C-1)s_C^2}{n}}$ and $s_T^2,s_C^2$ refer to the variances in the treatment and control groups respectively. The sampling variance of $g$ can be approximated by \citep{hedges2014statistical}

\begin{equation}
\label{eq:vars}
    v = \frac{1}{n_T}+\frac{1}{n_C}+\frac{g^2}{2(n_T+n_C)}.
\end{equation}

A mixed-effects meta-regression model with two covariates and their interaction is considered. The $y_i$ are assumed to be influenced by two covariates and their interaction. The interaction is modelled as $x_{i12} := x_{i1} x_{i2}$. Thus the model equation is given as

\begin{equation}
    y_i = \beta_1 x_{i1} + \beta_2 x_{i2} + \beta_{12} x_{i12} + u_i + \varepsilon_i.
\end{equation} 

The dependent variable $y_i$ is assumed to be the estimated SMD between an experimental and a control group in the $ith$ study for $i=1,\ldots,k$. There are four choices for the number of studies, $k \in \{6,10,20,50\}$. We note that test runs with $k=5$ frequently resulted in either a rank-deficient design matrix $\boldsymbol{X}$ or extremely wide confidence intervals. Therefore it cannot be recommended to use only $k = 5$ studies for a model with two covariates and interaction. We assume balanced study designs, i.e. $n_{T,i}=n_{C,i}=:n_i$ for each study. For each choice of $k \in \{6,10,20,50\}$ three different vectors of group sizes are considered. In the situation $k=6$, five studies contain the group sizes according to the following three vectors: $n_{15}=(6,8,9,10,42)',n_{25}=(16,18,19,20,52)'$ or $n_{50}=(41,43,44,45,77)'$. The size of the sixth study is set to the mean $\Bar{n}$ of the corresponding vector, either 15, 25 or 50. For $k \in \{10,20,50\}$ the vectors are repeated $k/5$ times and the resulting vector is used as the vector of study sizes. With this choice for the number of participants the study size vectors all have the same variance for a fixed $k$.

The covariates $x_{i1}$ and $x_{i2}$ are sampled from a joint normal distribution

\begin{equation*}
    \begin{pmatrix}
        x_{i1}\\
        x_{i2}
    \end{pmatrix} \sim \mathcal{N}\begin{pmatrix}
        1 & \varrho\\
        \varrho & 1
    \end{pmatrix},
\end{equation*}

\noindent
where $\varrho$ is the correlation between $x_{i1}$ and $x_{i2}$. We examined the settings of no correlation ($\varrho=0$), small correlation ($\varrho=0.2$), large correlation ($\varrho=0.5$) and large negative correlation ($\varrho=-0.5$). Possible adjustments for $\beta_1$, $\beta_2$ and $\beta_{12}$ are 0, 0.2 and 0.5. Additionally, the situation $\beta_{12}=-0.5$ is considered in order to check whether the estimates differ for a negative coefficient.

The random effects $u_i$ are chosen a $u_i = \tau q_i$, where $\tau^2 \in \{0.1,\ldots,0.9\}$ and the $q_i$'s are independently sampled from either a standard normal- or a standardized exponential-, Laplace-, log-normal- or $t_3$-distribution. Here $t_3$ denotes the $t$ distribution with three degrees of freedom. If $q_i$ is drawn from a standardized exponential distribution, then $q_i := a_i-1$ where $a_i \sim \exp(1)$. The $q_i$'s following a standardized Laplace distribution, are generated via $q_i = (a_i-b_i)/\sqrt{2}$ where $a_i,b_i \sim \exp(1)$ are sampled independently. For the $q_i$'s following a standardized log-normal distribution, $q_i$ is set to

\begin{equation*}
    q_i = \frac{\exp(z_i)-\exp(1/2)}{\sqrt{\exp(1)(\exp(1)-1)}},
\end{equation*}

\noindent
where $z_i \sim \mathcal{N}(0,1)$. Finally, $q_i$'s following a standardized $t_3$ distribution are set as $q_i = t_i/\sqrt{3}$ with $t_i \sim t_3$. The standardization of the $q_i$'s ensures that the corresponding $u_i$'s all have expectation $\mathbb{E}(u_i)=0$ and variance $\Var(u_i)=\tau^2$. Note, that if $u_i$ is not normally distributed, the $y_i$ are not normally distributed and the quantile $t_{k-m-1,1-\alpha/2}$ used in (\ref{eq:ki}) is not correct. However, results by \cite{kontopantelis2012performance} suggest that the distribution of the study outcomes has almost no impact on the resulting confidence intervals. Therefore the quantile of the $t$-distribution is used for this simulation as well.

The estimated effects (Hedges' g) $y_i$ are generated according to 

\begin{equation}
    g_i = \frac{\phi_i}{\sqrt{X_i/(2n_i-2)}},
\end{equation}

\noindent
where $\phi_i \sim \mathcal{N}(\theta_i,2/n_i)$ and $X_i \sim \chi^2_{(2n_i-2)}$ are sampled. The sampling variance $\sigma_i^2$ of $y_i$ is estimated using (\ref{eq:vars}). In total there are $77,760 = 3(\Bar{n}) \times 4(k) \times 9(\tau^2) \times 3(\beta_1) \times 3(\beta_2) \times 4(\beta_{12}) \times 5(u_i) \times 4(\varrho)$ different combinations of simulation parameters. For each combination the model is generated $N=10,000$ times. The confidence level is chosen as $1-\alpha=0.95$. For this choice of $N$ and $\alpha$ the Monte Carlo standard error of empirical coverage is approximately equal to $0.22\%$ \citep{morris2019using}. For each model the estimators $\bf{HC}_0$--$\bf{HC}_5$ and $\bf{HKSJ}$ are calculated and $\tau^2$ is estimated using the REML estimator, with a maximum of 5,000 iterations and a default step length of 0.5. Based on each estimator a $(1-\alpha)$ confidence interval is estimated for the coefficient $\beta_1$ of a single moderator and for the coefficient $\beta_{12}$ of the interaction term. Since $x_1$ and $x_2$ have the same distribution, intervals for $\beta_2$ are not considered. The proportion of estimated confidence intervals that cover the true coefficient is used as an estimate of the coverage probability. As an estimate of the interval length the average length of the estimated intervals is calculated.

\subsection*{Simulation Results}

In confidence interval estimation two properties are relevant, namely coverage and interval length. The actual coverage of the interval should be at least equal to the nominal confidence level $(1-\alpha)$. Second, we want to determine the interval, where the true parameter is included in with probability $(1-\alpha)\cdot 100\%$, as precisely as possible. This means of the intervals that have sufficient coverage, we choose the shortest one. Therefore, the coverage and lengths of the simulated intervals for $\beta_1$ (and $\beta_2$) as well as $\beta_\text{12}$ are compared in respect of the covariance estimators they are based on. Due to the high number of parameter adjustments not every adjustment is considered separately. Hence, the coverage and interval lengths of different settings are summarized by boxplots. That is, e.g., the boxplots in Section \ref{ssec:general} based upon the results for every adjustment of  $\beta_1, \beta_2, \beta_{12}, \rho, \tau^2$, $u_i$ and $\bar{n}$ and thus consider the overall performance of the estimators. 
 The aim of this section is to investigate, whether one estimator has a better overall performance compared to all other estimators. It is also of interest, whether there are any estimators that are outperformed by at least one other estimator in each situation. Because the intervals for $\beta_1$ and $\beta_{12}$ performed similarly for the most estimators and parameter adjustments, only the results for the confidence intervals of $\beta_1$ are shown in detail. The differences to the intervals for $\beta_{12}$ are highlighted in Section \ref{ssec:general}, the full results for the intervals for $\beta_{12}$ are shown in Section B the Supplement.

Since the number of studies $k$ strongly affects the coverage and interval lengths (Section \ref{ssec:details}), the results are compared separately for each $k$. How the adjustments of other simulation parameters affect the coverage and interval length is discussed in Section~ \ref{ssec:details}. There it is of interest, whether the performance of a certain estimator differs from its overall performance for a special adjustment. For example, it is analyzed whether there is an estimator whose intervals have the best performance but only for large correlations. For ease of presentation \enquote{confidence interval} is abbreviated with CI in this section. The CIs based on $\bf{HC}_0$ are abbreviated with $\bf{HC}_0$-CI, the CIs based on other estimators in an analogous manner. 

\subsubsection*{Overall Performance of the Estimators}
\label{ssec:general}
\begin{figure}[b!]
\includegraphics[width =\textwidth, page = 3]{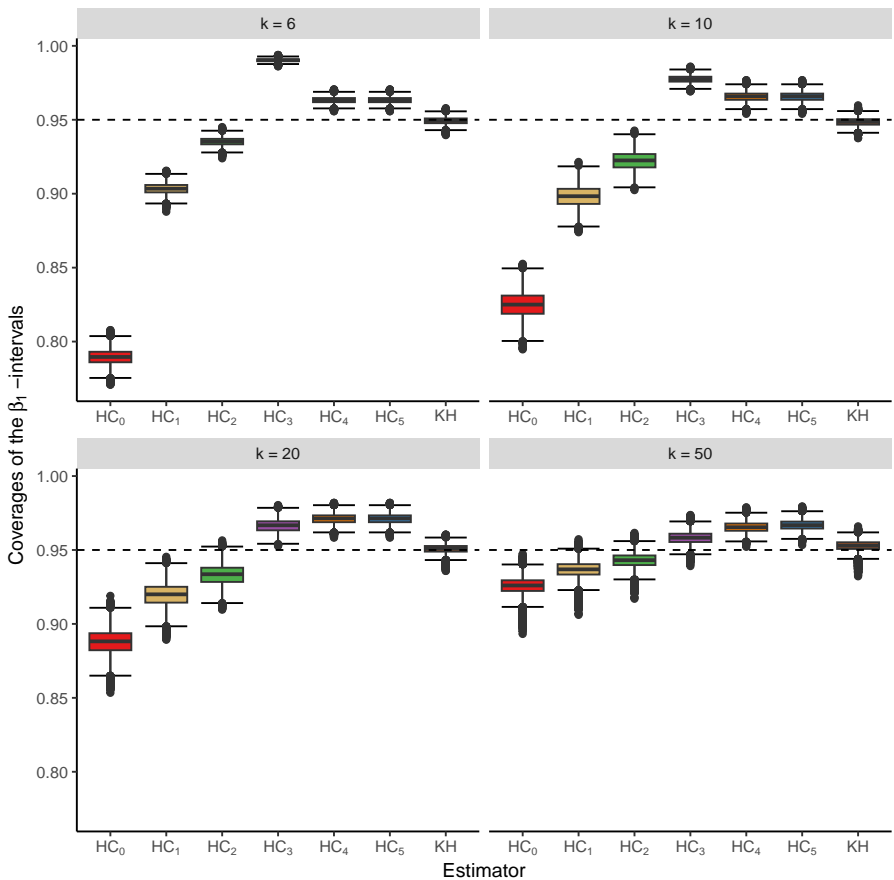}
\caption{Coverage probabilities of the confidence intervals for the regression parameter $\beta_1$ based on the estimators $\bf{HC}_0-\bf{HC}_5$ and $\bf{HKSJ}$ for different numbers of studies $k$.}
\label{b1_overall}
\end{figure}

\textbf{Confidence intervals for $\boldsymbol{\beta_1}$ -- Coverage Probability.}  In Figure \ref{b1_overall} the coverage of the CIs for $\beta_1$ are summarized using boxplots. 
Each plot reflects the results for a certain number of studies $k\in\{6,10,20,50\}$. The individual boxplots contain the coverage of all intervals based on the respective estimator and $k$.

The coverage of the $\bf{HC}_0$-CIs ranges from $0.7708$ to $0.8078$ for $k=6$. Although the coverage is growing in the number of studies $k$, the coverages of $\bf{HC}_0$-CIs are below the nominal confidence level $(1-\alpha)=0.95$ in every setting. For $k=50$ the coverage ranges from $0.8932$ to $0.9474$. The $\bf{HC}_1$-CIs have higher median coverages than the $\bf{HC}_0$-CIs for all $k$. But only for $1.76\%$ of the adjustments with $k=50$ the coverage is above the nominal confidence level. Similarly, $\bf{HC}_2$-CIs have a higher median coverage than $\bf{HC}_1$-CIs. Nevertheless, the coverage is below the nominal confidence level $(1-\alpha)=0.95$ for all adjustments with $k\in\{6, 10\}$ and in $94.68\%$ of the adjustments with $k\in\{20, 50\}$. Thus, $\bf{HC}_0-\bf{HC}_2$ seem to be inappropriate choices of estimators regarding their CI coverage.

$\bf{HC}_3-\bf{HC}_5$ based CIs have a higher median coverage than 0.95 for all number of studies $k$. For $k=6$ the $\bf{HC}_3$-CIs are the most conservative with coverages ranging from $0.9861$ to $0.9939$. However, the coverage of the $\bf{HC}_3$-CI is decreasing in $k$. Until $k=20$ the coverage is above the nominal level for every setting but for $k=50$ the coverage ranges from $0.9391$ to $0.9737$. But only in $0.06\%$ of the settings the coverages are below the nominal confidence level. Thus, the coverage of the $\bf{HC}_3$-CI for $\beta_1$ is quite accurate.

The coverages of the $\bf{HC}_4$-CIs and $\bf{HC}_5$-CIs range from $0.9521$ and $0.9533$ to $0.9819$ and $0.9819$, respectively and differ only slightly in respect of the number of studies. Thereby, the $\bf{HC}_4$-CIs and $\bf{HC}_5$-CIs are the only ones, whose coverages are above the nominal confidence level $(1-\alpha)=0.95$ for every adjustment. Thus, regarding coverage these estimators are suitable choices for all number of studies $k$.

Among all estimators $\bf{HKSJ}$-CIs show the closest coverages compared to the nominal confidence level $0.95$. The coverage tends to be slightly higher for larger number of studies $k$. For $k=6$ the actual coverage of the $\bf{HKSJ}$-CI is below the nominal confidence level in $60\%$ of the adjustments. In contrast, for $k=50$ the coverage is below $0.95$ for $18.55\%$ of the settings. However, only in $0.03\%$ of all settings the coverage of the $\bf{HKSJ}$-CI is below $0.94$. Although the $\bf{HKSJ}$-CIs have slightly smaller coverages than the $\bf{HC}_3-\bf{HC}_5$-CIs, their coverage is still acceptable for every adjustment. 
\begin{figure}[b!]
\includegraphics[width = \textwidth, page = 1]{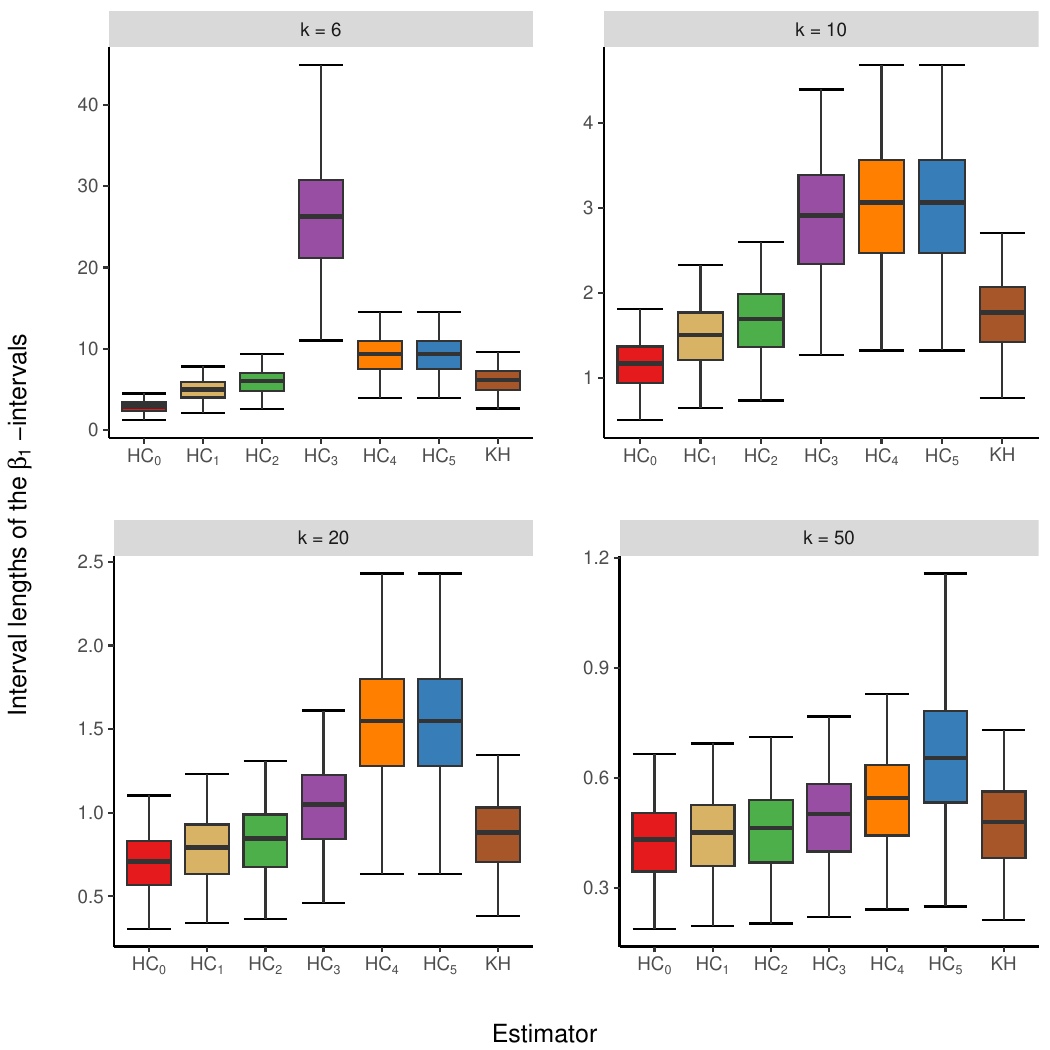}
\caption{Lengths of the confidence intervals for the regression parameter $\beta_1$ based on the estimators $\bf{HC}_0-\bf{HC}_5$ and $\bf{HKSJ}$ for different numbers of studies $k$ without outliers.}
\label{b1_length}
\end{figure}

\textbf{Confidence intervals for $\boldsymbol{\beta_1}$ -- Length.} Boxplots of the corresponding interval lengths are shown in Figure~\ref{b1_length}. 
Note, that for $k=6$ the lengths of the $\bf{HC}_3$-CIs have some extreme outliers with values up to 312.07, such as the lengths of the $\bf{HC}_5$-CIs for $k=50$ with values up to $48.38$. For a better visualization of the other results, outliers are not drawn in Figure \ref{b1_length}. The full results are shown 
in Section C of the Supplement. Moreover, the outliers are considered in Section~\ref{ssec:details} in more detail. 

The interval lengths of all estimators are monotonically decreasing in the number of studies $k$. Lengths of the $\bf{HKSJ}$-CIs range from $2.66$ to $9.62$ for $k=6$ and from $0.21$ to $0.73$ for $k=50$. Thereby, they are much shorter compared to the $\bf{HC}_4$- and $\bf{HC}_5$-CIs for all considered number of studies $k$. Except for $k=50$, where the lengths of the $\bf{HC}_5$-CIs tend to be longer, the lengths of the $\bf{HC}_4$- and $\bf{HC}_5$-CIs behave almost identically.  For $k=6$ the median length of the $\bf{HKSJ}$-CIs is equal to $6.20$, whereas it is equal to $9.34$ for the $\bf{HC}_4$- and $\bf{HC}_5$-CIs. In the situation of $k=50$ the median interval length of $\bf{HKSJ}$-CIs is 0.48, which is smaller than the $\bf{HC}_4$-CIs with $0.55$ and the $\bf{HC}_5$-CIs with $0.65$.

Lengths of the $\bf{HC}_3$-CIs are highly inflated for $k=6$. The lower quartile is equal to $21.17$ and the upper quartile's value is $30.76$. For $k=6$ the $\bf{HC}_4$- and $\bf{HC}_5$-CIs are shorter in the median than the $\bf{HC}_3$-CIs, for the other values of $k$ they are larger. With a value of $0.50$ the median interval length of the $\bf{HC}_3$-CIs is almost as short as the $\bf{HKSJ}$-CIs.

So in comparison of all estimators whose intervals have a suitable coverage, the $\bf{HKSJ}$-CIs are the shortest and therefore preferable. Since their CIs are much shorter for $k=6$, $\bf{HC}_4$ and $\bf{HC}_5$ have the second best performance for small $k$. If the number of studies is equal to $10$ or larger, $\bf{HC}_3$ is preferable compared to all other $\bf{HC}$ estimators. Due to the higher lengths of the $\bf{HC}_5$-CIs for $k=50$ compared to the $\bf{HC}_4$-CIs, $\bf{HC}_4$ should be preferred over $\bf{HC}_5$ for $k>20$. 

The $\bf{HC}_0-\bf{HC}_2$ based CIs tend to be shorter than the $\bf{HKSJ}$-CIs for all $k$, whereas the $\bf{HC}_0$-CIs have shorter median lengths than the $\bf{HC}_1$-CIs, which again are shorter in the median than the $\bf{HC}_2$-CIs. Nonetheless, due to their poor coverage they should not be used when calculating intervals of single parameters. 

\textbf{Performance of the intervals for $\boldsymbol{\beta_{12}}$}
Compared to the intervals for $\beta_1$ the intervals for $\beta_{12}$ tend to be longer for most estimators and adjustments of $k$. Exceptions are the lengths of the $\bf{HC}_0$, $\bf{HC}_1$, $\bf{HC}_3$ and $\bf{HKSJ}$ based CIs for $k=50$. For the $\bf{HC}_0$- and $\bf{HC}_1$-CIs this means, that they perform worse for intervals for $\beta_{12}$ compared to intervals for $\beta_1$, since also their coverages were lower for \bi. For the estimators $\bf{HC}_2-\bf{HC}_5$ it is arguable whether they perform worse for interaction coefficients, since the coverage is improved. However, the relatively big difference in the interval lengths of the $\bf{HC}_3- \bf{HC}_5$-CIs compared to the small gain of coverage indicates a worse performance compared to the intervals for $\beta_1$. When comparing CIs for $\beta_1$ to CIs for $\beta_{12}$, the coverage of the $\bf{HC}_2$-CIs increased from below the nominal confidence level to above in every situation. Therefore, the $\bf{HC}_2$-CIs perform better for interaction coefficients than for single coefficients, although their lengths are longer. The $\bf{HKSJ}$-CIs perform worse for $\beta_{12}$ than for \bo, since the share of coverages below 0.95 is higher for the intervals for $\beta_{12}$ and the intervals are longer. 

\begin{figure}[H]
\includegraphics[width =\textwidth, page = 4]{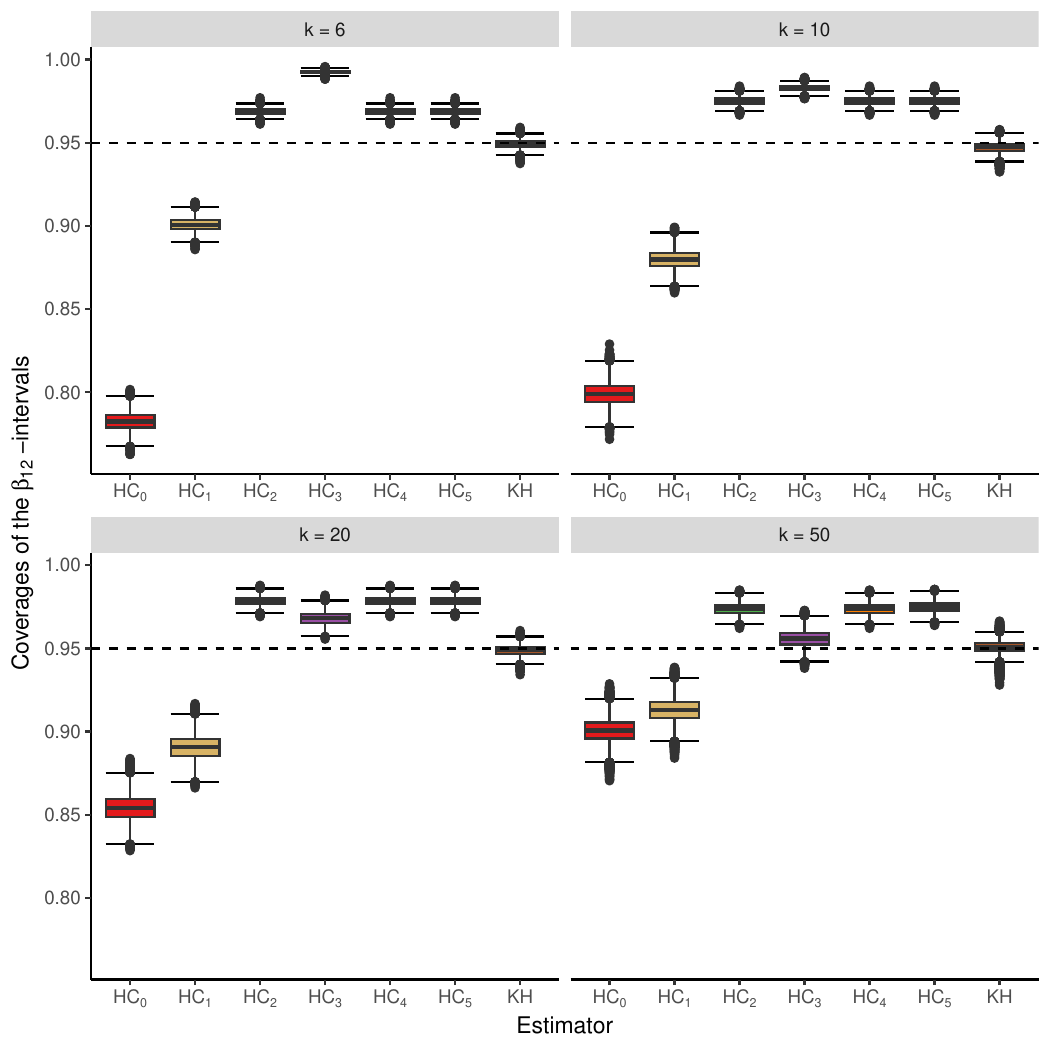}
\caption{Coverage probabilities of the confidence intervals for $\beta_{12}$ based on the estimators $\textbf{HC}_0-\textbf{HC}_5$ and $\textbf{KH}$ for different numbers of studies $k$.}
\label{bi_overall}
\end{figure}

\begin{figure}[H]
\includegraphics[width = \textwidth, page = 2]{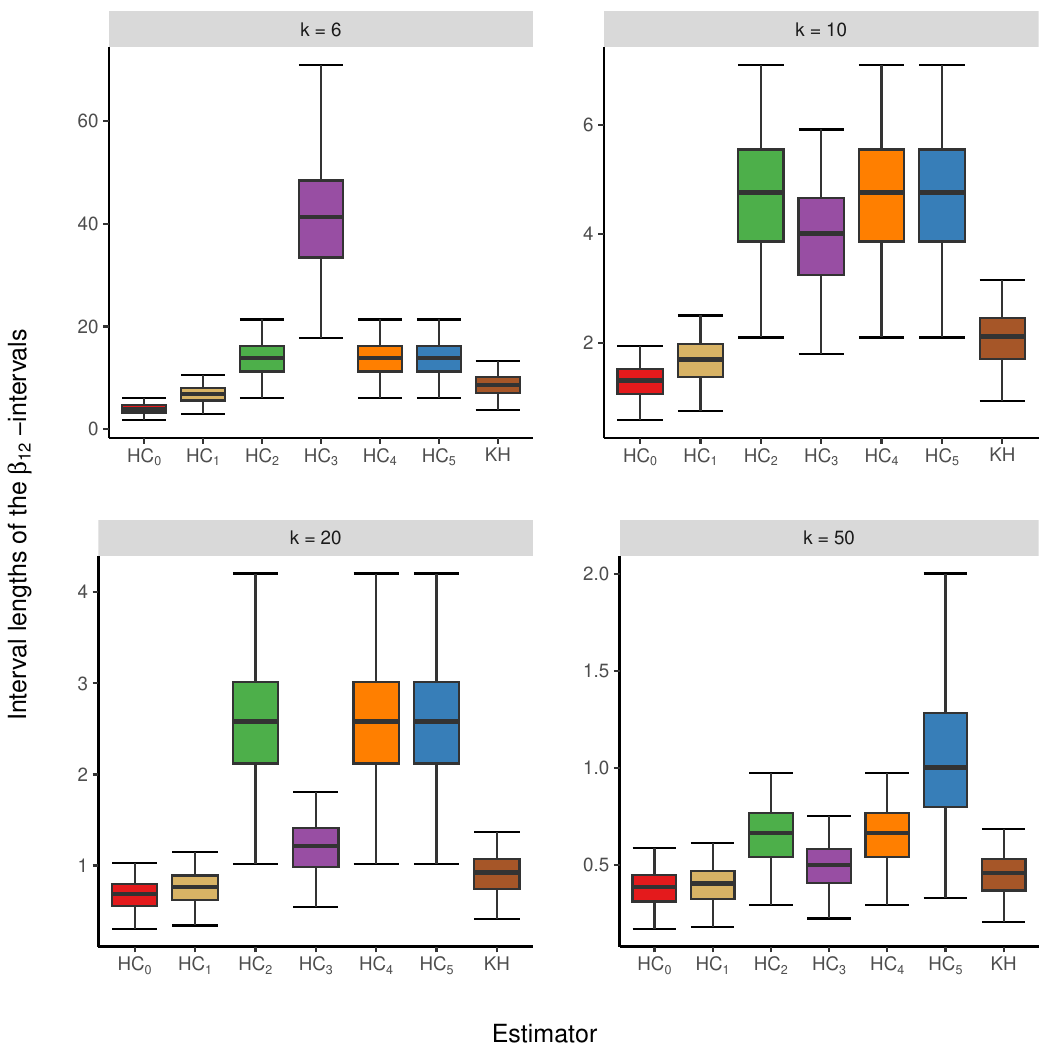}
\caption{Lengths of the confidence intervals for $\beta_{12}$ based on the estimators $\textbf{HC}_0-\textbf{HC}_5$ and $\textbf{KH}$ for different numbers of studies $k$ without outliers.}
\label{bi_length}
\end{figure}

\textbf{Summary of the overall performance} Overall the results are similar to the model with one covariate \citep{welzpauly2020}. Among all considered estimators the $\bf{HKSJ}$ estimator is the most appropriate for a model with an interaction term since it performed the best for the coefficient of the single moderator and the interaction term. Though its coverage is improvable in some situations. Focusing only on the $\bf{HC}$ estimators, $\bf{HC}_3$ is the best choice of estimator when the number of studies is not too small ($k\geq 10$). Otherwise $\bf{HC}_4$ or $\bf{HC}_5$ are preferable. For a larger number of studies $\bf{HC}_4$ is a better choice compared to $\bf{HC}_5$, due to the outliers that occur for the $\bf{HC}_5$-CIs. $\bf{HC}_2$ indeed performs as good as $\bf{HC}_4$ for the interaction coefficient, but in practice it is unlikely that only an interaction but not the single moderators are of interest. Therefore, $\bf{HC}_2$ is not recommendable for a model with interaction. $\bf{HC}_0$ and $\bf{HC}_1$ are not recommendable as well, since their performance is bad for both coefficients intervals.

\subsubsection*{Effects of Parameter Adjustments}\label{ssec:details}
This section will summarize how the coverages and interval lengths are affected by the adjustments of the flexible parameters. Since both coefficients are effected similar by most parameters they are considered together. We highlight the most important results and refer to Section D of the Supplement for complete results.

\textbf{Adjustments of the number of studies $\boldsymbol{k}$} Considered numbers of studies are 6, 10, 20 and 50.
The lengths of both coefficients intervals are monotonically decreasing in the number of studies $k$ 
This is comprehensible since the $t_{(k-p-1), (1-\alpha)}$-quantile in Equation (\ref{eq:ki}) is monotonically decreasing in $k$. The effect of the number of studies on coverage is not constant and depends on the considered covariance estimator. In general coverage tends towards the nominal level $1 - \alpha$ for increasing $k$.
Therefore, for all estimators a large number of studies is preferable. 

\textbf{Adjustments of study size}
Small ($\bar{n}=15$), medium ($\bar{n}=25$) and large ($\bar{n}=50$) group sizes are compared. For most covariance estimators the median coverage is slightly increasing in the study size. 
The corresponding interval lengths are decreasing as the study sizes increase for all $k$ and estimators. 
This trend may be caused by the impact of $n_i$ on $v_i$ in Equation (\ref{eq:vars}), which leads to decreasing standard errors in equation (\ref{eq:ki}).
Thus, overall larger studies lead to better confidence intervals, since both coverages and interval lengths are improved for larger study sizes.



\textbf{Adjustments of $\boldsymbol{\tau^2}$}
Coverages of both coefficients intervals are increasing slightly in the heterogeneity parameter $\tau^2$ for all estimators and $k$. 
For a larger number of studies, the effect is stronger. The increasing coverages in $\tau^2$ show that the model used in the simulation is adequate to model a study effect. On the other hand the interval lengths are increasing in $\tau^2$ strongly. This result is explicable by the direct impact the value of $\tau^2$ has on the variances of the coefficients and thus on the interval bounds.

In order to provide the reader with an idea of the amount of heterogeneity relative to the sampling variance, which we considered in the simulations, we considered an $I^2$ statistic of sorts by calculating $\frac{\tau^2}{\tau^2 + \sigma_i^2}$. This ratio represents the amount of heterogeneity variance relative to the total variation in study $i$. At the extremes this value lies between 42\% and 72\% for $\tau^2=0.1$ and between 87\% and 96\% for $\tau^2=0.9$, depending on the other parameter adjustments.

\textbf{Adjustments of $\boldsymbol{\beta_1}$} Examined adjustments of $\beta_1$ are 0, 0.2 and 0.5. The CIs for $\beta_{12}$ were not affected by these adjustments of $\beta_1$, whereas the CIs for $\beta_1$ have slightly lower coverage for $k\in\{20,50\}$ studies and all estimators 
. Adjustments of $\beta_1$ had no influence on the interval lengths of the CIs for $\beta_1$ and $\beta_{12}$ 
.


\textbf{Adjustments of $\boldsymbol{\beta_2}$} For $\beta_2$ the adjustments 0, 0.2 and 0.5 were considered as well. None of the intervals was affected by the adjustment of $\beta_2$ regarding the coverage or length 
. Noteworthy is that the extreme interval lengths of the $\bf{HC}_3$-CIs for $k=6$ only occur for small values of $\beta_2$.

\textbf{Adjustments of $\boldsymbol{\beta_{12}}$} Besides the adjustments 0, 0.2 and 0.5 for $\beta_{12}$ the adjustment -0.5 was simulated as well, to check whether it differs from the 0.5 adjustment. This is neither the case for the interval lengths nor for the coverages of the CIs for $\beta_1$ and the CIs for $\beta_{12}$. However, the $\bf{HC}_4$- and $\bf{HC}_5$-CIs for $\beta_1$ have slightly lower coverage for a high absolute value of $\beta_{12}$. There were no clear general trends, but it is noticeable that most of the extreme outliers of $\bf{HC}_5$-CIs occur for $\beta_{12}\in\{0,0.2\}$ 
.


Altogether the true values of the considered parameters do not have a strong impact on the intervals of any estimator. Therefore, there is no coefficient for which an estimator performs better or worse compared to the other estimators than in the overall results.

\textbf{Adjustments of the correlation $\boldsymbol{\rho}$} Examined adjustments of $\rho$ are 0, 0.2, 0.5 and -0.5. In additional simulations we also considered $\rho=0.9$. The findings did not alter the results. For details see Section E of the supplement.

The sign of the correlation affects neither the coverages nor the interval lengths. Intervals for $\beta_1$ that are based on $\bf{HC}_3-\bf{HC}_5$ tend to have a lower coverage for higher correlations, whereas CIs based on  $\bf{HC}_0-\bf{HC}_2$ tend to have higher coverages for $\lvert\rho\rvert=0.5$. For $\bf{HC}_2$ and $\bf{HC}_3$ the respective effect is only marginal. Large correlations induce longer CIs for $\beta_1$ for all number of studies and estimators 
. There is no consistent impact of the correlation on the CIs for $\beta_{12}$. The changes depend on both the estimator and number of studies $k$. However, these changes are only slight.
 $\bf{HC}_0$, $\bf{HC}_1$, $\bf{HC}_3$ and $\bf{HKSJ}$ based CIs have shorter lengths for larger values of $\lvert\rho\rvert$ and all $k$. Intervals based on $\bf{HC}_2$, $\bf{HC}_4$ and $\bf{HC}_5$ have marginally decreasing lengths in $\lvert\rho\rvert$ for $k=6$, slightly increasing lengths for $k\in\{10,20\}$ and again marginally decreasing lengths for $k=50$ 
 . It is also interesting to note, that most of the extreme outliers of $\bf{HC}_5$ occur for high correlations 
 .

\textbf{Adjustments of the random effect distribution}
Simulated random effect distributions are the standard normal distribution and standardized Laplace-, exponential, $t_3$- and log-normal-distributions. In comparison with the other simulated distributions, the coverages of CIs for $\beta_1$ based on $\bf{HC}_0-\bf{HC}_5$ are on average the lowest with normal distributed $u_i$ and highest with log-normal distributed random effects. The coverages do not differ much in respect of the other random effect. The $\bf{HKSJ}$-CIs for $k\in\{6, 10\}$ have the highest coverage with normal distributed random effects and the lowest with log-normal random effects. Especially for $k=10$ the coverages of the $\bf{HKSJ}$-CIs with non-normal random effects tend to be lower. In $73.71\%$ of the adjustments with non-normal random effects the coverages of the $\bf{HKSJ}$-CIs are below 0.95. For $k=20$ the $\bf{HKSJ}$-CIs show no observable differences between the random effect distributions, whereas for $k=50$ the order of the median coverages is the same as for the other estimators 
.

\thispagestyle{empty}
\begin{figure}
\vspace{-1.3cm}
\includegraphics[scale=0.7, page=2]{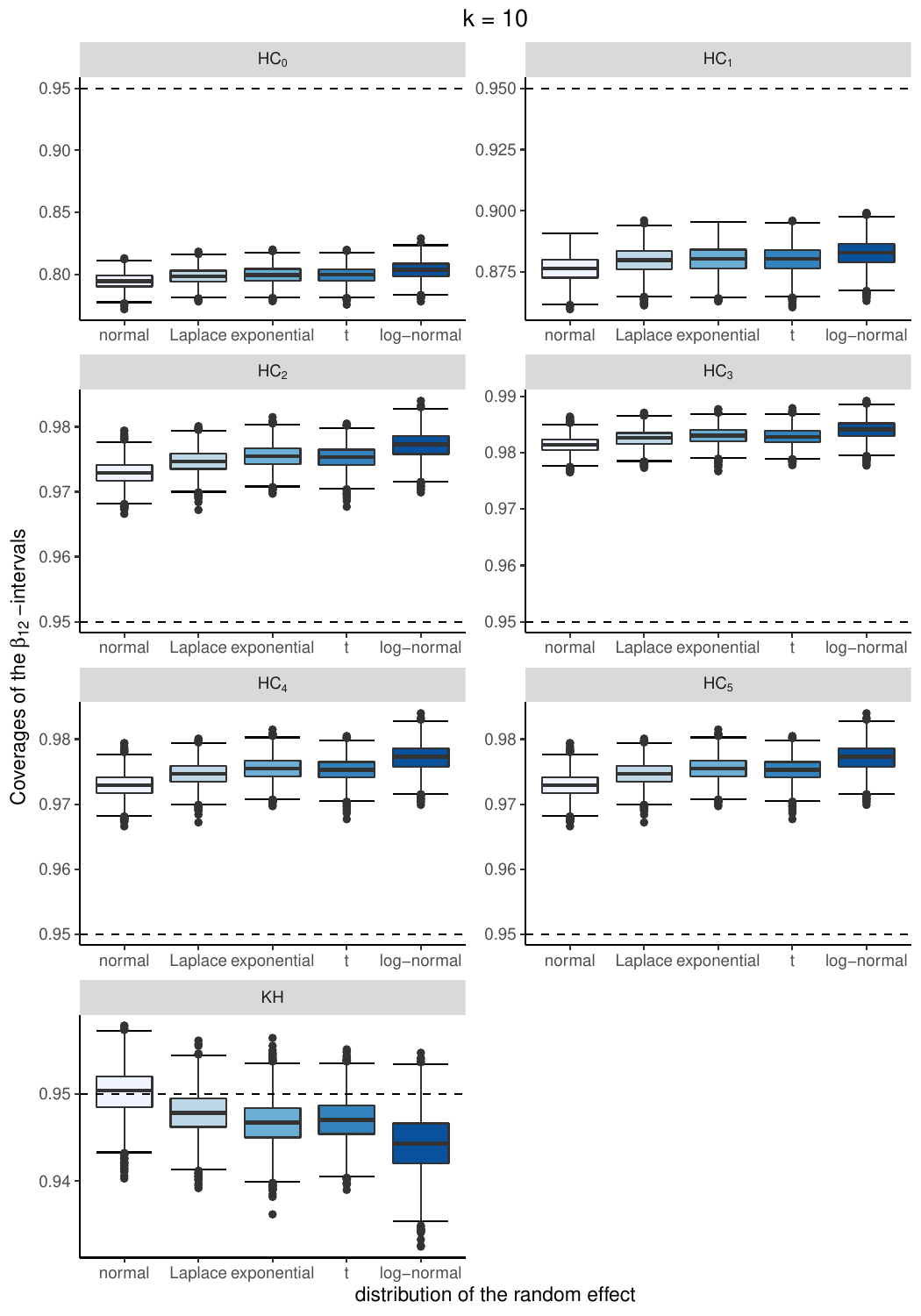}
\caption{Coverages of the \bi -intervals compared regarding the adjustments of $u_i$ for the estimators $HC_0-HC_5$ and $KH$ with $k=10$.}
\label{bi_err_2}
\end{figure}

\thispagestyle{empty}
\begin{figure}
\vspace{-1.3cm}
\includegraphics[scale=0.7, page=3]{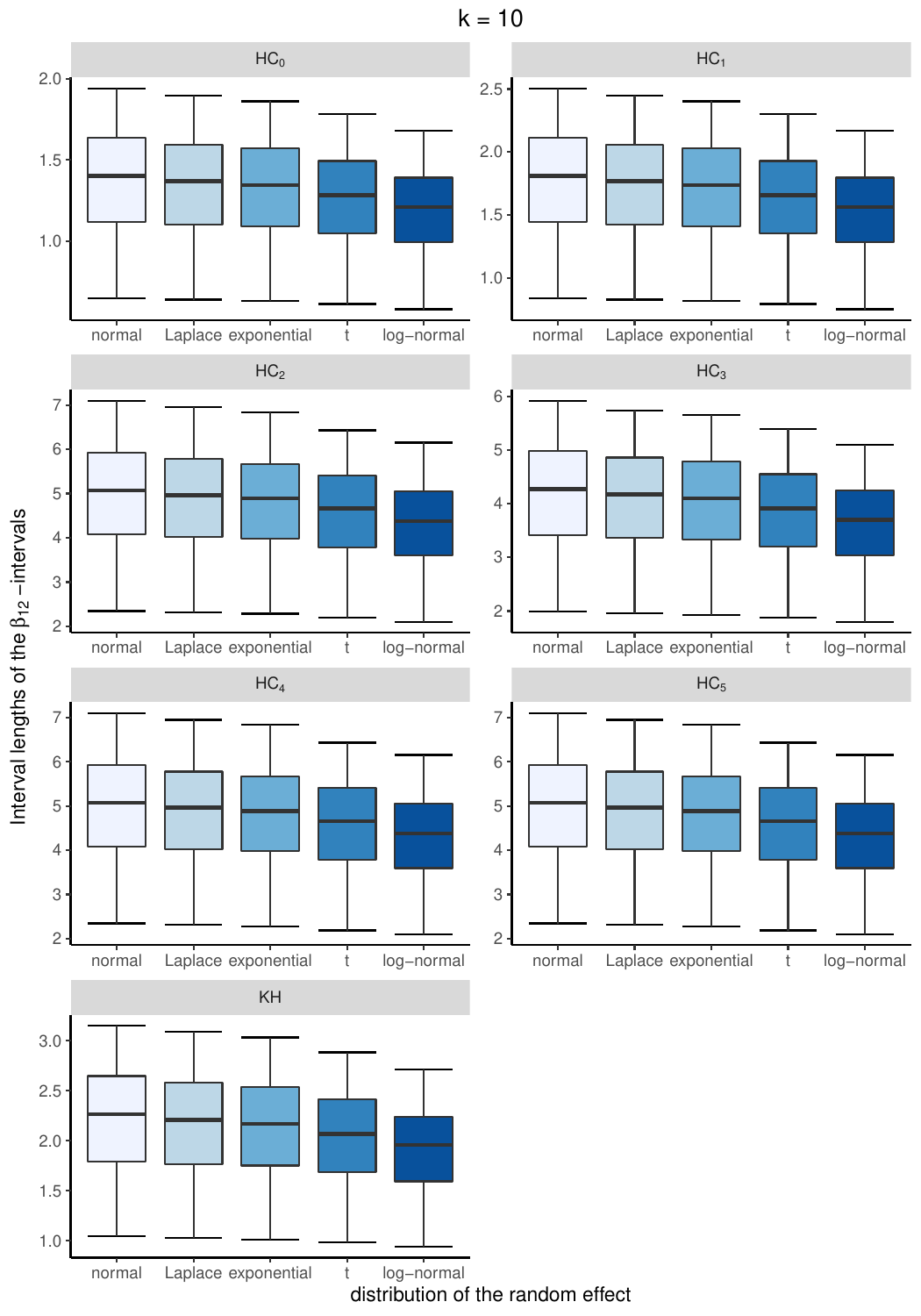}
\caption{Lengths of the \bi -intervals compared regarding the adjustments of $u_i$ for the estimators $HC_0-HC_5$ and $KH$ with $k=10$.}
\label{bi_err_l_2}
\end{figure}


The coverage of the $\bf{HC}_0-\bf{HC}_5$ CIs for $\beta_{12}$ are affected similarly by the random effect distribution for $k\in\{10,20,50\}$. 
For $k\in\{6, 10, 20\}$ the coverage of the $\bf{HKSJ}$-CIs for $\beta_{12}$ are below 0.95 in most of the adjustments with non-normal random effects.
Thus, in this situation the coverages of the $\bf{HKSJ}$-CIs for $\beta_{12}$ are even less adequate than for the intervals for $\beta_1$. If $k=50$, the $\bf{HKSJ}$-CIs are not affected by the random effects distribution 
.


The median lengths of both coefficients CIs depends on the underlying distribution can can be ordered in the following way for all $k$ and estimators: normal $>$ Laplace $>$ exponential $>$ $t_3$ $>$ log-normal 
. Thus, for $\bf{HC}_0-\bf{HC}_5$ the confidence intervals have better properties, when the random effect distribution is different from a normal distribution. Therefore, the quantile used as critical value is suitable, even if the distribution of the $u_i$ is not normal. In contrast, the $\bf{HKSJ}$-CIs depend more on the normality assumption  for smaller numbers of studies ($k\in\{6,10, 20\}$), especially for $k=10$. Due to the high share of coverage of the $\bf{HKSJ}$-CIs below the nominal confidence level, for non-normal and particularly log-normal random effects it is arguable whether $\bf{HKSJ}$ is the best estimator in this situation. If a precise control of the nominal confidence level is required $\bf{HC}_3$ (for $k\in{10, 20}$) or $\bf{HC}_4/\bf{HC}_5$ (for $k=6$) may be preferable
. For $k=50$ the performance of the $\bf{HKSJ}$-CIs is still the best for all distributions of $u_i$.


In sum, the estimators are affected by most parameter adjustments in the same way or a similar manner. Only the number of studies $k$ shows a strong varying effect on the coverage of some estimators. Besides the number of studies, the group size and the heterogeneity parameter $\tau^2$ have impact on the interval lengths. However, the trend is the same for all estimators and reducible to the direct impact of these parameters on components of the confidence interval in equation (\ref{eq:ki}). The results of the different random effect distributions indicate that the $\bf{HC}$ estimators are more robust against deviations from the normal distribution. For small numbers of studies it is questionable whether the coverage of the $\bf{HKSJ}$-CIs for non-normal random effects are still adequate. In this situation $\bf{HC}_3-\bf{HC}_5$ might be more suitable compared to $\bf{HKSJ}$. Otherwise, there is no situation where any estimator performs superior compared to its overall performance.

{\bf Additional Simulations}
In additional simulations we investigated the effect of omitted and redundant interactions. The results were the same as for usual omitted or redundant regressors: Omitted interactions caused lower coverages of the intervals, especially of $\beta_0$-CIs, whereas redundant interactions did not result in lower coverages but higher interval lengths. For details see Section E of the Supplement.

\section*{Discussion}
\label{sec:discussion}

Here we compared different confidence intervals for a mixed-effects meta-regression model with two moderators and an interaction term.
The confidence intervals were based on one of the six different heteroscedasticity consistent covariance estimators $\bf{HC}_0, ..., \bf{HC}_5$ or the Hartung-Knapp-Sidik-Jonkman covariance estimator $\bf{HKSJ}$. In a simulation study the confidence intervals based on these estimators were compared regarding their coverage and lengths for numerous combinations of simulation parameters. The simulation settings varied in the number of studies, the study sizes, a heterogeneity parameter, the coefficients of the moderators, the correlation between the covariates and the distribution of the random effect. A total of $77,760$ combinations was simulated $10,000$ times.

The coverage of the confidence intervals based on $\bf{HC}_0$ and $\bf{HC}_1$ turned out to be below the nominal confidence level $(1-\alpha)=0.95$ for almost every setting and are therefore not adequate. Although the coverage of the confidence intervals based on $\bf{HC}_2$ ($\bf{HC}_2$-CIs) for $\beta_{12}$ were suitable, $\bf{HC}_2$ is not recommended due to the low coverage of the intervals for $\beta_1$. The CIs based on the estimators $\bf{HC}_3-\bf{HC}_5$ and $\bf{HKSJ}$ showed suitable coverage for both parameters. Concerning the interval lengths the $\bf{HKSJ}$-CIs performed the best for all settings among all estimators with adequate coverage. For a small number of studies $(k=6)$ the lengths of the $\bf{HC}_3$-CIs were highly inflated. Therefore, in this situation the $\bf{HC}_4$ and $\bf{HC}_5$ should be preferred over $\bf{HC}_3$. For larger numbers of studies ($k\geq10$) the lengths of the $\bf{HC}_3$-CIs are shorter compared to the $\bf{HC}_4$ and $\bf{HC}_5$ intervals. Thus, for $k\geq10$ $\bf{HC}_3$-CIs are preferable compared to $\bf{HC}_4$- and $\bf{HC}_5$-CIs.

The results for single parameter adjustments differ only slightly from the overall results. The interval lengths were shown to be increasing in the amount of heterogeneity $\tau^2$, whereas they were decreasing in the number of studies $k$ and the mean study sizes $\bar{n}$ for  all estimators. Coverages were mostly decreasing in $\bar{n}$ and $\tau^2$. The confidence intervals were only slightly affected by the values of the true coefficients. Only high values of $\beta_1$ and strong interactions ($\lvert\beta_{12}\rvert=0.5$) reduced the coverage of some intervals. For all different estimators the CIs for $\beta_1$ had coverage closer to the nominal confidence level for higher correlations $\rho$ but their lengths were increasing in $\rho$. Concerning coverage and lengths of the CIs for $\beta_{12}$ no such trend was observable. Surprisingly, all estimators but $\bf{HKSJ}$ performed better for non-normal distributed random effects regarding their coverage and lengths. For small numbers of studies $k\in\{6,10\}$ the coverage of the $\bf{HKSJ}$-CIs tend to be below the nominal confidence level $(1-\alpha)=0.95$. In these situations the coverage of the  $\bf{HKSJ}$-CIs are still close to 0.95, but if an exact control of the significance level is required $\bf{HC}_3$ is more suitable.

Altogether, for the most parts the results of this work are in accordance with the results of \cite{welzpauly2020}. The superior performance of the $\bf{HKSJ}$ estimator and the behavior of the $\bf{HC}$ estimators observed in the model with one moderator also holds for most situations when studying the model with two covariates and an interaction. However, some aspects of this work indicate that $\bf{HKSJ}$ may not be the best estimator for more complex models or more extreme parameter adjustments. In fact, the $\bf{HKSJ}$-estimator performed worse compared to the model with one covariate in the work of \cite{welzpauly2020}. In their simulation study the Type 1 random error of the $\bf{HKSJ}$ based tests was below the nominal significance level $\alpha=0.05$ in more than 50$\%$ of the adjustments for all $k$. In this work situations were observed where the coverage of the $\bf{HKSJ}$-CIs was below the nominal confidence level $(1-\alpha)=0.95$ in almost $80\%$ of the adjustments.  This may be due to its worse performance for non-normal distributed random effects compared to the other estimators, which was observed especially for small numbers of studies. Additionally, the coverage of all estimators were lower for high values of $\beta_1$. Furthermore, the coverage of the $\bf{HKSJ}$-CIs for $\beta_{12}$ tended to be lower.

As the model examined in this work still has a simple structure, in further research it might be of interest to consider the performance of $\bf{HKSJ}$ for more complex models. Interesting settings are interaction terms of higher order, other random effect distributions and more extreme coefficients. Based on this simulation study it seems plausible that the good performance of $\bf{HKSJ}$ might be limited to models of simpler structure. For more complex models coverage of $\bf{HKSJ}$-CIs may be inadequate and alternative estimators may be required. For such situations $\bf{HC}_4$ and $\bf{HC}_5$ may be suitable choices of estimators for small and large number of studies, since their CIs controlled the nominal confidence level well in every situation and had shorter interval lengths compared to $\bf{HC}_3$-CIs for $k=6$. For a medium number of studies $\bf{HC}_3$ might be the most suitable, since its intervals held the nominal confidence level in every situation with $k\in\{10,20\}$ and were shorter compares to the $\bf{HC}_4$- and $\bf{HC}_5$-CIs. In further research it may also be of interest to analyze the situations where highly inflated interval lengths of the $\bf{HC}_3$- and $\bf{HC}_5$-CIs occurred in detail, because they cannot be explained by the results of this work. A limitation of our research regarding the estimator $\bf{HC}_5$ is that we did not optimize the tuning parameter $\eta$, relying on the recommendation of $\eta=0.7$ by \cite{neto2007}. The question whether and how the optimal choice of $\eta$ depends on a given context remains an open question for further research.

Concluding, meta-regression remains an important field of statistical research. The Hartung-Knapp-Sidik-Jonkman estimator is generally a good choice, especially for simple models and situations where a normality assumption for the distribution of effect estimates is appropriate. Among the various \textbf{HC} estimators, $\bf{HC}_3$ and $\bf{HC}_4$ appear to be the best choices, depending on the model assumptions. However, in most cases researchers would do well, to use the Hartung-Knapp-Sidik-Jonkman estimator.

\section*{Declarations}
\backmatter

\bmhead{Supplementary information}

This manuscript has an accompanying supplement, which contains detailed simulation results and some relevant mathematical theory.

\bmhead{Conflicts of Interest}

The authors have declared no conflict of interest.

\bmhead{Funding}

This work was supported by the German Research Foundation: project Grant no. PA-2409 7-1 (Markus Pauly) and FR 3070/3-1 (Tim Friede).

\bmhead{Acknowledgements}

The authors gratefully acknowledge the computing time provided on the
Linux HPC cluster at Technical University Dortmund (LiDO3), partially
funded in the course of the Large-Scale Equipment Initiative by the German
Research Foundation (DFG) as project 271512359.

\bibliography{literature}

\clearpage
\setcounter{page}{1}
\thispagestyle{empty}
\vspace{2em}

\begin{center}
\LARGE{\textbf{Supplement to:\\
Robust Confidence Intervals for Meta-Regression with Interaction Effects\\[1cm]
}}
\end{center}

\begin{center}
\large{Eric S. Knop\footnote{
Department of Statistics, TU Dortmund University}, Markus Pauly\footnotemark[1], Tim Friede\footnote{Department of Medical Statistics, University of Göttingen}\footnote{DZHK (German Center for Cardiovascular Research), partner site Göttingen} and Thilo Welz\footnotemark[1]}
\end{center}

\vspace{2em}
\begin{center}
 \today  
\end{center}

\vspace{1em}

\begin{small}
    \begin{center}
    \textbf{Abstract}
    \end{center}
    This is a supplement to the main paper "Robust Confidence Intervals for Meta-Regression with Interaction Effects". The representation of $\textbf{HC}_0$ given in Equation (7) of the main paper is derived in section A. In Section B the full results of the confidence intervals for $\beta_{12}$ are presented. Versions of Figures 3 \& 4 of the main paper where outliers are shown are presented in Section C. In Section D the effects of the parameter adjustments are considered in more detail and corresponding boxplots are shown. Finally, the results of the additional simulations are presented in Section E.
    \end{small}

\vspace{2em}


\addtocounter{page}{-1}

\clearpage

\addtocounter{page}{-1}
\tableofcontents

\clearpage


\renewcommand{\thesection}{\Alph{section}}

\section{Derivation of \textbf{HC}$_0$ for Weighted Least Squares Estimators}
For $\textbf{HC}_0$ the form of the estimator given in Equation (7)
in the main paper can be obtained  by considering the transformed model \begin{equation}
\tilde{\vy}=\tilde{\vX}\vb+\tilde{\vu}+\tilde{\ve},
\label{eq:transmodel}
\end{equation}
where $\tilde{\vy}=\vW^{\frac{1}{2}}\vy, \tilde{\vX}=\vW^{\frac{1}{2}}\vX, \tilde{\vu}=\vW^{\frac{1}{2}}\vu$ and $\tilde{\ve}=\vW^{\frac{1}{2}}\ve$. The transformed model satisfies at least asymptotically all assumptions of the classical linear regression model, hence the OLS estimator of the transformed model $\hat{\vb}_{OLS}=(\tilde{\vX}^\top\tilde{\vX})^{-1}\tilde{\vX}^\top\vy$ is an appropriate estimator for $\vb$. Since the OLS estimator of the transformed model is $\hat{\vb}_{OLS}=\hat{\vb}$ \citep{hayashi2000} one simply has to consider the estimator given in equation (5) of \cite{mackinnonwhite1985} for the transformed model: \begin{eqnarray*}
 \textbf{HC}_0&=&(\tilde{\vX}^\top\tilde{\vX})^{-1}\tilde{\vX}^\top\tilde{\vO}\tilde{\vX}(\tilde{\vX}^\top\tilde{\vX})^{-1}\\&=&(\vX^\top\vW\vX)^{-1}\vX^\top\vW^{\frac{1}{2}}\tilde{\vO}\vW^{\frac{1}{2}}\vX(\vX^\top\vW\vX)^{-1}\\&=&(\vX^\top\vW\vX)^{-1}\vX^\top\vW\vE\vD_0\vD_0^\top\vE^\top\vW\vX(\vX^\top\vW\vX)^{-1}
\end{eqnarray*} as 
\begin{eqnarray*}
\tilde{\vO}&=&diag(\tilde{e_1}^2,...,\tilde{e}_k^2)\\&=&diag\bigl(\vD_0(\tilde{\vy}-\tilde{\vX}(\tilde{\vX}^\top\tilde{\vX})^{-1}\tilde{\vX}^\top\tilde{\vy})\bigr)diag\bigl(\vD_0(\tilde{\vy}-\tilde{\vX}(\tilde{\vX}^\top\tilde{\vX})^{-1}\tilde{\vX}^\top\tilde{\vy})\bigr)^\top\\&=&diag\bigl(\vD_0\vW^{\frac{1}{2}}(\vy-\vX\hat{\vb})\bigr)diag\bigl(\vD_0\vW^{\frac{1}{2}}(\vy-\vX\hat{\vb})\bigr)^\top
\\&=&\vD_0\vW^{\frac{1}{2}}\vE\vE^\top\vW^{\frac{1}{2}}\vD_0\\&=&\vW^{\frac{1}{2}}\vD_0\vE\vE^\top\vD_0^\top\vW^{\frac{1}{2}}\\&=&\vW^{\frac{1}{2}}\vE\vD_0\vD_0^\top\vE^\top\vW^{\frac{1}{2}}.
\end{eqnarray*}
Note that $\vD_0$, $\vW^{\frac{1}{2}}$ and $\vE_0$ are interconvertible in the calculation above, given that they are all diagonal matrices.\\
The form of the other $\textbf{HC}$ estimators $\textbf{ HC}_1-\textbf{HC}_5$ can be derived by analogous transformations using the suitable residuals and $\vD_i$, $i=1,...,5$.
\clearpage

\section{Full results of the intervals for $\beta_{12}$}
\textbf{Confidence intervals for $\boldsymbol{\beta_{12}}$ -- Coverage Probabilities.} Boxplots, which summarize the coverage of the confidence intervals for $\beta_{12}$ are shown in Figure \ref{bi_overall}.
\begin{figure}[b!]
\includegraphics[width =\textwidth, page = 4]{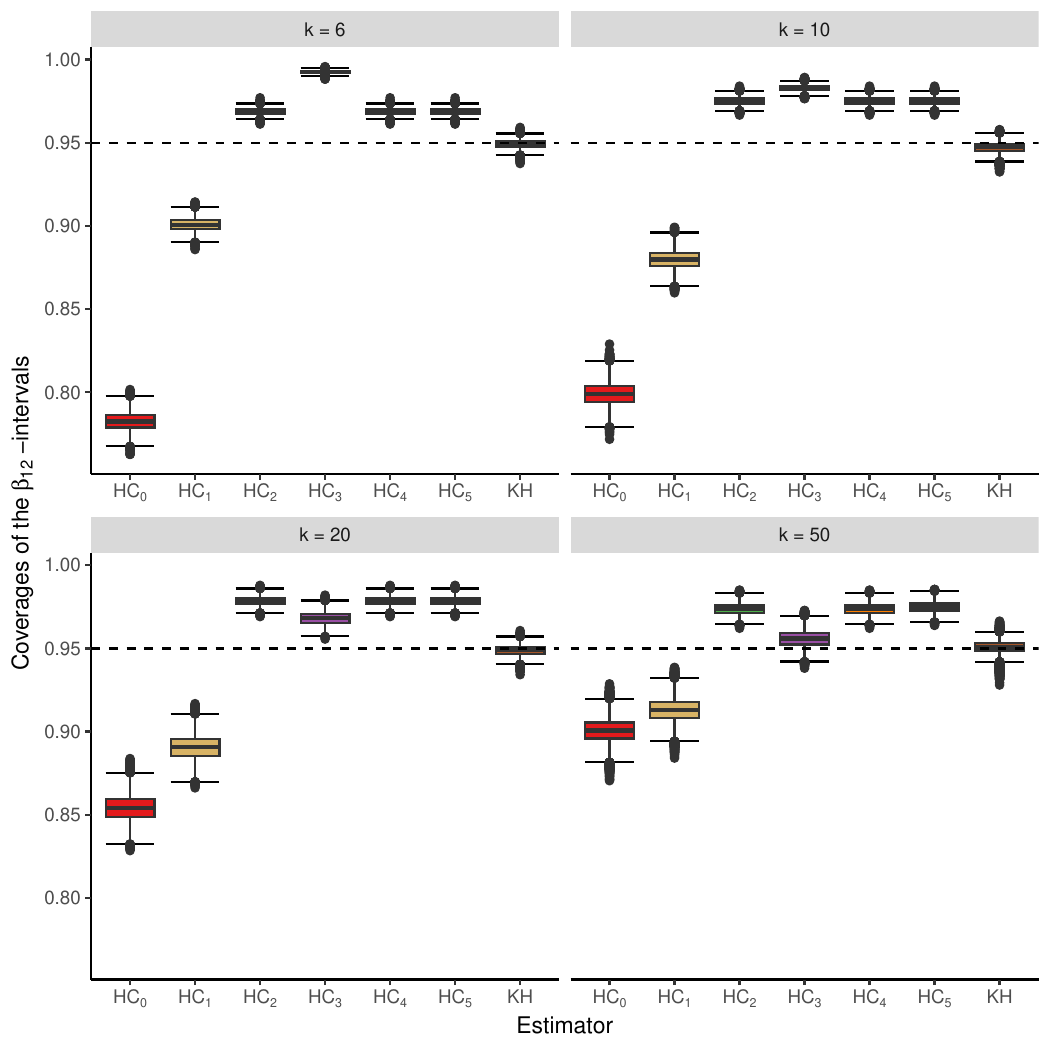}
\caption{Coverage probabilities of the confidence intervals for $\beta_{12}$ based on the estimators $\textbf{HC}_0-\textbf{HC}_5$ and $\textbf{KH}$ for different numbers of studies $k$.}
\label{bi_overall}
\end{figure}
The biggest difference regarding the coverage of those intervals compared to the intervals for $\beta_1$, is that all $\textbf{HC}_2$-CIs are conservative for $\beta_{12}$. Other than for the CIs for $\beta_1$, the coverage of the $\textbf{HC}_2$-CIs for $\beta_{12}$ are higher than the nominal confidence level for all adjustments. The median coverages of the $\textbf{HC}_2$-CIs are equal to the median coverages of the $\textbf{HC}_4$-CIs for all considered number of studies $k$, which again are equal to the median coverage of the $\textbf{HC}_5$-CIs for $k\in\{6, 10, 20\}$.  For $k=50$ the median coverage of the $\textbf{HC}_5$-CIs is slightly higher than the median coverages of the $\textbf{HC}_2$- and $\textbf{HC}_4$-CIs. 

Except for the $\textbf{HC}_2$-CIs the coverages of the CIs for $\beta_{12}$ resemble the ones of the CIs for $\beta_1$ in the corresponding situation. $\textbf{HC}_0$- and $\textbf{HC}_1$-CIs are liberal, whereas $\textbf{HC}_4$- and $\textbf{HC}_5$-CIs are conservative. Again the $\textbf{HC}_3$-CIs are extremely conservative for $k=6$ and only slightly conservative for $k=50$. Also the coverage of the $\textbf{KH}$-CIs are very close to the nominal confidence level $(1-\alpha)=0.95$ for most settings. But there are still some interesting minor differences for the other estimators, which need to be considered in detail. 

Median coverage of the coefficient of interest $\beta_{12}$ differs more from the nominal confidence level compared to the CIs for $\beta_1$ for most estimators and number of studies. Exceptions are, besides the $\textbf{HC}_2$-CIs, the $\textbf{HC}_3$- and $\textbf{KH}$-CIs for $k=50$. This results in even more liberal $\textbf{HC}_0$- and $\textbf{HC}_1$-CIs, as well as more conservative $\textbf{HC}_3$ - $\textbf{HC}_5$ based CIs for $\beta_{12}$ compared to the CIs for $\beta_1$. For the $\textbf{KH}$-CIs the boxplots of the CIs for $\beta_{12}$ are very similar to the CIs for $\beta_1$.

However, for the $\textbf{KH}$-CIs for $\beta_{12}$ the ratio of coverage below the nominal confidence level is larger compared to the CIs for $\beta_1$ for all $k$. For $k=10$ in $79.88\%$ of the adjustments $\beta_{12}$ is included in less than $95\%$ of the intervals. For $k=50$ the ratio is still $38.45\%$. Nevertheless, only in $1.12\%$ of all adjustments the coverage is below 0.94. \\In conclusion, the performance of the estimators can be assessed similar to the CIs for $\beta_1$. The only major exception is the performance of the $\textbf{HC}_2$-CI, which is as good as the performance of $\textbf{HC}_4$-CI. Moreover, the coverage of the $\textbf{KH}$-CI are still improvable.

\textbf{Confidence intervals for $\boldsymbol{\beta_{12}}$ -- Length.} 
Boxplots containing the interval lengths of the CIs for $\beta_{12}$ are shown in Figure \ref{bi_length}. Regarding the lengths, all intervals except for $\textbf{HC}_2$-CIs behave similar compared to the CIs for \bo. 
\begin{figure}[h!]
\includegraphics[width = \textwidth, page = 2]{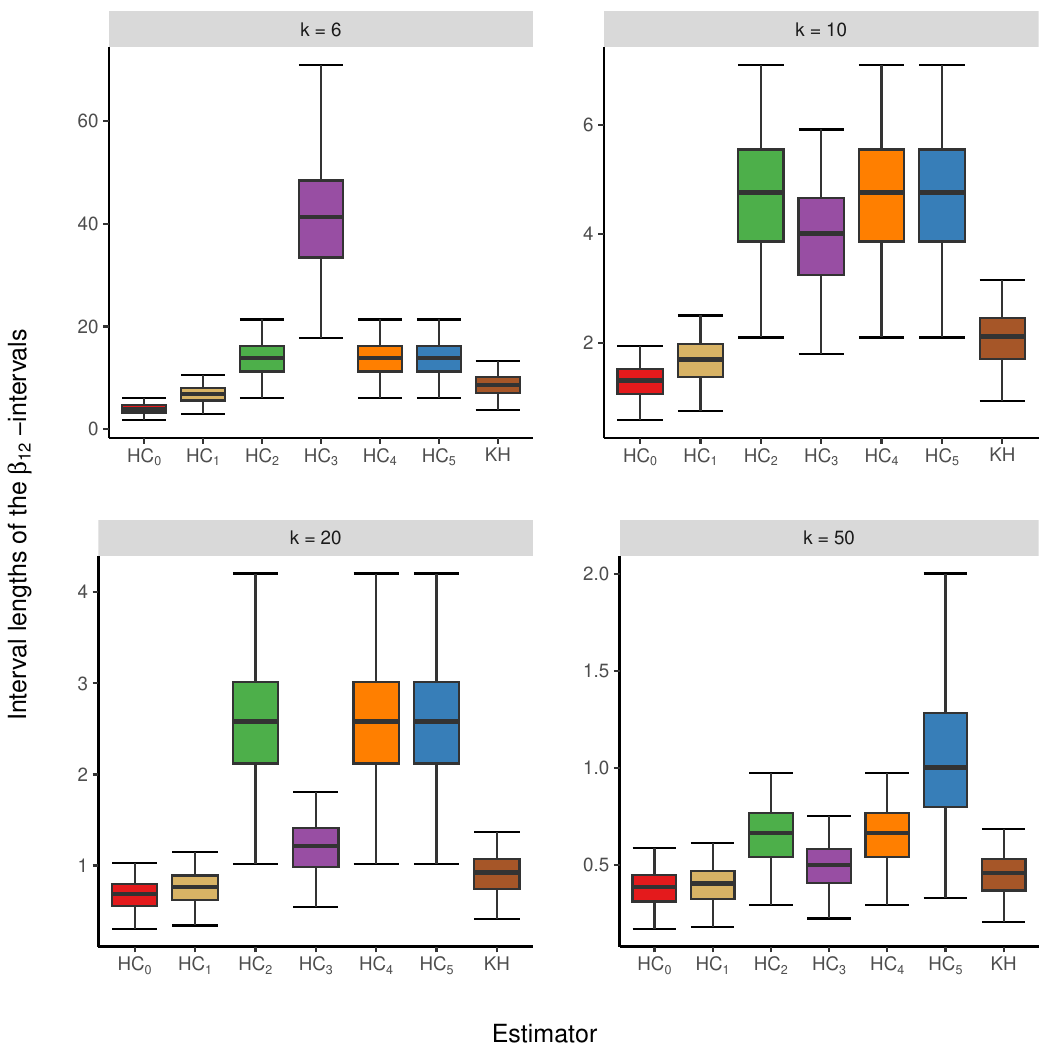}
\caption{Lengths of the confidence intervals for $\beta_{12}$ based on the estimators $\textbf{HC}_0-\textbf{HC}_5$ and $\textbf{KH}$ for different numbers of studies $k$ without outliers.}
\label{bi_length}
\end{figure}
Again there occurred heavy outliers for the $\textbf{HC}_3$-CIs with $k=6$ and the $\textbf{HC}_5$-CIs with  $k=50$. In this case the $\textbf{HC}_3$-CIs have a maximum length of 393.1 and the $\textbf{HC}_5$-CIs of 239.1. For a better visualization in Figure \ref{bi_length} the outliers are not shown. Full results are shown in 
Section D of this Supplement. 

For $k=6$ the $\textbf{HC}_3$-CIs are the longest with a median length of 41.30. This is almost triple of the median lengths of the $\textbf{HC}_2$, $\textbf{HC}_4$ and $\textbf{HC}_5$ based CIs, which are 13.81. The shortest median length of the intervals, that have a coverage above 0.95, is portrayed by the $\textbf{KH}$-CIs with 8.64. $\textbf{HC}_0$- and $\textbf{HC}_1$- CIs have a median length of 3.96 and 7.87 respectively for $k=6$. For all estimators the median interval length is decreasing in the number of studies $k$. Except for the $\textbf{HC}_3$-CIs the rank order of the lengths does not change for the different $k$. The $\textbf{HC}_3$-CIs are only the longest for $k= 6$. For $k=50$ their median length is 0.50 and therefore close to the $\textbf{KH}$-CIs median length of 0.46. This was also the case for the CIs of \bo. Again the $\textbf{HC}_0$- and $\textbf{HC}_1$-CIs are the shortest for all numbers of studies $k$. Regarding the lengths except for $\textbf{HC}_2$ the same conclusions are drawn as for the intervals for $\beta_1$. Of all intervals with adequate coverage, the $\textbf{KH}$-CIs perform the best. The performance of $\textbf{HC}_2$ is as good as the performance of $\textbf{HC}_4$ while estimating intervals for the interaction term. 

\clearpage
\section{Boxplots of Lengths for all Results with Outliers}

\begin{figure}[h!]
\includegraphics[width = \textwidth, page = 1]{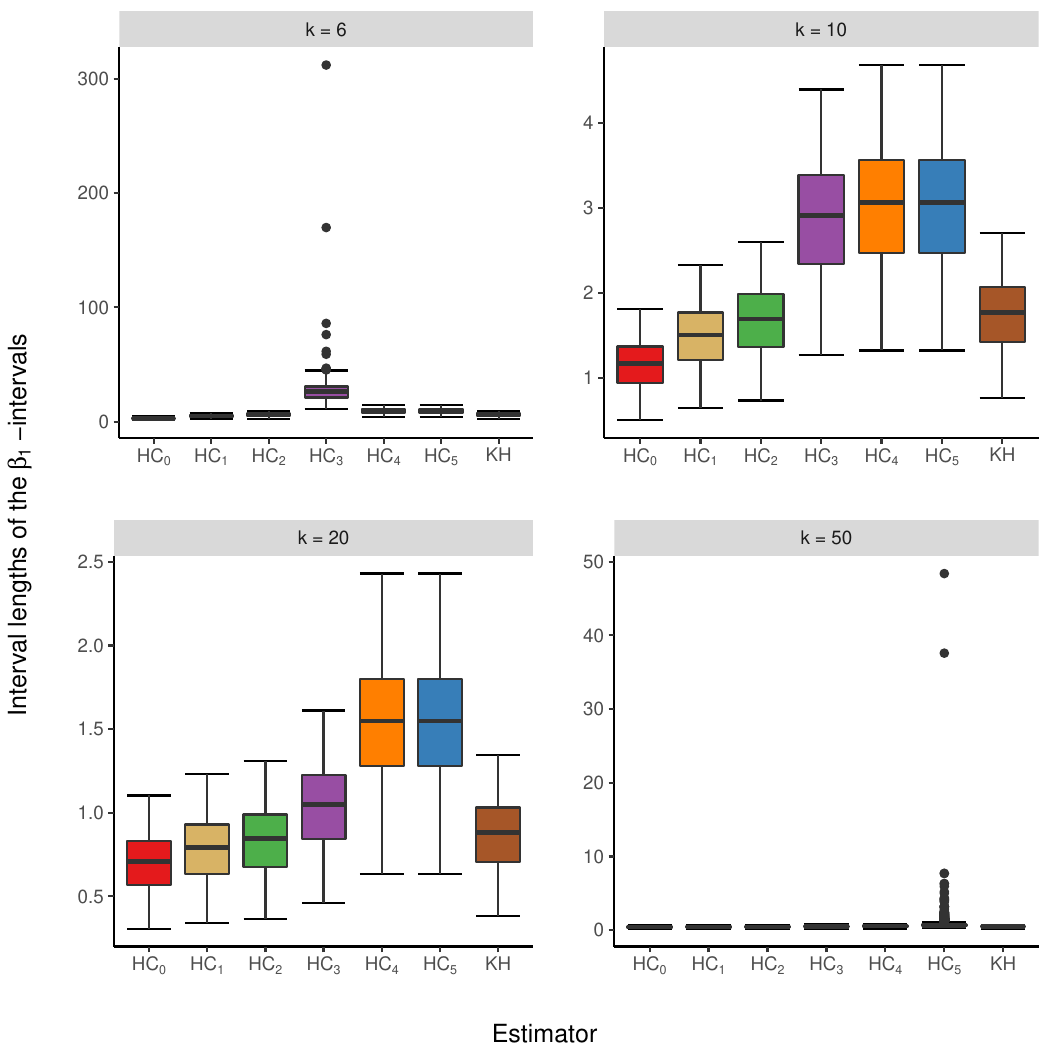}
\caption{Lengths of the confidence intervals for $\beta_1$ based on the estimators ${HC}_0-{HC}_5$ and $KH$ for different numbers of studies $k$.}
\label{b1_length_outliers}
\end{figure}

\begin{figure}[b!]
\includegraphics[width = \textwidth, page = 2]{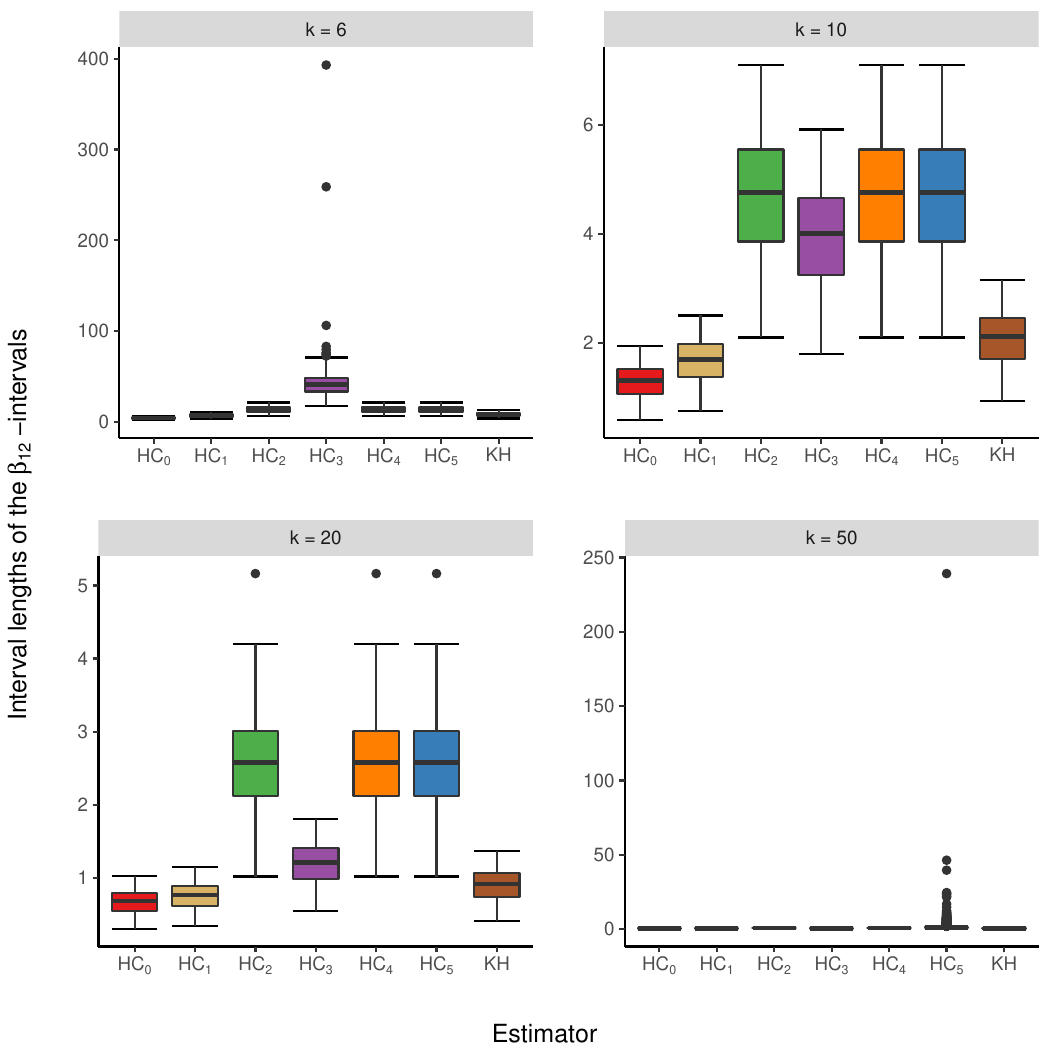}
\caption{Lengths of the confidence intervals for $\beta_{12}$ based on the estimators ${HC}_0-{HC}_5$ and $KH$ for different numbers of studies $k$.}
\label{bi_length_outliers}
\end{figure}

In Figures \ref{b1_length_outliers} and \ref{bi_length_outliers} the lengths of all confidence intervals are shown. Due to the large outliers for the $\textbf{HC}_3$-CIs with $k=6$ and the $\textbf{HC}_5$-CIs with $k=50$ the estimators are not comparable to the others. Therefore in the main paper the lengths are shown without outliers. In Section 5.2 of the main paper and D of this supplement the outliers are considered in more detail.

\clearpage

\section{Boxplots for Separate Parameter Adjustments}

This section will focus on how the coverages and interval lengths are affected by the adjustments of the flexible parameters. Since both coefficients are affected similarly by most parameters they are considered together. Situations where the influences differ are emphasized. To assess effects of a flexible parameter each boxplot in Section 5.1 of the main paper
is split into parallel boxplots that contain all situations with the considered parameter adjustment. The boxplots of the interval lengths with heavy outliers ($\textbf{HC}_3$-CIs for $k=6$ and $\textbf{HC}_5$-CIs for $k=50$) are shown twice, once without outliers and once with outliers.\\

\begin{figure}
\includegraphics[scale=0.7, page=1]{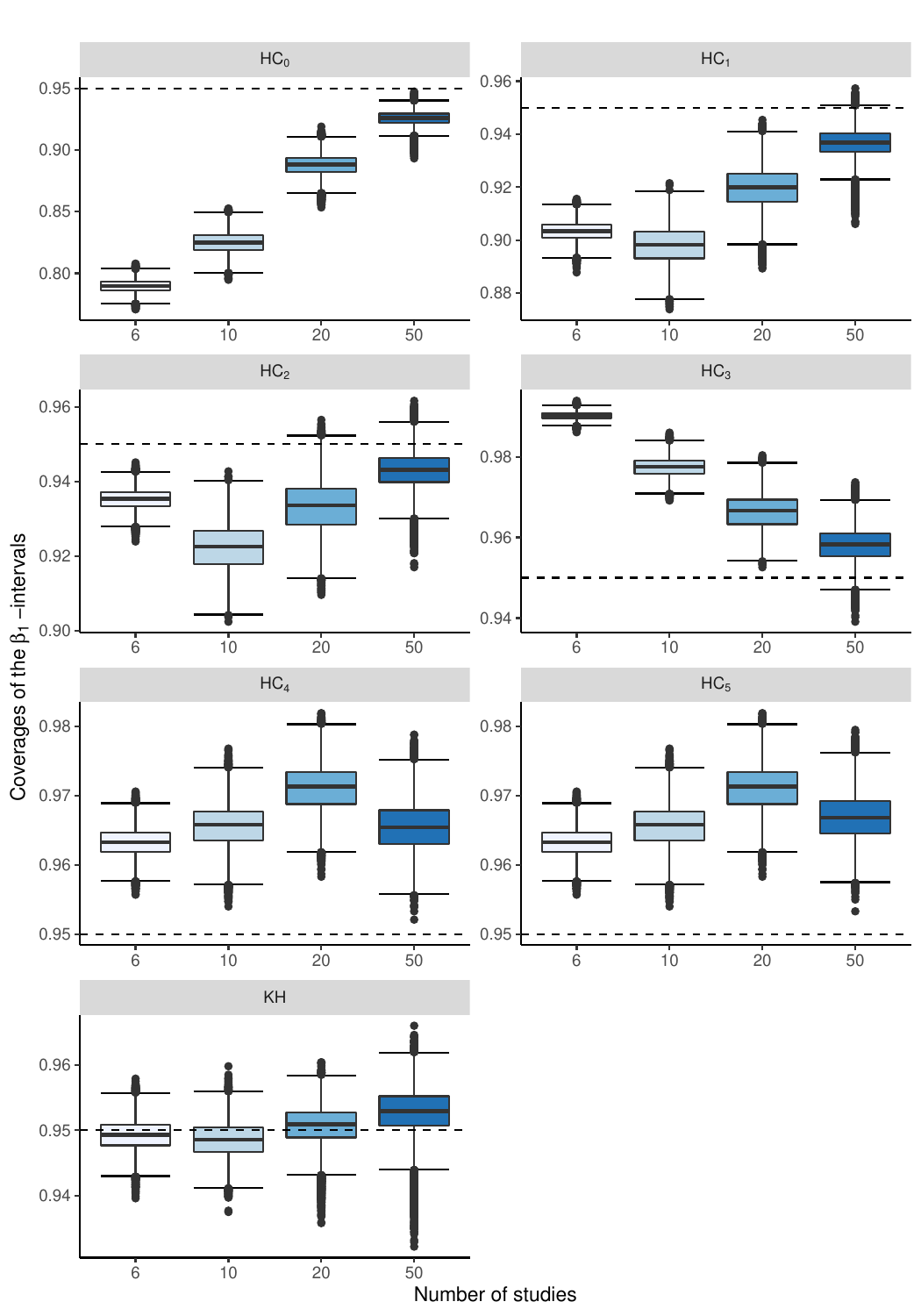}
\caption{Coverages of the \bo -intervals compared regarding the adjustments of $k$ for the estimators $HC_0-HC_5$ and $KH$.}
\label{b1_k}
\end{figure}

\begin{figure}
\includegraphics[scale=0.7, page=2]{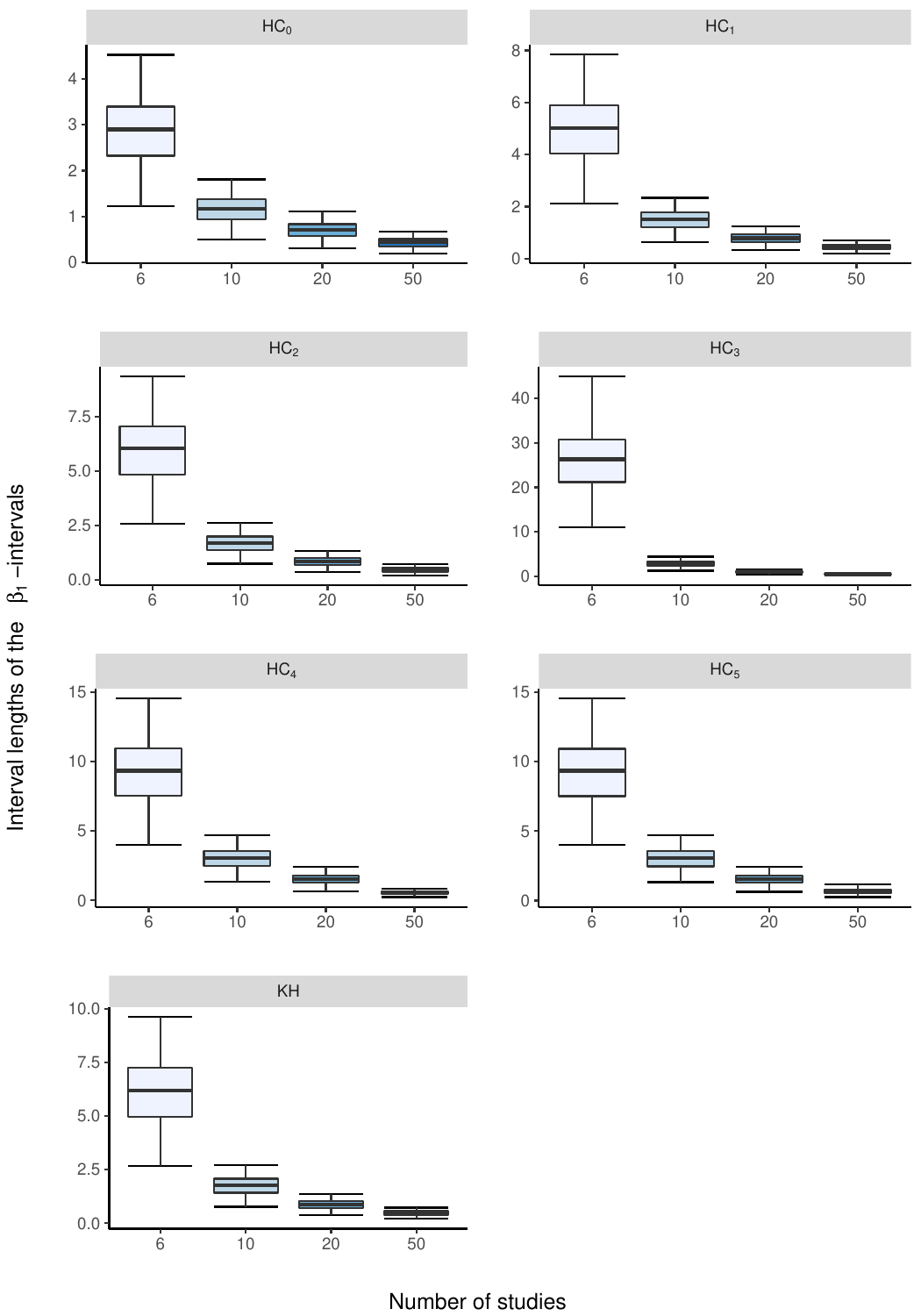}
\caption{Lengths of the \bo -intervals compared regarding the adjustments of $k$ for the estimators $HC_0-HC_5$ and $KH$ without outliers.}
\label{b1_k_l}
\end{figure}

\begin{figure}
\includegraphics[scale=0.7, page=1]{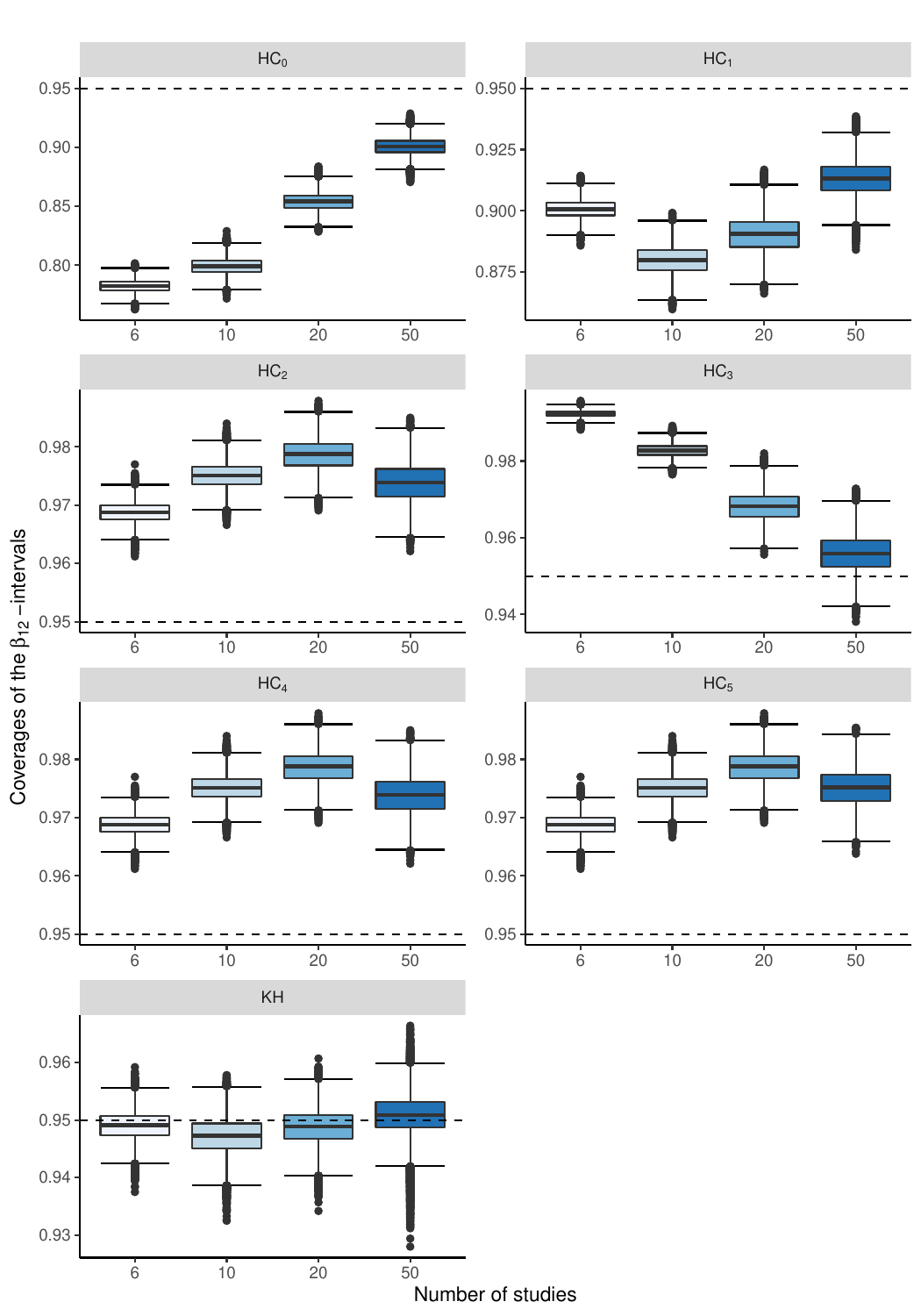}
\caption{Coverages of the \bi -intervals compared regarding the adjustments of $k$ for the estimators $HC_0-HC_5$ and $KH$.}
\label{bi_k}
\end{figure}

\begin{figure}
\includegraphics[scale=0.7, page=2]{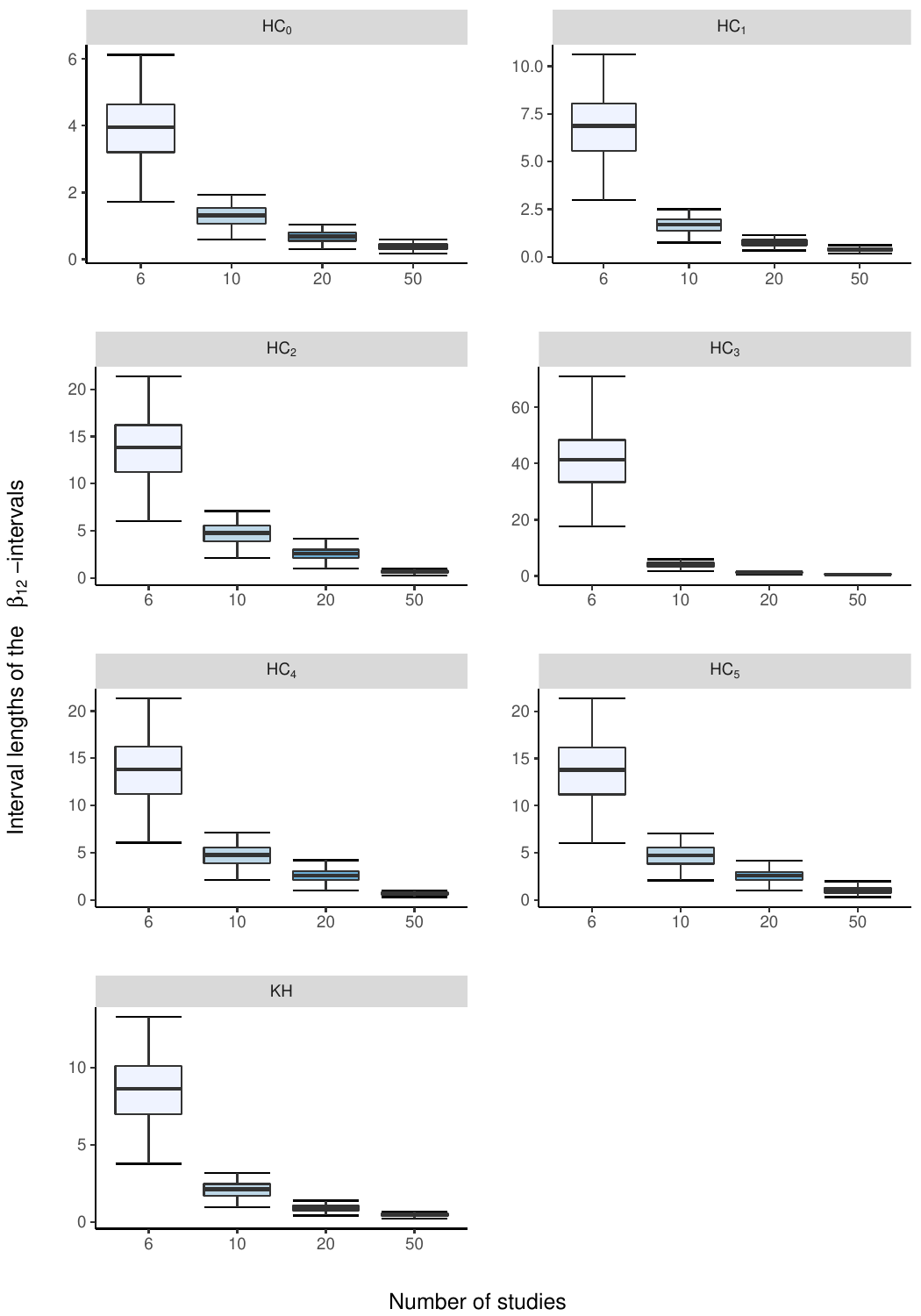}
\caption{Lengths of the \bi -intervals compared regarding the adjustments of $k$ for the estimators $HC_0-HC_5$ and $KH$ without outliers.}
\label{bi_k_l}
\end{figure}

\textbf{Adjustments of the number of studies $\boldsymbol{k}$} Considered numbers of studies are 6, 10, 20 and 50.
The lengths of both coefficients' intervals are monotonically decreasing in the number of studies $k$ (Figures \ref{b1_k_l} and \ref{bi_k_l}). This is comprehensible since the $t_{(k-p-1), (1-\alpha)}$-quantile in Equation (6)
of the main paper is monotonically decreasing in $k$.\\ Effects of the number of studies on the coverage depend on the covariance estimator used. The coverages of the $\textbf{HC}_3$-CIs for $\beta_1$ are monotonically decreasing in $k$, whereas the $\textbf{HC}_0$-CIs are monotonically increasing in $k$. Also coverages of the CIs based on $\textbf{HC}_1$, $\textbf{HC}_2$ and $\textbf{KH}$ are increasing in $k$, but only for $k\geq10$. For $k=6$ their coverages are higher than for $k=10$. Moreover, coverages of $\textbf{HC}_4$- and $\textbf{HC}_5$-CIs are increasing in $k$, but only for $k\leq20$. For $k=50$ coverages are lower than for $k=20$ (Figure \ref{b1_k}).\\ Except for the $\textbf{HC}_2$-CIs, the effect of the number of studies on the coverage of the CIs for $\beta_{12}$ is the same. In this situation the $\textbf{HC}_2$-CIs are affected by $k$ similarly to the $\textbf{HC}_4$- and $\textbf{HC}_5$-CIs (Figure \ref{bi_k}).\\ In sum the considered intervals depend highly on the number of studies $k$. Obviously, for all estimators a large number of studies is preferable. 

\begin{figure}
\includegraphics[scale=0.7, page=1]{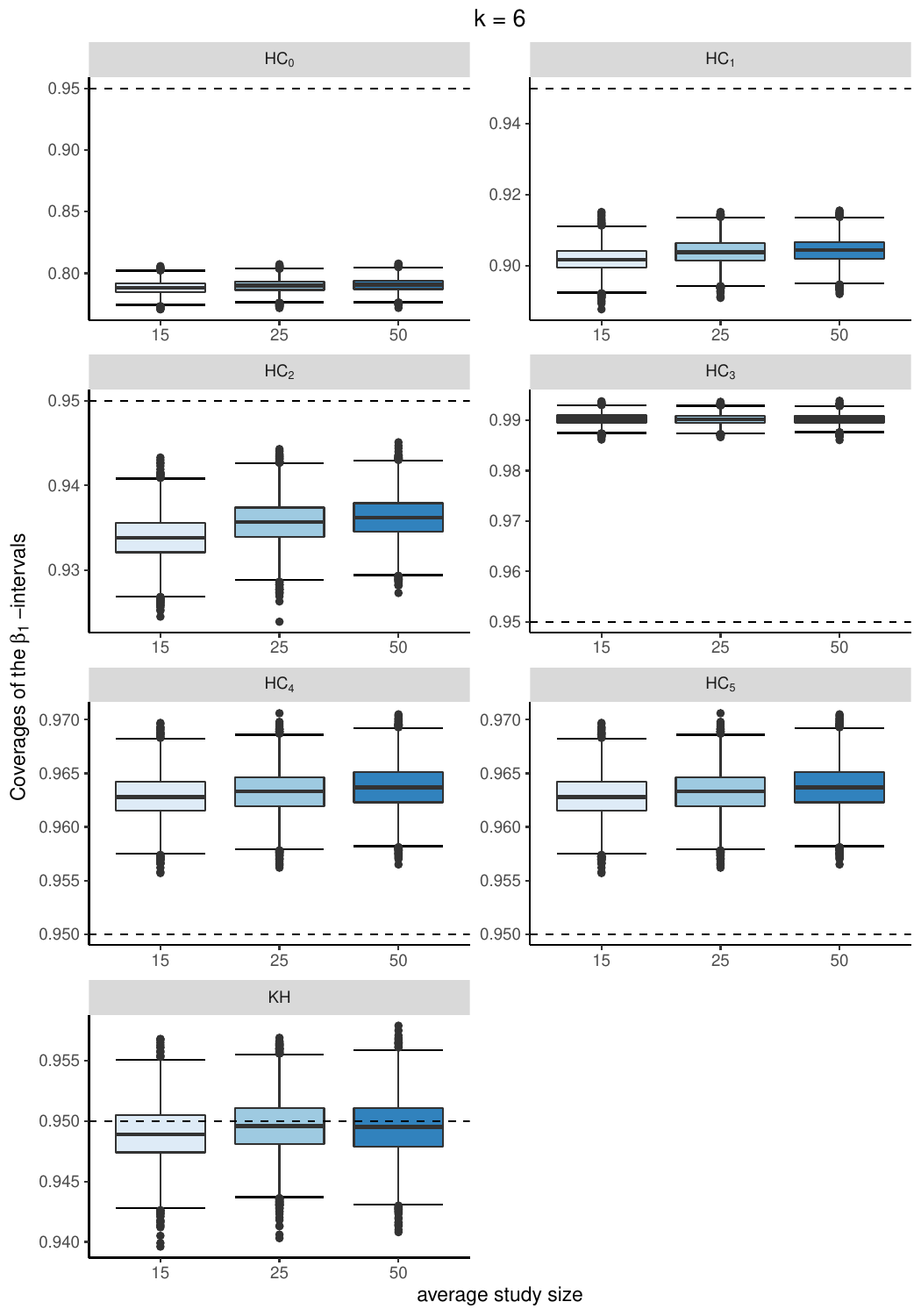}
\caption{Coverages of the \bo -intervals compared regarding the adjustments of group size ($n_i$) for the estimators $HC_0-HC_5$ and $KH$ with $k=6$.}
\label{b1_ni_1}
\end{figure}

\begin{figure}
\includegraphics[scale=0.7, page=2]{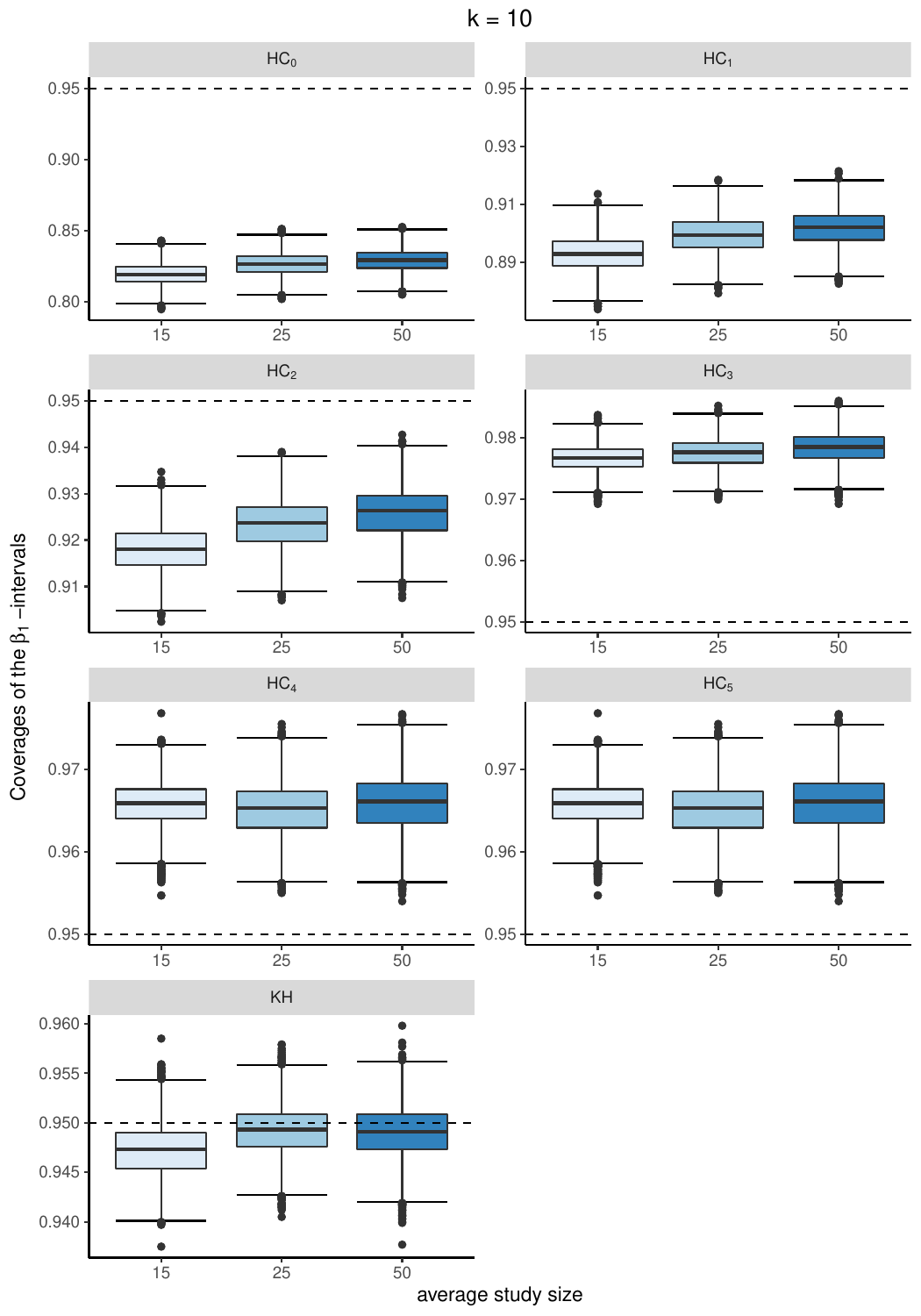}
\caption{Coverages of the \bo -intervals compared regarding the adjustments of group size ($n_i$) for the estimators $HC_0-HC_5$ and $KH$ with $k=10$.}
\label{b1_ni_2}
\end{figure}

\begin{figure}
\includegraphics[scale=0.7, page=3]{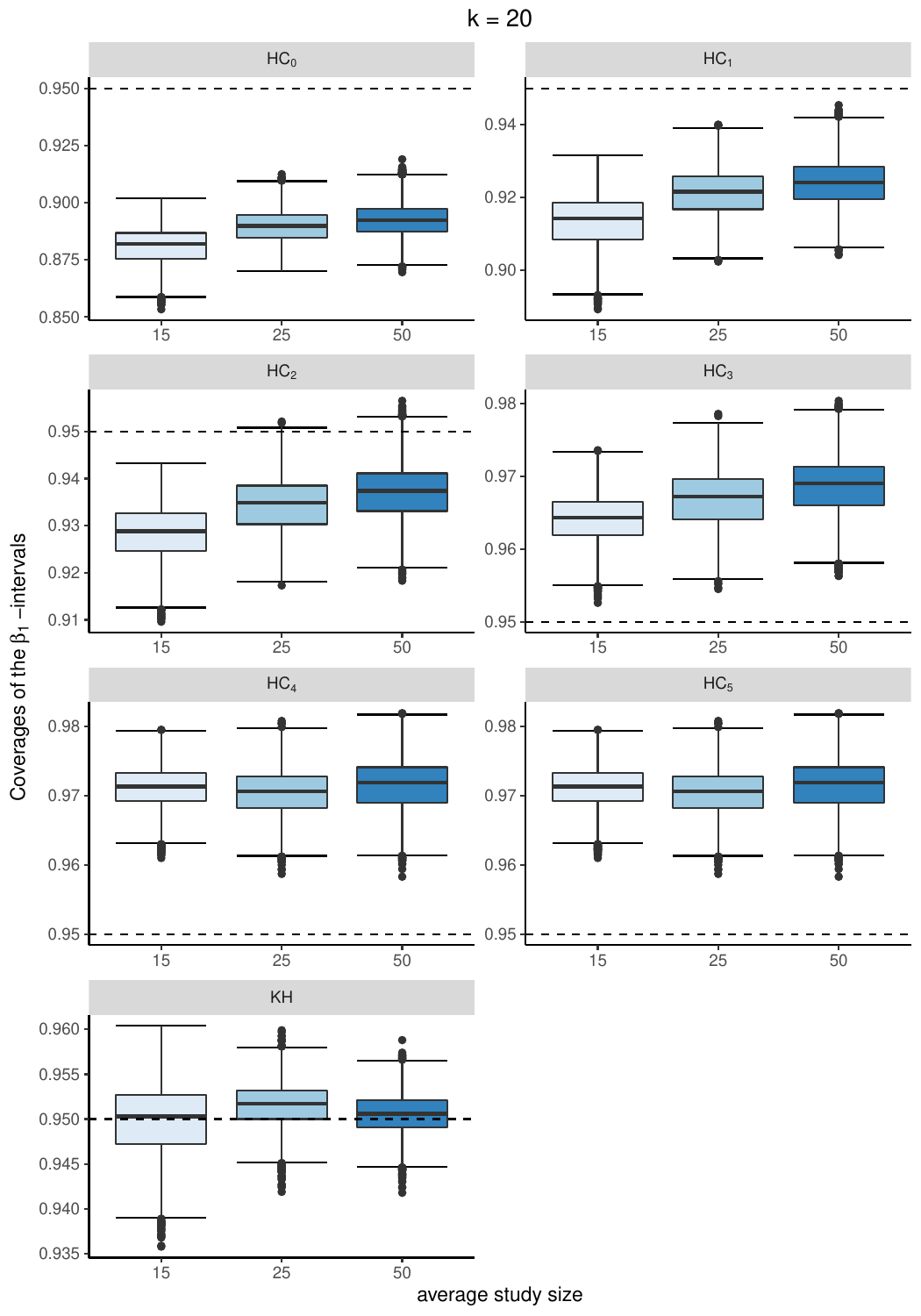}
\caption{Coverages of the \bo -intervals compared regarding the adjustments of group size ($n_i$) for the estimators $HC_0-HC_5$ and $KH$ with $k=20$.}
\label{b1_ni_3}
\end{figure}

\begin{figure}
\includegraphics[scale=0.7, page=4]{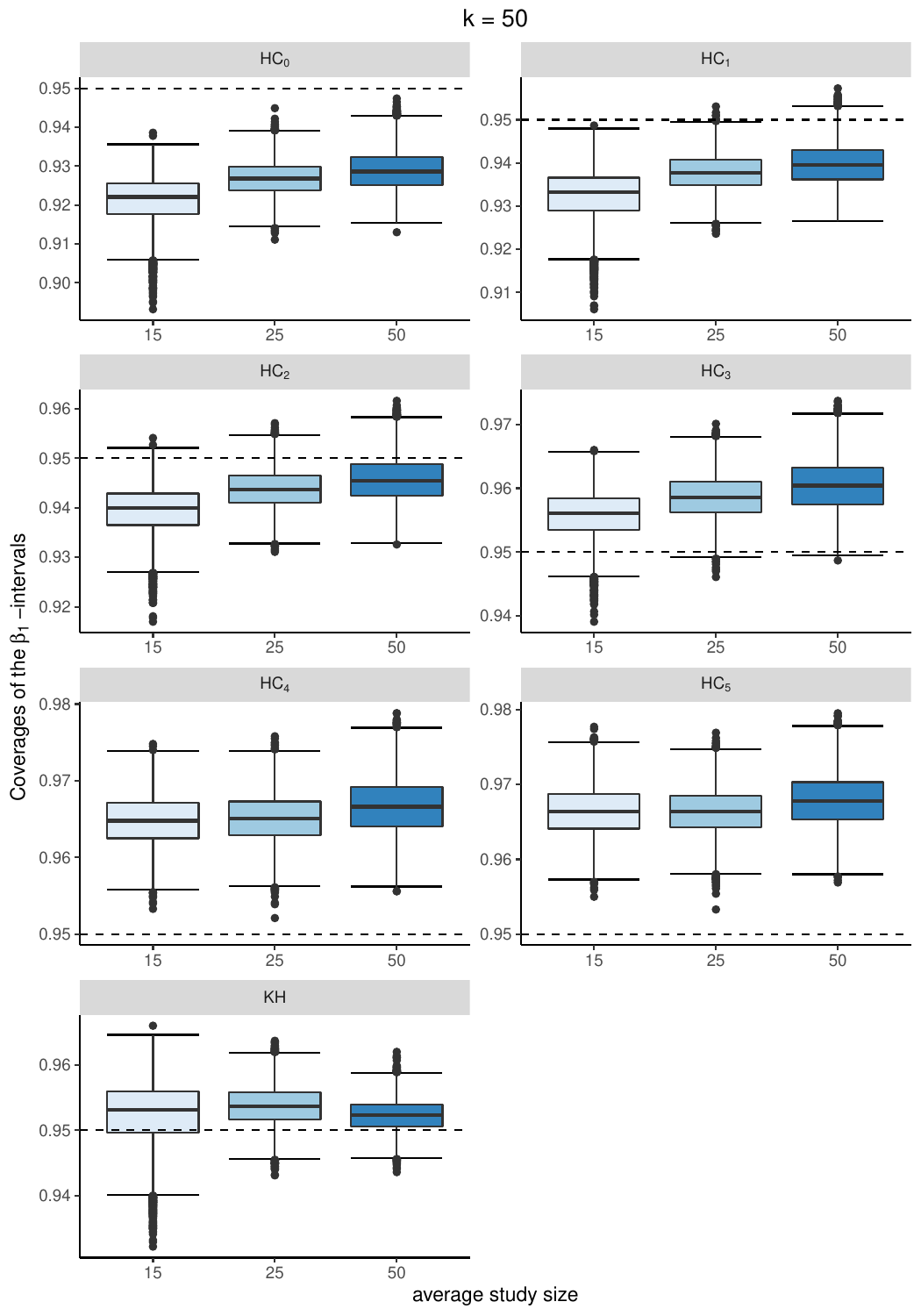}
\caption{Coverages of the \bo -intervals compared regarding the adjustments of group size ($n_i$) for the estimators $HC_0-HC_5$ and $KH$ with $k=50$.}
\label{b1_ni_4}
\end{figure}

\begin{figure}
\includegraphics[scale=0.7, page=1]{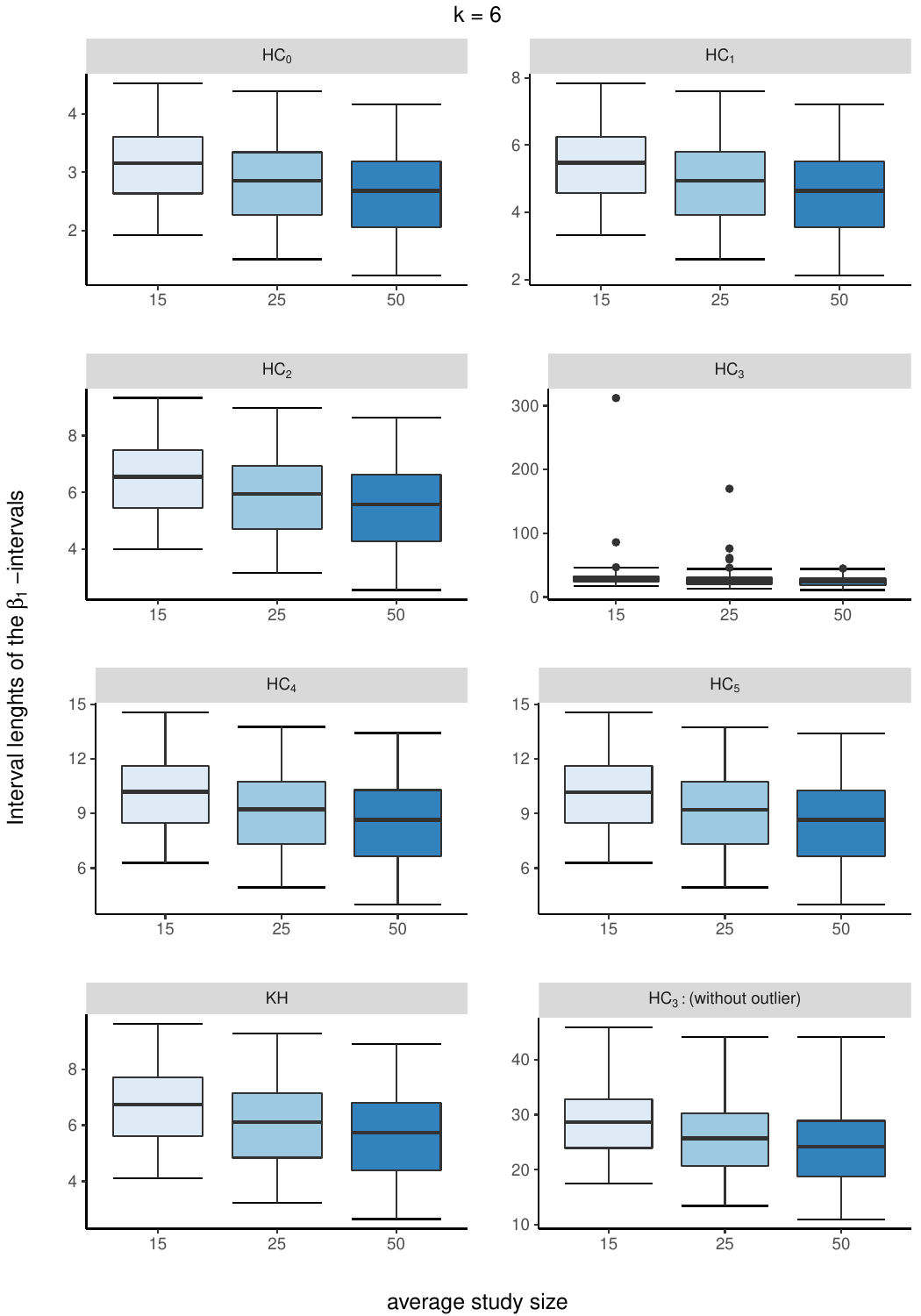}
\caption{Lengths of the \bo -intervals compared regarding the adjustments of group size ($n_i$) for the estimators $HC_0-HC_5$ and $KH$ with $k=6$.}
\label{b1_ni_l_1}
\end{figure}

\begin{figure}
\includegraphics[scale=0.7, page=3]{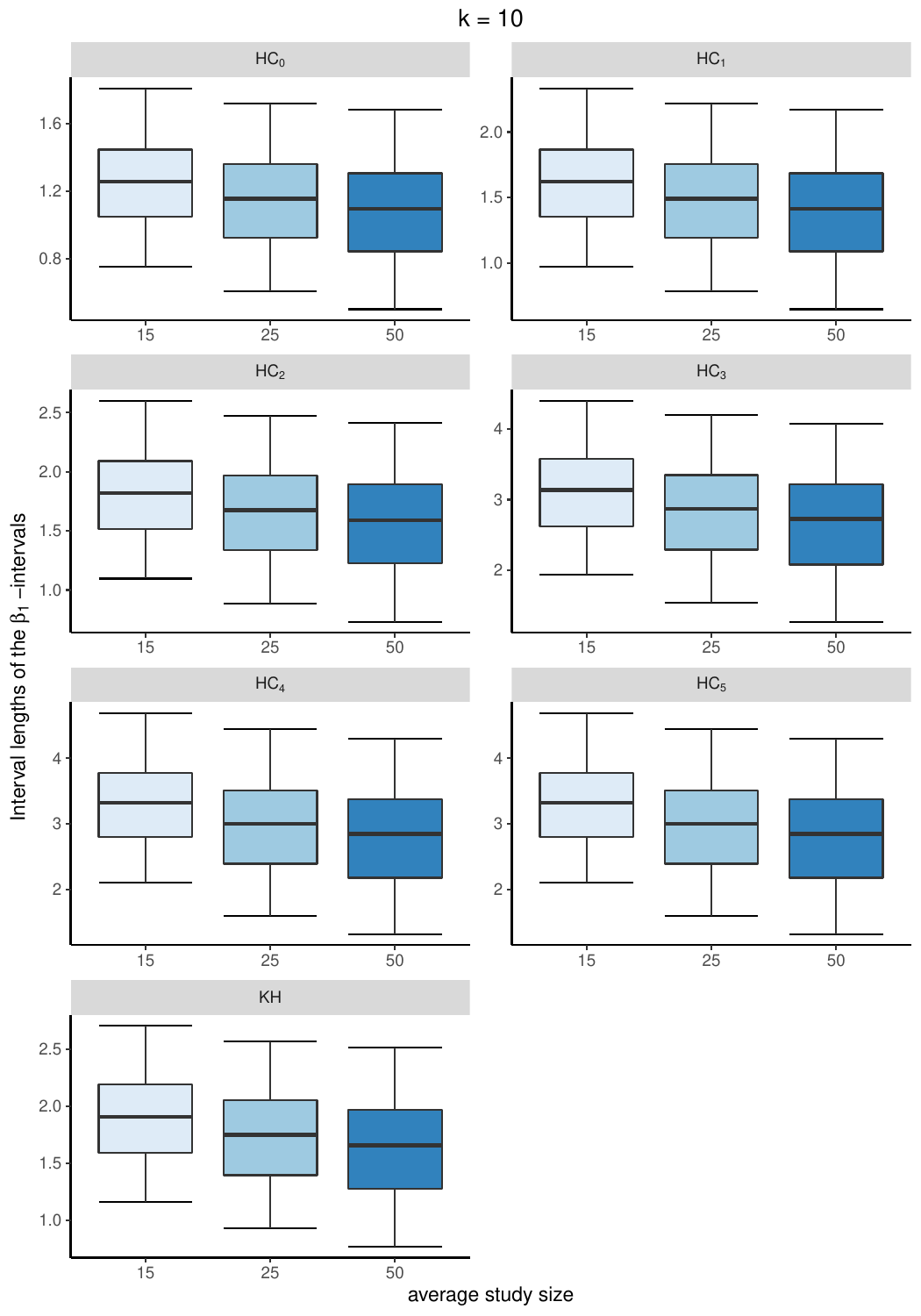}
\caption{Lengths of the \bo -intervals compared regarding the adjustments of group size ($n_i$) for the estimators $HC_0-HC_5$ and $KH$ with $k=10$.}
\label{b1_ni_l_2}
\end{figure}

\begin{figure}
\includegraphics[scale=0.7, page=4]{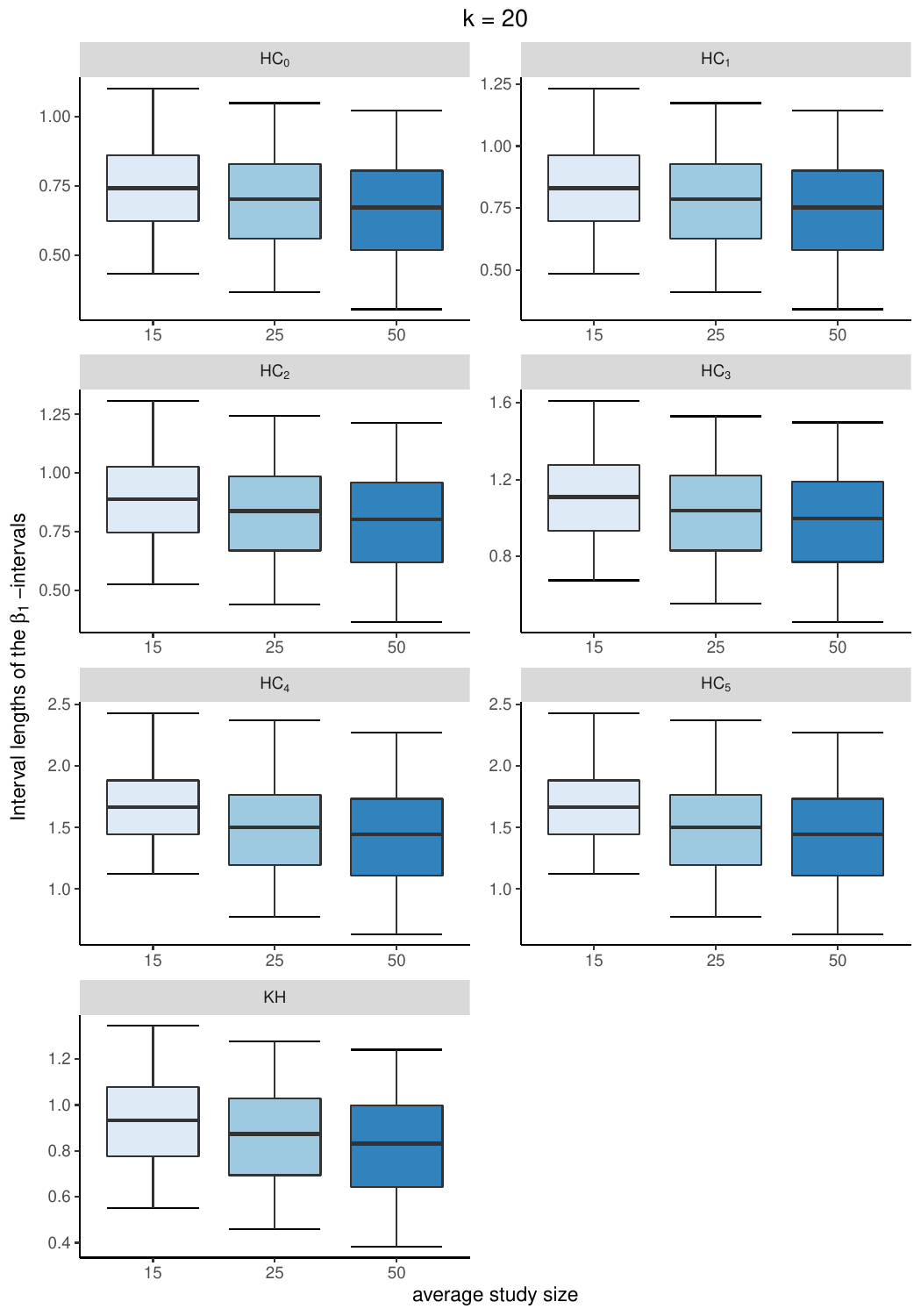}
\caption{Lengths of the \bo -intervals compared regarding the adjustments of group size ($n_i$) for the estimators $HC_0-HC_5$ and $KH$ with $k=20$.}
\label{b1_ni_l_3}
\end{figure}

\begin{figure}
\includegraphics[scale=0.7, page=2]{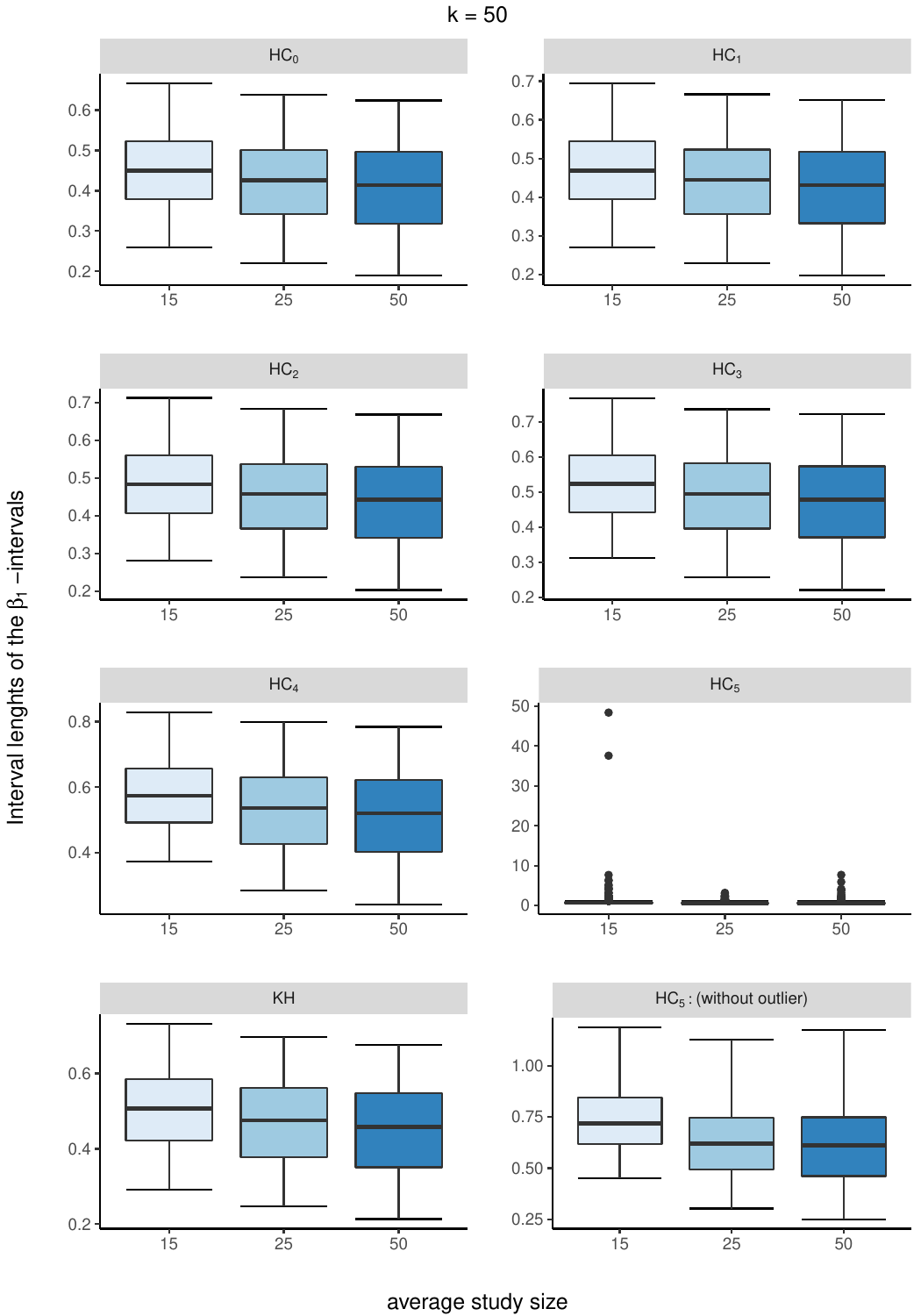}
\caption{Lengths of the \bo -intervals compared regarding the adjustments of group size ($n_i$) for the estimators $HC_0-HC_5$ and $KH$ with $k=50$.}
\label{b1_ni_l_4}
\end{figure}

\begin{figure}
\includegraphics[scale=0.7, page=1]{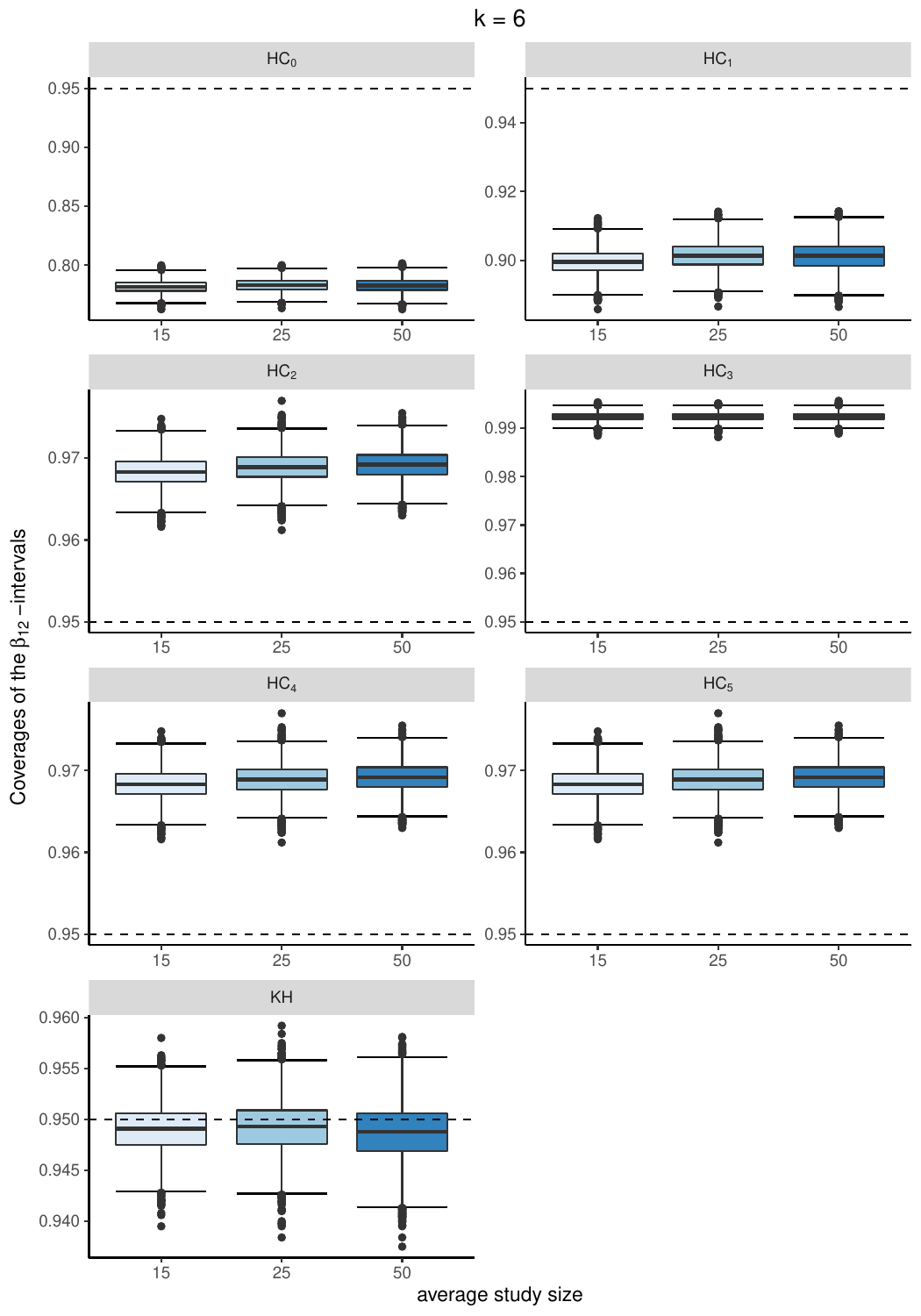}
\caption{Coverages of the \bi -intervals compared regarding the adjustments of group size ($n_i$) for the estimators $HC_0-HC_5$ and $KH$ with $k=6$.}
\label{bi_ni_1}
\end{figure}

\begin{figure}
\includegraphics[scale=0.7, page=2]{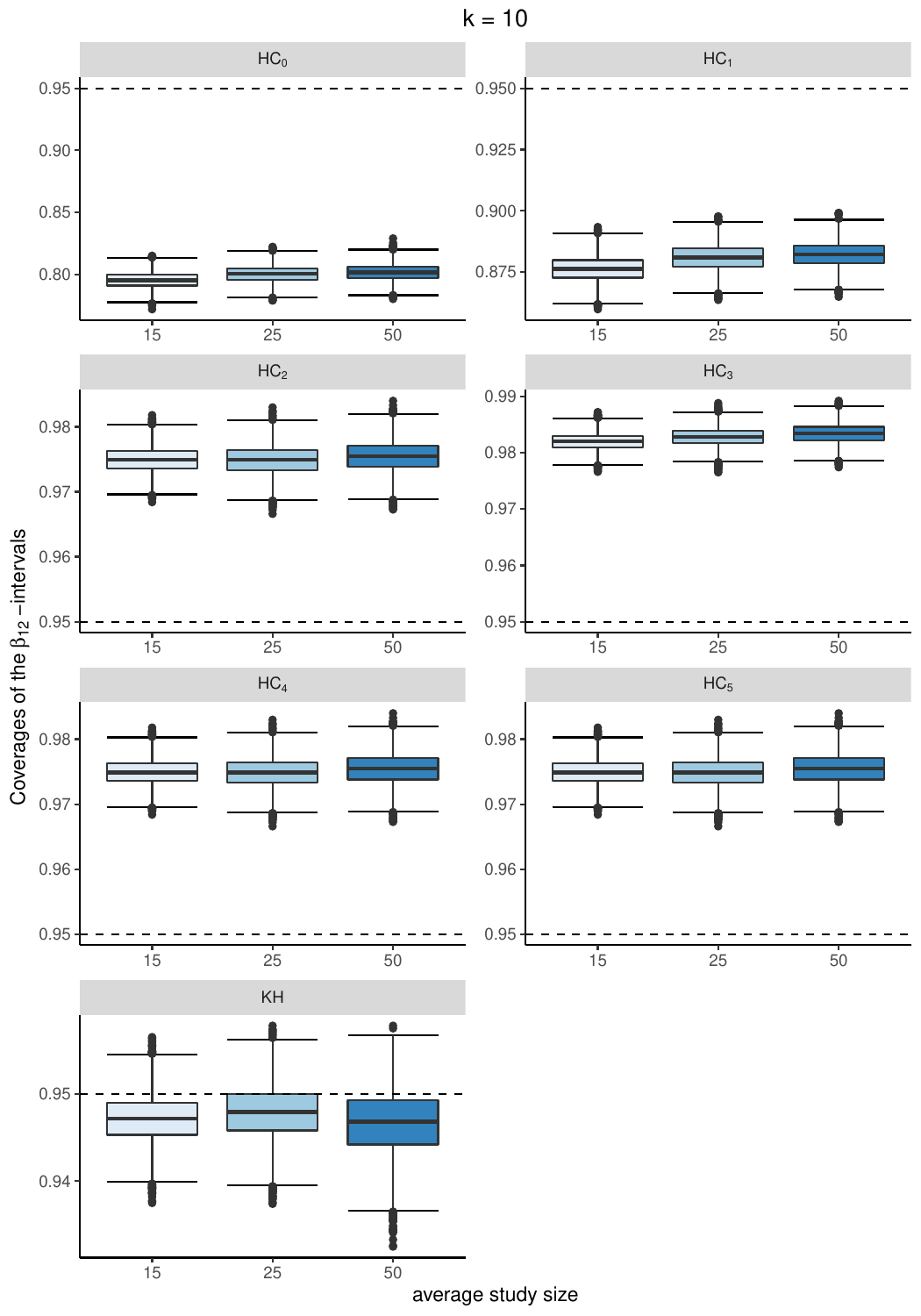}
\caption{Coverages of the \bi -intervals compared regarding the adjustments of group size ($n_i$) for the estimators $HC_0-HC_5$ and $KH$ with $k=10$.}
\label{bi_ni_2}
\end{figure}

\begin{figure}
\includegraphics[scale=0.7, page=3]{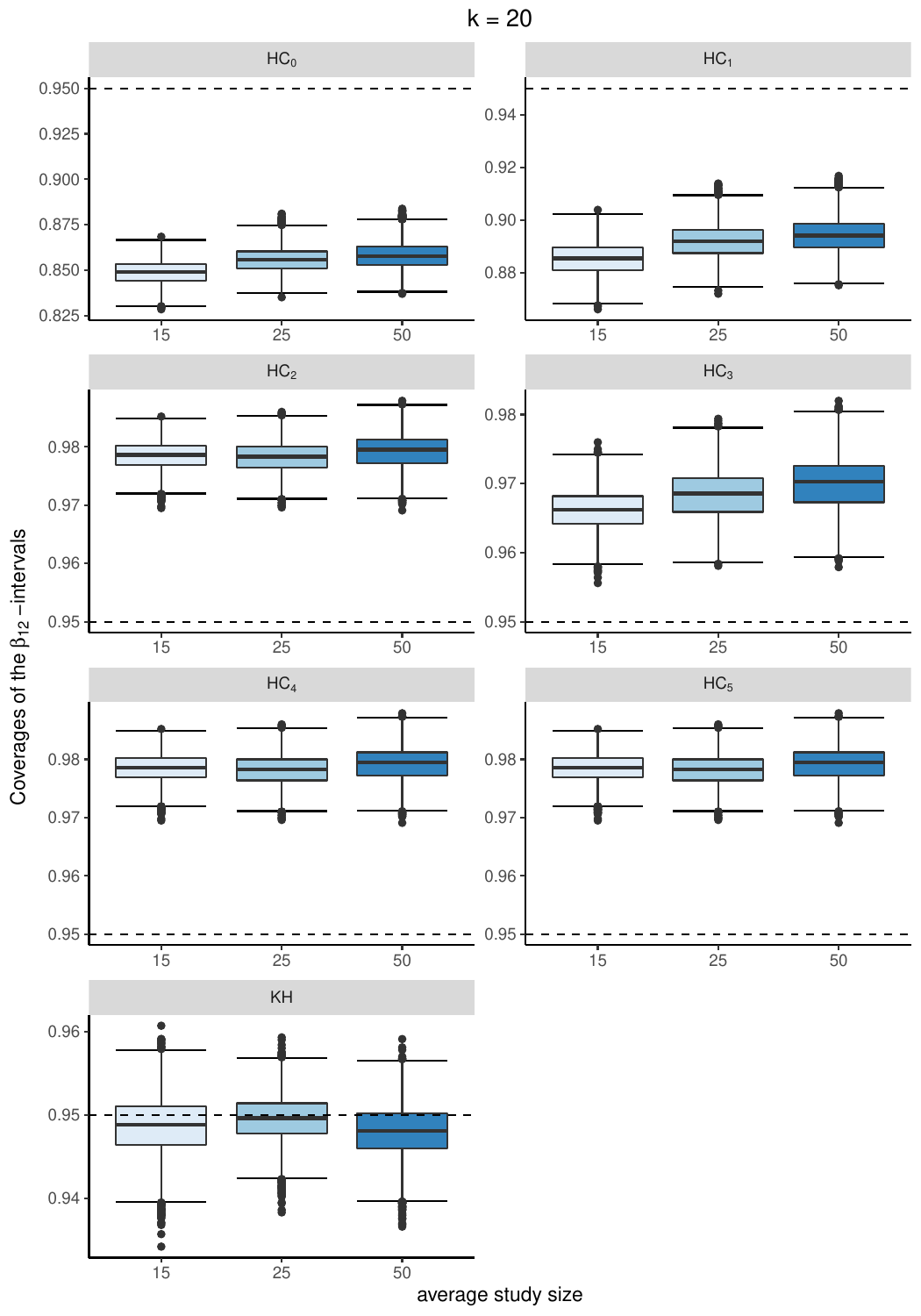}
\caption{Coverages of the \bi -intervals compared regarding the adjustments of group size ($n_i$) for the estimators $HC_0-HC_5$ and $KH$ with $k=20$.}
\label{bi_ni_3}
\end{figure}

\begin{figure}
\includegraphics[scale=0.7, page=4]{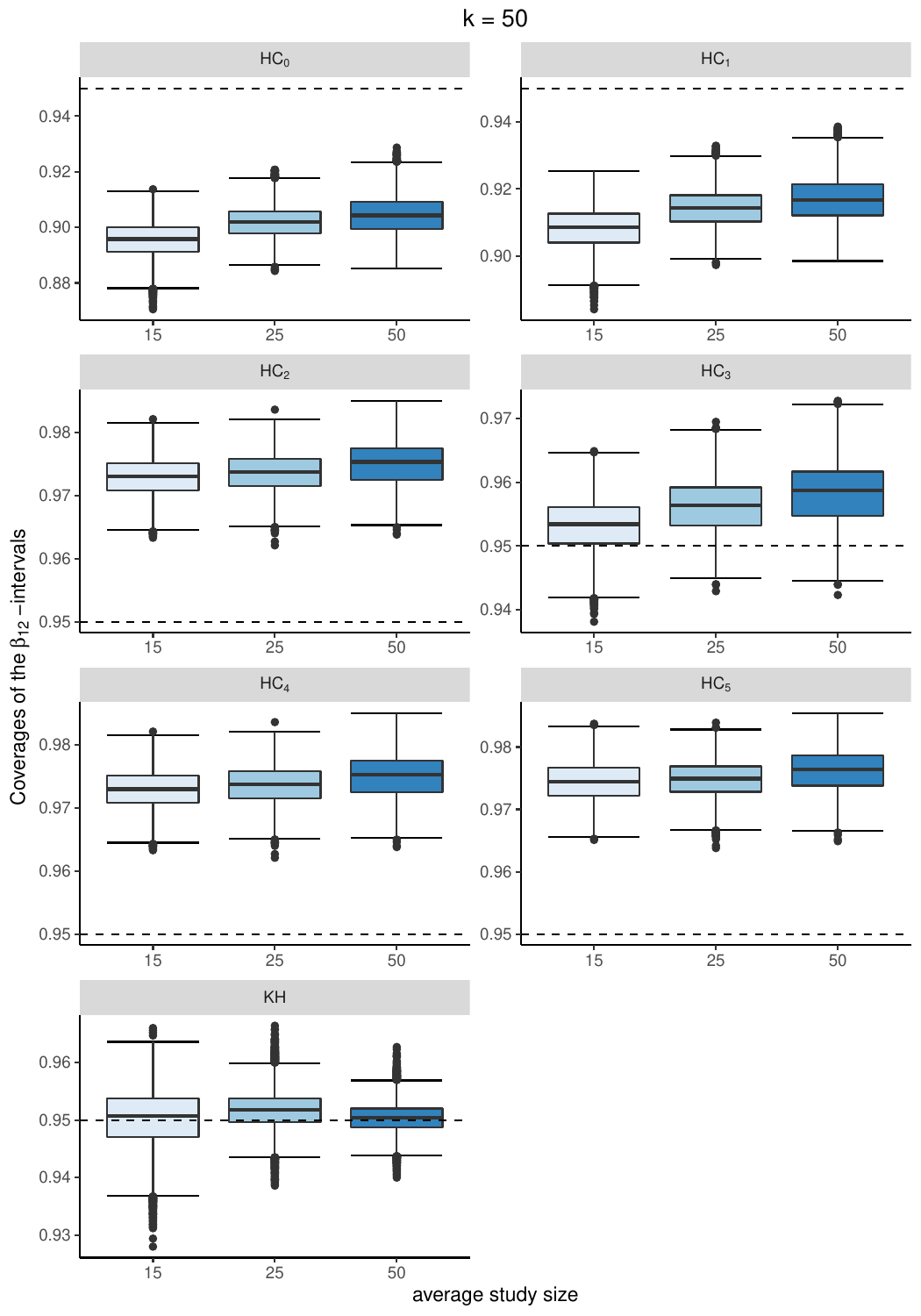}
\caption{Coverages of the \bi -intervals compared regarding the adjustments of group size ($n_i$) for the estimators $HC_0-HC_5$ and $KH$ with $k=50$.}
\label{bi_ni_4}
\end{figure}

\begin{figure}
\includegraphics[scale=0.7, page=1]{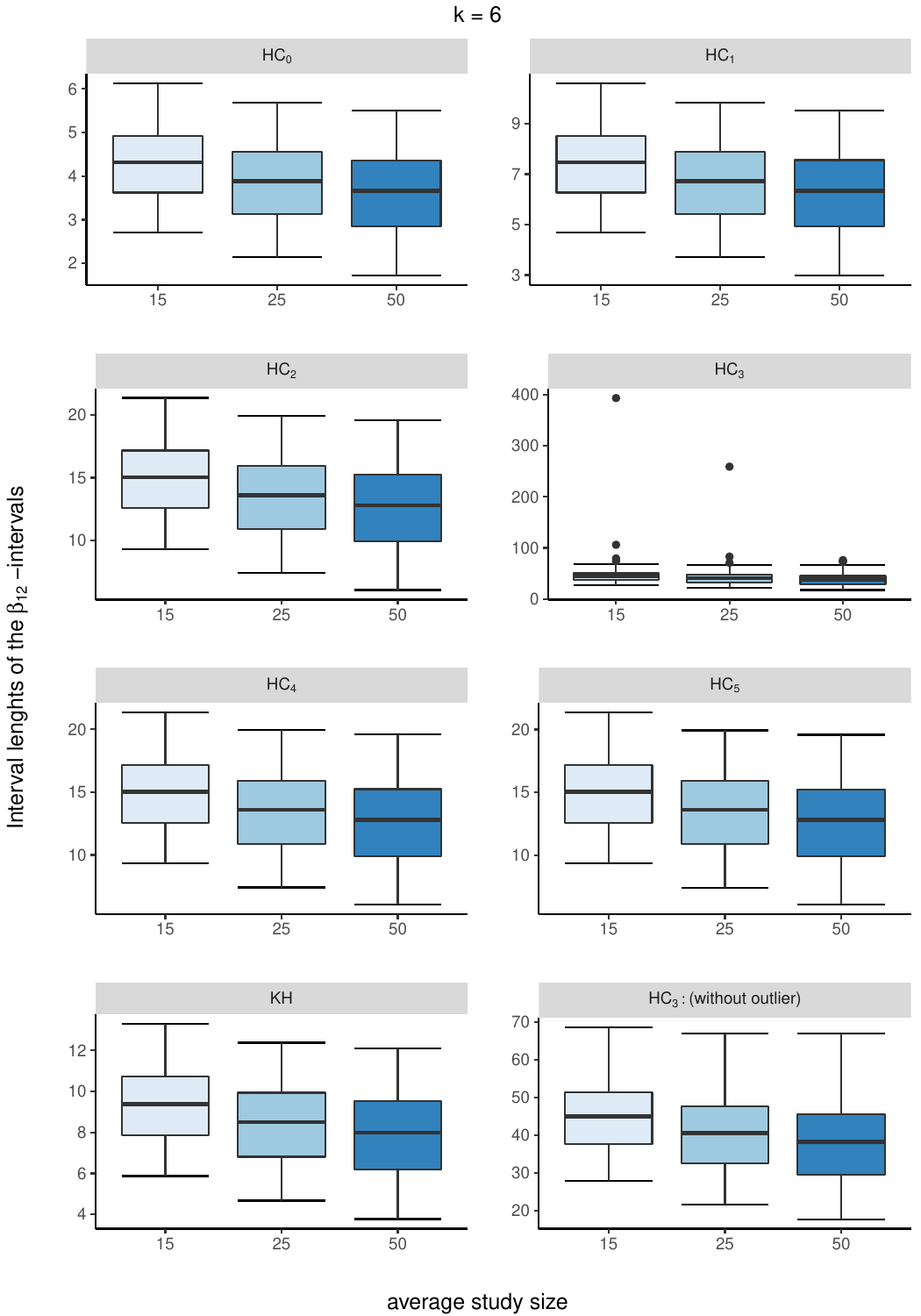}
\caption{Lengths of the \bi -intervals compared regarding the adjustments of group size ($n_i$) for the estimators $HC_0-HC_5$ and $KH$ with $k=6$.}
\label{bi_ni_l_1}
\end{figure}

\begin{figure}
\includegraphics[scale=0.7, page=3]{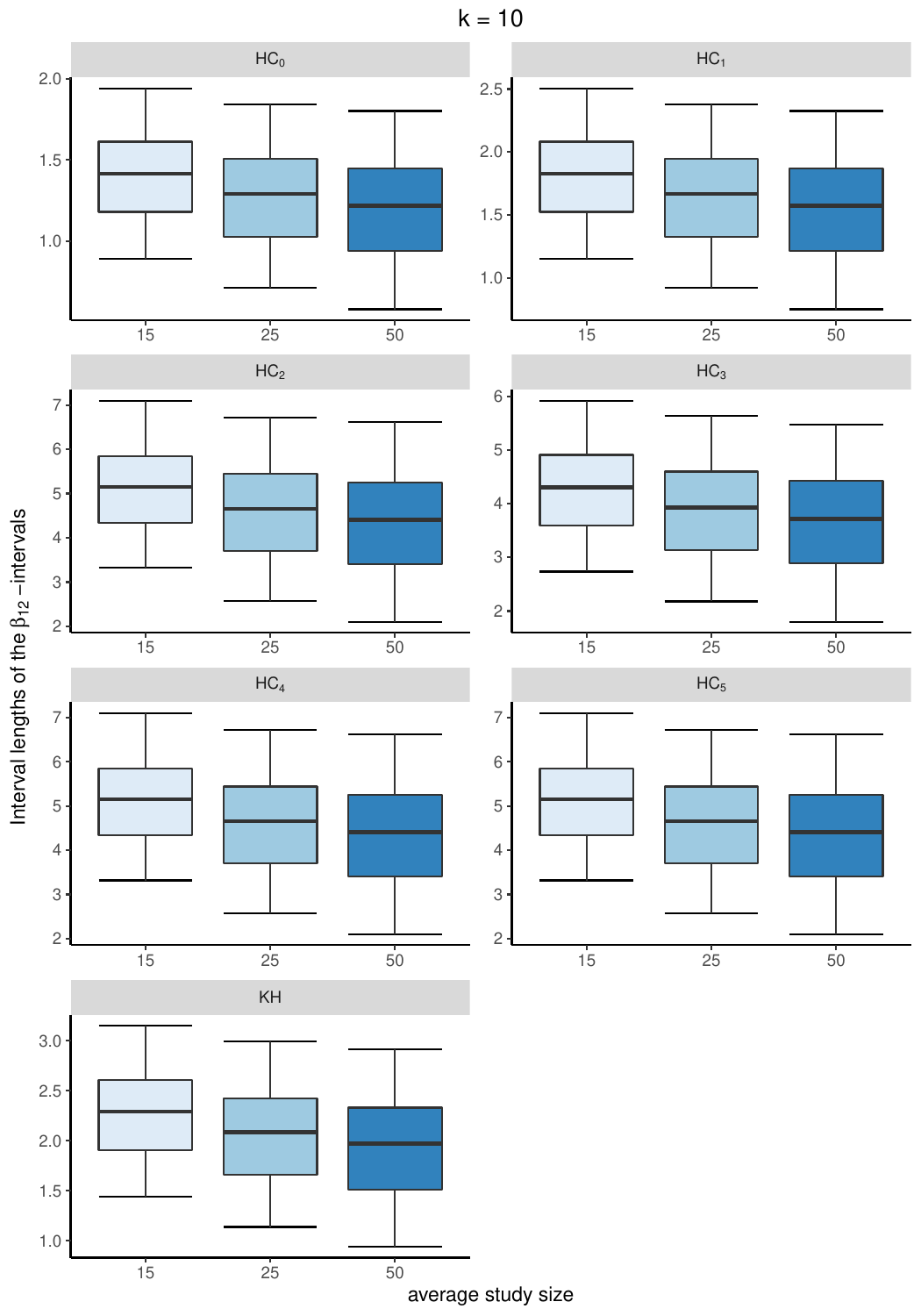}
\caption{Lengths of the \bi -intervals compared regarding the adjustments of group size ($n_i$) for the estimators $HC_0-HC_5$ and $KH$ with $k=10$.}
\label{bi_ni_l_2}
\end{figure}

\begin{figure}
\includegraphics[scale=0.7, page=4]{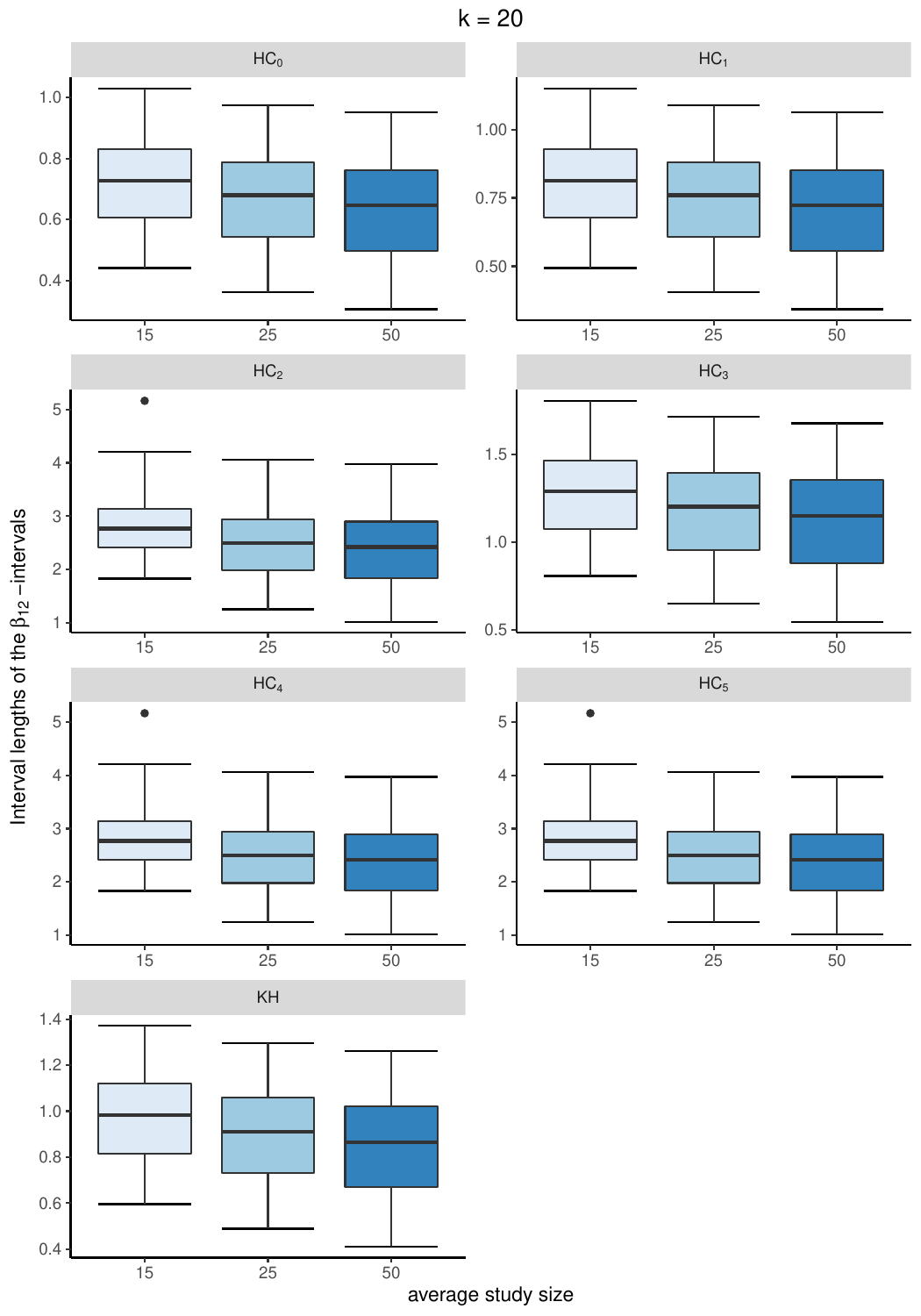}
\caption{Lengths of the \bi -intervals compared regarding the adjustments of group size ($n_i$) for the estimators $HC_0-HC_5$ and $KH$ with $k=20$.}
\label{bi_ni_l_3}
\end{figure}

\begin{figure}
\includegraphics[scale=0.7, page=2]{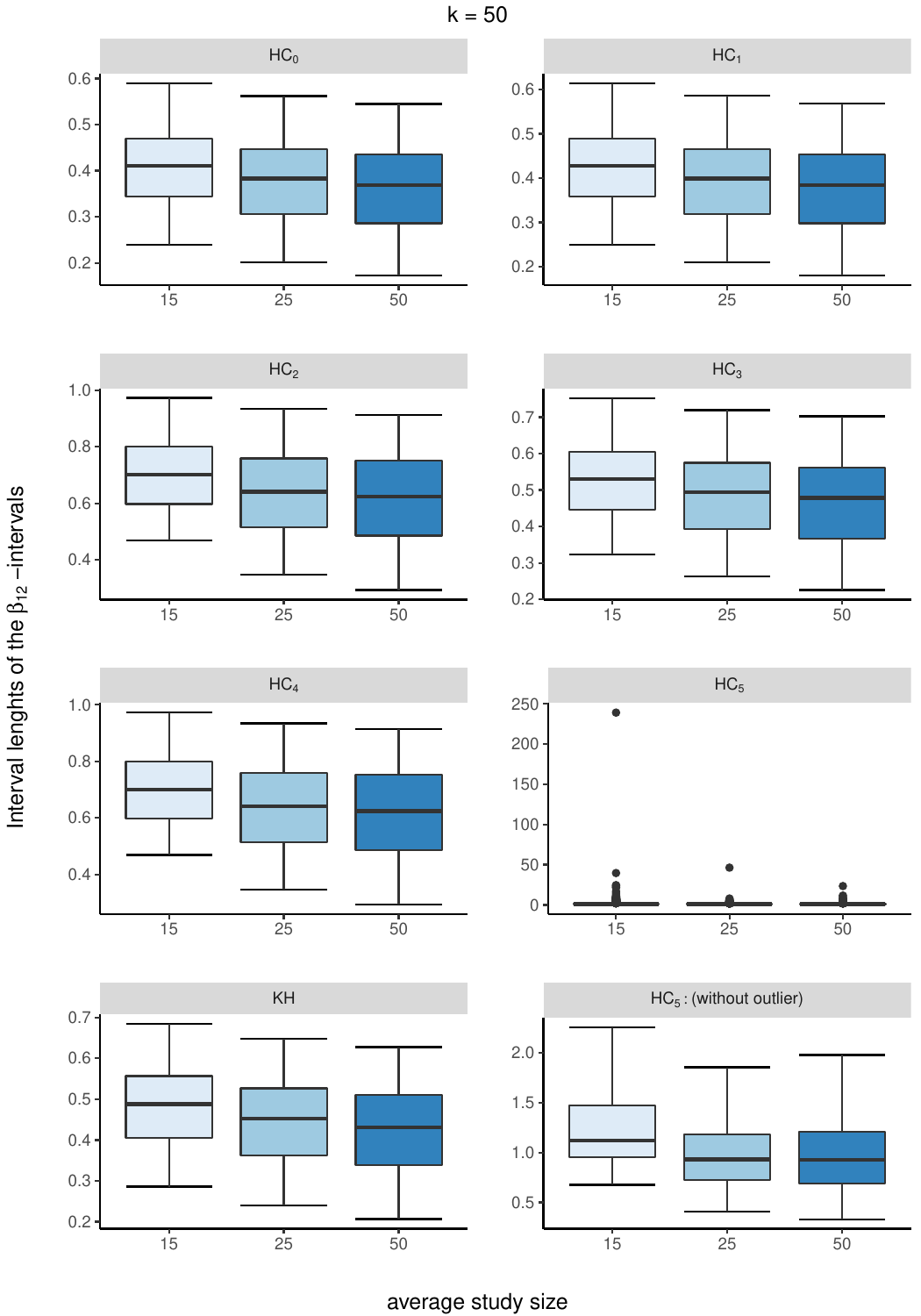}
\caption{Lengths of the \bi -intervals compared regarding the adjustments of group size ($n_i$) for the estimators $HC_0-HC_5$ and $KH$ with $k=50$.}
\label{bi_ni_l_4}
\end{figure}

\textbf{Adjustments of study size}
Small ($\bar{n}=15$), medium ($\bar{n}=25$) and large ($\bar{n}=50$) group sizes are compared. For the most numbers of studies $k$ and covariance estimators the median coverage is slightly increasing in the study size (Figures \ref{b1_ni_1}-\ref{b1_ni_4} and \ref{bi_ni_1}-\ref{bi_ni_4}). For the CIs for $\beta_1$ exceptions are the $\textbf{HC}_4$, $\textbf{HC}_5$ and $\textbf{KH}$ based CIs for $k\in\{10,20\}$ and the $\textbf{KH}$-CIs for $k=50$. The coverages of the CIs for $\beta_{12}$ have no increasing trend for the $\textbf{KH}$-CIs for all $k$, nor the coverages of the  $\textbf{HC}_0, \textbf{HC}_1$ and $\textbf{HC}_3$ based CIs for $k=6$ and the $\textbf{HC}_2$, $\textbf{HC}_4$ and $\textbf{HC}_5$ based CIs for $k=20$.\\ The corresponding interval lengths are decreasing in the study sizes for all $k$ and estimators (Figures \ref{b1_ni_l_1}-\ref{b1_ni_l_4} and \ref{bi_ni_l_1}-\ref{bi_ni_l_4}). This trend may be caused by the impact of $n_i$ on $v_i$ in Equation (9) 
of the main paper, which leads to decreasing standard errors in equation (6)
.\\ Thus, overall larger studies lead to better confidence intervals, since both coverages and interval lengths are improved for larger study sizes. Among the considered study sizes there is no size where an estimator has a different performance compared to the other estimators.\\ It is noticeable that the extreme lengths of $\textbf{HC}_3$-CIs for $k=6$ appear mostly for small and medium group sizes (Figures \ref{b1_ni_l_1} and \ref{bi_ni_l_1}).\\

\begin{figure}
\includegraphics[scale=0.7, page=1]{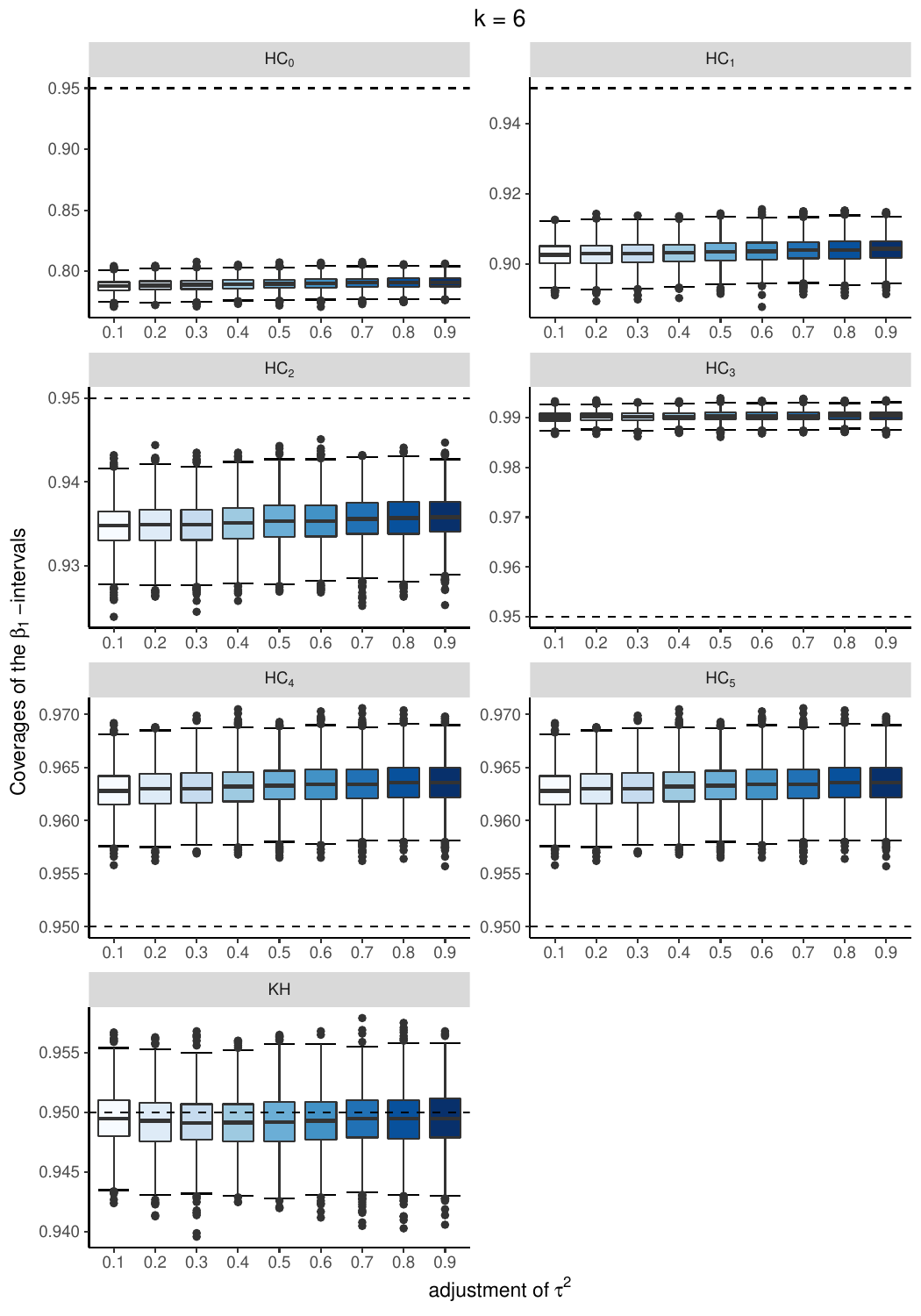}
\caption{Coverages of the \bo -intervals compared regarding the adjustments of the heterogeneity parameter $\tau^2$ for the estimators $HC_0-HC_5$ and $KH$ with $k=6$.}
\label{b1_tau_1}
\end{figure}

\begin{figure}
\includegraphics[scale=0.7, page=2]{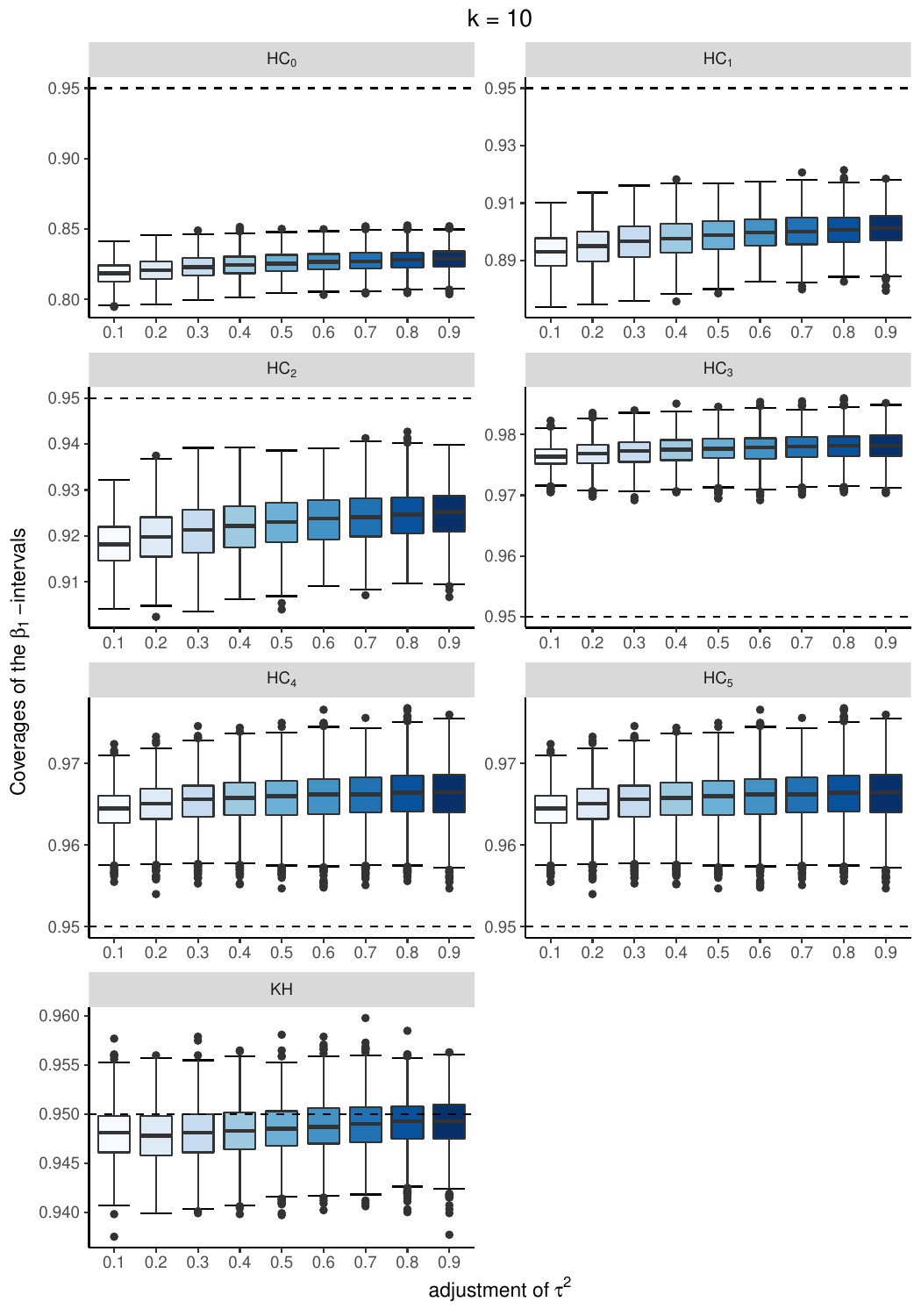}
\caption{Coverages of the \bo -intervals compared regarding the adjustments of the heterogeneity parameter $\tau^2$ for the estimators $HC_0-HC_5$ and $KH$ with $k=10$.}
\label{b1_tau_2}
\end{figure}

\begin{figure}
\includegraphics[scale=0.7, page=3]{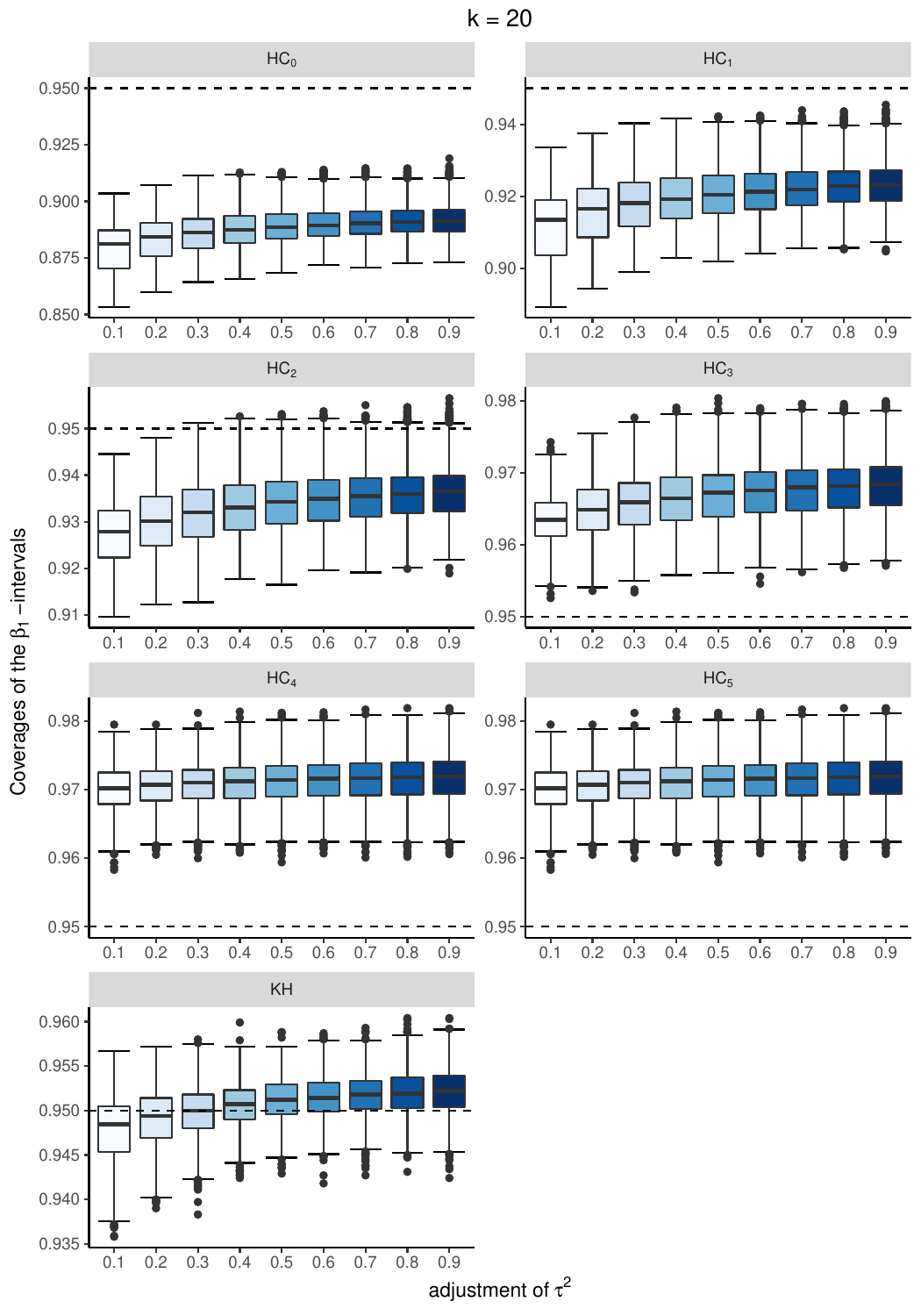}
\caption{Coverages of the \bo -intervals compared regarding the adjustments of the heterogeneity parameter $\tau^2$ for the estimators $HC_0-HC_5$ and $KH$ with $k=20$.}
\label{b1_tau_3}
\end{figure}

\begin{figure}
\includegraphics[scale=0.7, page=4]{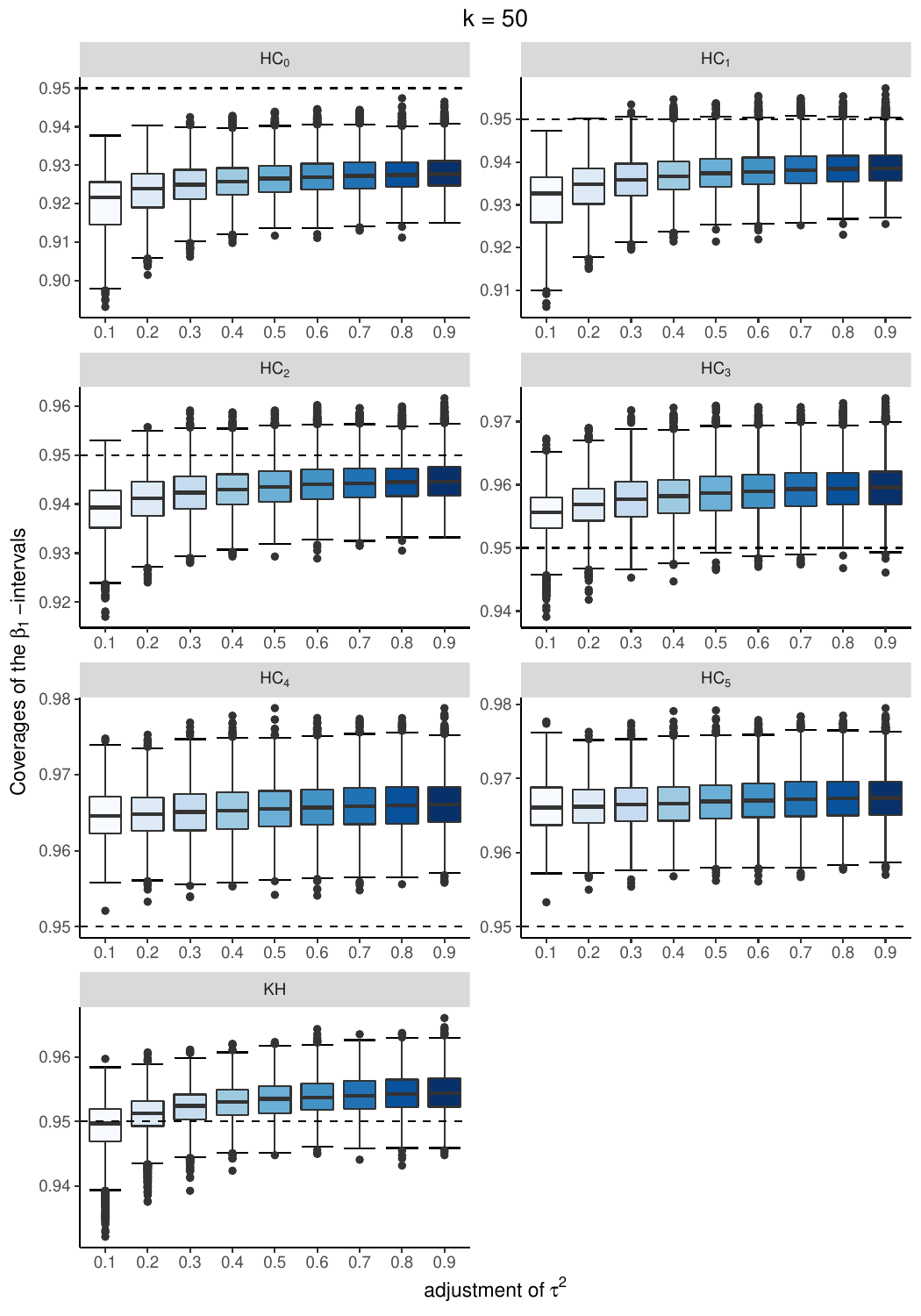}
\caption{Coverages of the \bo -intervals compared regarding the adjustments of the heterogeneity parameter $\tau^2$ for the estimators $HC_0-HC_5$ and $KH$ with $k=50$.}
\label{b1_tau_4}
\end{figure}

\begin{figure}
\includegraphics[scale=0.7, page=1]{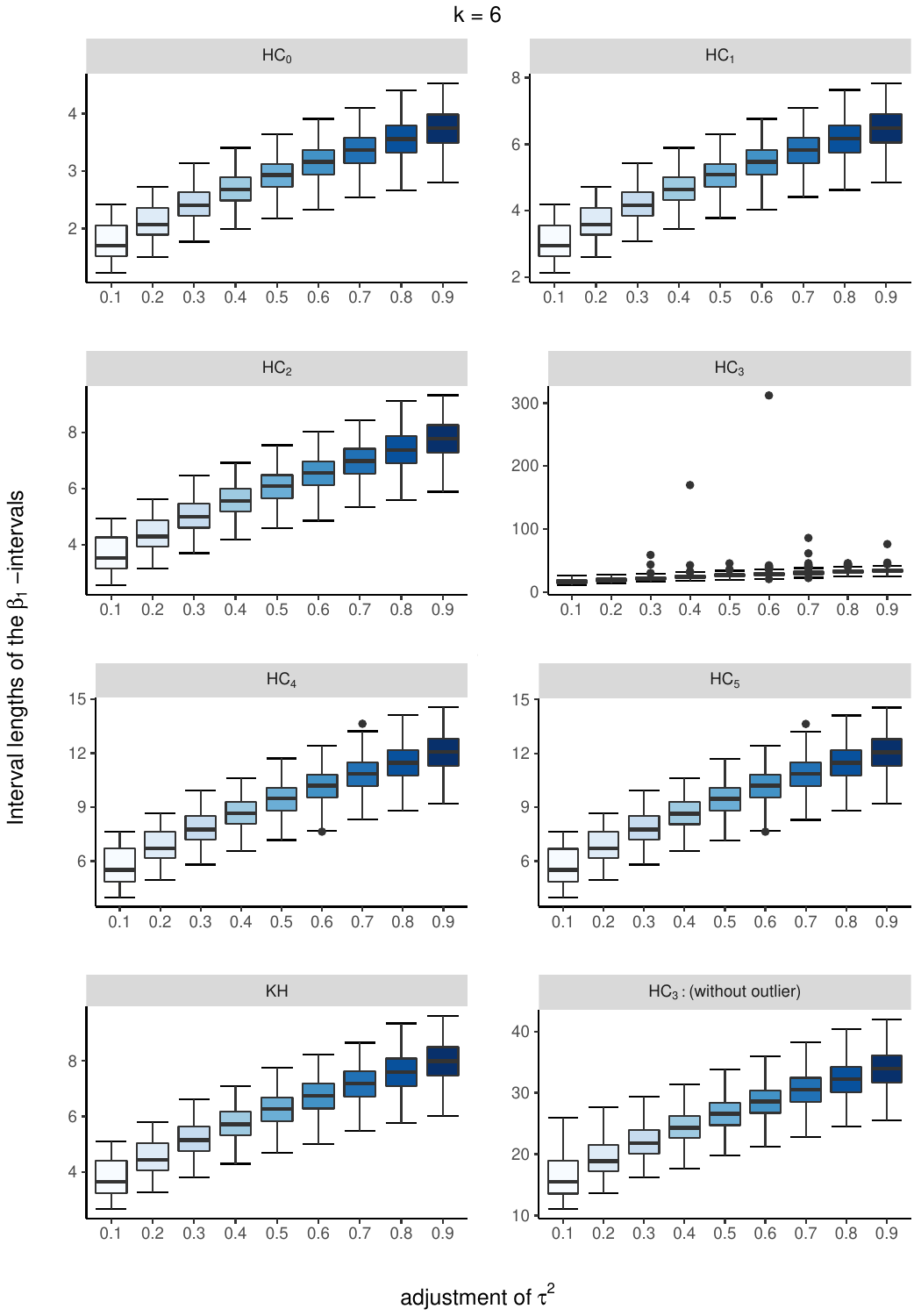}
\caption{Lengths of the \bo -intervals compared regarding the adjustments of the heterogeneity parameter $\tau^2$ for the estimators $HC_0-HC_5$ and $KH$ with $k=6$.}
\label{b1_tau_l_1}
\end{figure}

\begin{figure}
\includegraphics[scale=0.7, page=3]{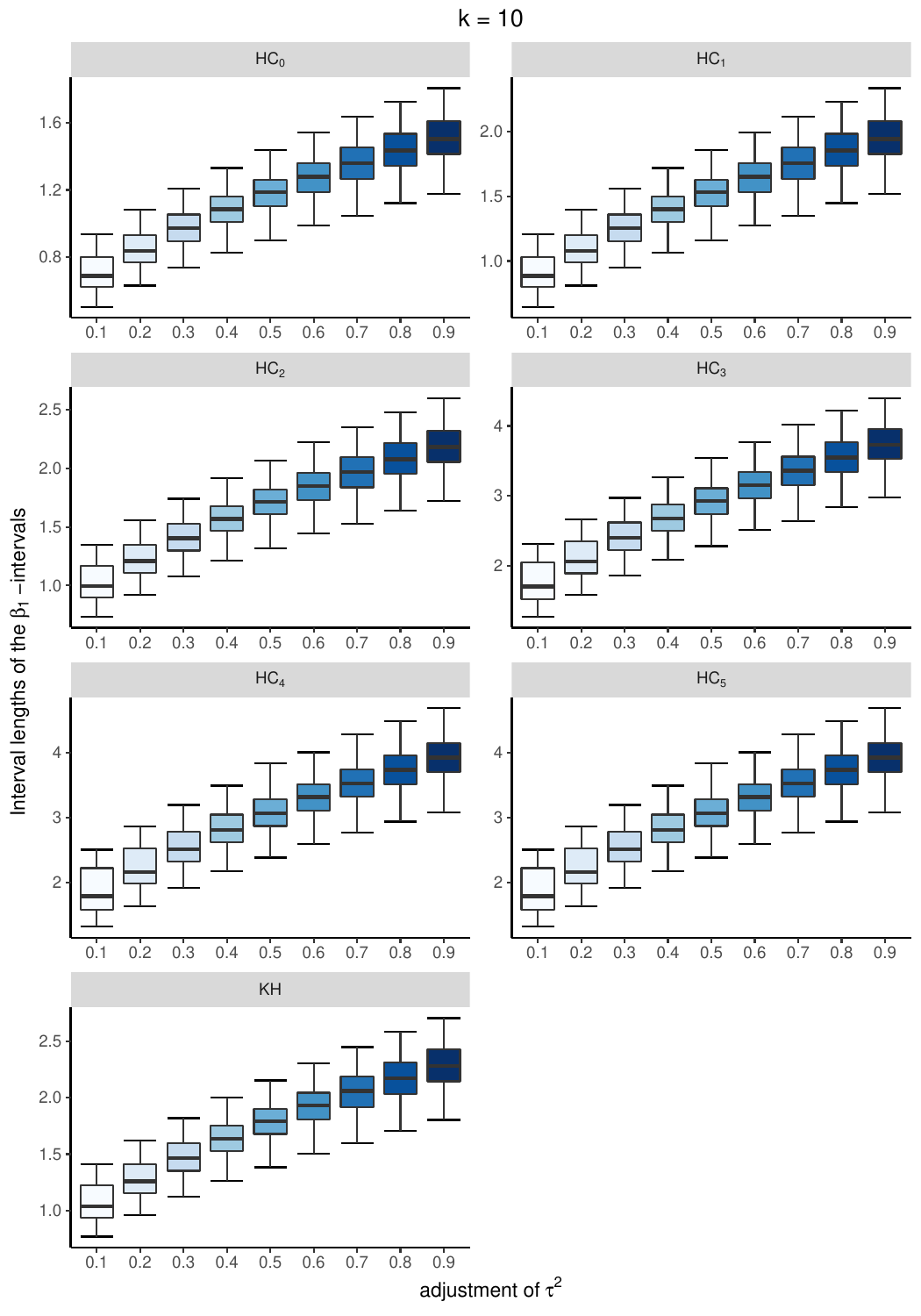}
\caption{Lengths of the \bo -intervals compared regarding the adjustments of the heterogeneity parameter $\tau^2$ for the estimators $HC_0-HC_5$ and $KH$ with $k=10$.}
\label{b1_tau_l_2}
\end{figure}

\begin{figure}
\includegraphics[scale=0.7, page=4]{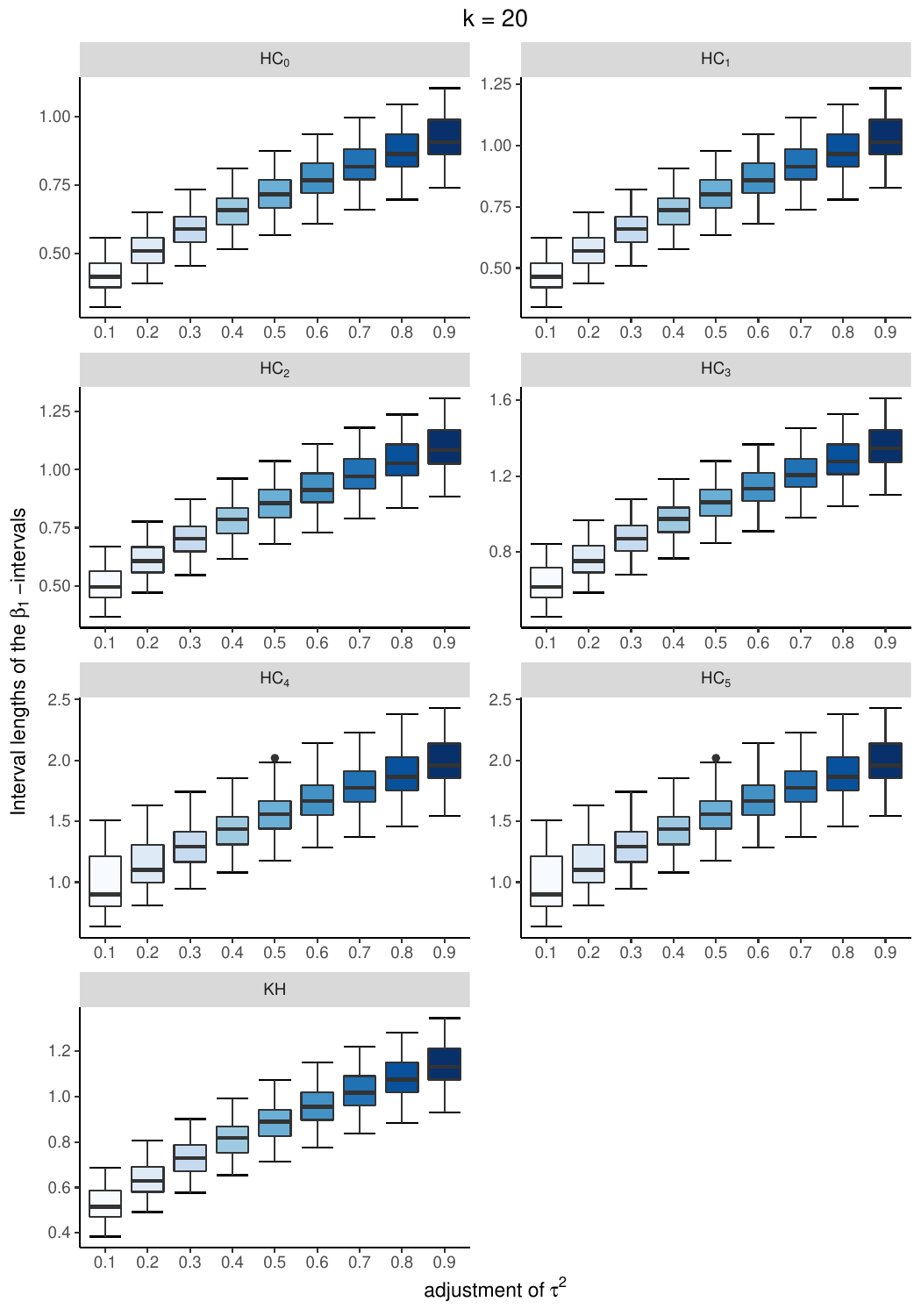}
\caption{Lengths of the \bo -intervals compared regarding the adjustments of the heterogeneity parameter $\tau^2$ for the estimators $HC_0-HC_5$ and $KH$ with $k=20$.}
\label{b1_tau_l_3}
\end{figure}

\begin{figure}
\includegraphics[scale=0.7, page=2]{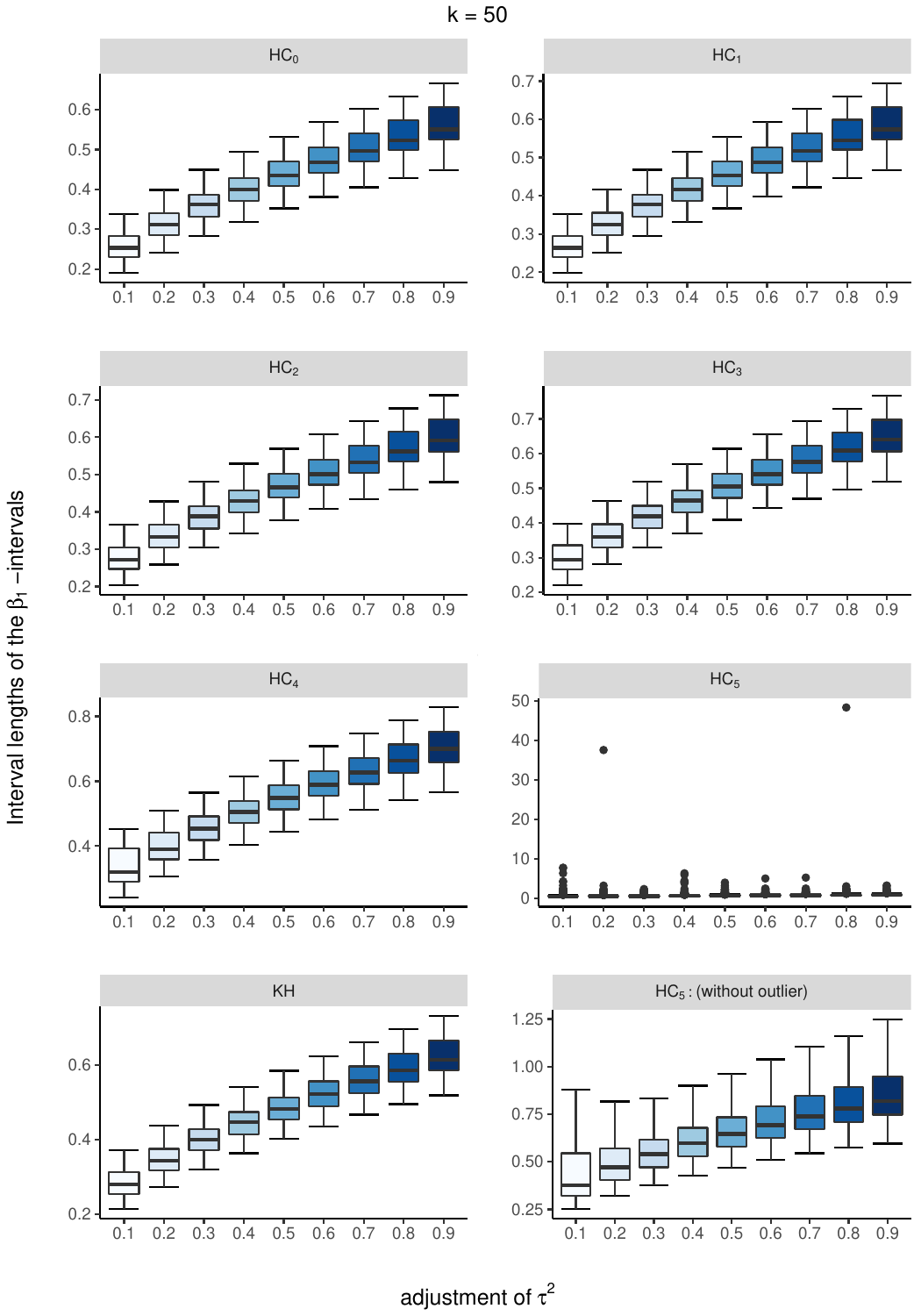}
\caption{Lengths of the \bo -intervals compared regarding the adjustments of the heterogeneity parameter $\tau^2$ for the estimators $HC_0-HC_5$ and $KH$ with $k=50$.}
\label{b1_tau_l_4}
\end{figure}

\begin{figure}
\includegraphics[scale=0.7, page=1]{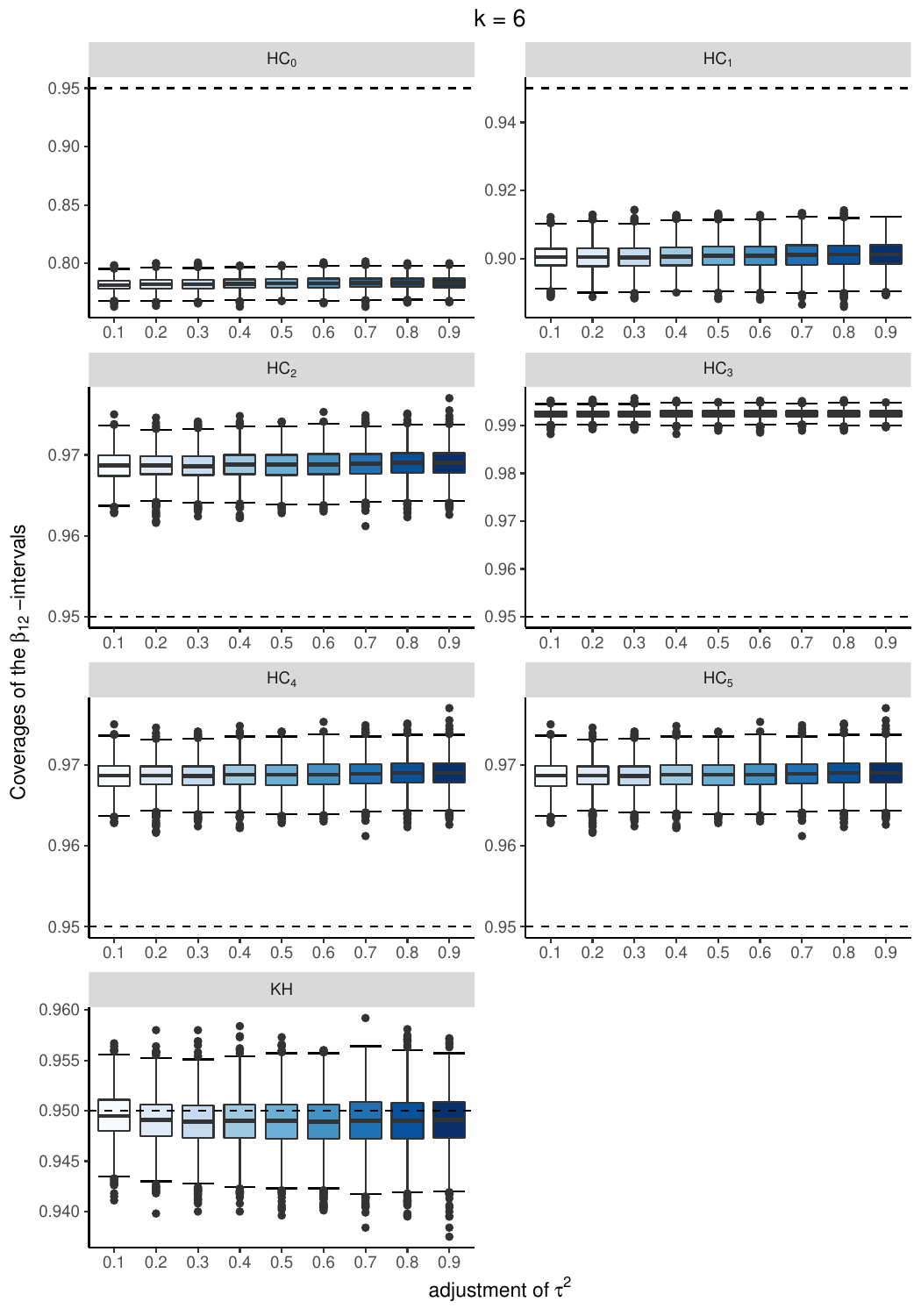}
\caption{Coverages of the \bi -intervals compared regarding the adjustments of the heterogeneity parameter $\tau^2$ for the estimators $HC_0-HC_5$ and $KH$ with $k=6$.}
\label{bi_tau_1}
\end{figure}

\begin{figure}
\includegraphics[scale=0.7, page=2]{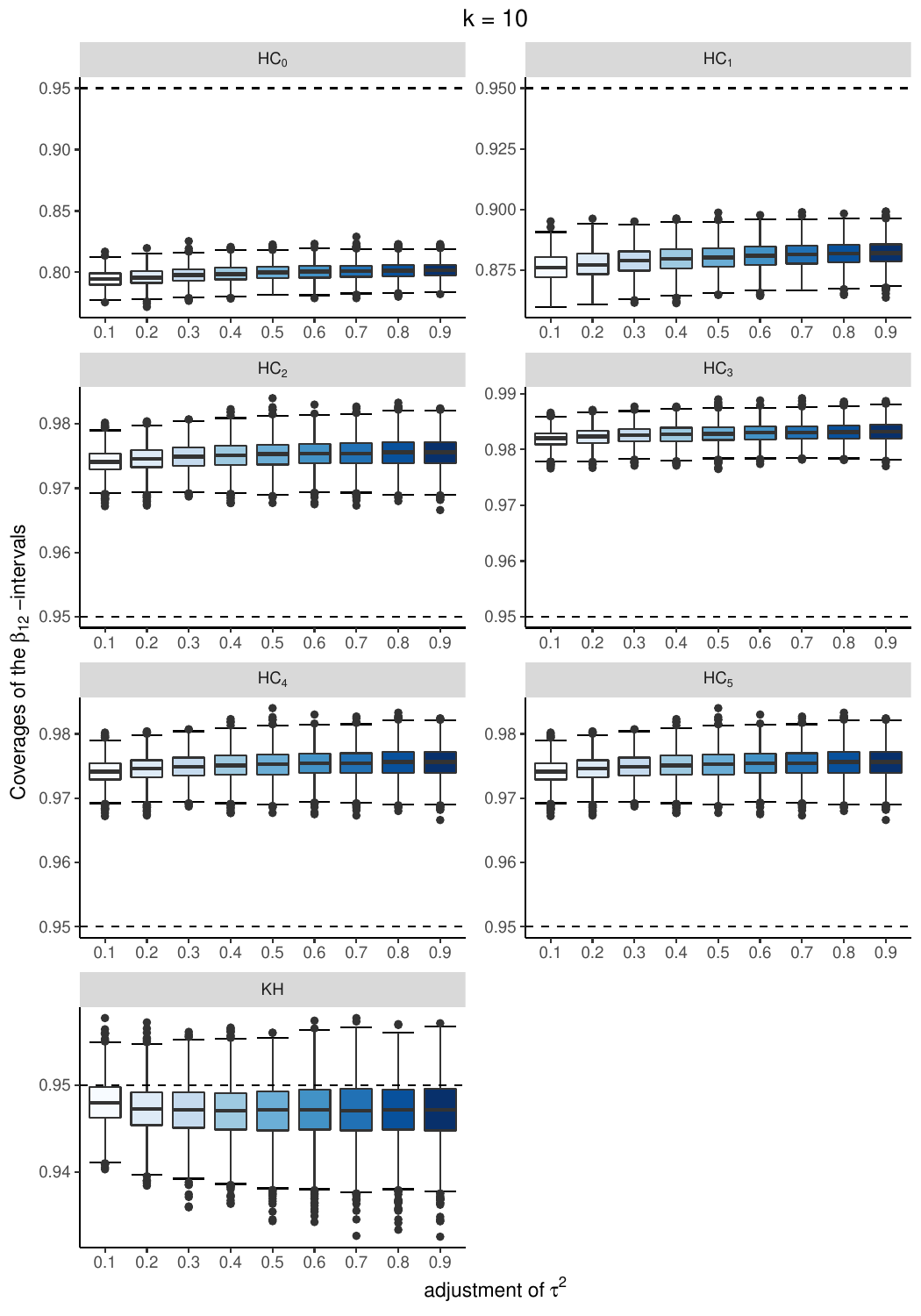}
\caption{Coverages of the \bi -intervals compared regarding the adjustments of the heterogeneity parameter $\tau^2$ for the estimators $HC_0-HC_5$ and $KH$ with $k=10$.}
\label{bi_tau_2}
\end{figure}

\begin{figure}
\includegraphics[scale=0.7, page=3]{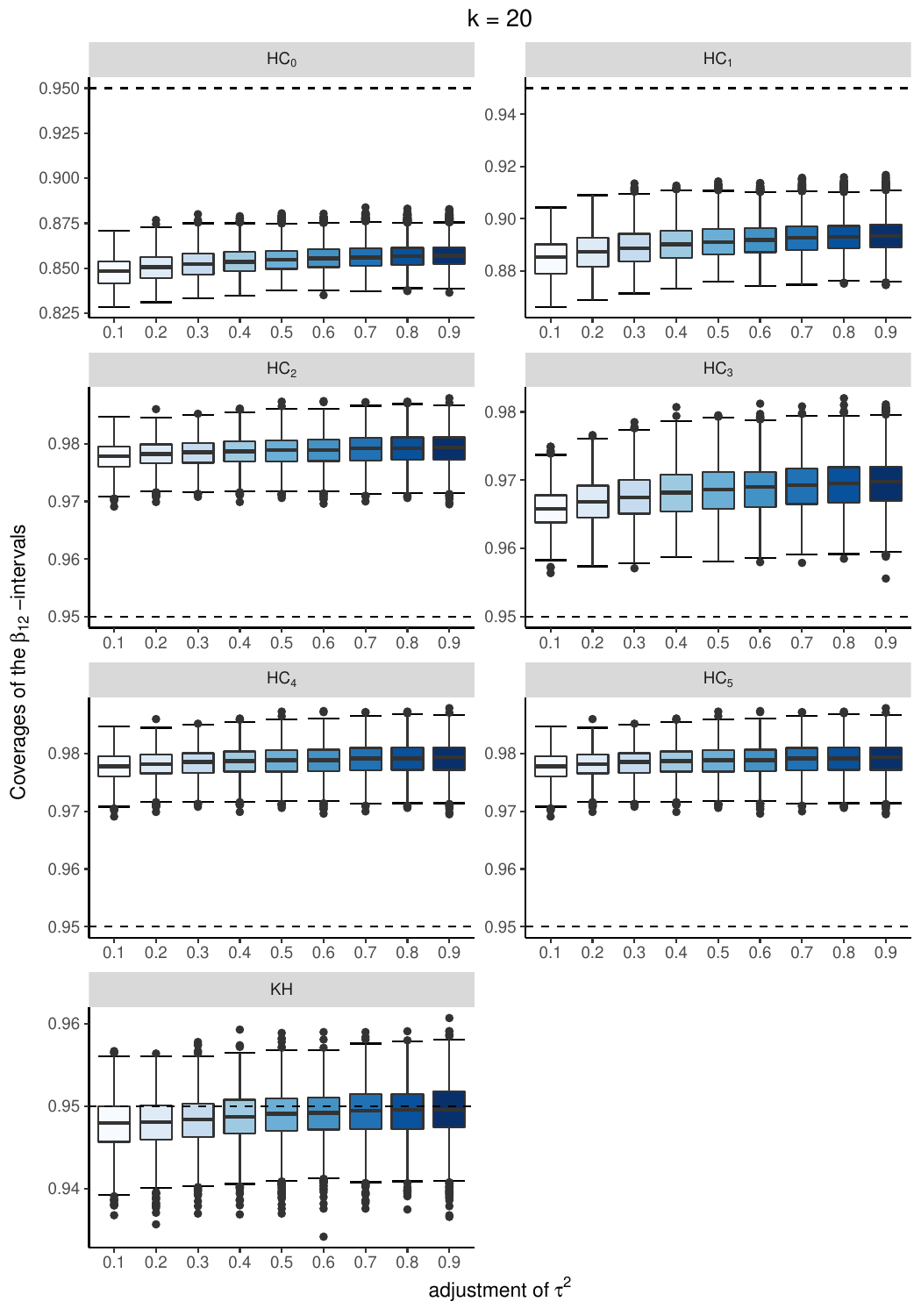}
\caption{Coverages of the \bi -intervals compared regarding the adjustments of the heterogeneity parameter $\tau^2$ for the estimators $HC_0-HC_5$ and $KH$ with $k=20$.}
\label{bi_tau_3}
\end{figure}

\begin{figure}
\includegraphics[scale=0.7, page=4]{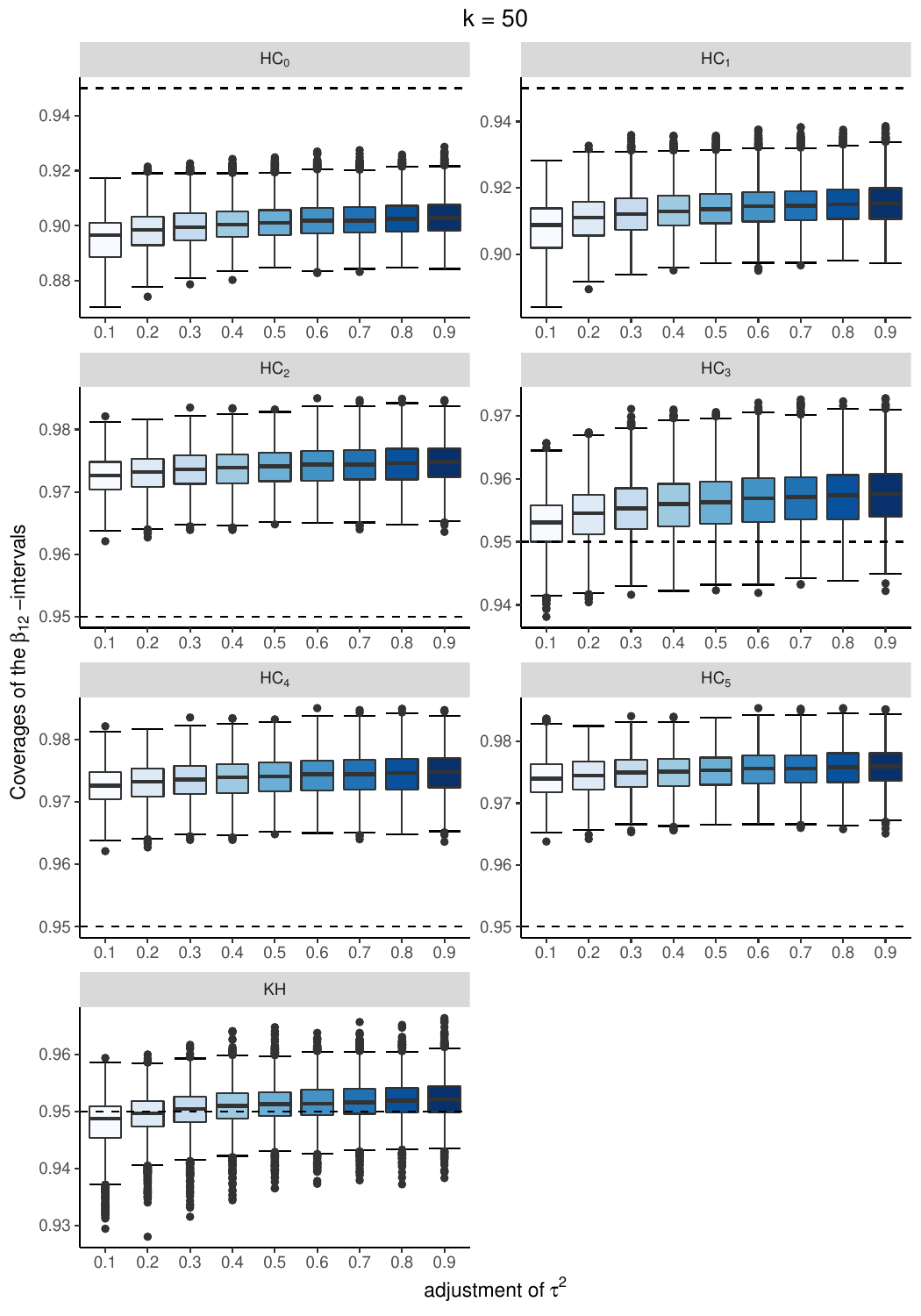}
\caption{Coverages of the \bi -intervals compared regarding the adjustments of the heterogeneity parameter $\tau^2$ for the estimators $HC_0-HC_5$ and $KH$ with $k=50$.}
\label{bi_tau_4}
\end{figure}

\begin{figure}
\includegraphics[scale=0.7, page=1]{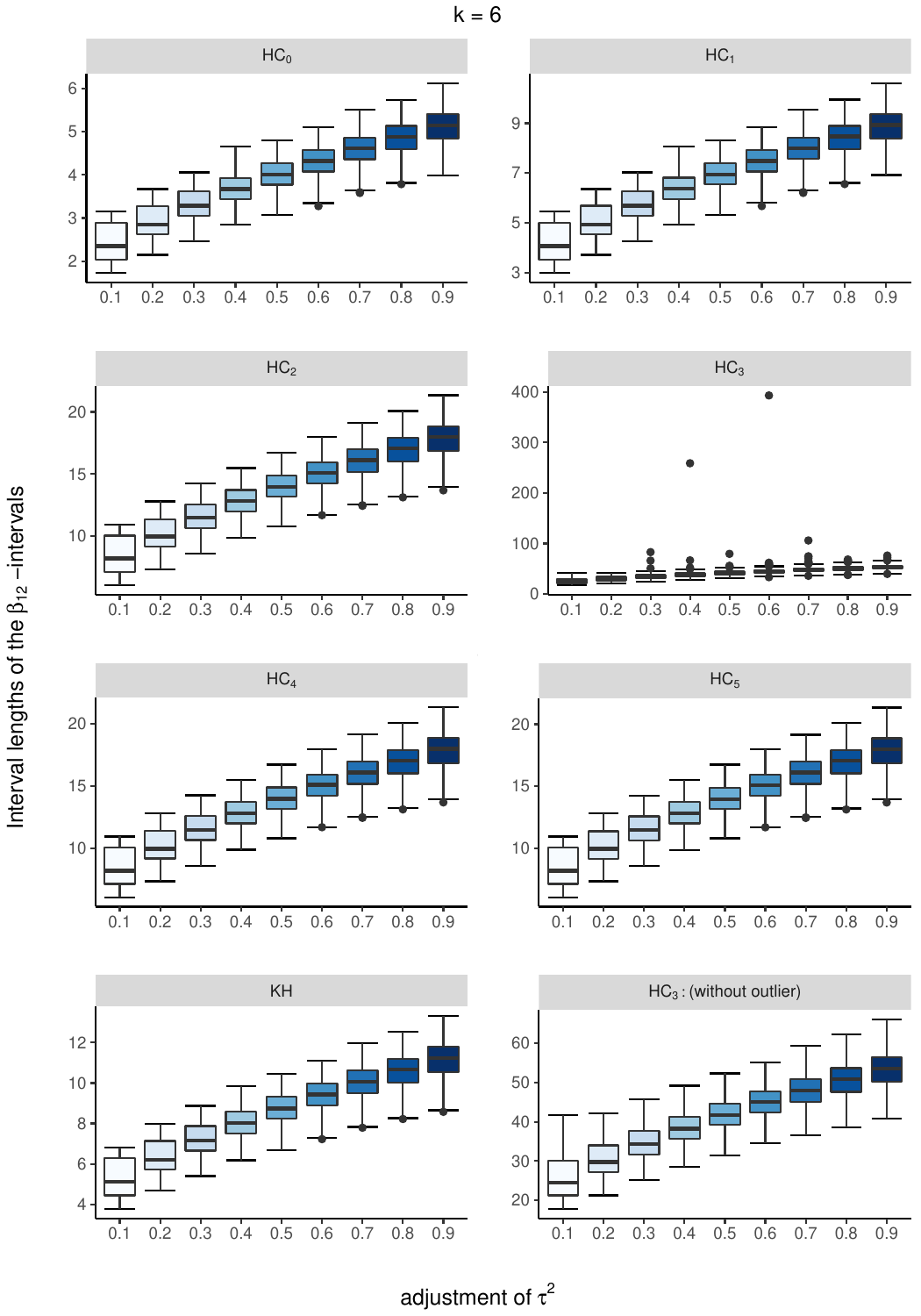}
\caption{Lengths of the \bi -intervals compared regarding the adjustments of the heterogeneity parameter $\tau^2$ for the estimators $HC_0-HC_5$ and $KH$ with $k=6$.}
\label{bi_tau_l_1}
\end{figure}

\begin{figure}
\includegraphics[scale=0.7, page=3]{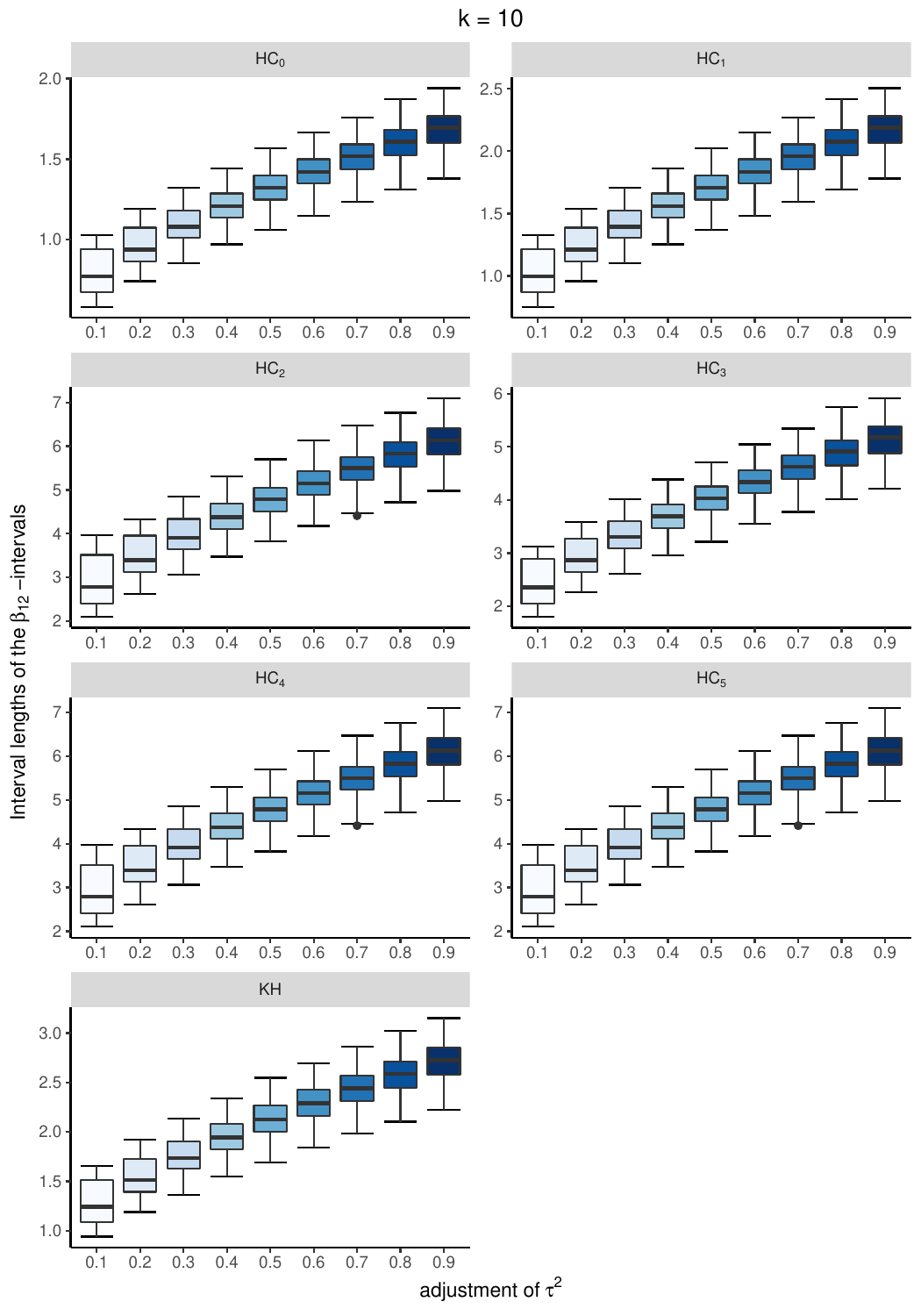}
\caption{Lengths of the \bi -intervals compared regarding the adjustments of the heterogeneity parameter $\tau^2$ for the estimators $HC_0-HC_5$ and $KH$ with $k=10$.}
\label{bi_tau_l_2}
\end{figure}

\begin{figure}
\includegraphics[scale=0.7, page=4]{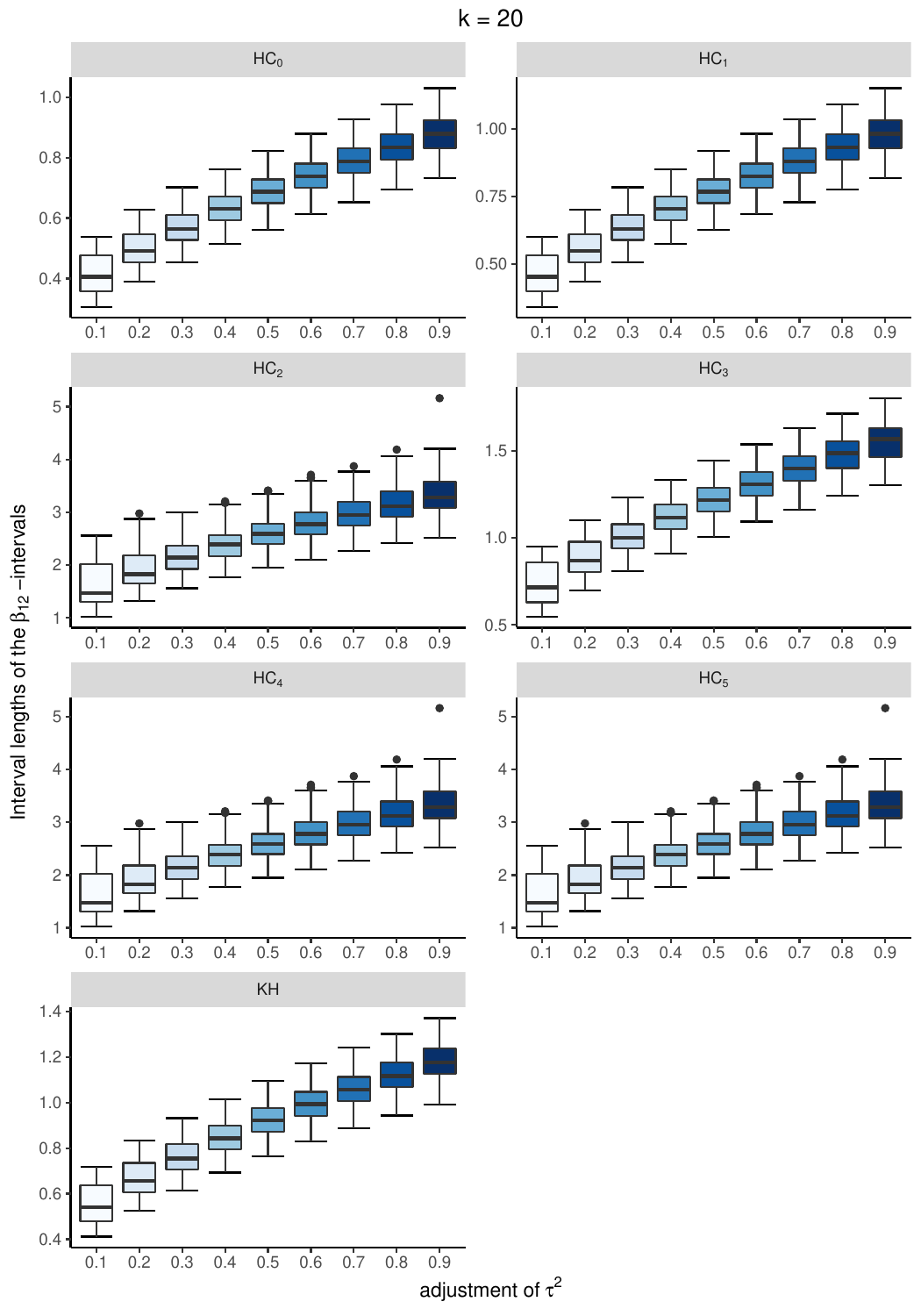}
\caption{Lengths of the \bi -intervals compared regarding the adjustments of the heterogeneity parameter $\tau^2$ for the estimators $HC_0-HC_5$ and $KH$ with $k=20$.}
\label{bi_tau_l_3}
\end{figure}

\begin{figure}
\includegraphics[scale=0.7, page=2]{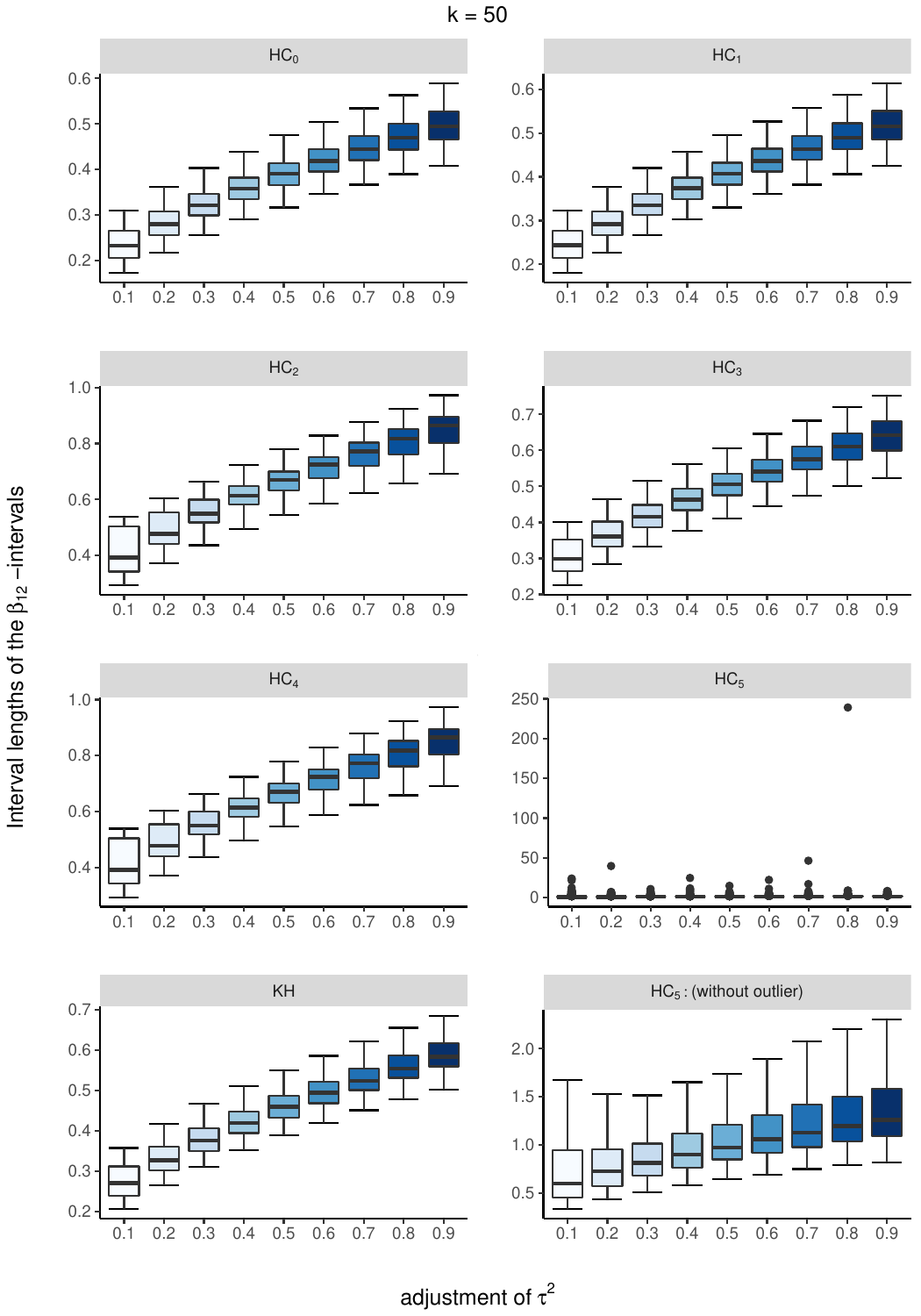}
\caption{Lengths of the \bi -intervals compared regarding the adjustments of the heterogeneity parameter $\tau^2$ for the estimators $HC_0-HC_5$ and $KH$ with $k=50$.}
\label{bi_tau_l_4}
\end{figure}

\textbf{Adjustments of $\boldsymbol{\tau^2}$}
Coverages of both coefficients' intervals are increasing slightly in the heterogeneity parameter $\tau^2$ for all estimators and $k$ (Figures \ref{b1_tau_1}-\ref{b1_tau_4} and \ref{bi_tau_1}-\ref{bi_tau_4}). The only exception are the $\textbf{KH}$-CIs for $k\in\{6,10\}$. For a larger number of studies, the effect is stronger. The increasing coverages in $\tau^2$ show that the model used in the simulation is adequate to model a study effect.\\ On the other hand the interval lengths are increasing in $\tau^2$ strongly. Intervals for the adjustment $\tau^2=0.9$ have more than double median length compared to the intervals with $\tau^2=0.1$ (Figures \ref{b1_tau_l_1}-\ref{b1_tau_l_4} and \ref{bi_tau_l_1}-\ref{bi_tau_l_4}). This result is explicable by the direct impact the value of $\tau^2$ has on the variances of the coefficients and thus on the interval bounds. \\ The recommendation for the choice of estimator does not differ from the overall recommendation for any observed $\tau^2$.\\

\begin{figure}
\includegraphics[scale=0.7, page=1]{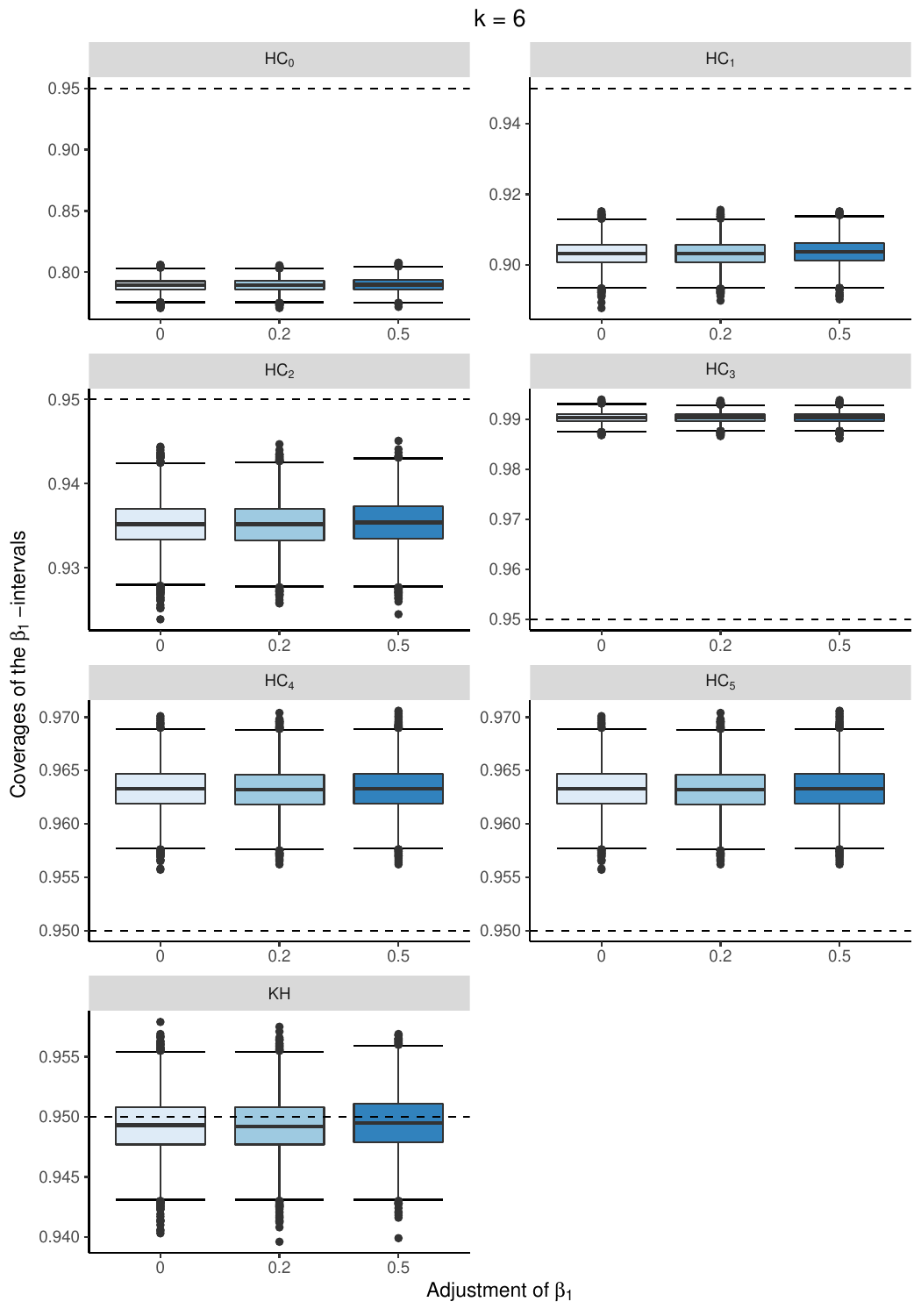}
\caption{Coverages of the \bo -intervals compared regarding the adjustments of $\beta_1$ for the estimators $HC_0-HC_5$ and $KH$ with $k=6$.}
\label{b1_b1_1}
\end{figure}

\begin{figure}
\includegraphics[scale=0.7, page=2]{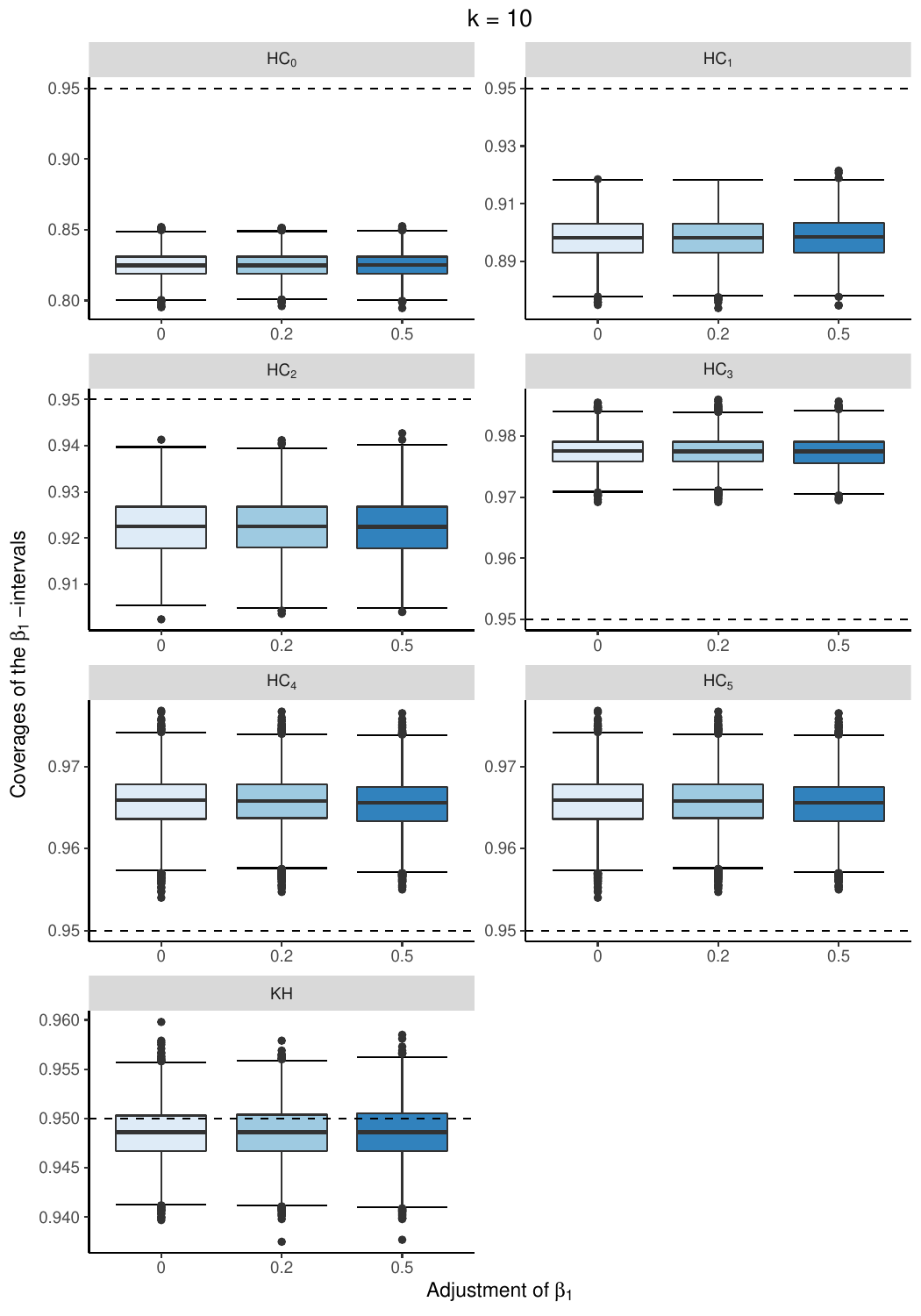}
\caption{Coverages of the \bo -intervals compared regarding the adjustments of $\beta_1$ for the estimators $HC_0-HC_5$ and $KH$ with $k=10$.}
\label{b1_b1_2}
\end{figure}

\begin{figure}
\includegraphics[scale=0.7, page=3]{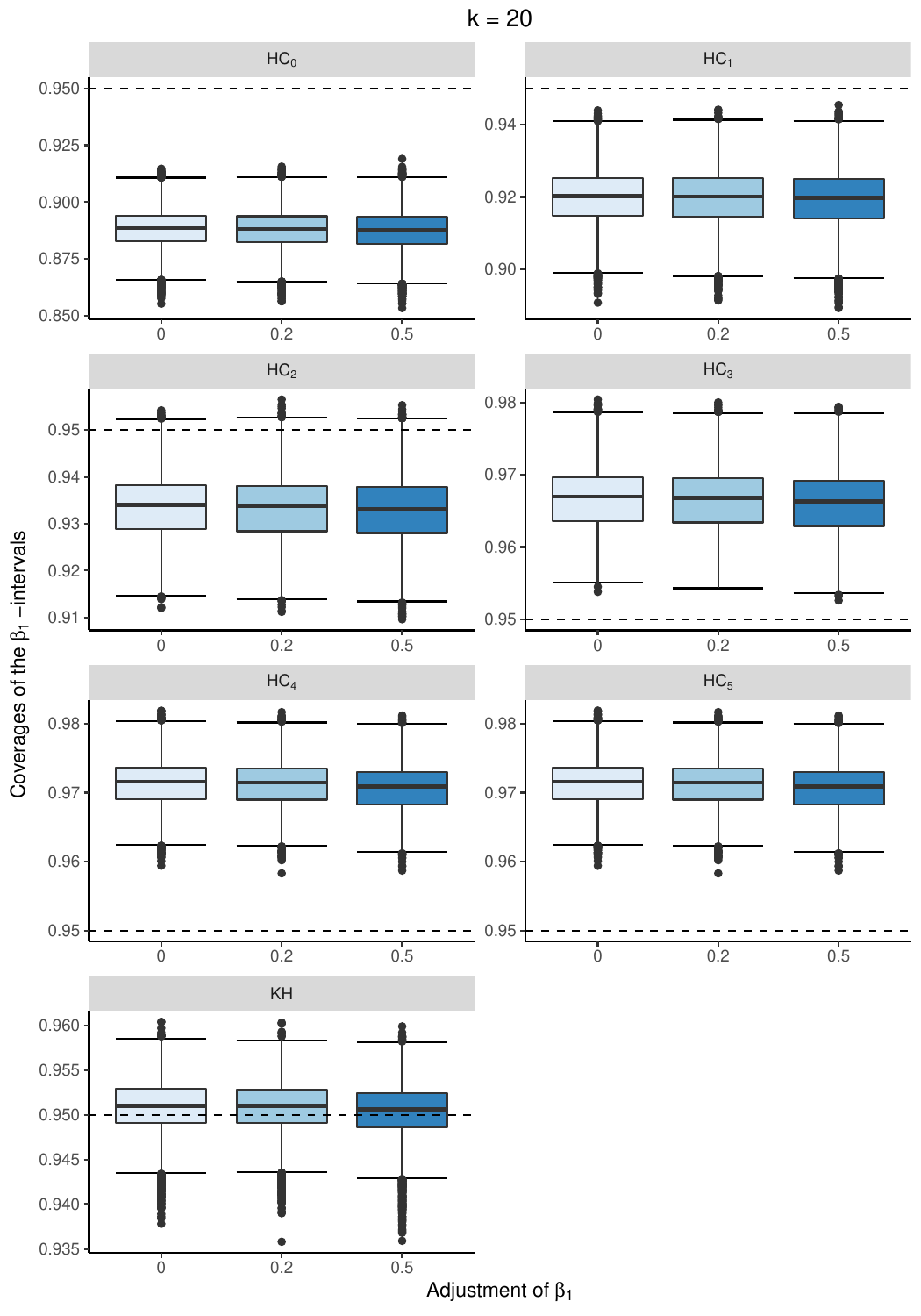}
\caption{Coverages of the \bo -intervals compared regarding the adjustments of $\beta_1$ for the estimators $HC_0-HC_5$ and $KH$ with $k=20$.}
\label{b1_b1_3}
\end{figure}

\begin{figure}
\includegraphics[scale=0.7, page=4]{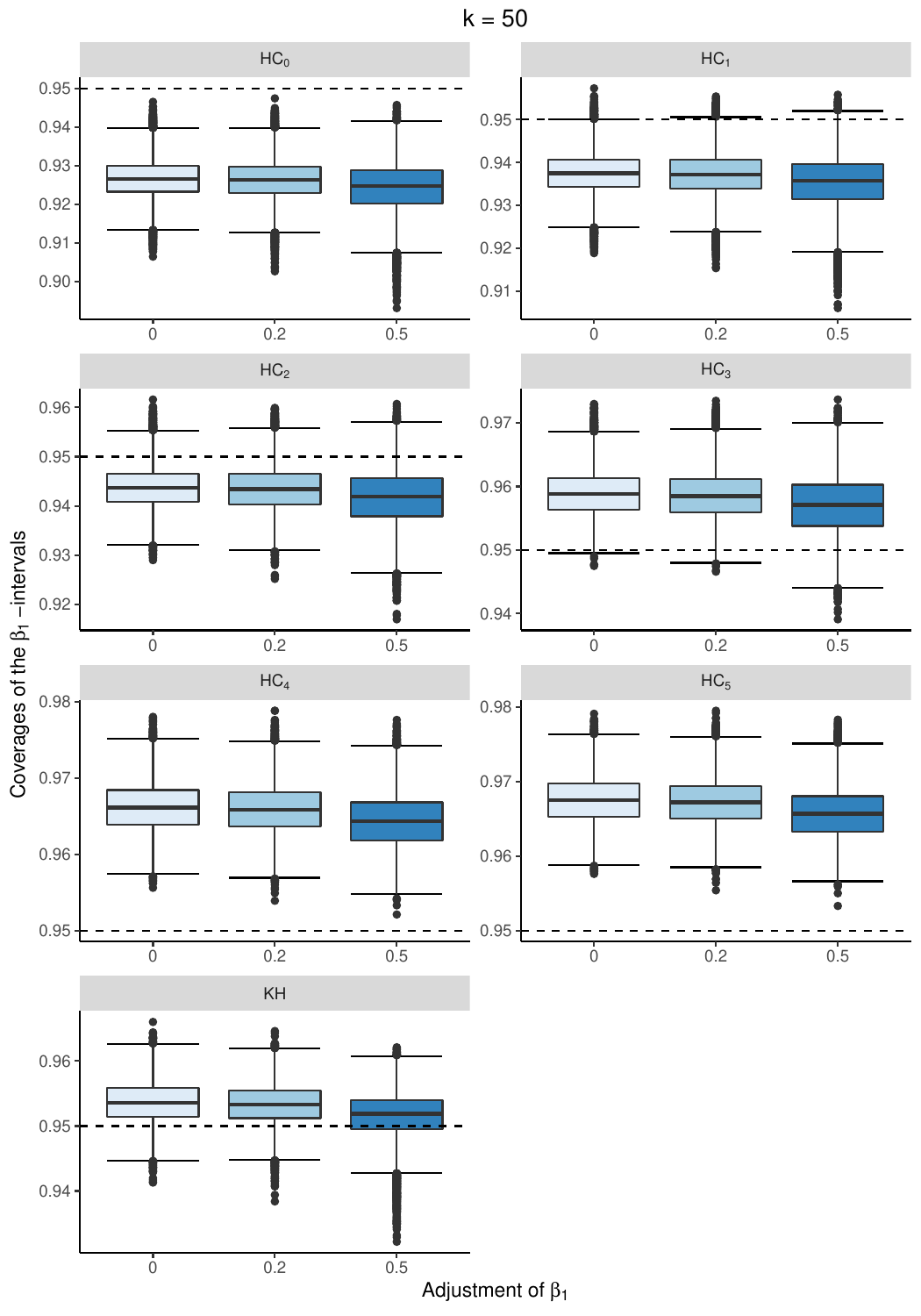}
\caption{Coverages of the \bo -intervals compared regarding the adjustments of $\beta_1$ for the estimators $HC_0-HC_5$ and $KH$ with $k=50$.}
\label{b1_b1_4}
\end{figure}

\begin{figure}
\includegraphics[scale=0.7, page=1]{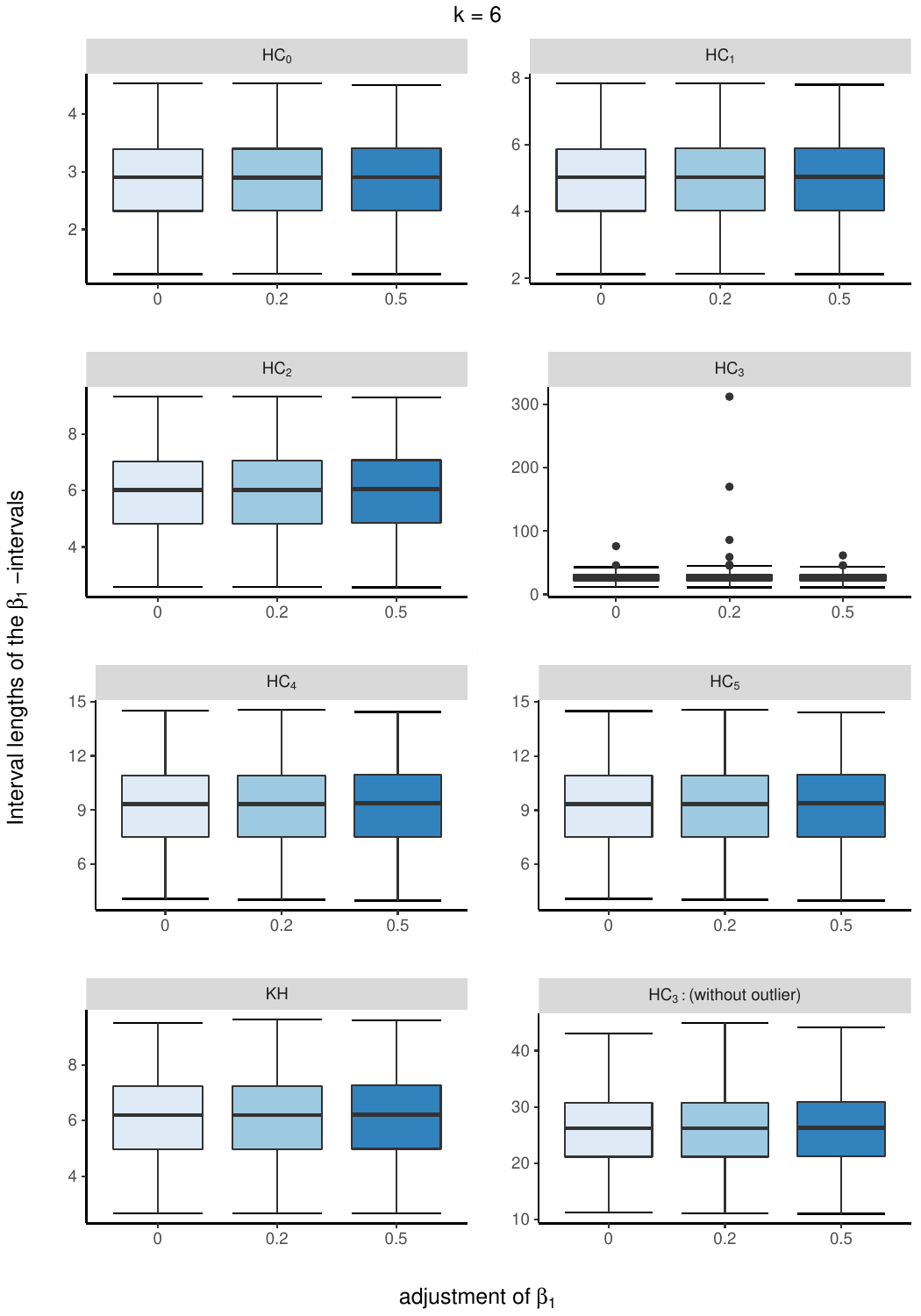}
\caption{Lengths of the \bo -intervals compared regarding the adjustments of $\beta_1$ for the estimators $HC_0-HC_5$ and $KH$ with $k=6$.}
\label{b1_b1_l_1}
\end{figure}

\begin{figure}
\includegraphics[scale=0.7, page=3]{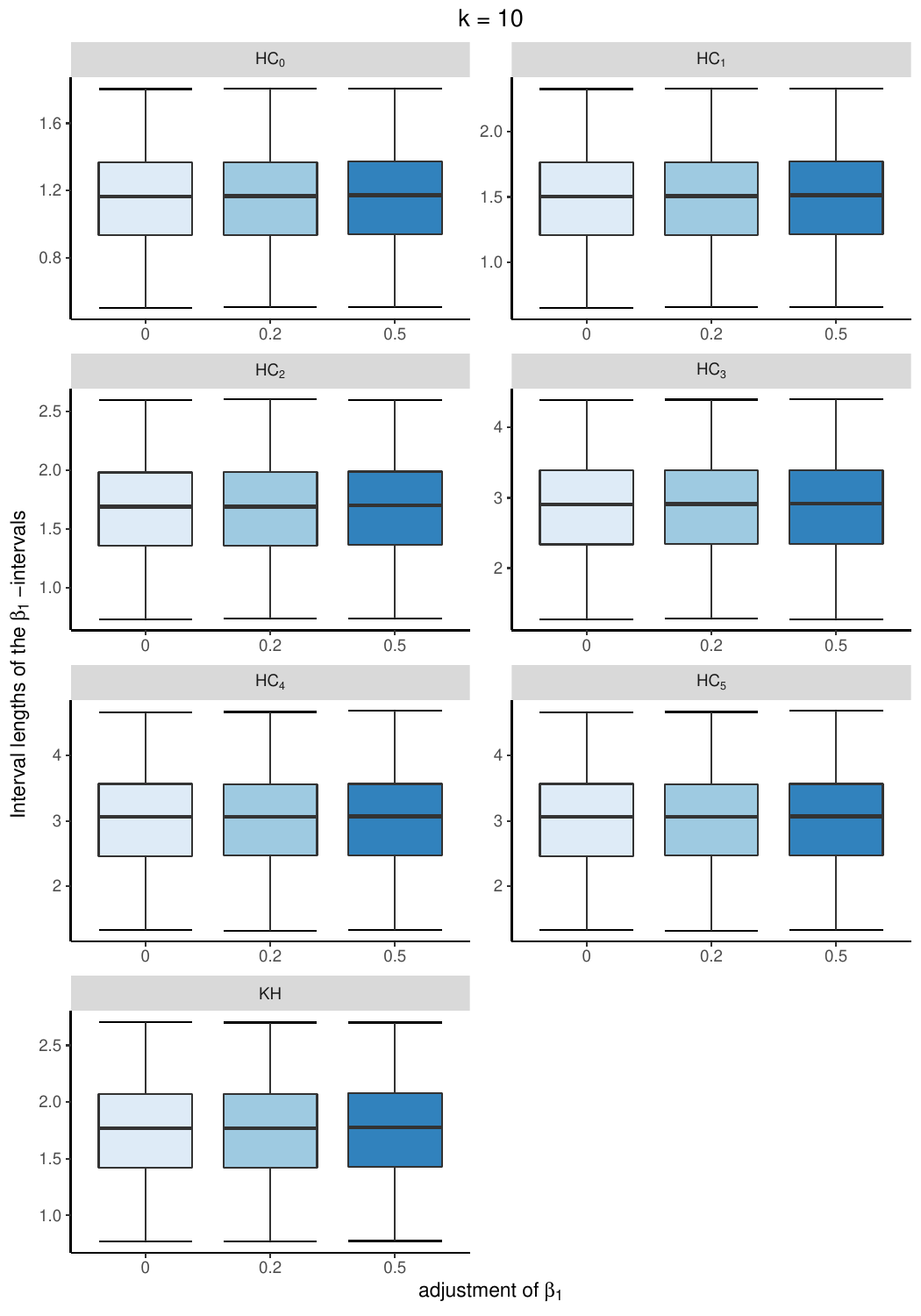}
\caption{Lengths of the \bo -intervals compared regarding the adjustments of $\beta_1$ for the estimators $HC_0-HC_5$ and $KH$ with $k=10$.}
\label{b1_b1_l_2}
\end{figure}

\begin{figure}
\includegraphics[scale=0.7, page=4]{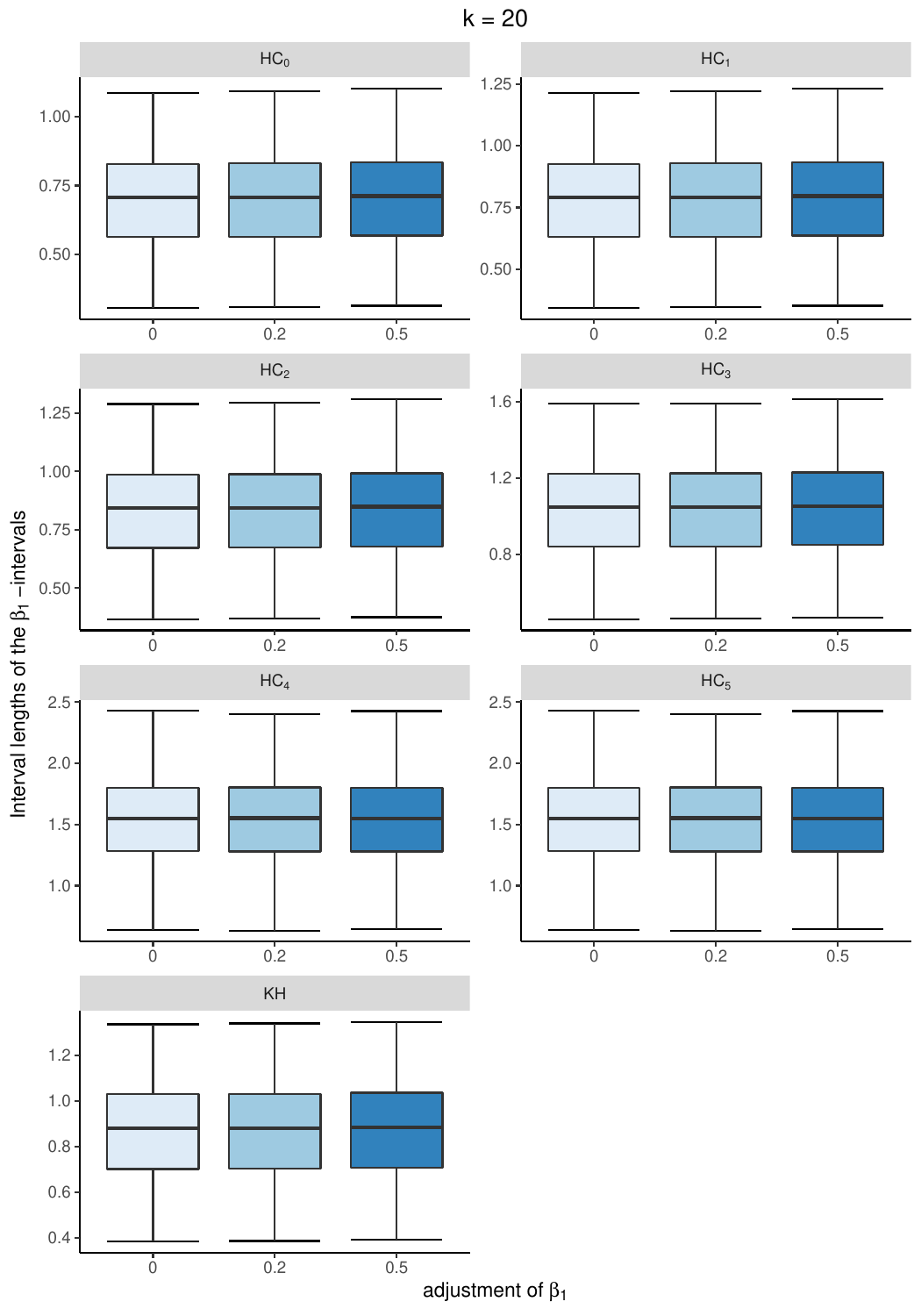}
\caption{Lengths of the \bo -intervals compared regarding the adjustments of $\beta_1$ for the estimators $HC_0-HC_5$ and $KH$ with $k=20$.}
\label{b1_b1_l_3}
\end{figure}

\begin{figure}
\includegraphics[scale=0.7, page=2]{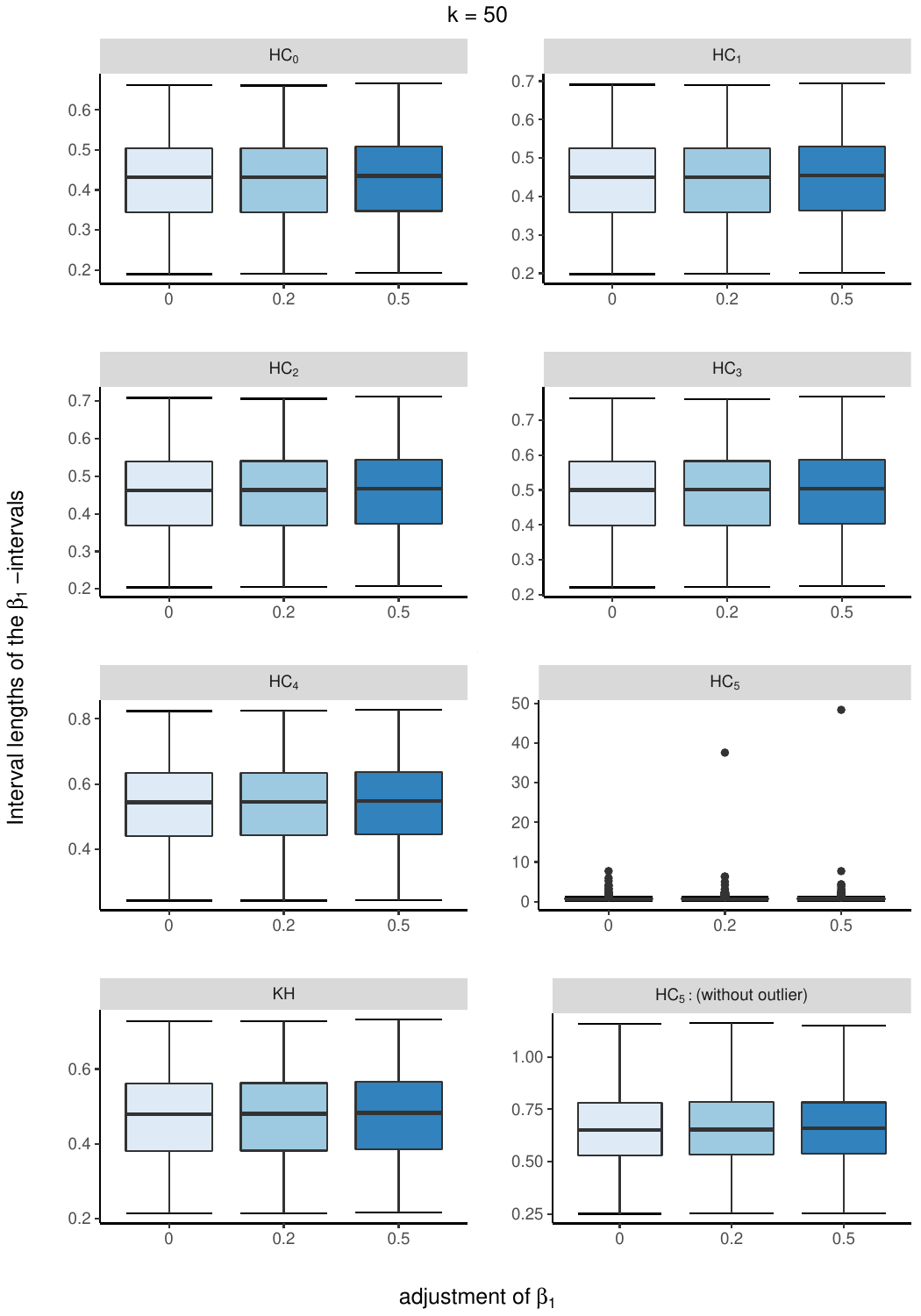}
\caption{Lengths of the \bo -intervals compared regarding the adjustments of $\beta_1$ for the estimators $HC_0-HC_5$ and $KH$ with $k=50$.}
\label{b1_b1_l_4}
\end{figure}

\begin{figure}
\includegraphics[scale=0.7, page=1]{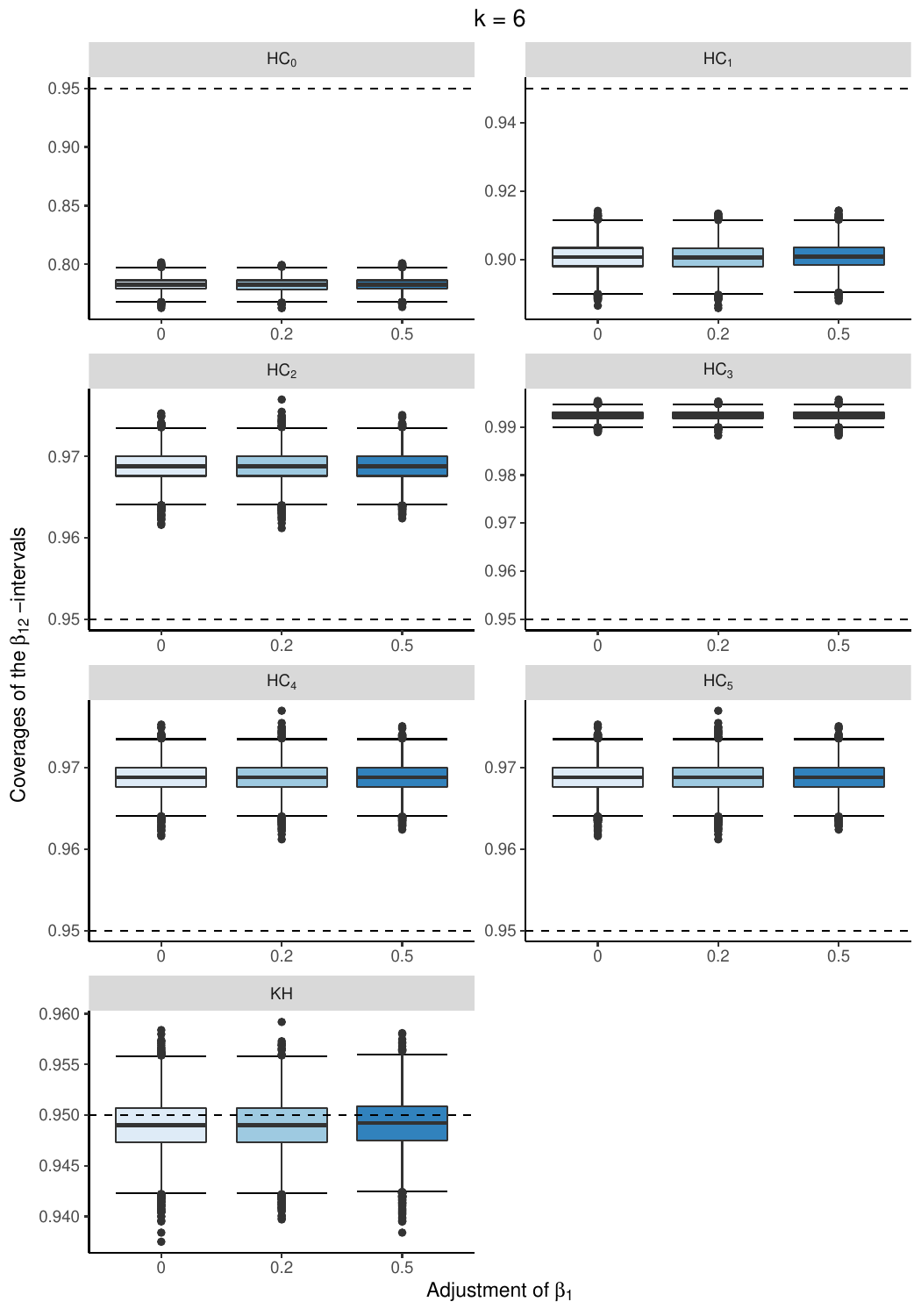}
\caption{Coverages of the \bi -intervals compared regarding the adjustments of $\beta_1$ for the estimators $HC_0-HC_5$ and $KH$ with $k=6$.}
\label{bi_b1_1}
\end{figure}

\begin{figure}
\includegraphics[scale=0.7, page=2]{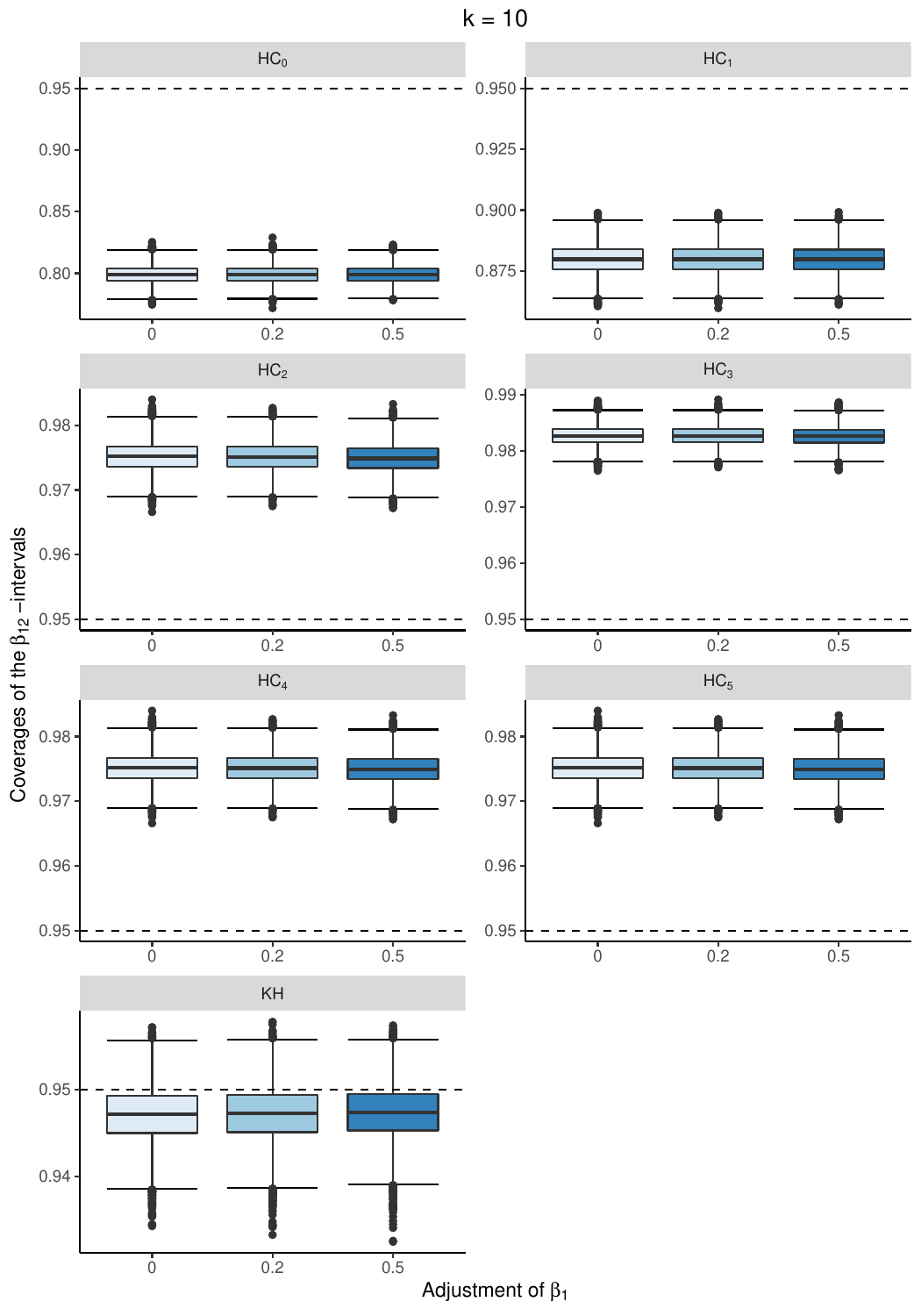}
\caption{Coverages of the \bi -intervals compared regarding the adjustments of $\beta_1$ for the estimators $HC_0-HC_5$ and $KH$ with $k=10$.}
\label{bi_b1_2}
\end{figure}

\begin{figure}
\includegraphics[scale=0.7, page=3]{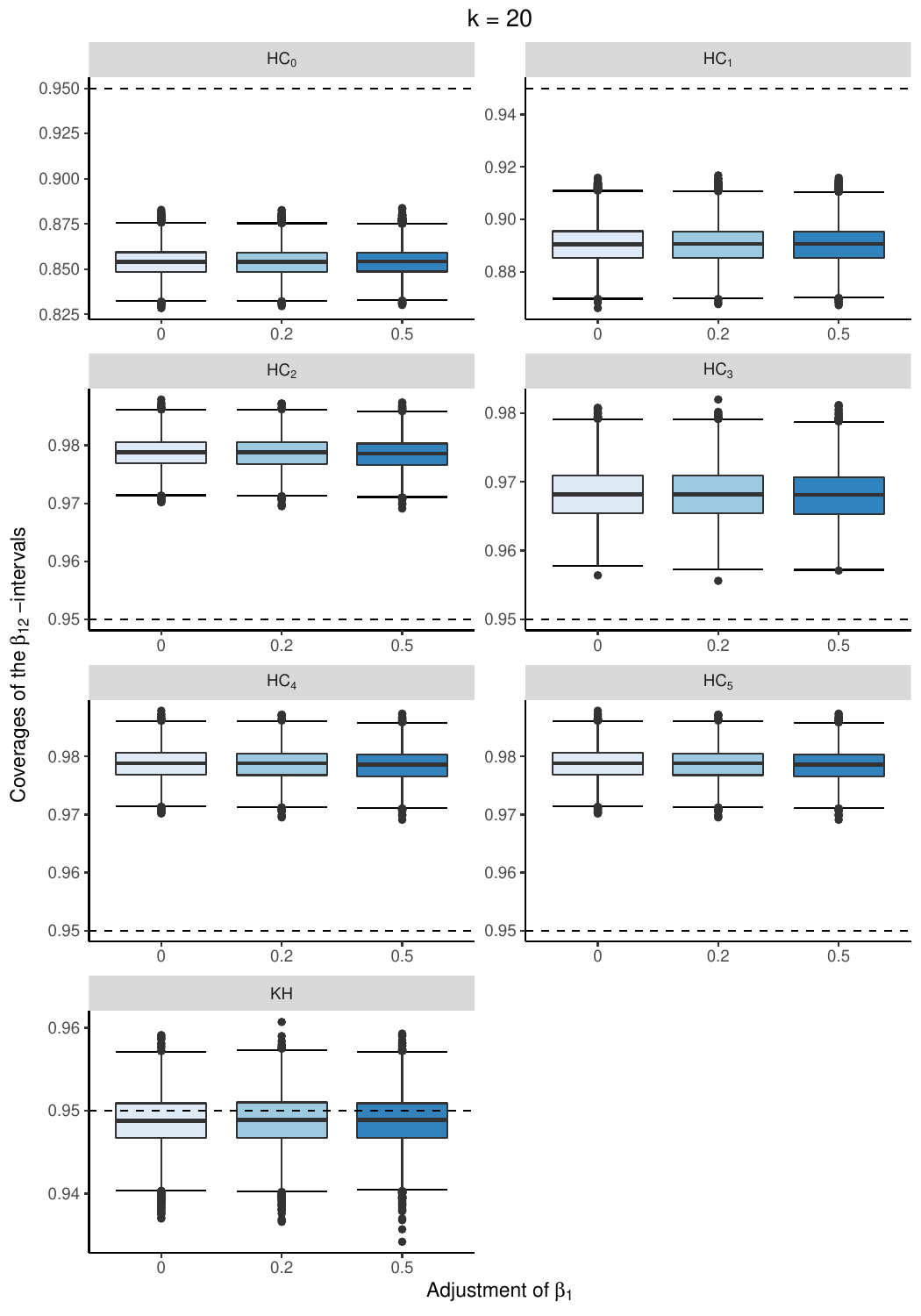}
\caption{Coverages of the \bi -intervals compared regarding the adjustments of $\beta_1$ for the estimators $HC_0-HC_5$ and $KH$ with $k=20$.}
\label{bi_b1_3}
\end{figure}

\begin{figure}
\includegraphics[scale=0.7, page=4]{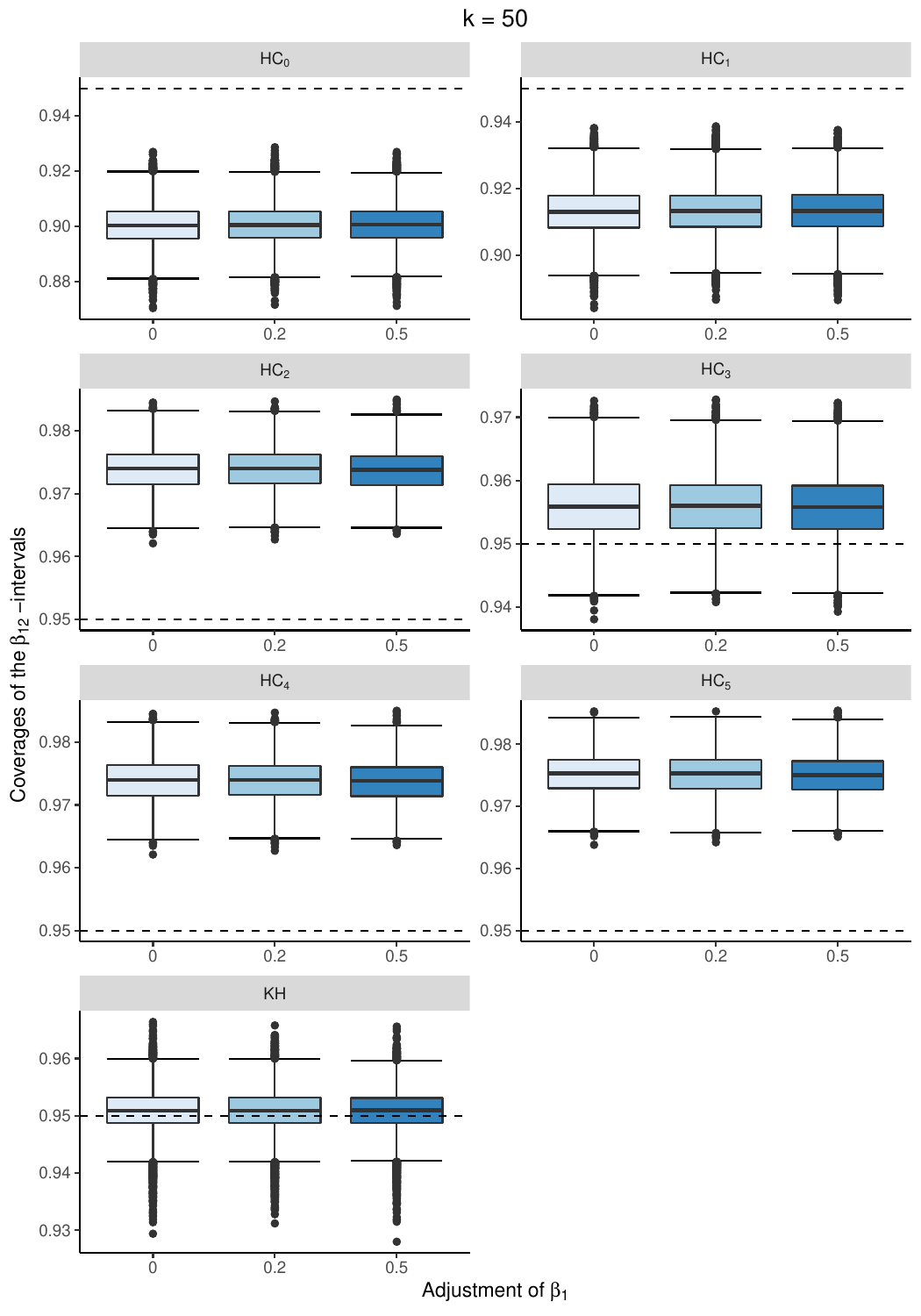}
\caption{Coverages of the \bi -intervals compared regarding the adjustments of $\beta_1$ for the estimators $HC_0-HC_5$ and $KH$ with $k=50$.}
\label{bi_b1_4}
\end{figure}

\begin{figure}
\includegraphics[scale=0.7, page=1]{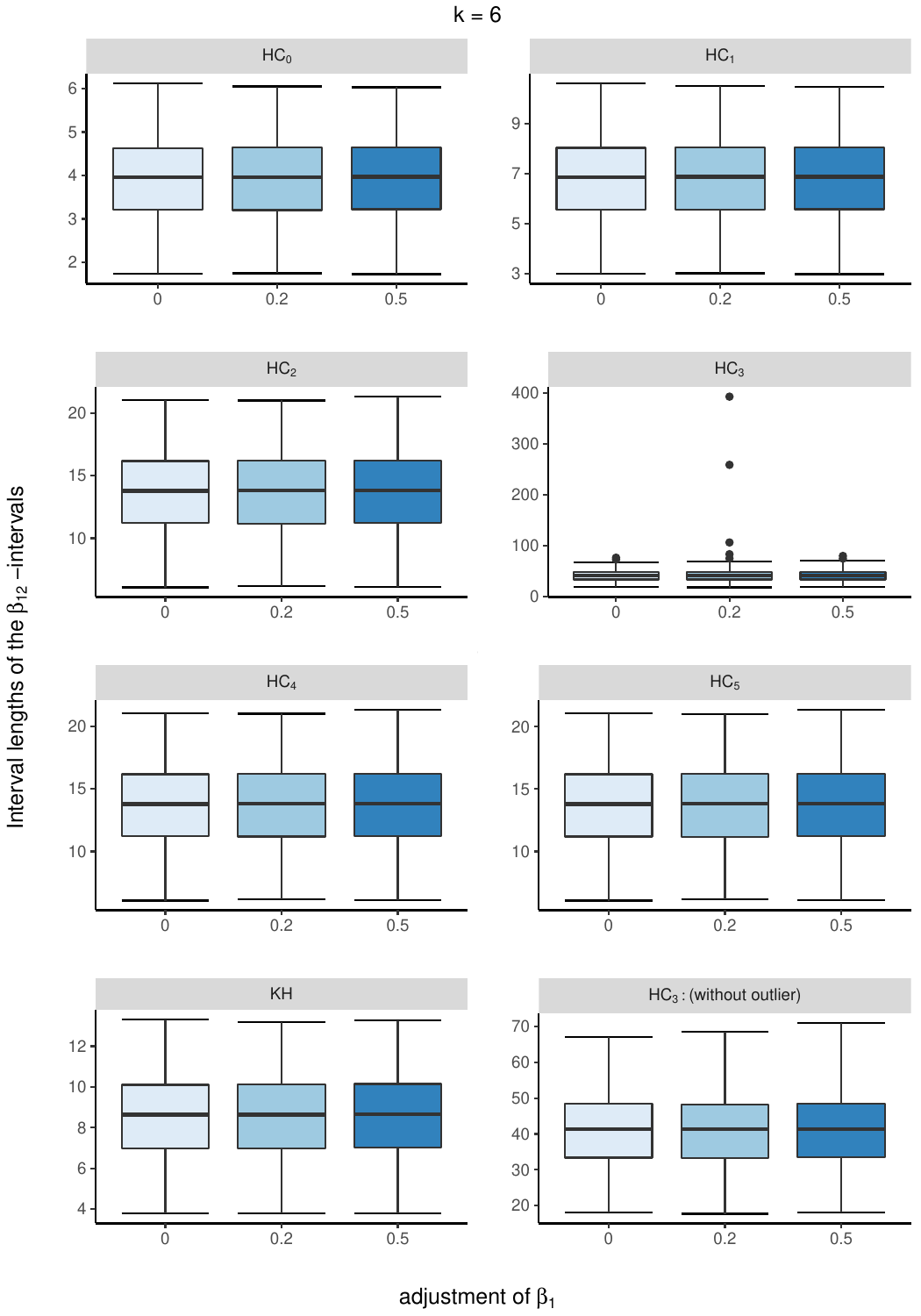}
\caption{Lengths of the \bi -intervals compared regarding the adjustments of $\beta_1$ for the estimators $HC_0-HC_5$ and $KH$ with $k=6$.}
\label{bi_b1_l_1}
\end{figure}

\begin{figure}
\includegraphics[scale=0.7, page=3]{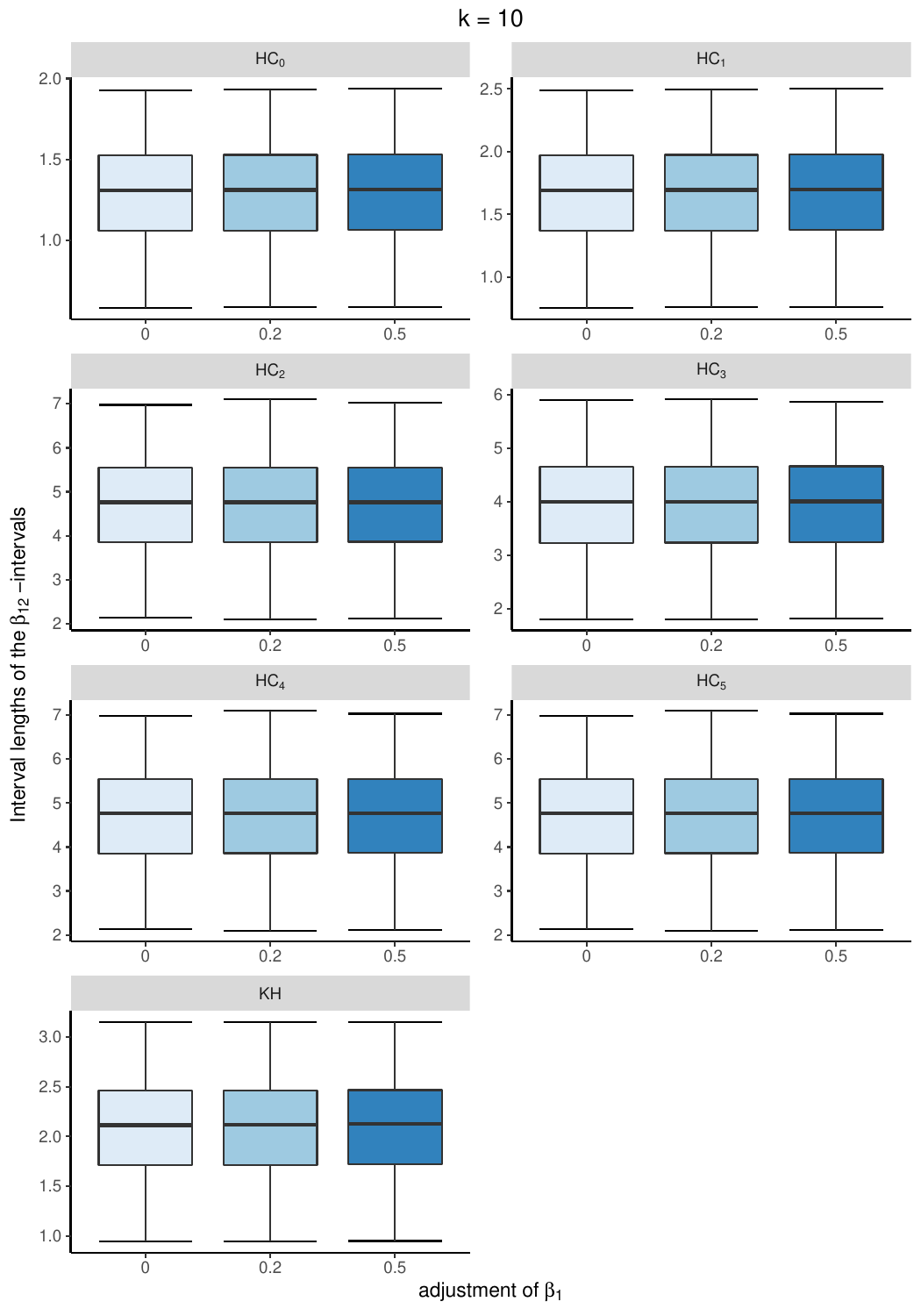}
\caption{Lengths of the \bi -intervals compared regarding the adjustments of $\beta_1$ for the estimators $HC_0-HC_5$ and $KH$ with $k=10$.}
\label{bi_b1_l_2}
\end{figure}

\begin{figure}
\includegraphics[scale=0.7, page=4]{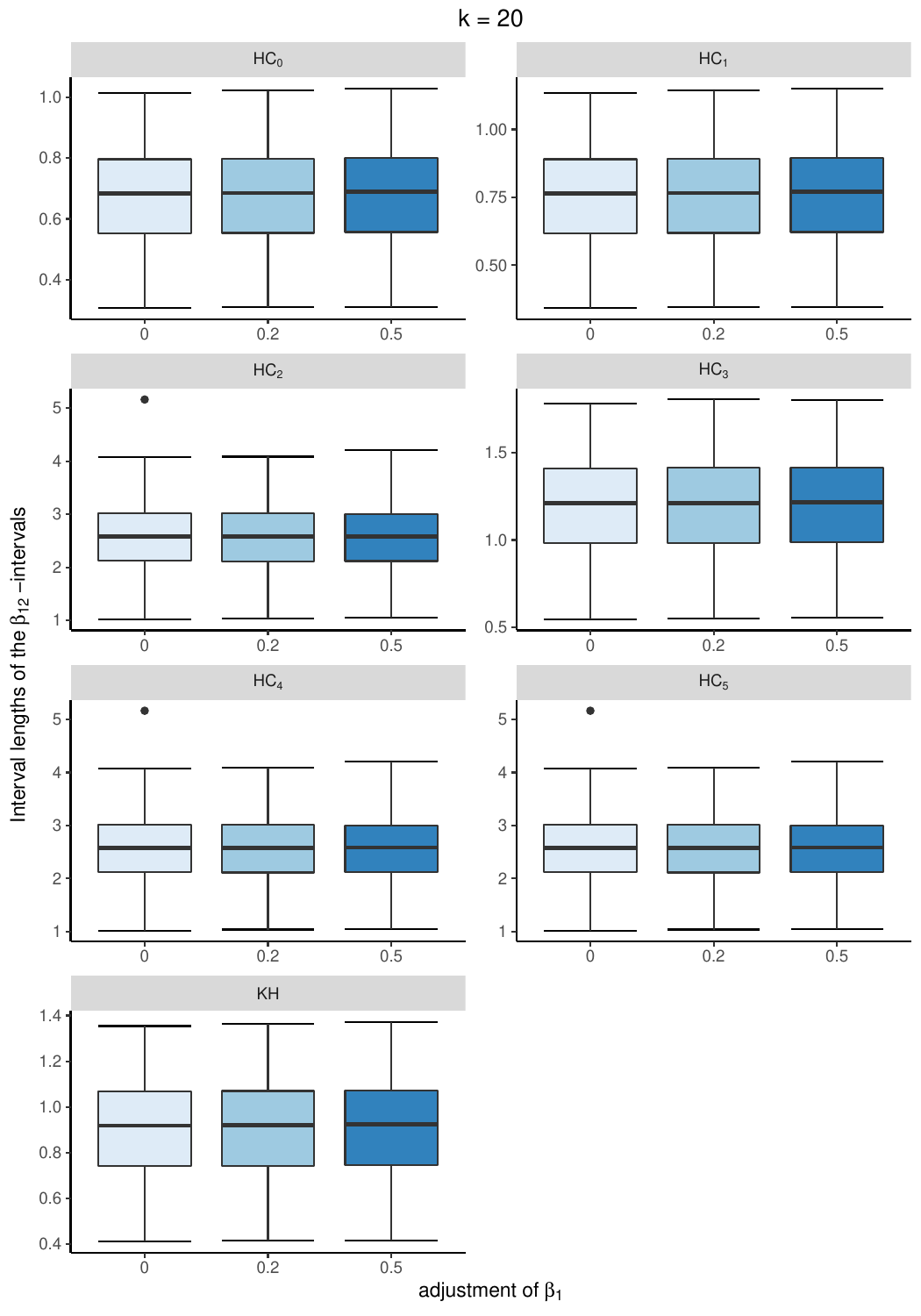}
\caption{Lengths of the \bi -intervals compared regarding the adjustments of $\beta_1$ for the estimators $HC_0-HC_5$ and $KH$ with $k=20$.}
\label{bi_b1_l_3}
\end{figure}

\begin{figure}
\includegraphics[scale=0.7, page=2]{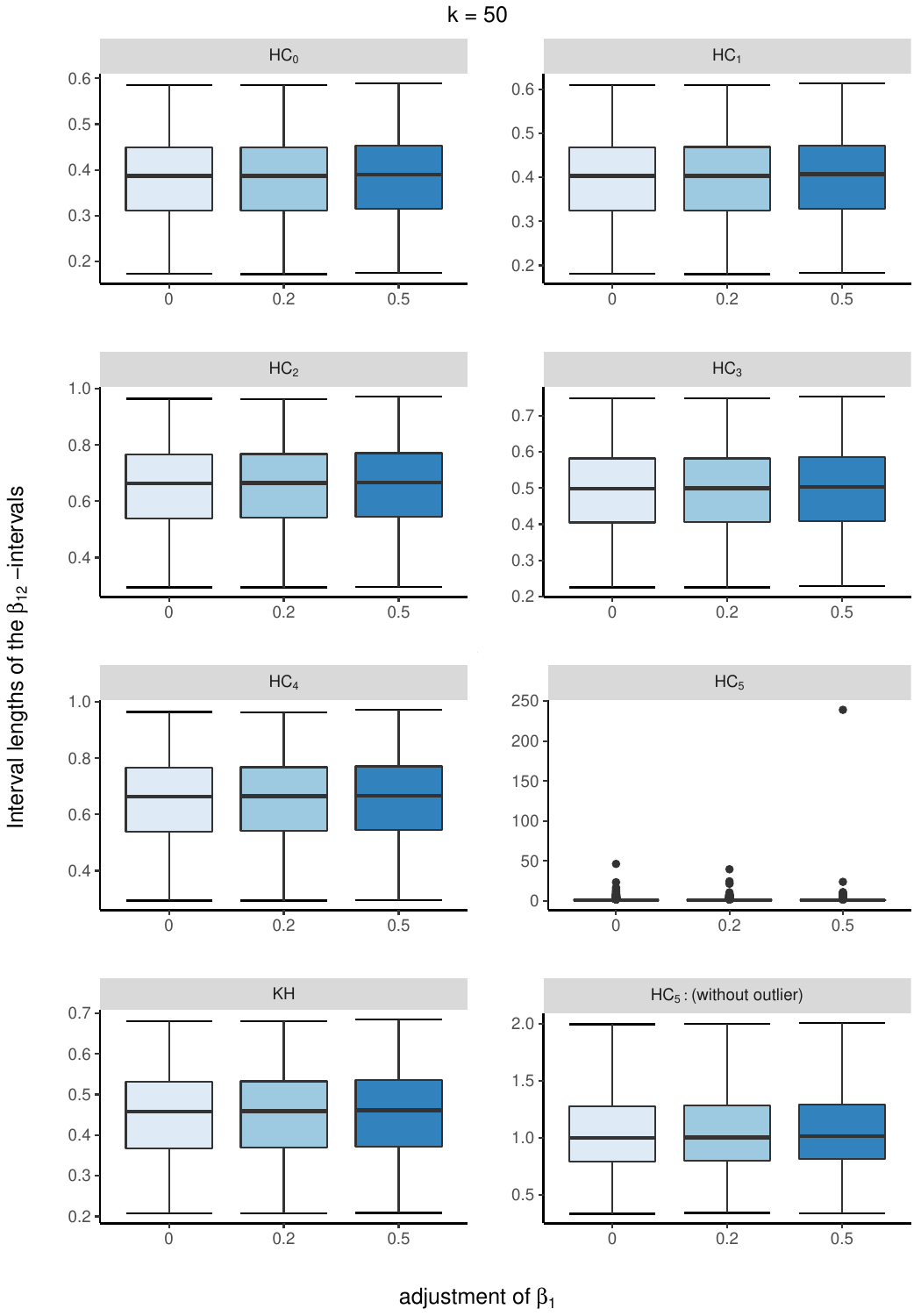}
\caption{Lengths of the \bi -intervals compared regarding the adjustments of $\beta_1$ for the estimators $HC_0-HC_5$ and $KH$ with $k=50$.}
\label{bi_b1_l_4}
\end{figure}

\textbf{Adjustments of $\boldsymbol{\beta_1}$} Examined adjustments of $\beta_1$ are 0, 0.2 and 0.5. The coverages of CIs for $\beta_{12}$ are not affected by the adjustment of $\beta_1$ (Figures \ref{bi_b1_1}-\ref{bi_b1_4}), whereas the CIs for $\beta_1$ have a slightly lower coverage for a number of studies $k\in\{20,50\}$ and all estimators (Figures \ref{b1_b1_1}-\ref{b1_b1_4}). Neither the interval lengths of the CIs for $\beta_1$ nor of the CIs for $\beta_{12}$ differ regarding the adjustment of $\beta_1$ (Figures \ref{b1_b1_l_1}-\ref{b1_b1_l_4} and \ref{bi_b1_l_1}-\ref{bi_b1_l_4}).

\begin{figure}
\includegraphics[scale=0.7, page=1]{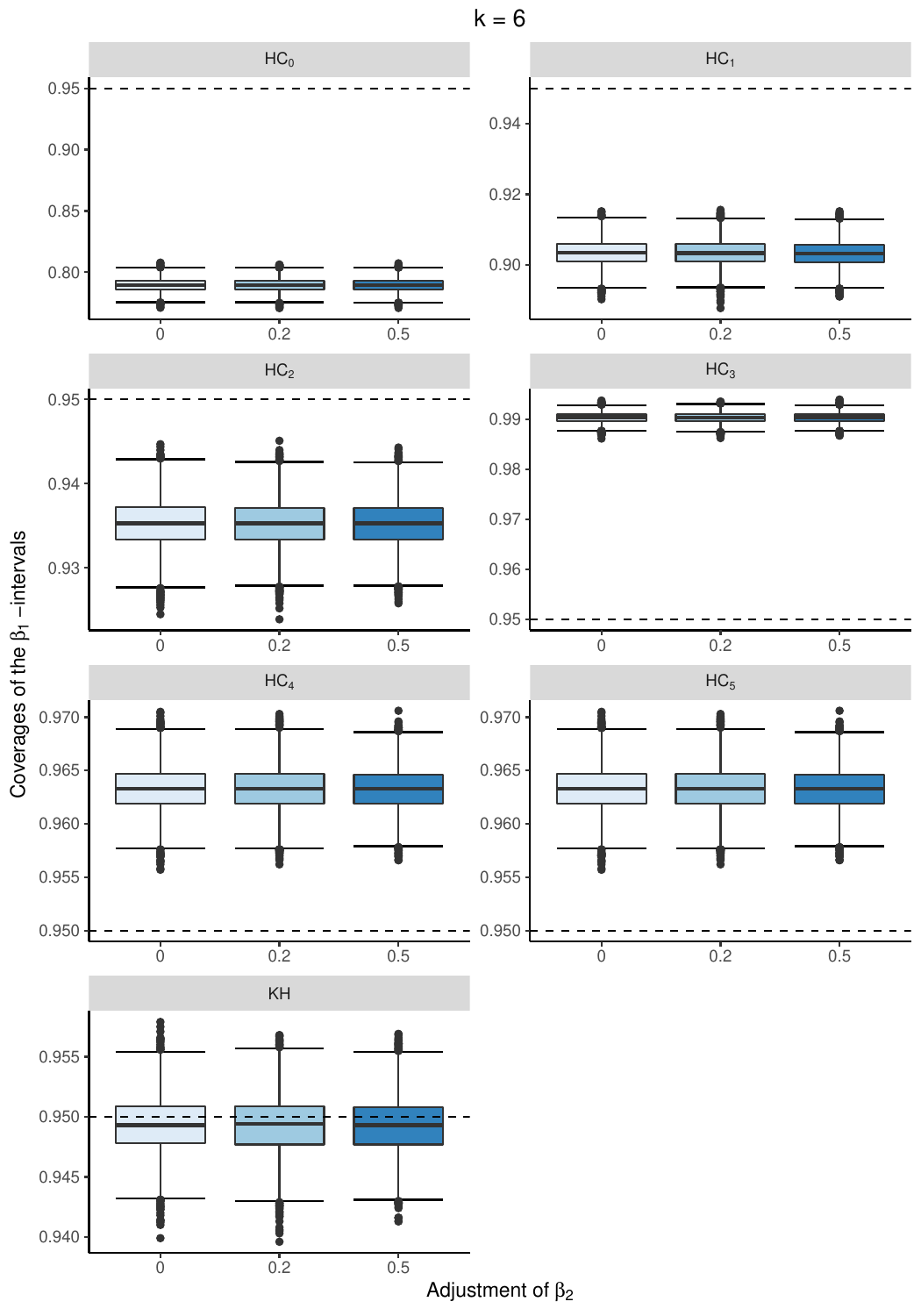}
\caption{Coverages of the \bo -intervals compared regarding the adjustments of $\beta_2$ for the estimators $HC_0-HC_5$ and $KH$ with $k=6$.}
\label{b1_b2_1}
\end{figure}

\begin{figure}
\includegraphics[scale=0.7, page=2]{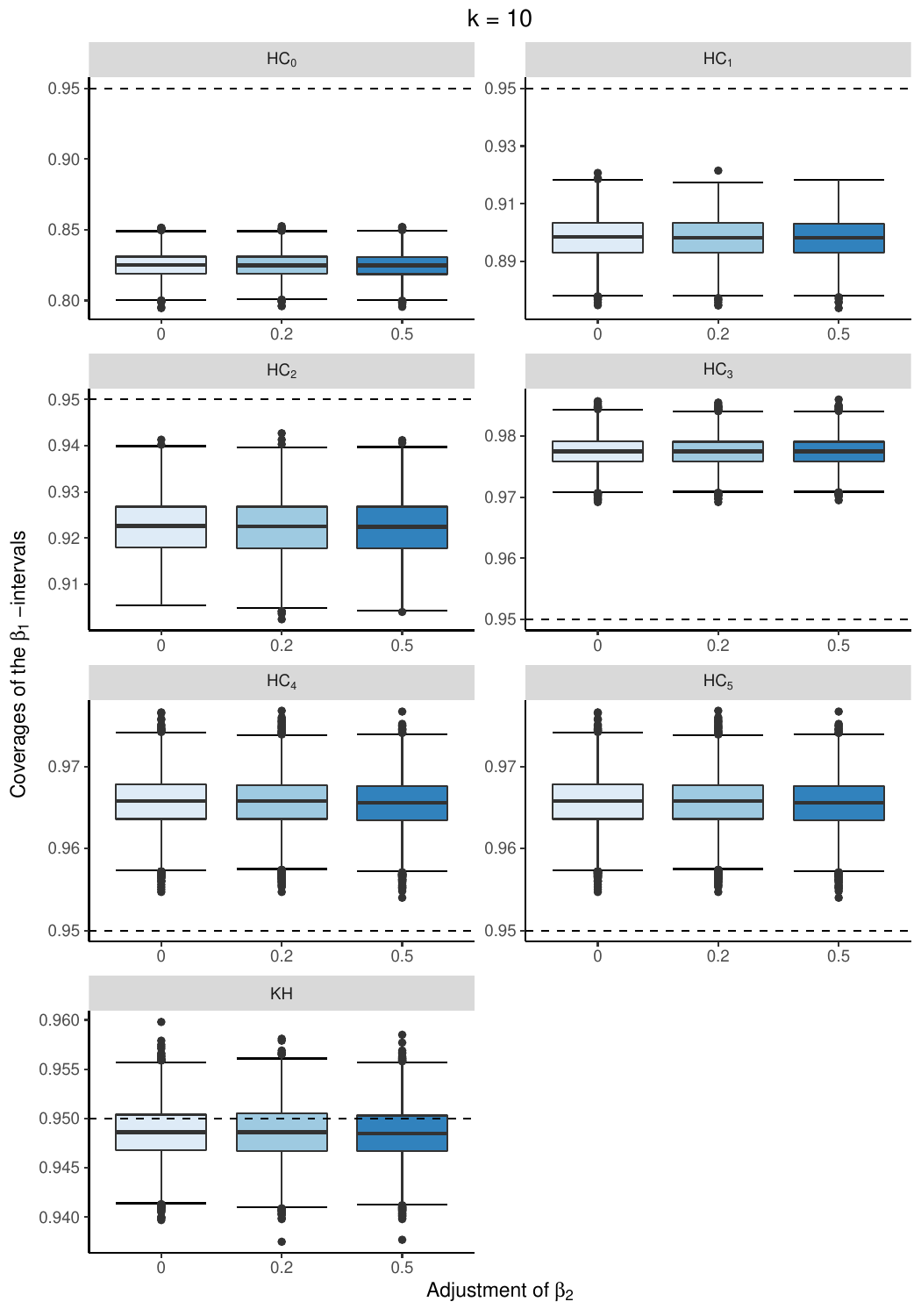}
\caption{Coverages of the \bo -intervals compared regarding the adjustments of $\beta_2$ for the estimators $HC_0-HC_5$ and $KH$ with $k=10$.}
\label{b1_b2_2}
\end{figure}

\begin{figure}
\includegraphics[scale=0.7, page=3]{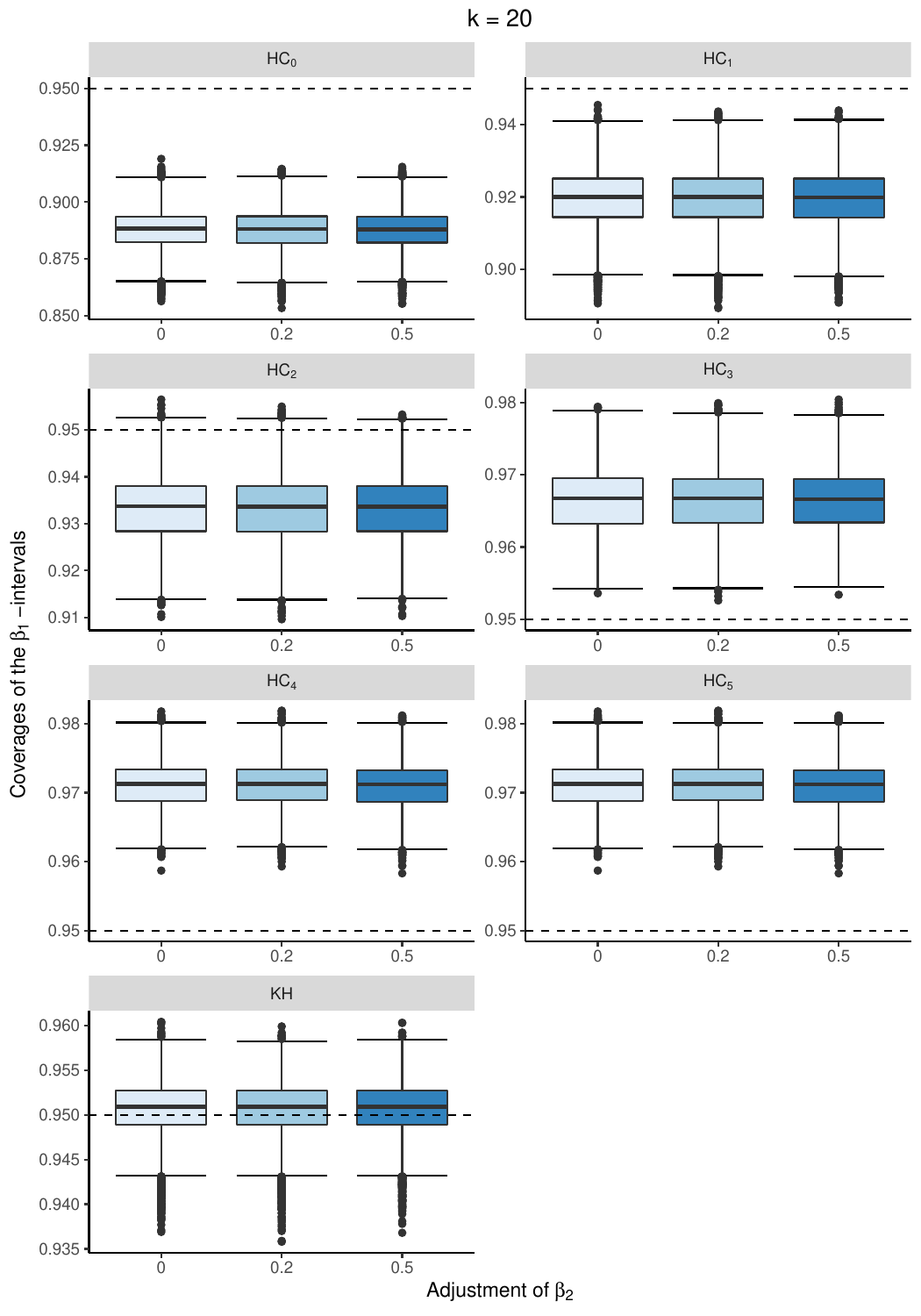}
\caption{Coverages of the \bo -intervals compared regarding the adjustments of $\beta_2$ for the estimators $HC_0-HC_5$ and $KH$ with $k=20$.}
\label{b1_b2_3}
\end{figure}

\begin{figure}
\includegraphics[scale=0.7, page=4]{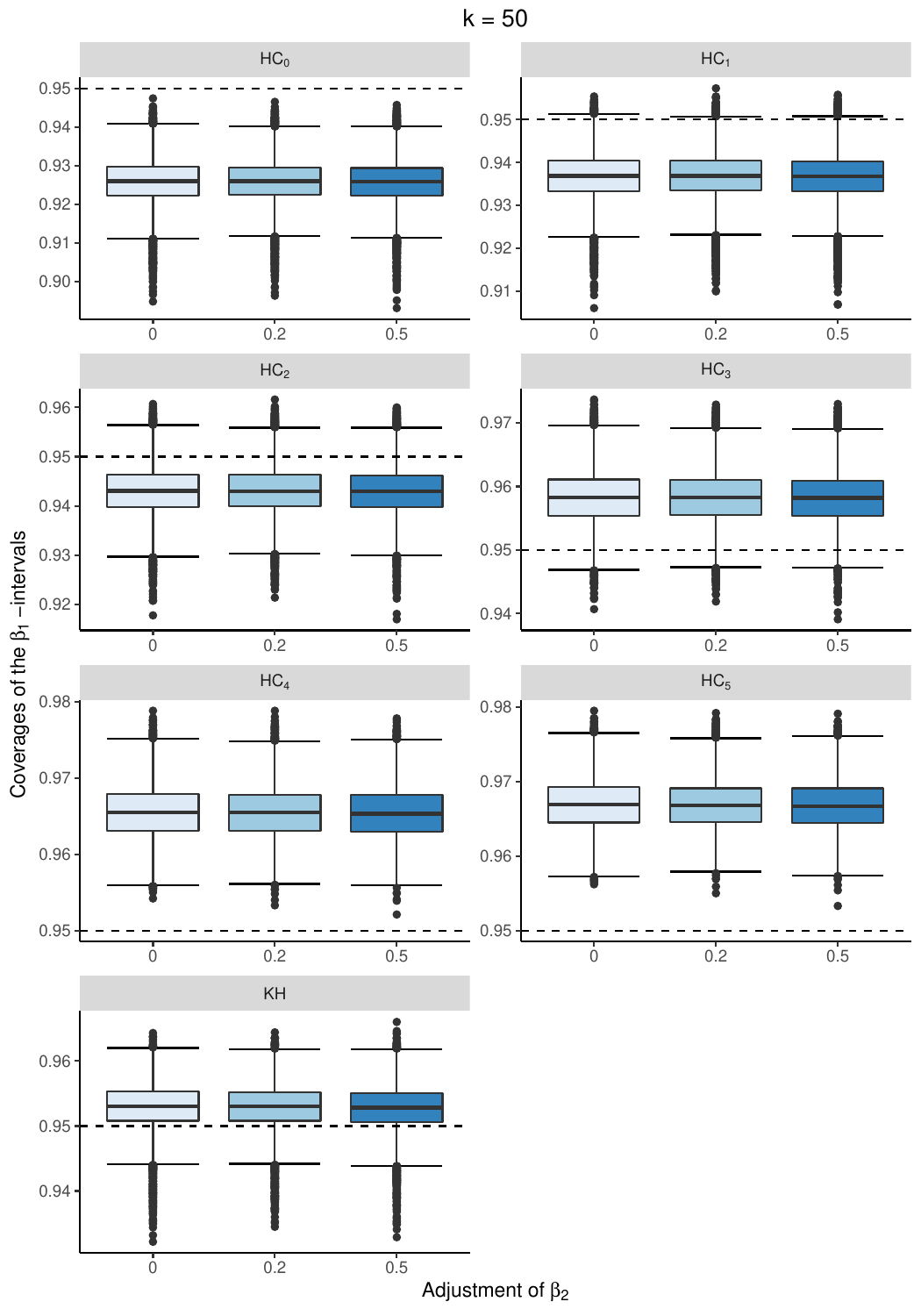}
\caption{Coverages of the \bo -intervals compared regarding the adjustments of $\beta_2$ for the estimators $HC_0-HC_5$ and $KH$ with $k=50$.}
\label{b1_b2_4}
\end{figure}

\begin{figure}
\includegraphics[scale=0.7, page=1]{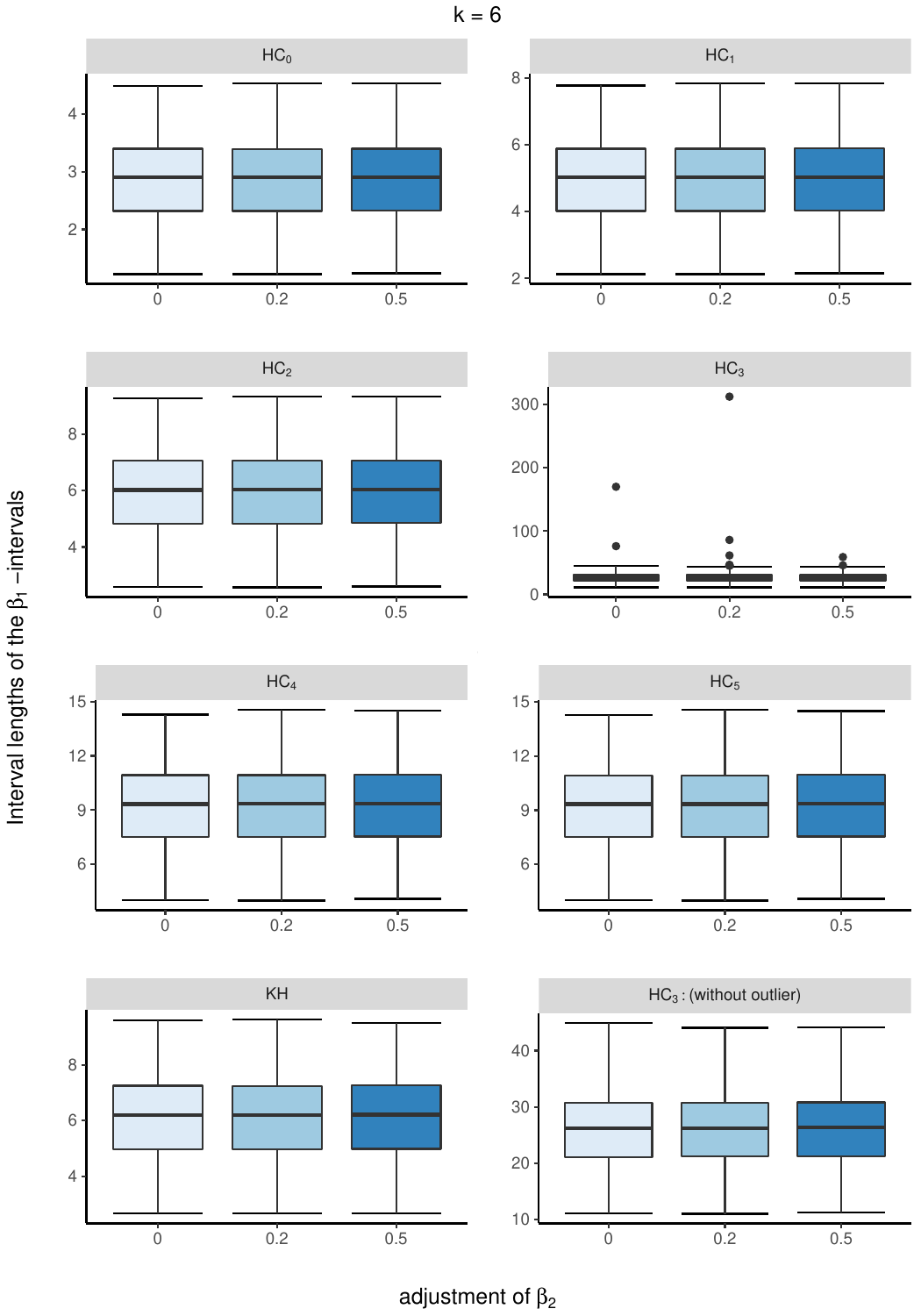}
\caption{Lengths of the \bo -intervals compared regarding the adjustments of $\beta_2$ for the estimators $HC_0-HC_5$ and $KH$ with $k=6$.}
\label{b1_b2_l_1}
\end{figure}

\begin{figure}
\includegraphics[scale=0.7, page=3]{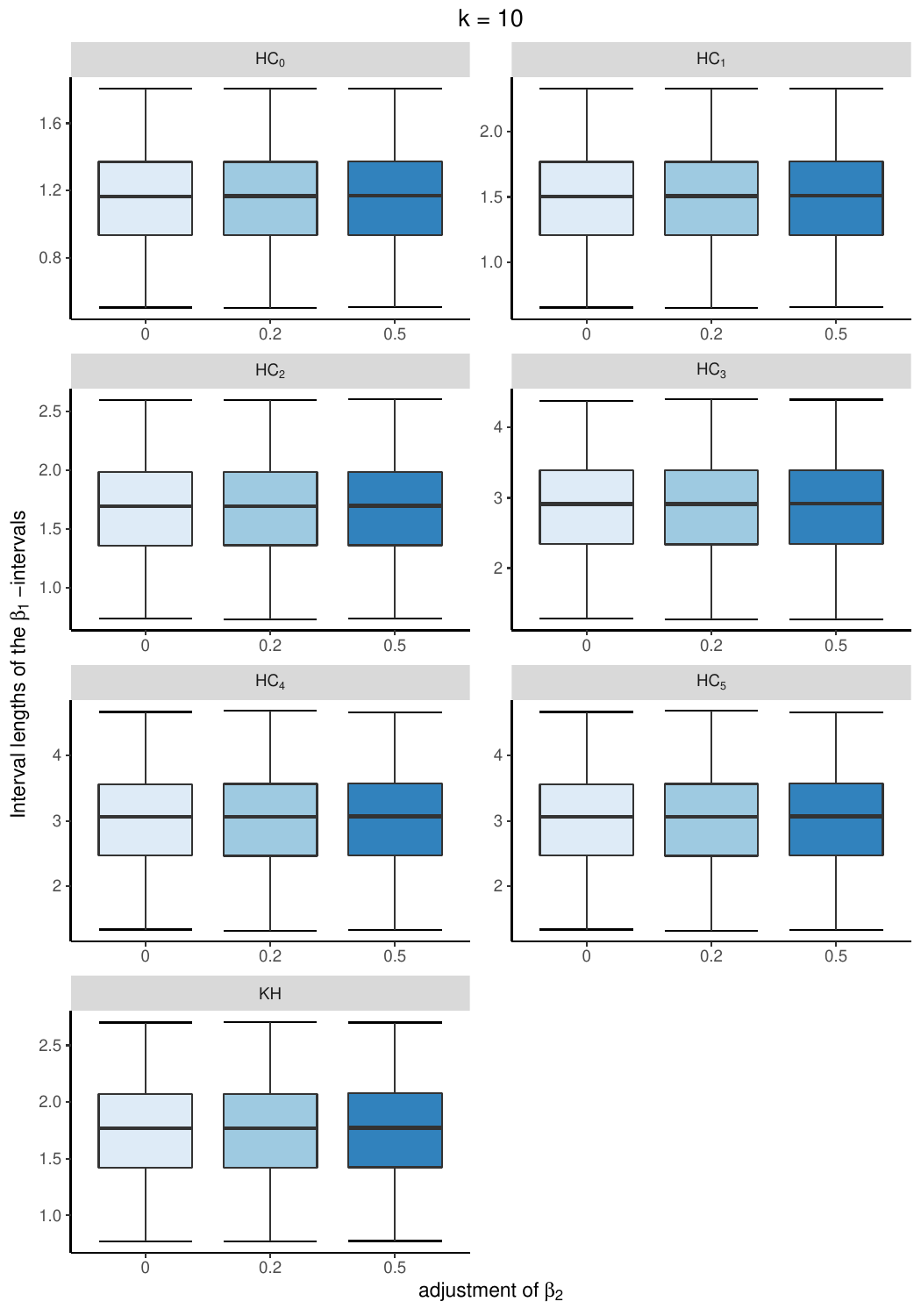}
\caption{Lengths of the \bo -intervals compared regarding the adjustments of $\beta_2$ for the estimators $HC_0-HC_5$ and $KH$ with $k=10$.}
\label{b1_b2_l_2}
\end{figure}

\begin{figure}
\includegraphics[scale=0.7, page=4]{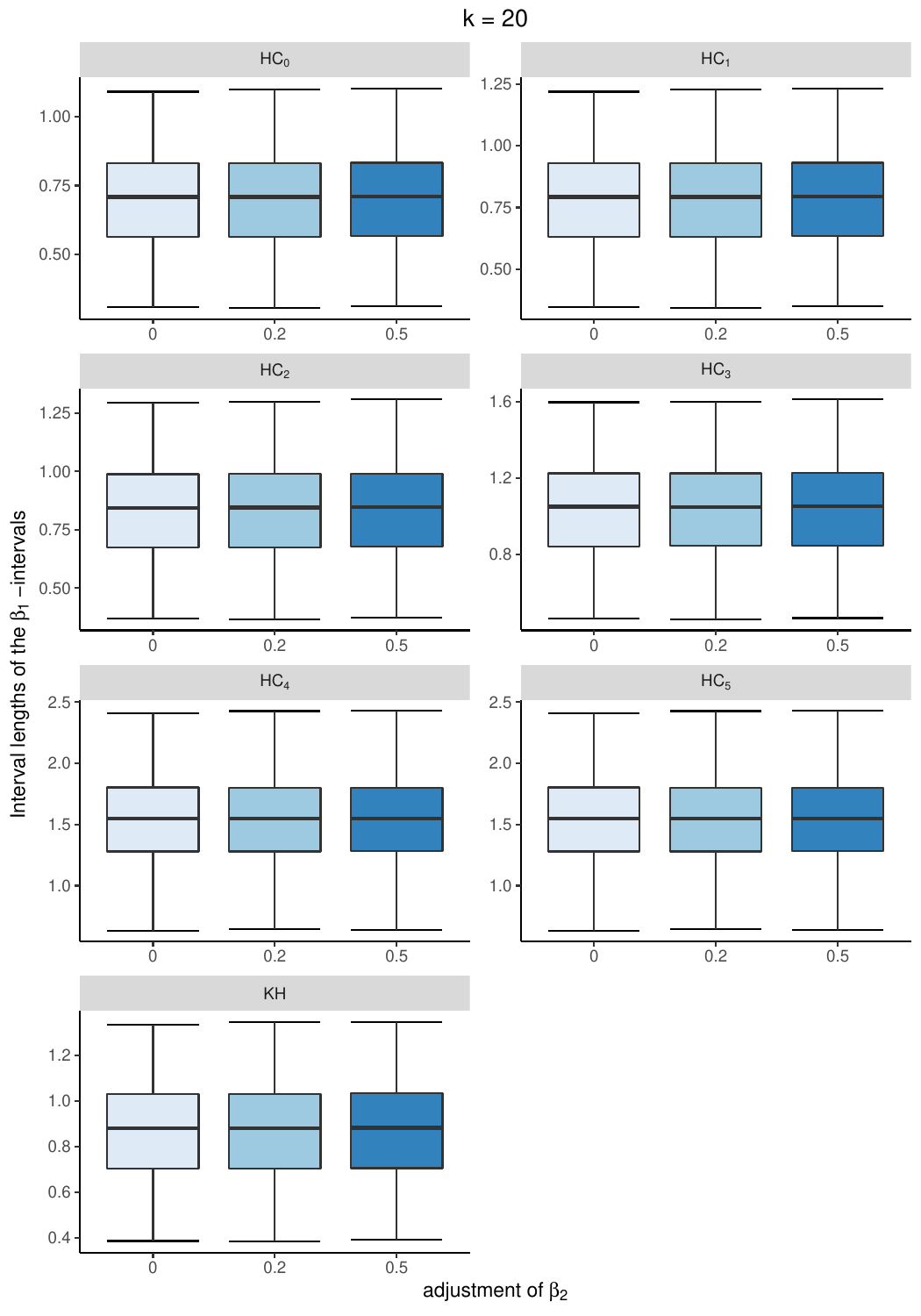}
\caption{Lengths of the \bo -intervals compared regarding the adjustments of $\beta_2$ for the estimators $HC_0-HC_5$ and $KH$ with $k=20$.}
\label{b1_b2_l_3}
\end{figure}

\begin{figure}
\includegraphics[scale=0.7, page=2]{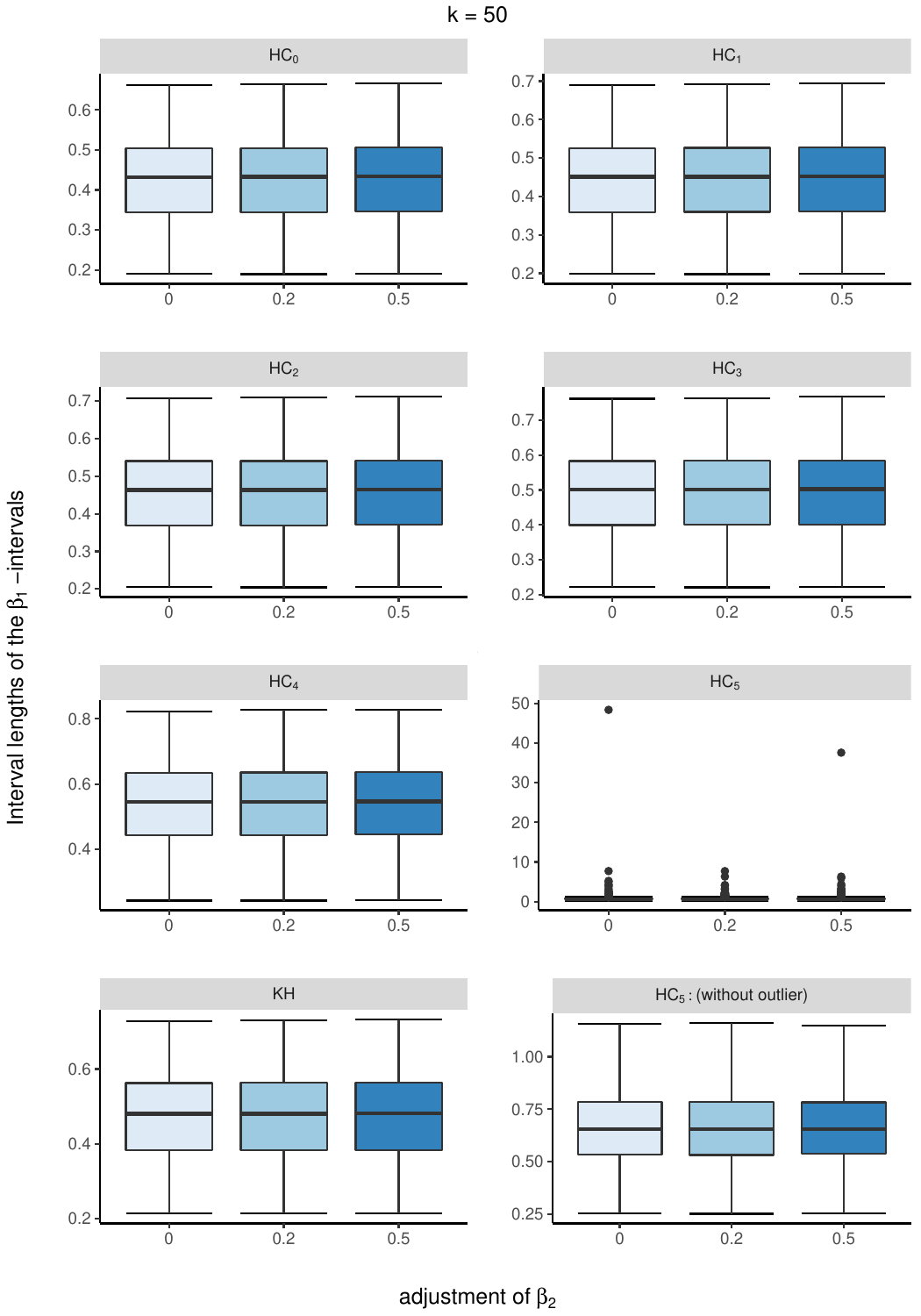}
\caption{Lengths of the \bo -intervals compared regarding the adjustments of $\beta_2$ for the estimators $HC_0-HC_5$ and $KH$ with $k=50$.}
\label{b1_b2_l_4}
\end{figure}

\begin{figure}
\includegraphics[scale=0.7, page=1]{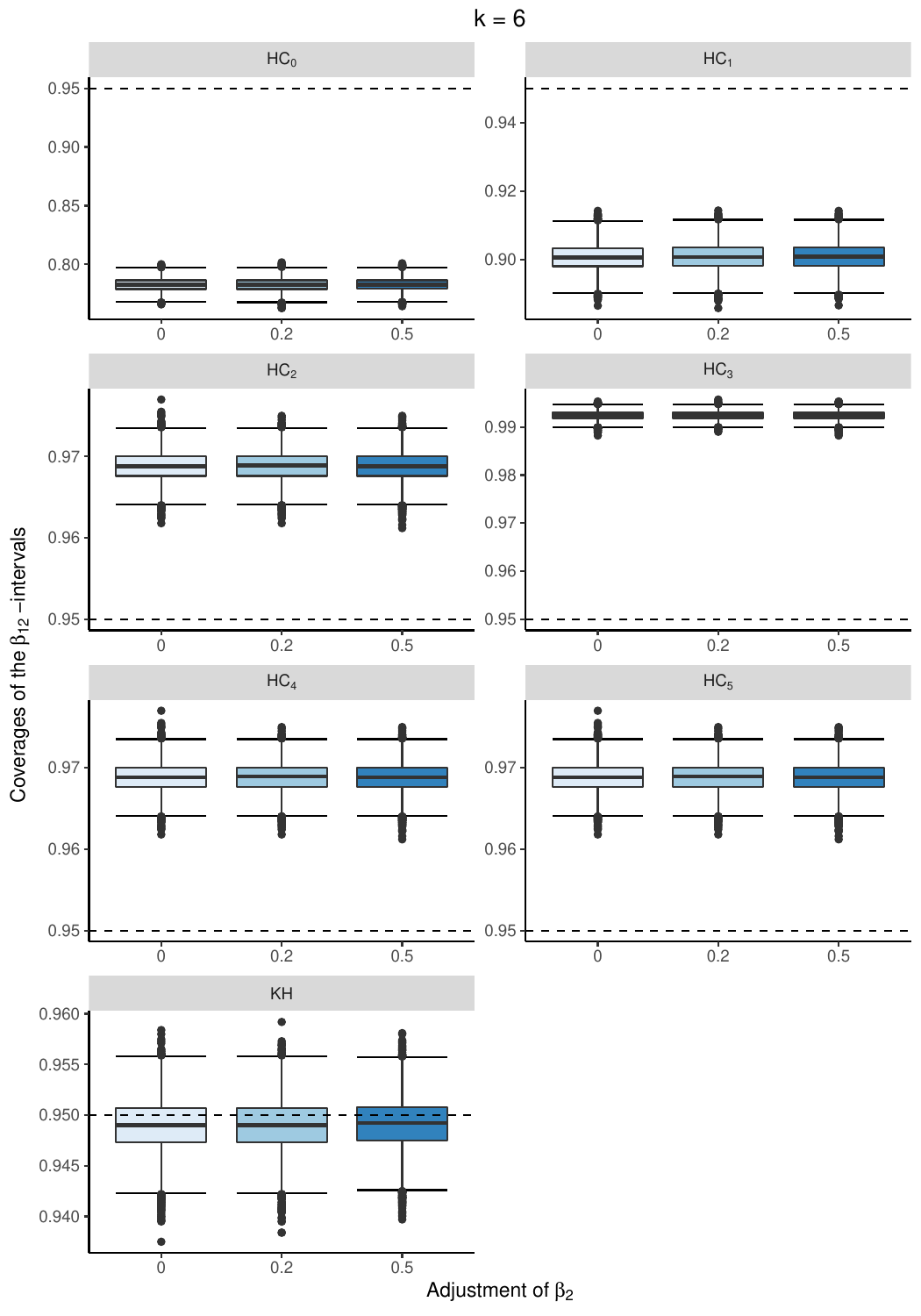}
\caption{Coverages of the \bi -intervals compared regarding the adjustments of $\beta_2$ for the estimators $HC_0-HC_5$ and $KH$ with $k=6$.}
\label{bi_b2_1}
\end{figure}

\begin{figure}
\includegraphics[scale=0.7, page=2]{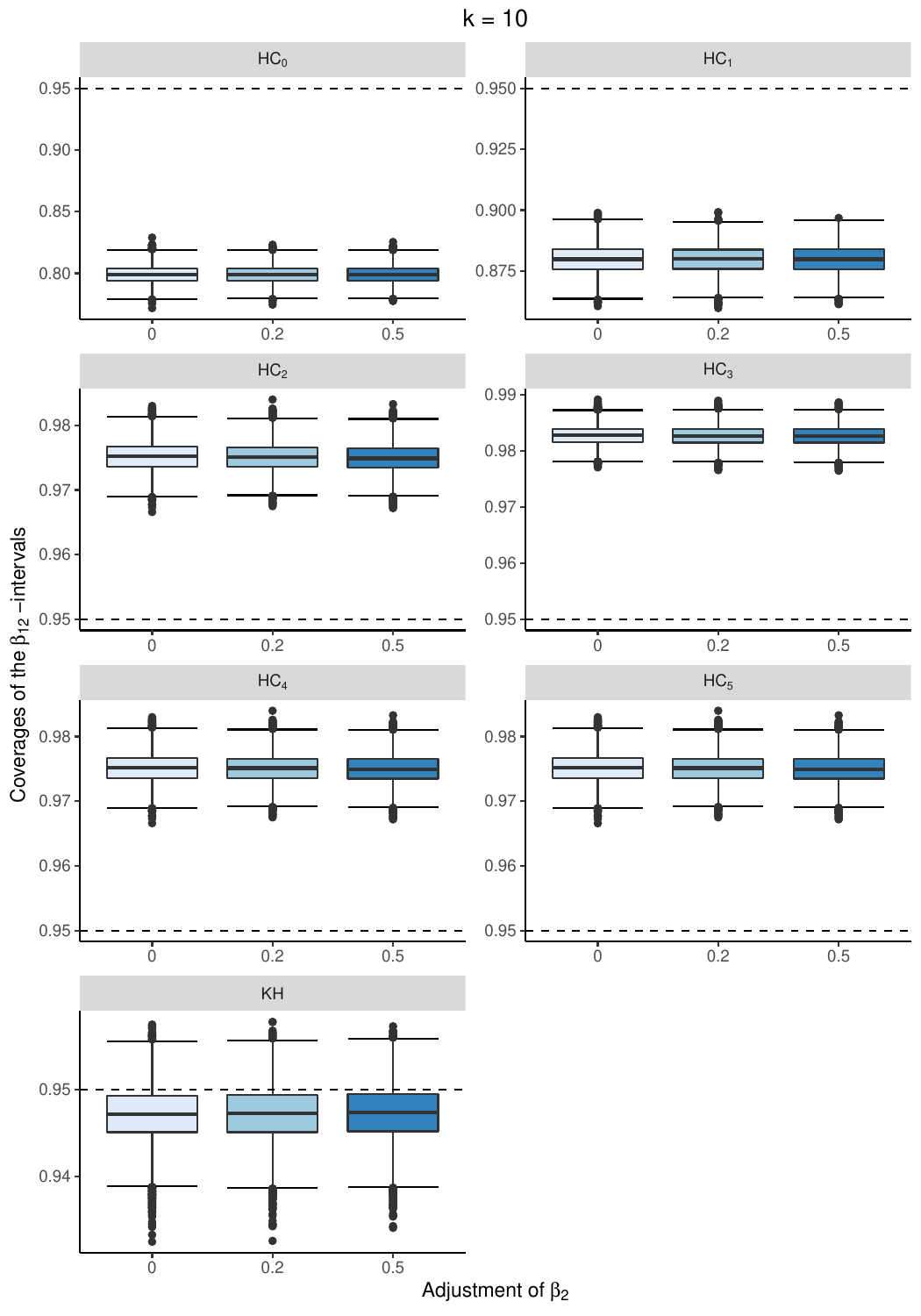}
\caption{Coverages of the \bi -intervals compared regarding the adjustments of $\beta_2$ for the estimators $HC_0-HC_5$ and $KH$ with $k=10$.}
\label{bi_b2_2}
\end{figure}

\begin{figure}
\includegraphics[scale=0.7, page=3]{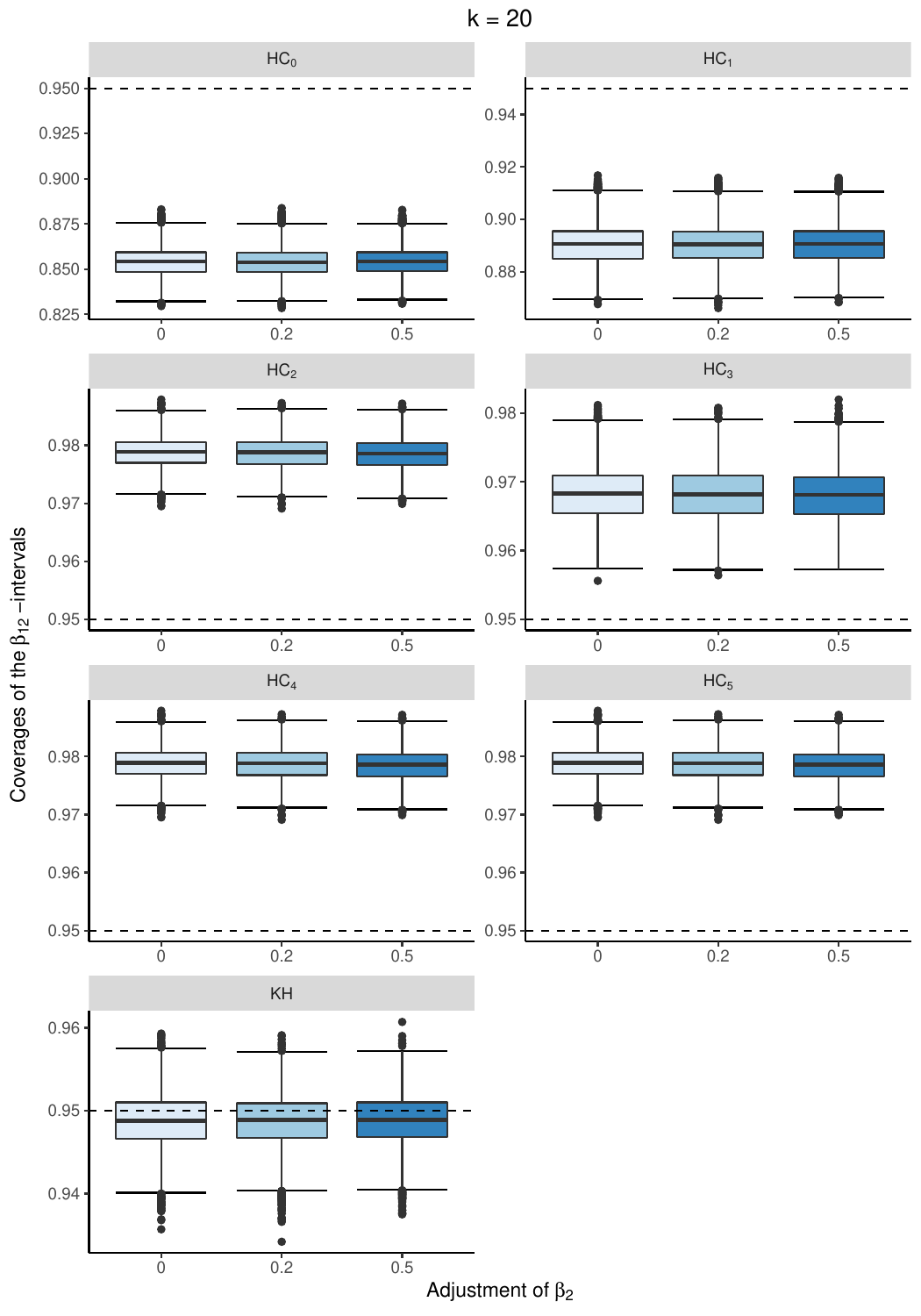}
\caption{Coverages of the \bi -intervals compared regarding the adjustments of $\beta_2$ for the estimators $HC_0-HC_5$ and $KH$ with $k=20$.}
\label{bi_b2_3}
\end{figure}

\begin{figure}
\includegraphics[scale=0.7, page=4]{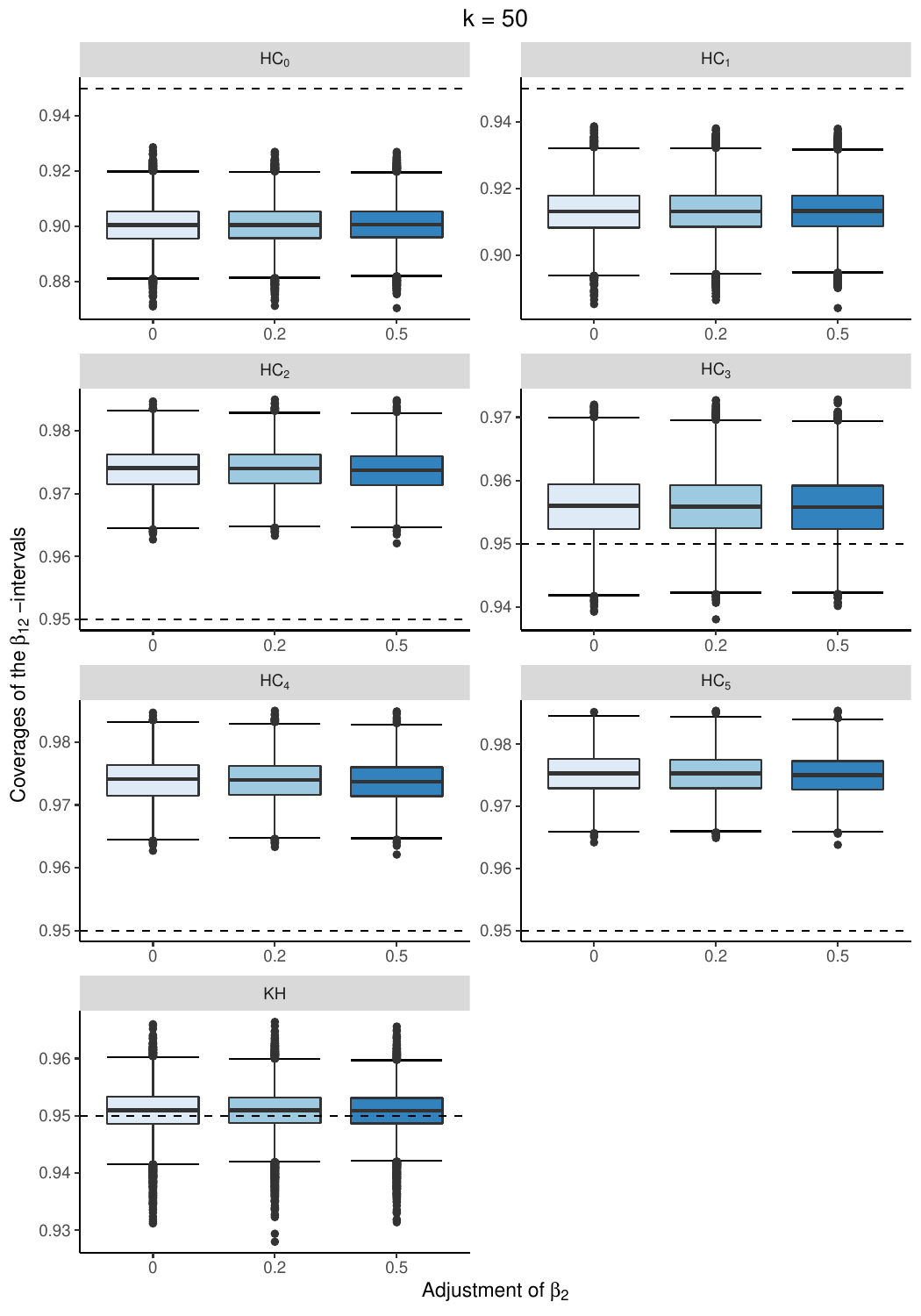}
\caption{Coverages of the \bi -intervals compared regarding the adjustments of $\beta_2$ for the estimators $HC_0-HC_5$ and $KH$ with $k=50$.}
\label{bi_b2_4}
\end{figure}

\begin{figure}
\includegraphics[scale=0.7, page=1]{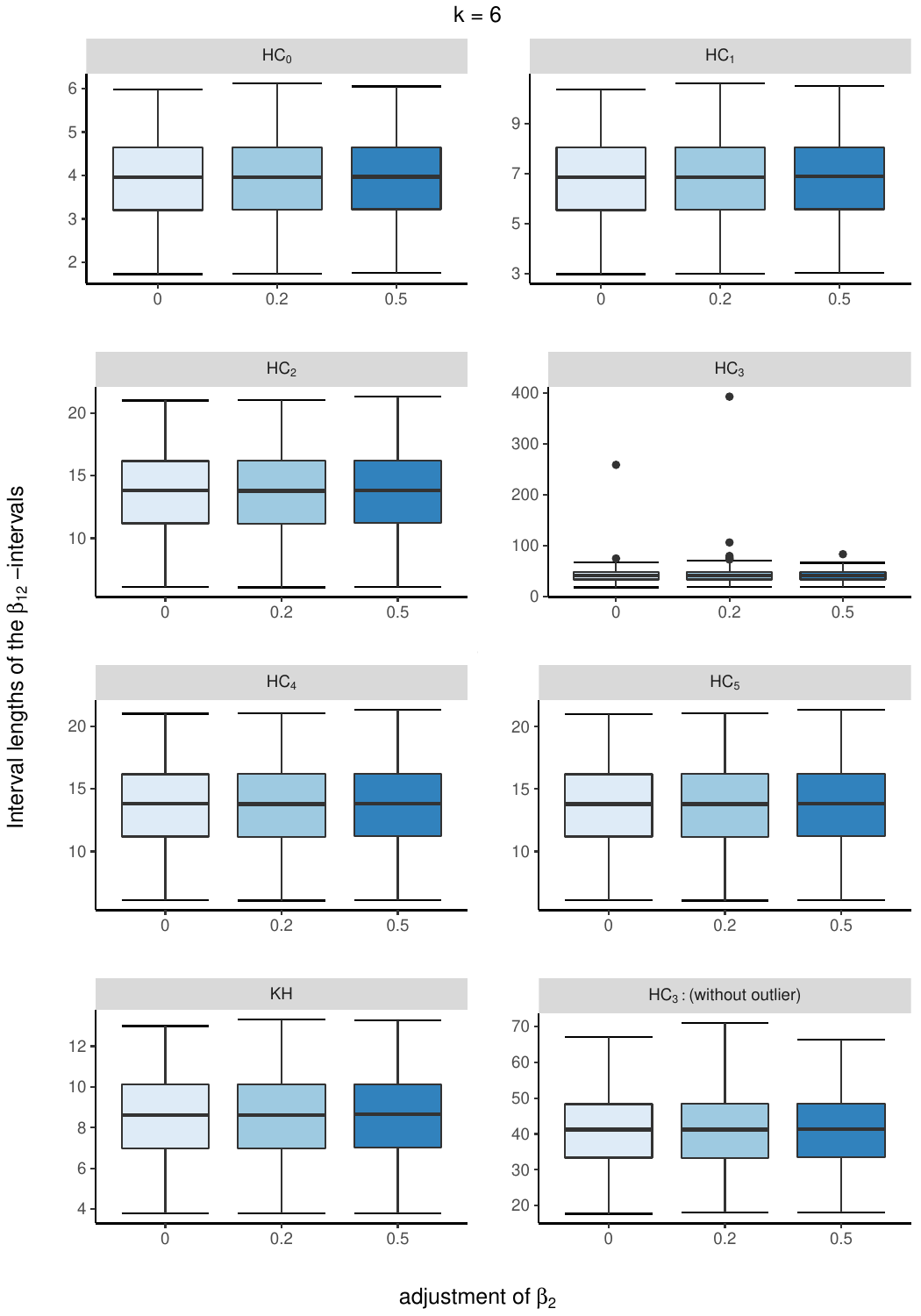}
\caption{Lengths of the \bi -intervals compared regarding the adjustments of $\beta_2$ for the estimators $HC_0-HC_5$ and $KH$ with $k=6$.}
\label{bi_b2_l_1}
\end{figure}

\begin{figure}
\includegraphics[scale=0.7, page=3]{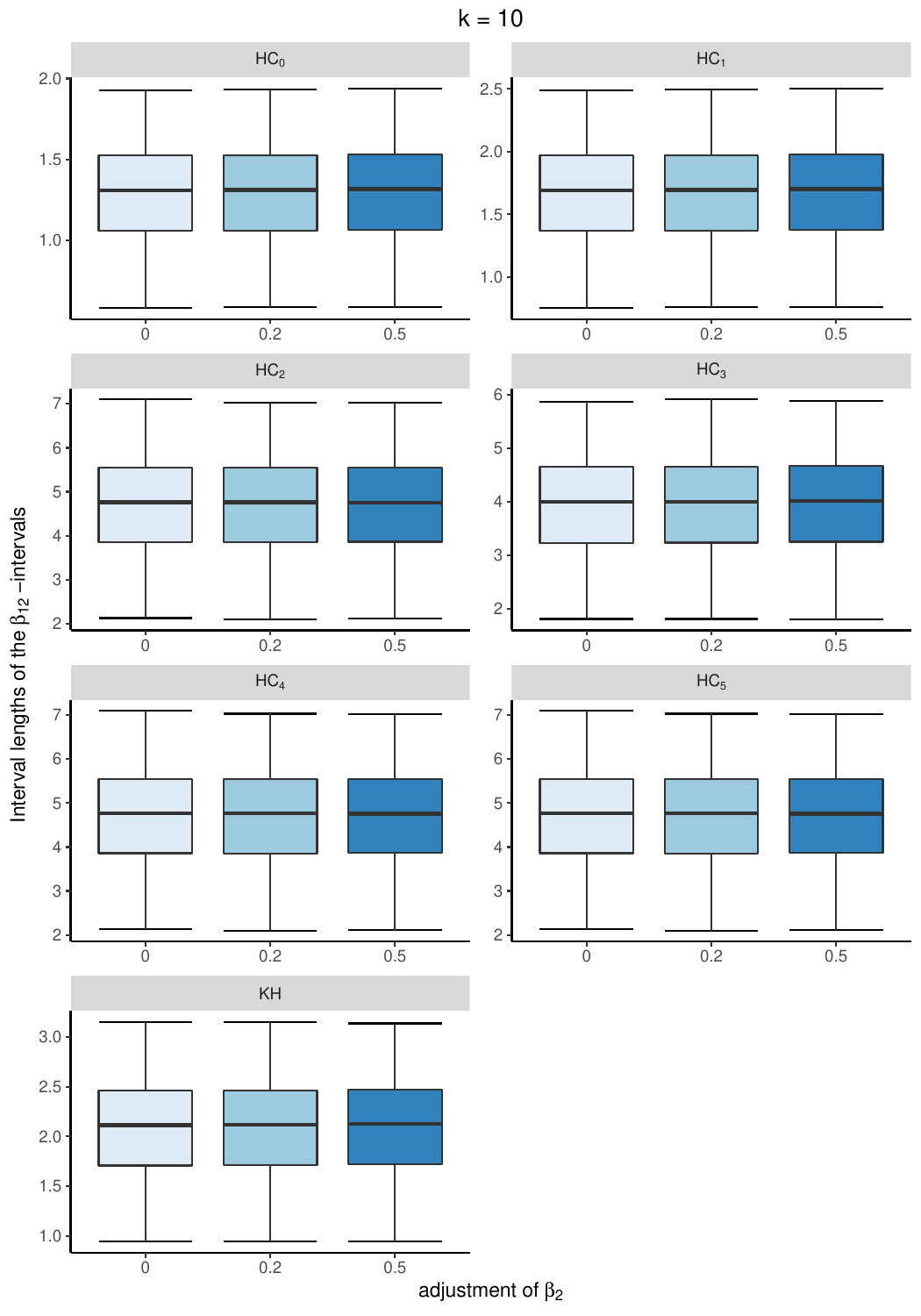}
\caption{Lengths of the \bi -intervals compared regarding the adjustments of $\beta_2$ for the estimators $HC_0-HC_5$ and $KH$ with $k=10$.}
\label{bi_b2_l_2}
\end{figure}

\begin{figure}
\includegraphics[scale=0.7, page=4]{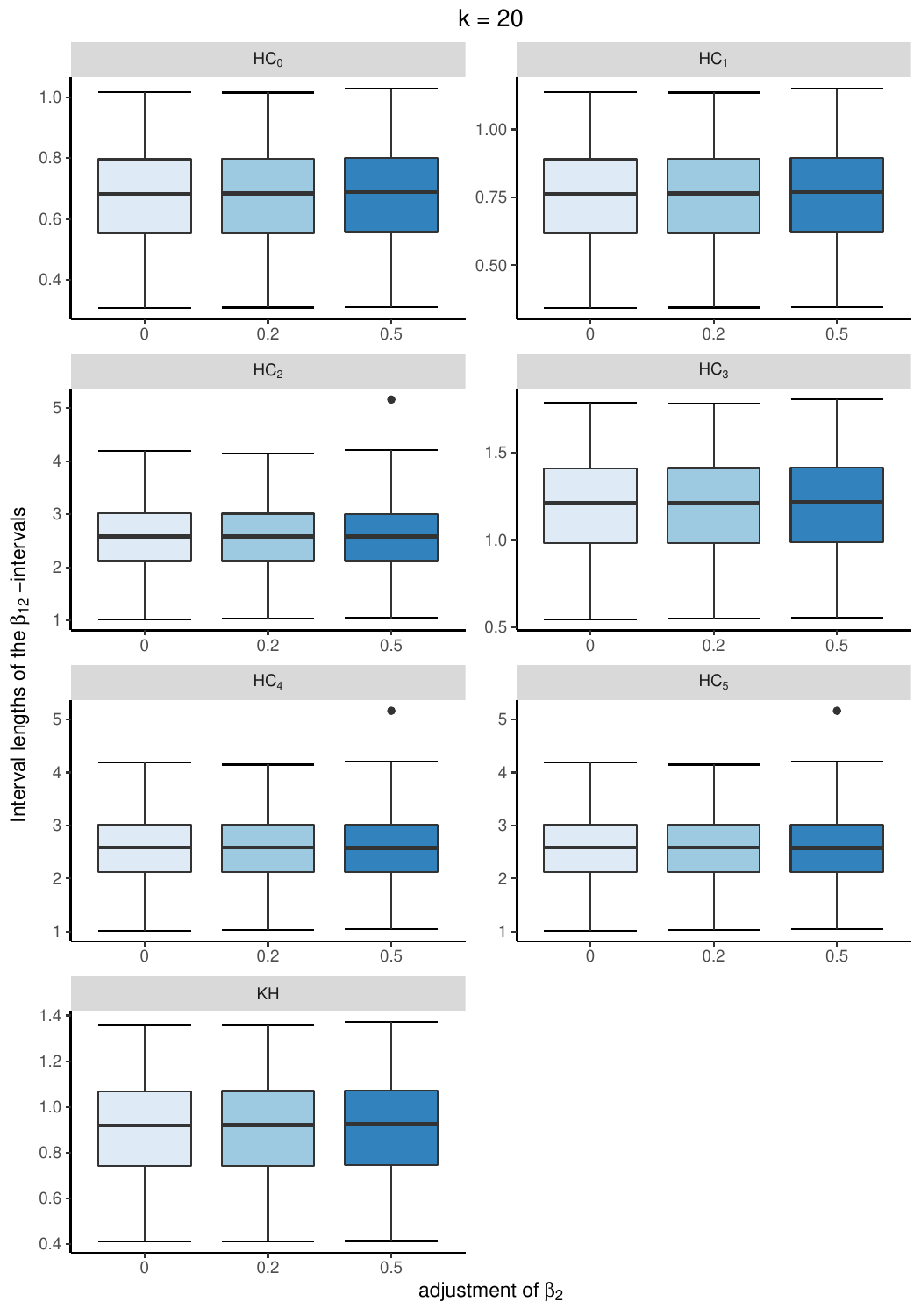}
\caption{Lengths of the \bi -intervals compared regarding the adjustments of $\beta_2$ for the estimators $HC_0-HC_5$ and $KH$ with $k=20$.}
\label{bi_b2_l_3}
\end{figure}

\begin{figure}
\includegraphics[scale=0.7, page=2]{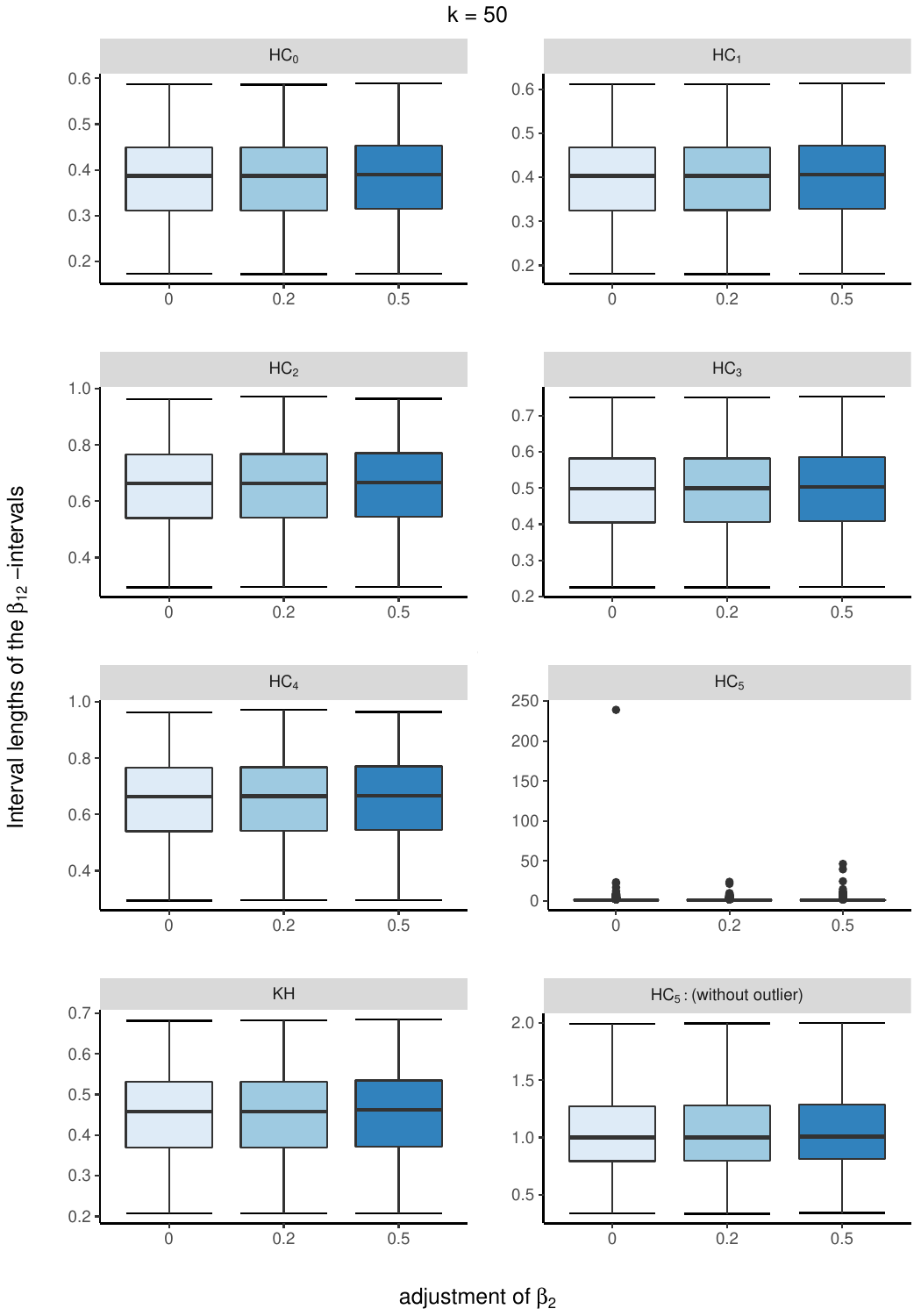}
\caption{Lengths of the \bi -intervals compared regarding the adjustments of $\beta_2$ for the estimators $HC_0-HC_5$ and $KH$ with $k=50$.}
\label{bi_b2_l_4}
\end{figure}

\textbf{Adjustments of $\boldsymbol{\beta_2}$} For $\beta_2$ the adjustments 0, 0.2 and 0.5 were considered as well. None of the intervals were affected by the adjustment of $\beta_2$ regarding coverage or length (Figures \ref{b1_b2_1}-\ref{bi_b2_l_4}). Mentionable is that the extreme interval lengths of the $\textbf{HC}_3$-CIs for $k=6$ only occur for small values of $\beta_2$.

\begin{figure}
\includegraphics[scale=0.7, page=1]{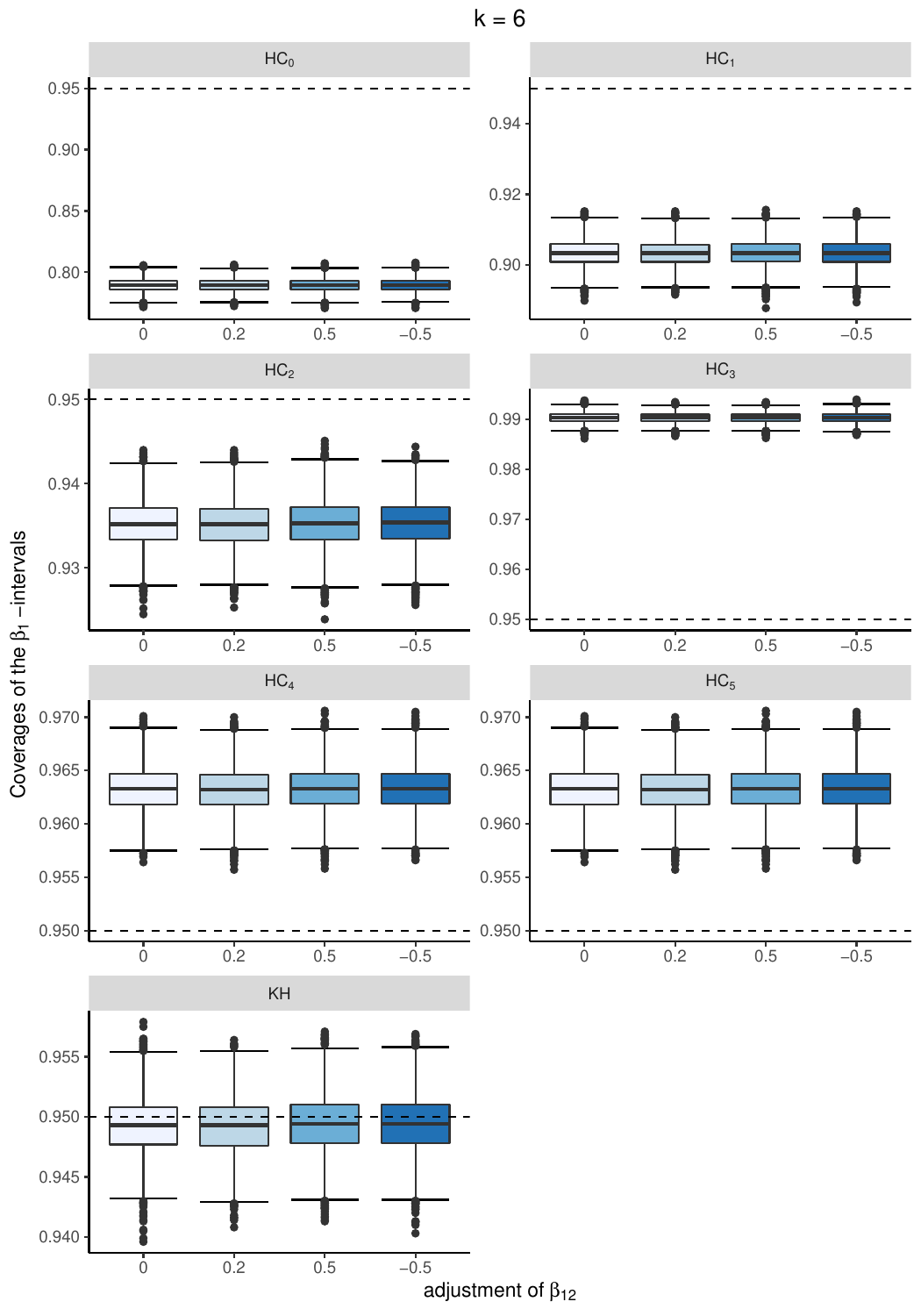}
\caption{Coverages of the \bo -intervals compared regarding the adjustments of $\beta_{12}$ for the estimators $HC_0-HC_5$ and $KH$ with $k=6$.}
\label{b1_b12_1}
\end{figure}

\begin{figure}
\includegraphics[scale=0.7, page=2]{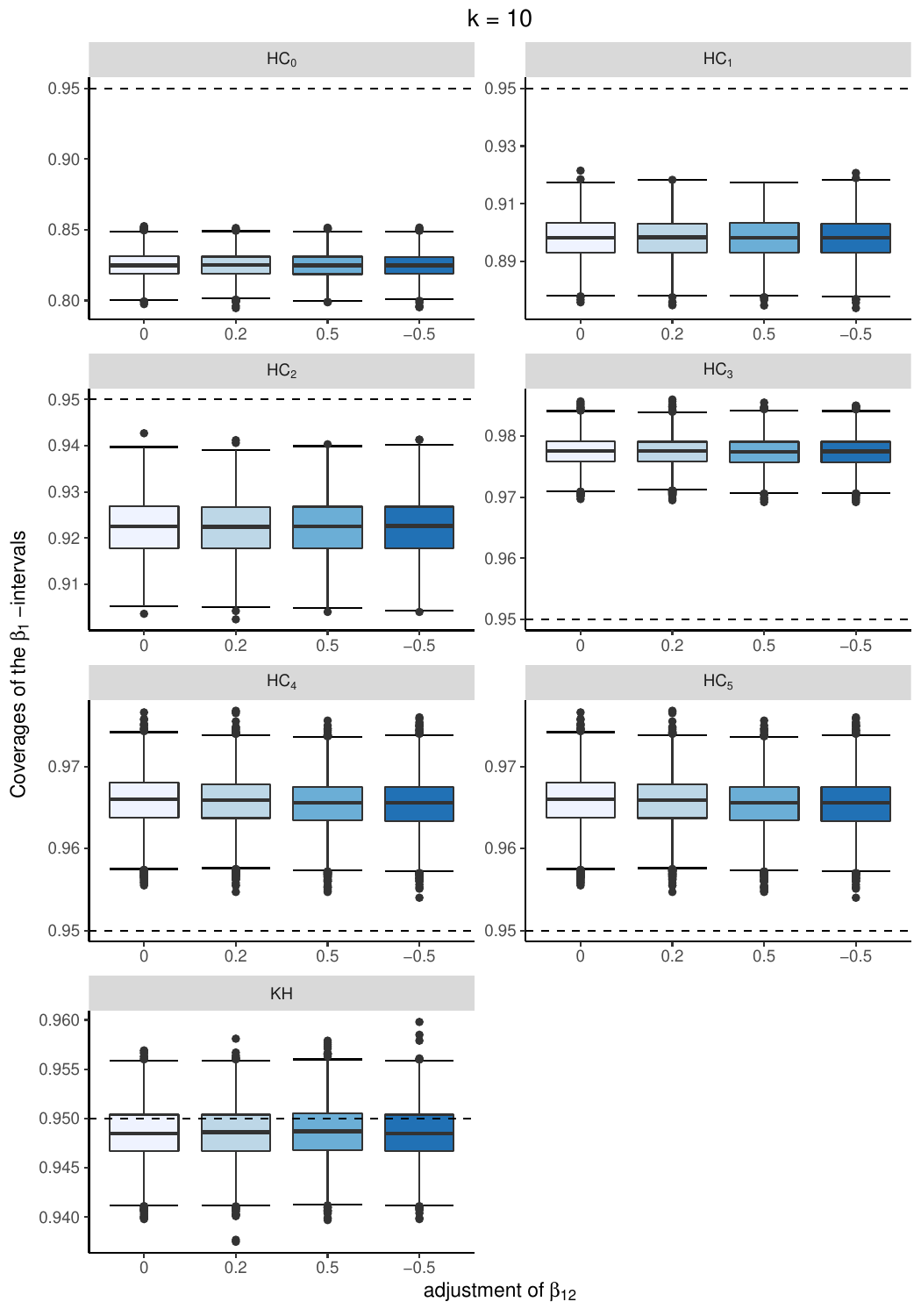}
\caption{Coverages of the \bo -intervals compared regarding the adjustments of $\beta_{12}$ for the estimators $HC_0-HC_5$ and $KH$ with $k=10$.}
\label{b1_b12_2}
\end{figure}

\begin{figure}
\includegraphics[scale=0.7, page=3]{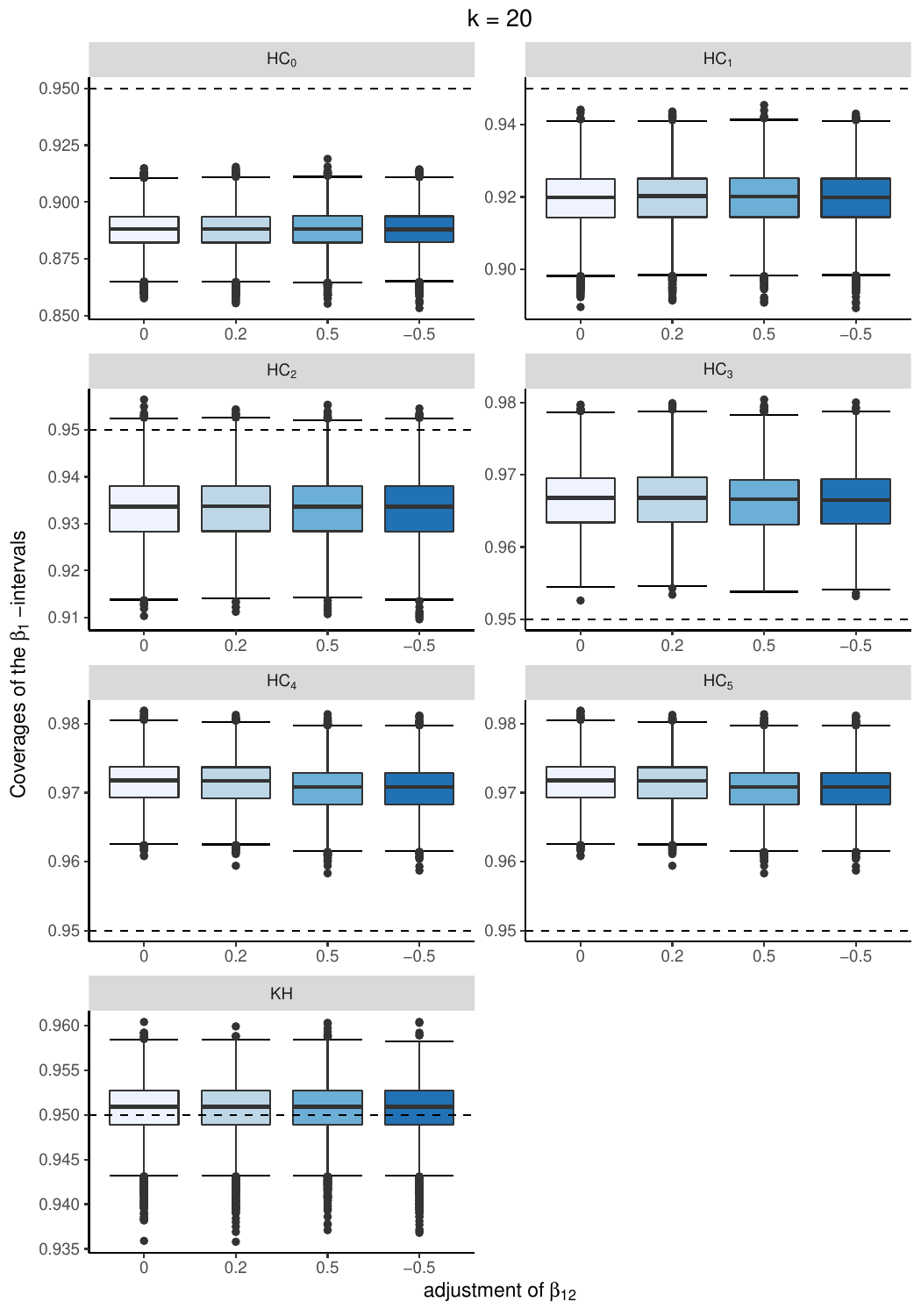}
\caption{Coverages of the \bo -intervals compared regarding the adjustments of $\beta_{12}$ for the estimators $HC_0-HC_5$ and $KH$ with $k=20$.}
\label{b1_b12_3}
\end{figure}

\begin{figure}
\includegraphics[scale=0.7, page=4]{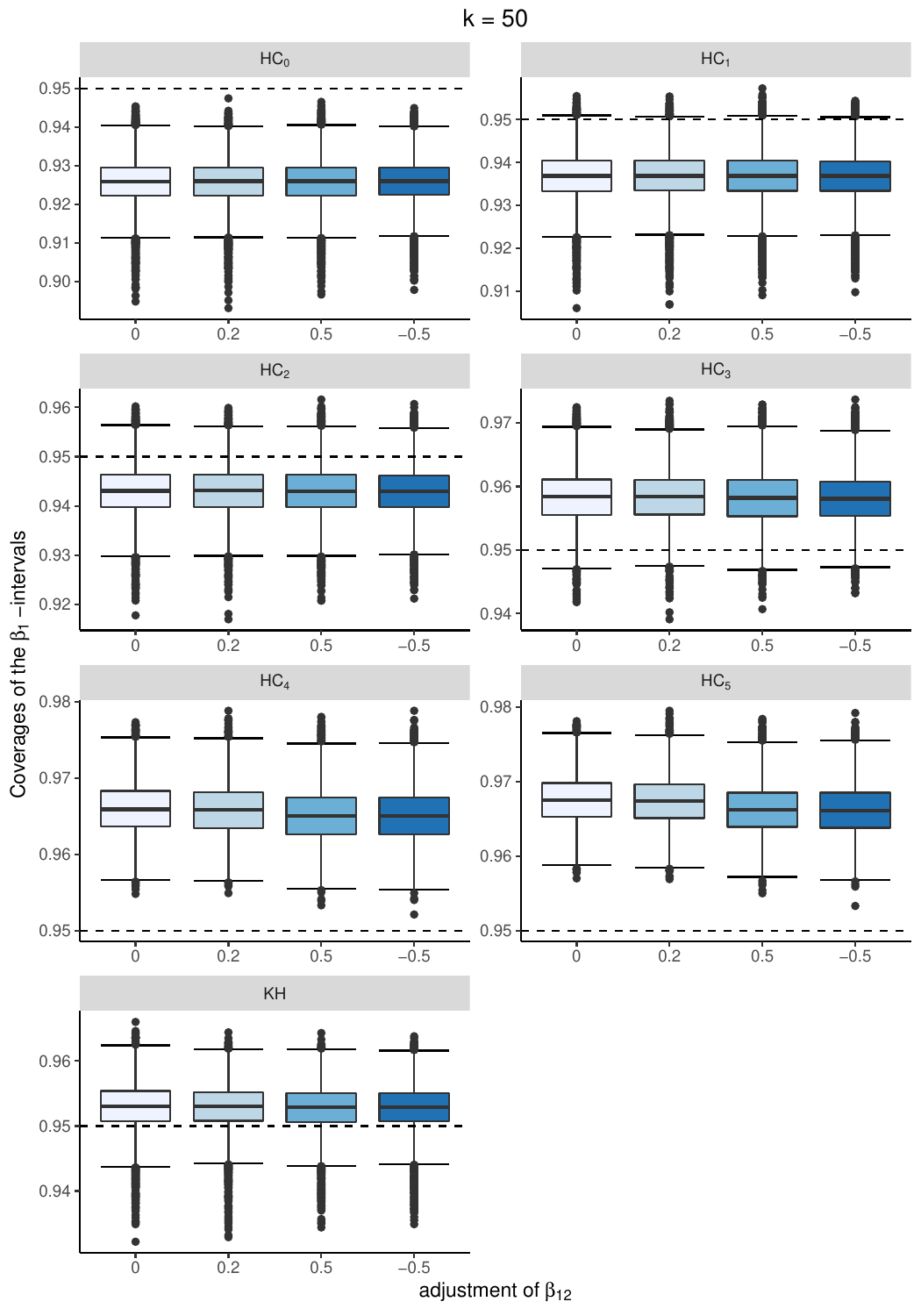}
\caption{Coverages of the \bo -intervals compared regarding the adjustments of $\beta_{12}$ for the estimators $HC_0-HC_5$ and $KH$ with $k=50$.}
\label{b1_b12_4}
\end{figure}

\begin{figure}
\includegraphics[scale=0.7, page=1]{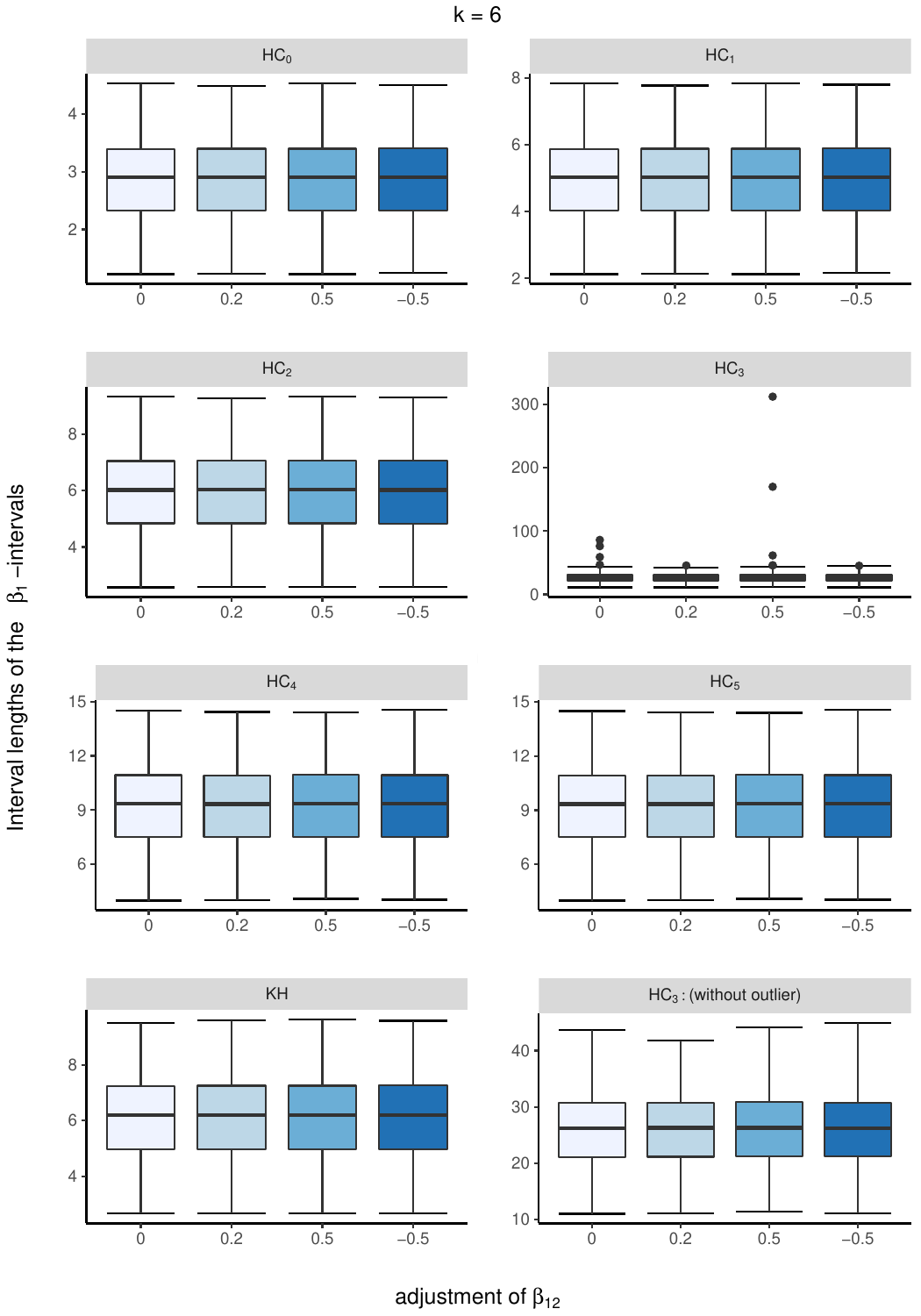}
\caption{Lengths of the \bo -intervals compared regarding the adjustments of $\beta_{12}$ for the estimators $HC_0-HC_5$ and $KH$ with $k=6$.}
\label{b1_b12_l_1}
\end{figure}

\begin{figure}
\includegraphics[scale=0.7, page=3]{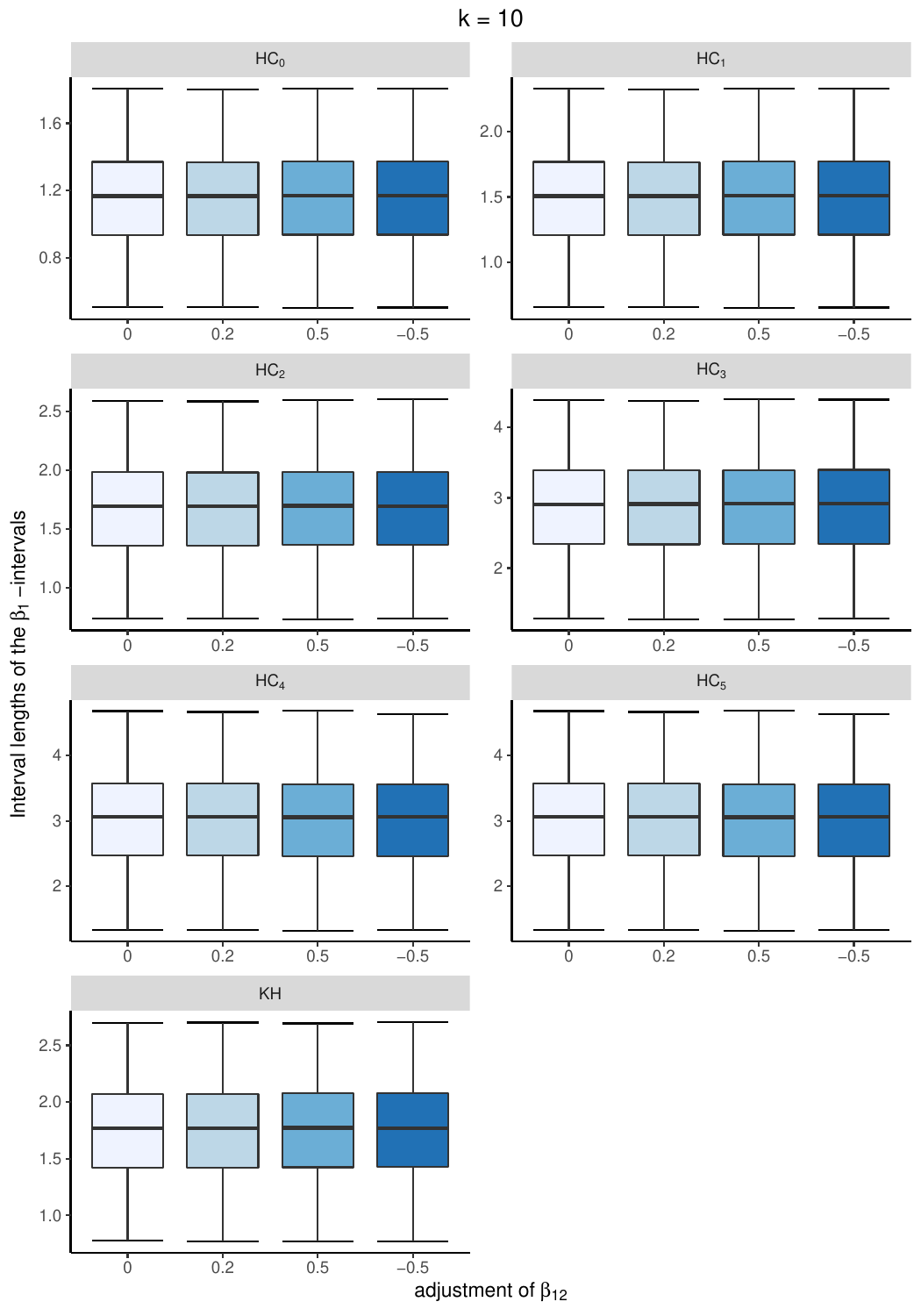}
\caption{Lengths of the \bo -intervals compared regarding the adjustments of $\beta_{12}$ for the estimators $HC_0-HC_5$ and $KH$ with $k=10$.}
\label{b1_b12_l_2}
\end{figure}

\begin{figure}
\includegraphics[scale=0.7, page=4]{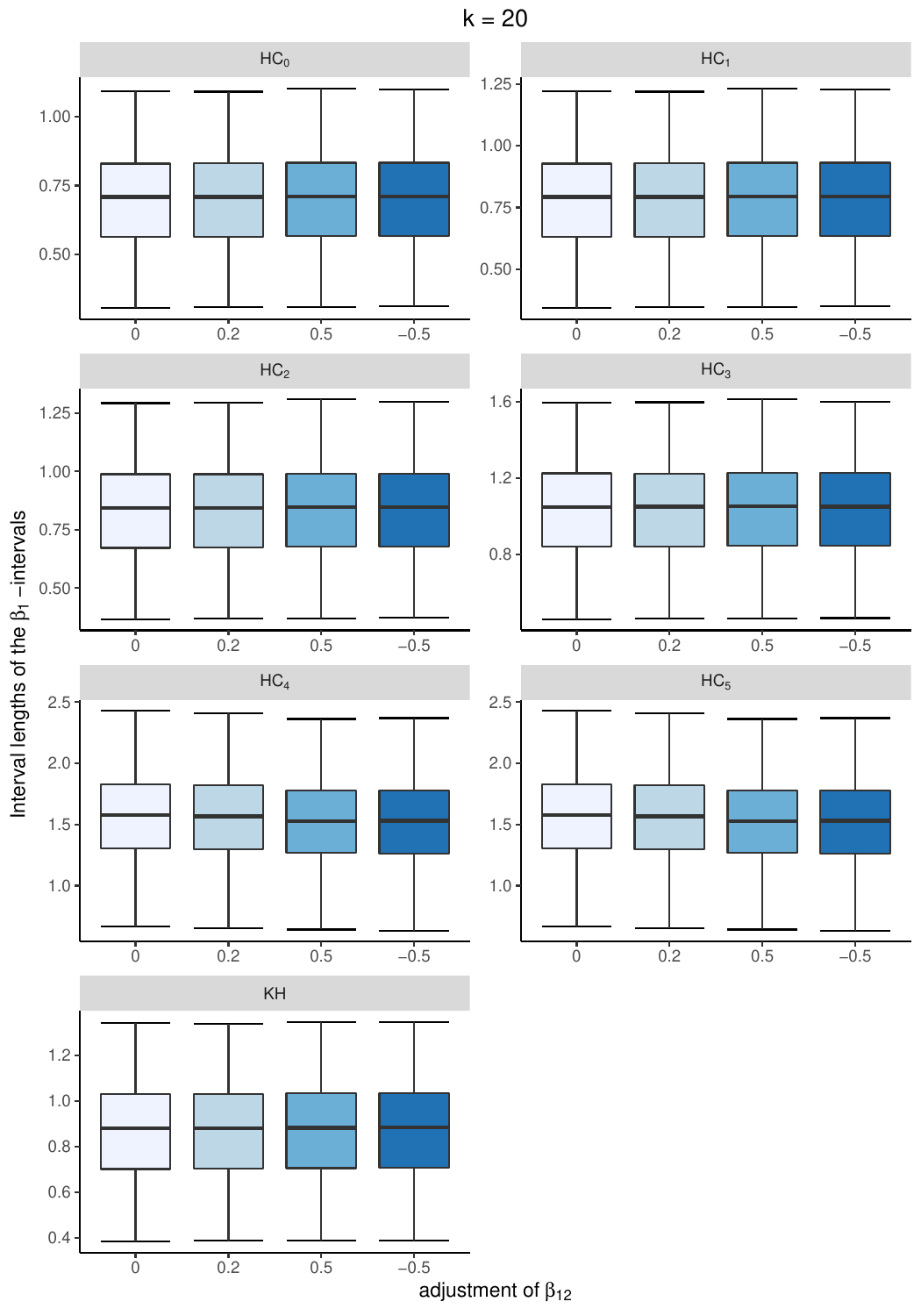}
\caption{Lengths of the \bo -intervals compared regarding the adjustments of $\beta_{12}$ for the estimators $HC_0-HC_5$ and $KH$ with $k=20$.}
\label{b1_b12_l_3}
\end{figure}

\begin{figure}
\includegraphics[scale=0.7, page=2]{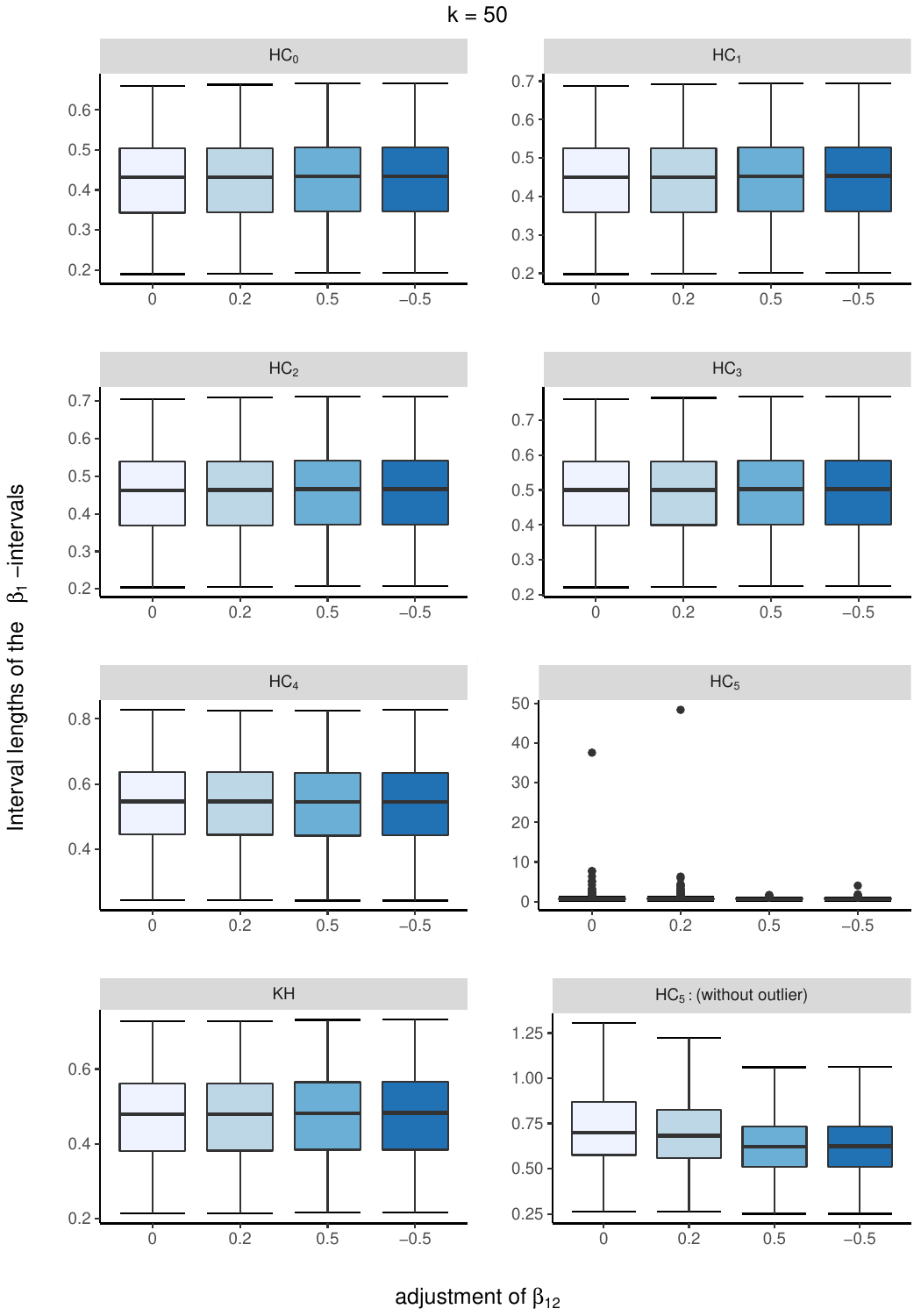}
\caption{Lengths of the \bo -intervals compared regarding the adjustments of $\beta_{12}$ for the estimators $HC_0-HC_5$ and $KH$ with $k=50$.}
\label{b1_b12_l_4}
\end{figure}

\begin{figure}
\includegraphics[scale=0.7, page=1]{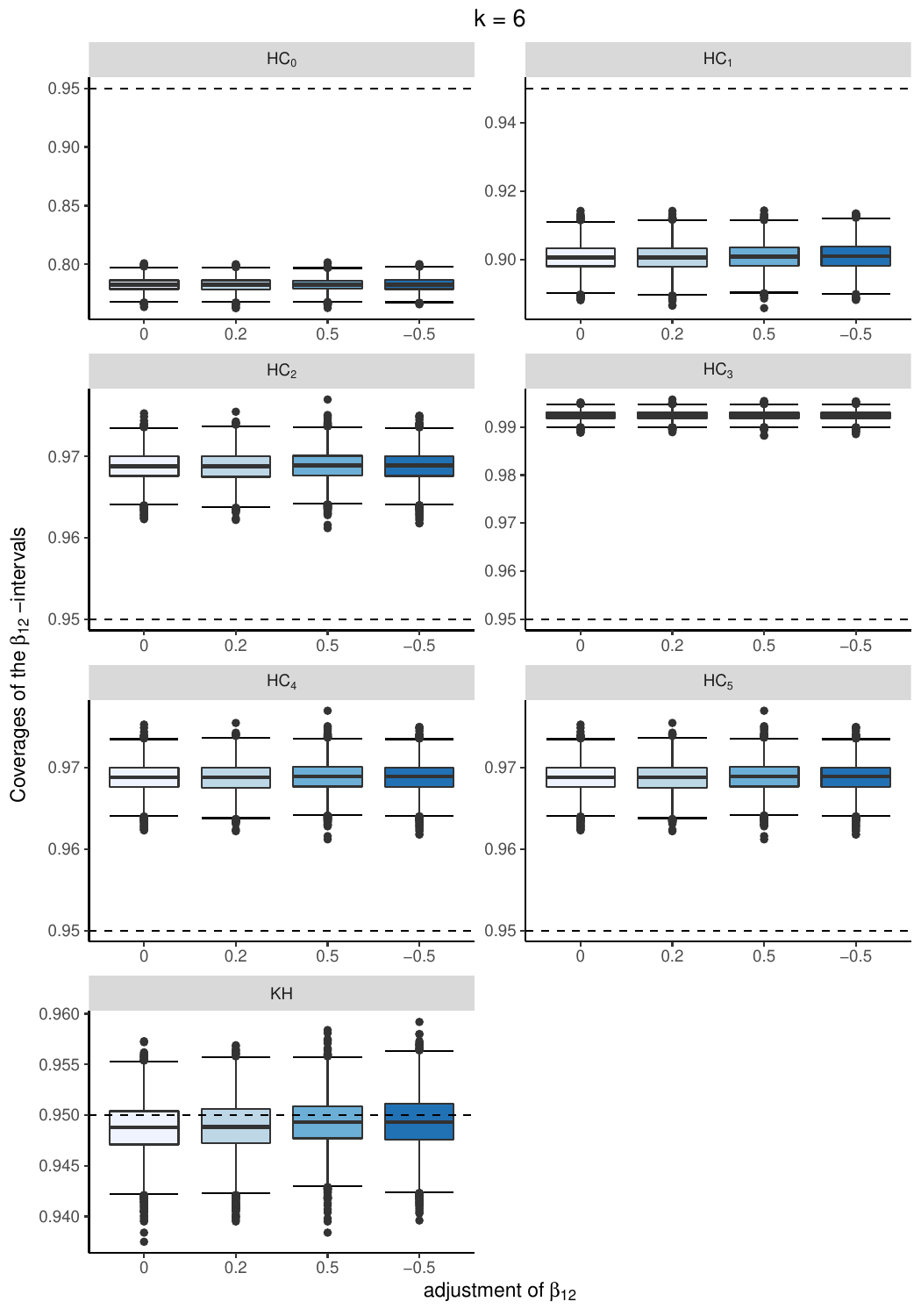}
\caption{Coverages of the \bi -intervals compared regarding the adjustments of $\beta_{12}$ for the estimators $HC_0-HC_5$ and $KH$ with $k=6$.}
\label{bi_b12_1}
\end{figure}

\begin{figure}
\includegraphics[scale=0.7, page=2]{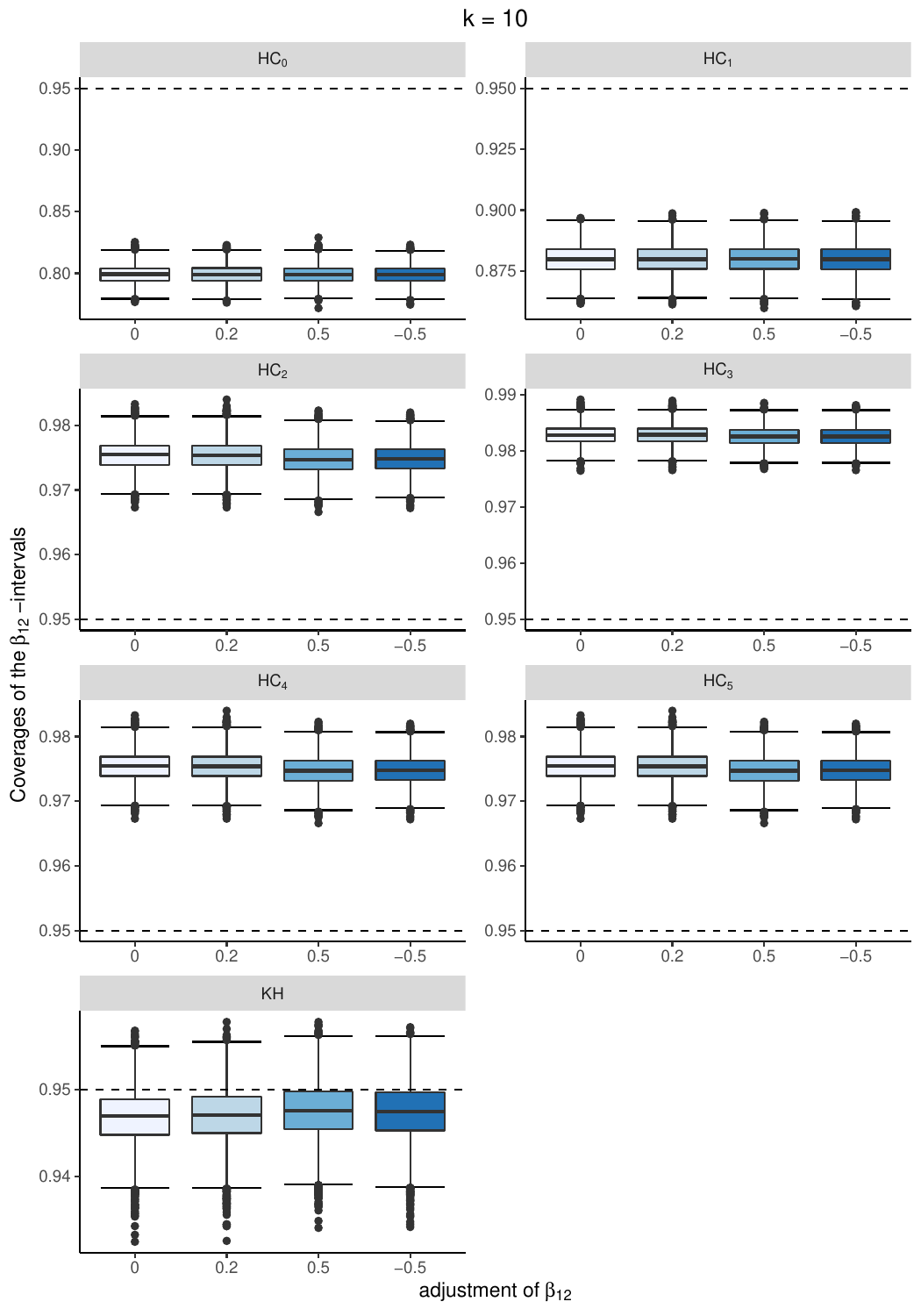}
\caption{Coverages of the \bi -intervals compared regarding the adjustments of $\beta_{12}$ for the estimators $HC_0-HC_5$ and $KH$ with $k=10$.}
\label{bi_b12_2}
\end{figure}

\begin{figure}
\includegraphics[scale=0.7, page=3]{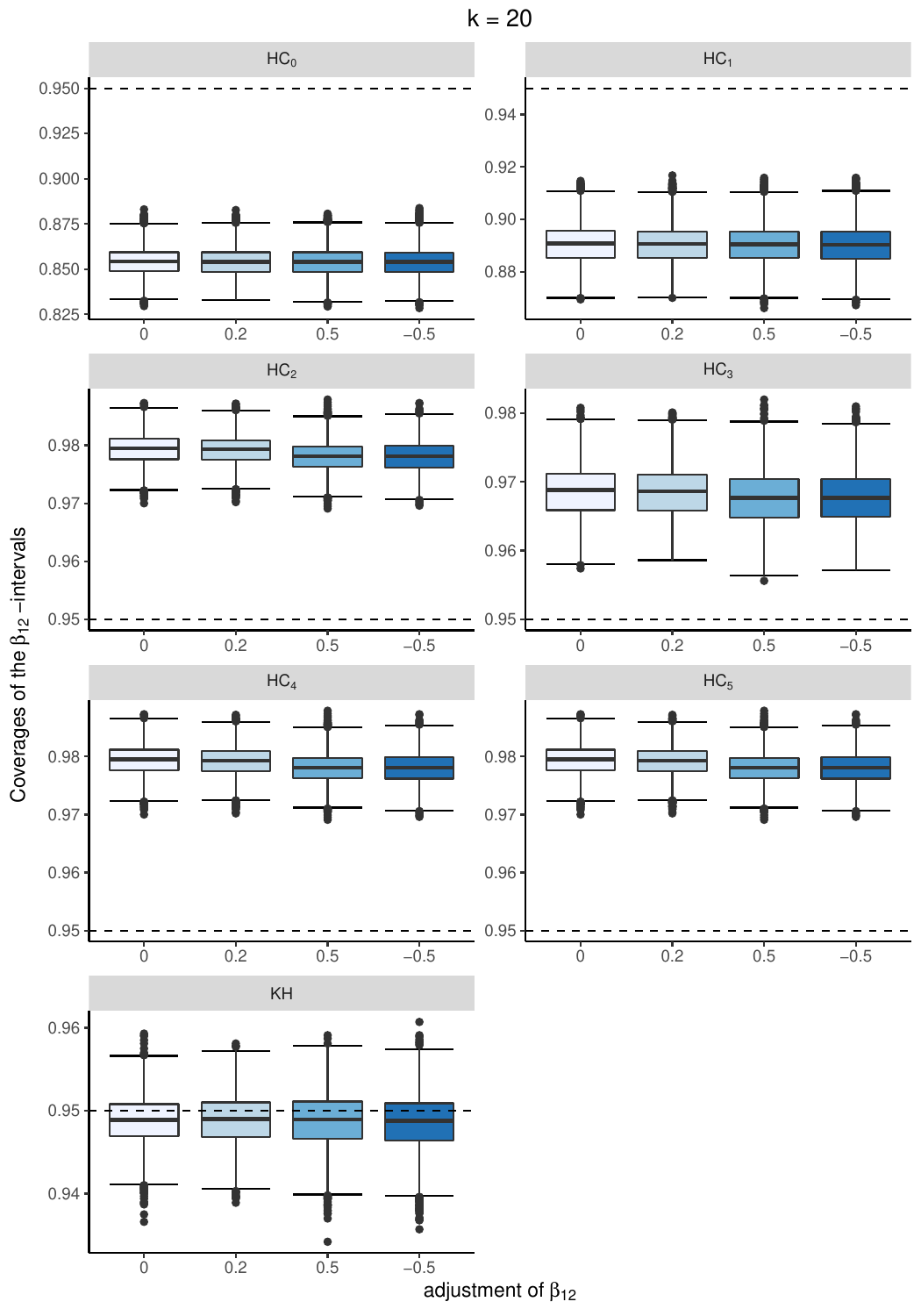}
\caption{Coverages of the \bi -intervals compared regarding the adjustments of $\beta_{12}$ for the estimators $HC_0-HC_5$ and $KH$ with $k=20$.}
\label{bi_b12_3}
\end{figure}

\begin{figure}
\includegraphics[scale=0.7, page=4]{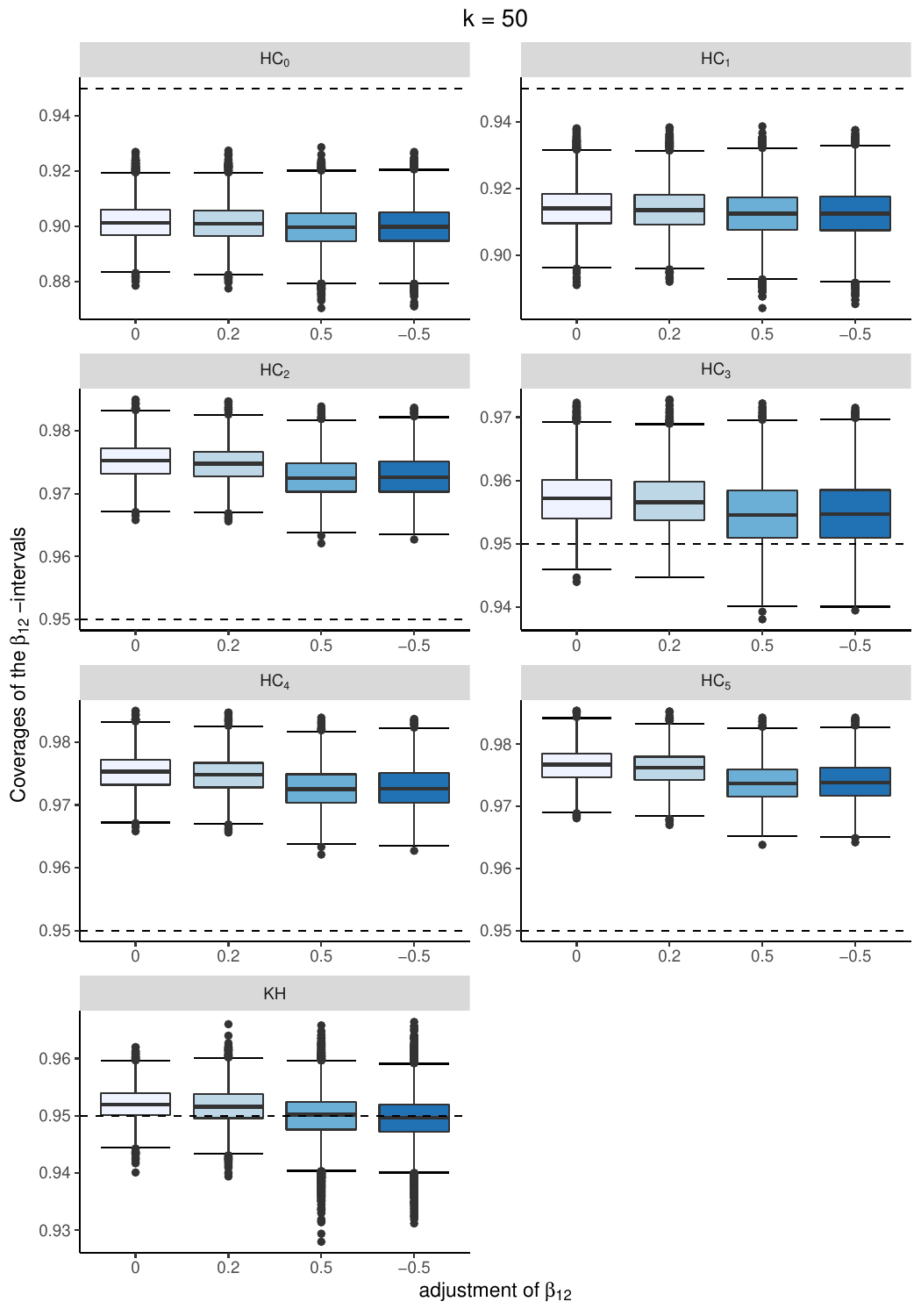}
\caption{Coverages of the \bi -intervals compared regarding the adjustments of $\beta_{12}$ for the estimators $HC_0-HC_5$ and $KH$ with $k=50$.}
\label{bi_b12_4}
\end{figure}

\begin{figure}
\includegraphics[scale=0.7, page=1]{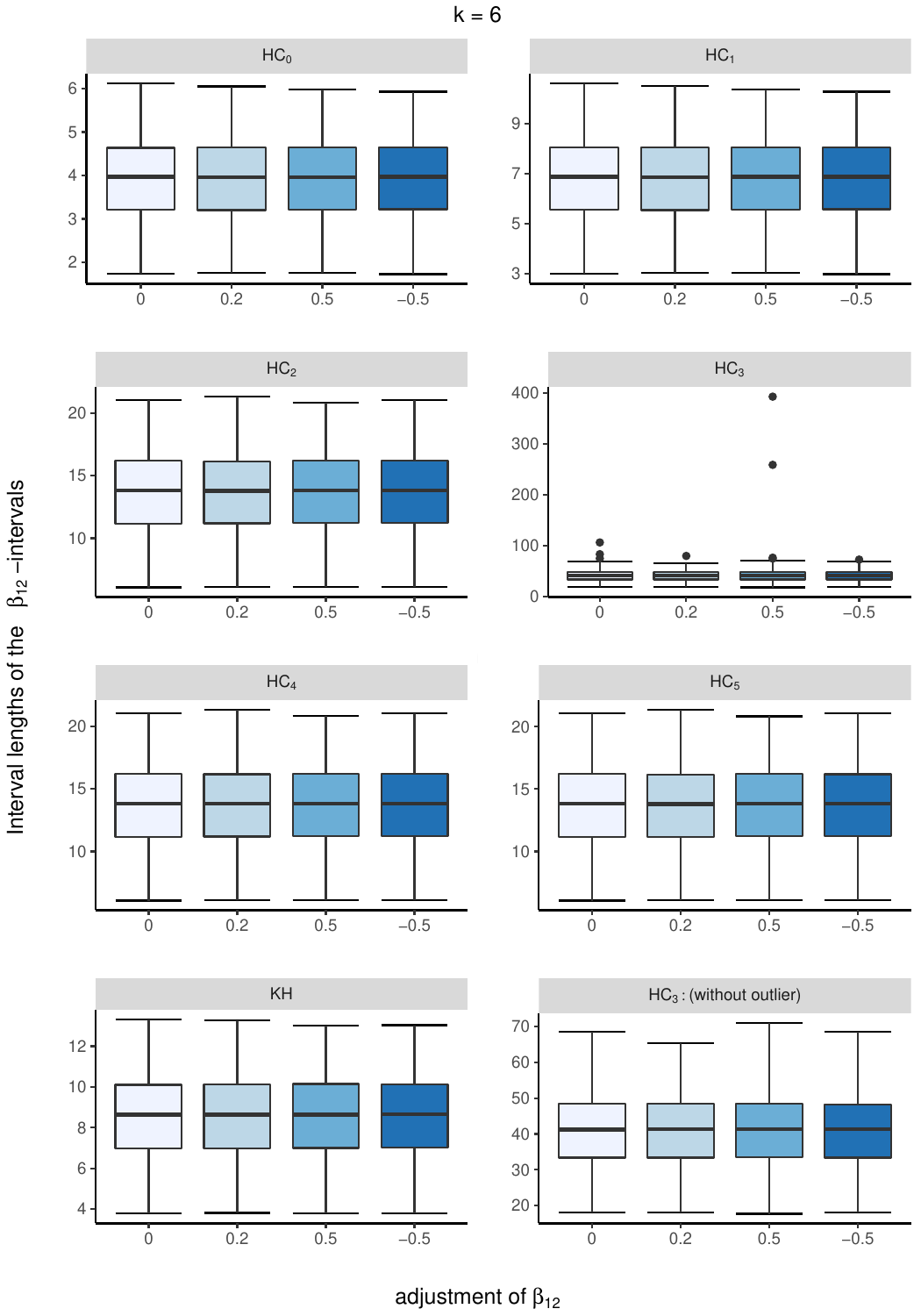}
\caption{Lengths of the \bi -intervals compared regarding the adjustments of $\beta_{12}$ for the estimators $HC_0-HC_5$ and $KH$ with $k=6$.}
\label{bi_b12_l_1}
\end{figure}

\begin{figure}
\includegraphics[scale=0.7, page=3]{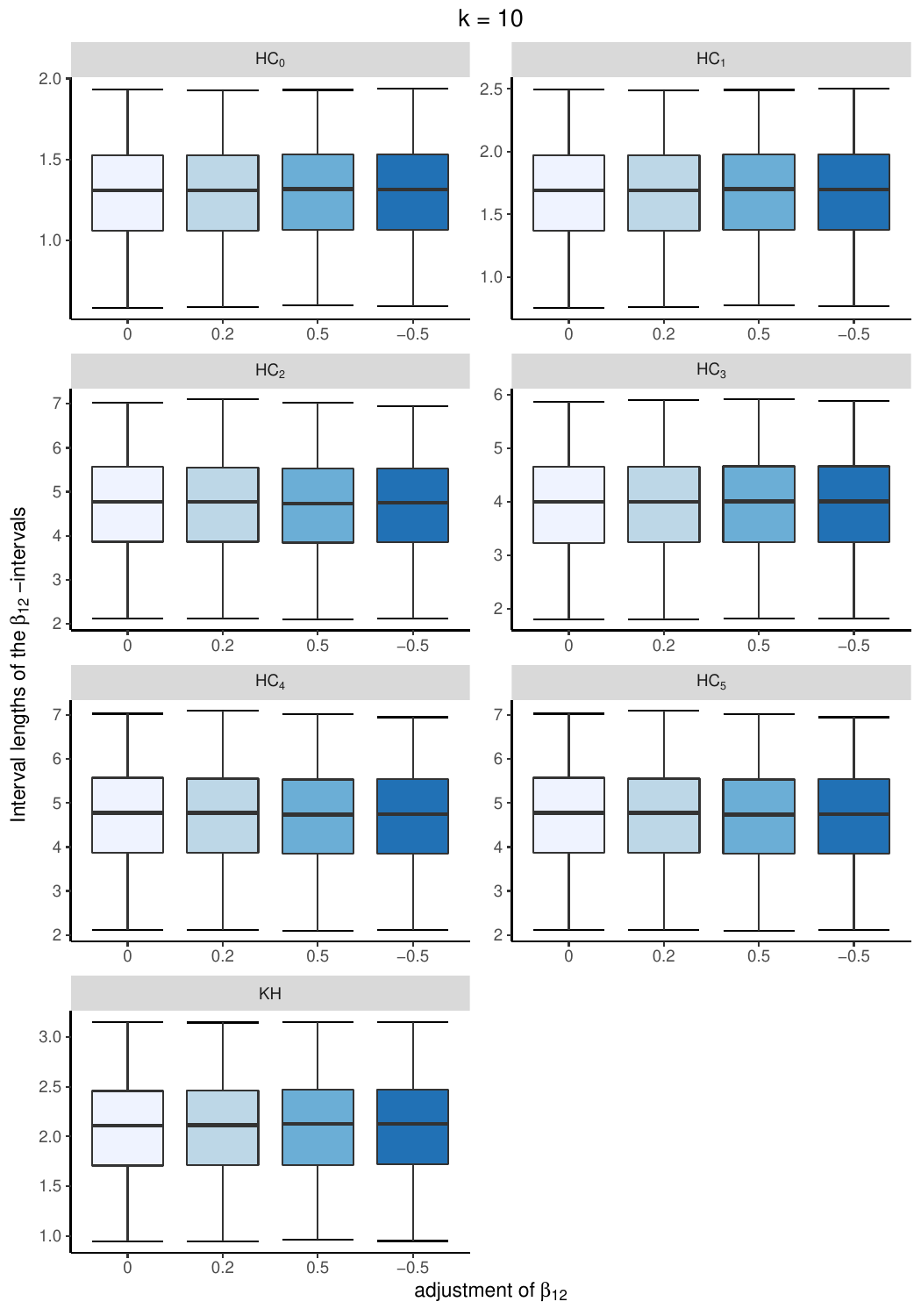}
\caption{Lengths of the \bi -intervals compared regarding the adjustments of $\beta_{12}$ for the estimators $HC_0-HC_5$ and $KH$ with $k=10$.}
\label{bi_b12_l_2}
\end{figure}

\begin{figure}
\includegraphics[scale=0.7, page=4]{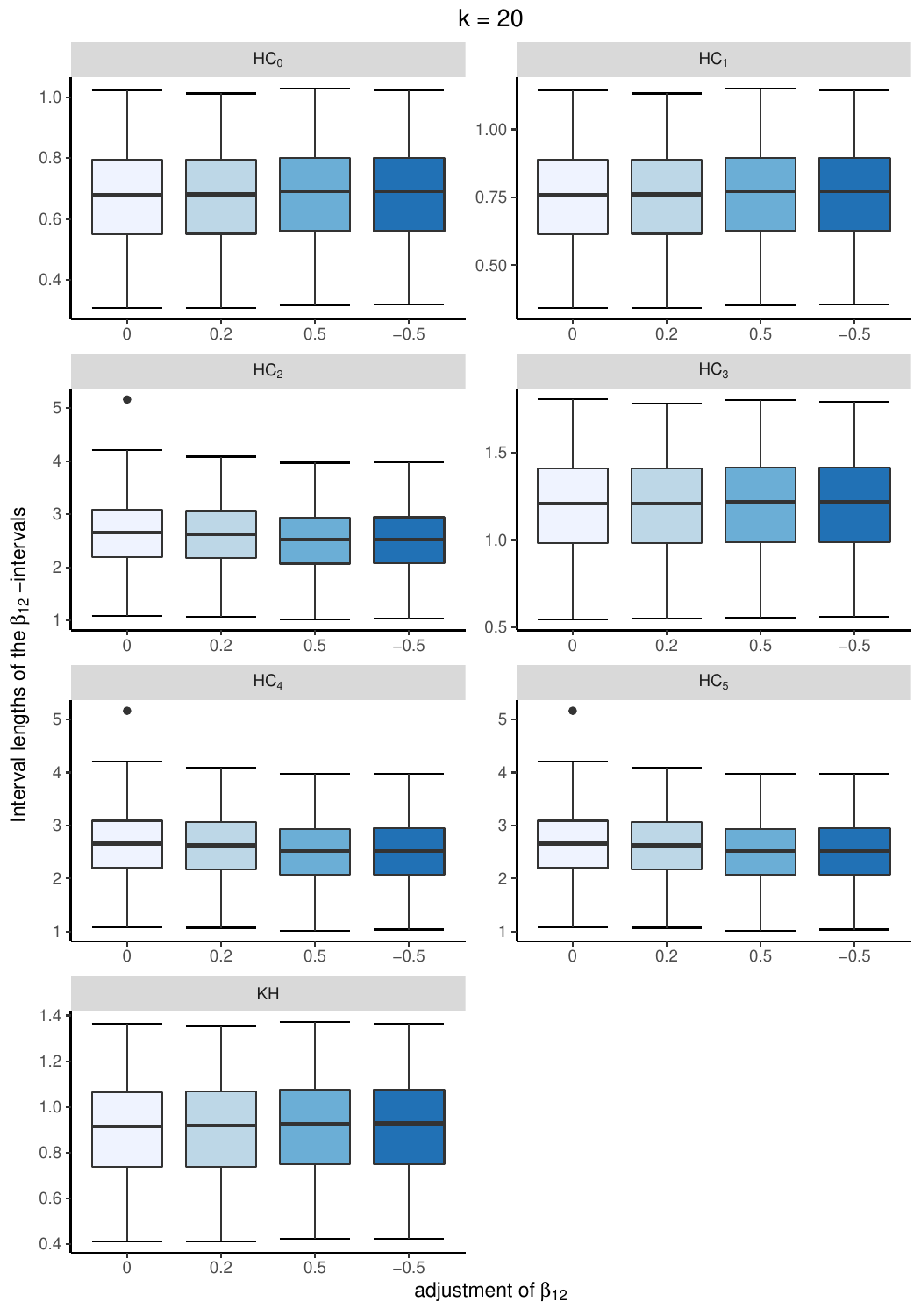}
\caption{Lengths of the \bi -intervals compared regarding the adjustments of $\beta_{12}$ for the estimators $HC_0-HC_5$ and $KH$ with $k=20$.}
\label{bi_b12_l_3}
\end{figure}

\begin{figure}
\includegraphics[scale=0.7, page=2]{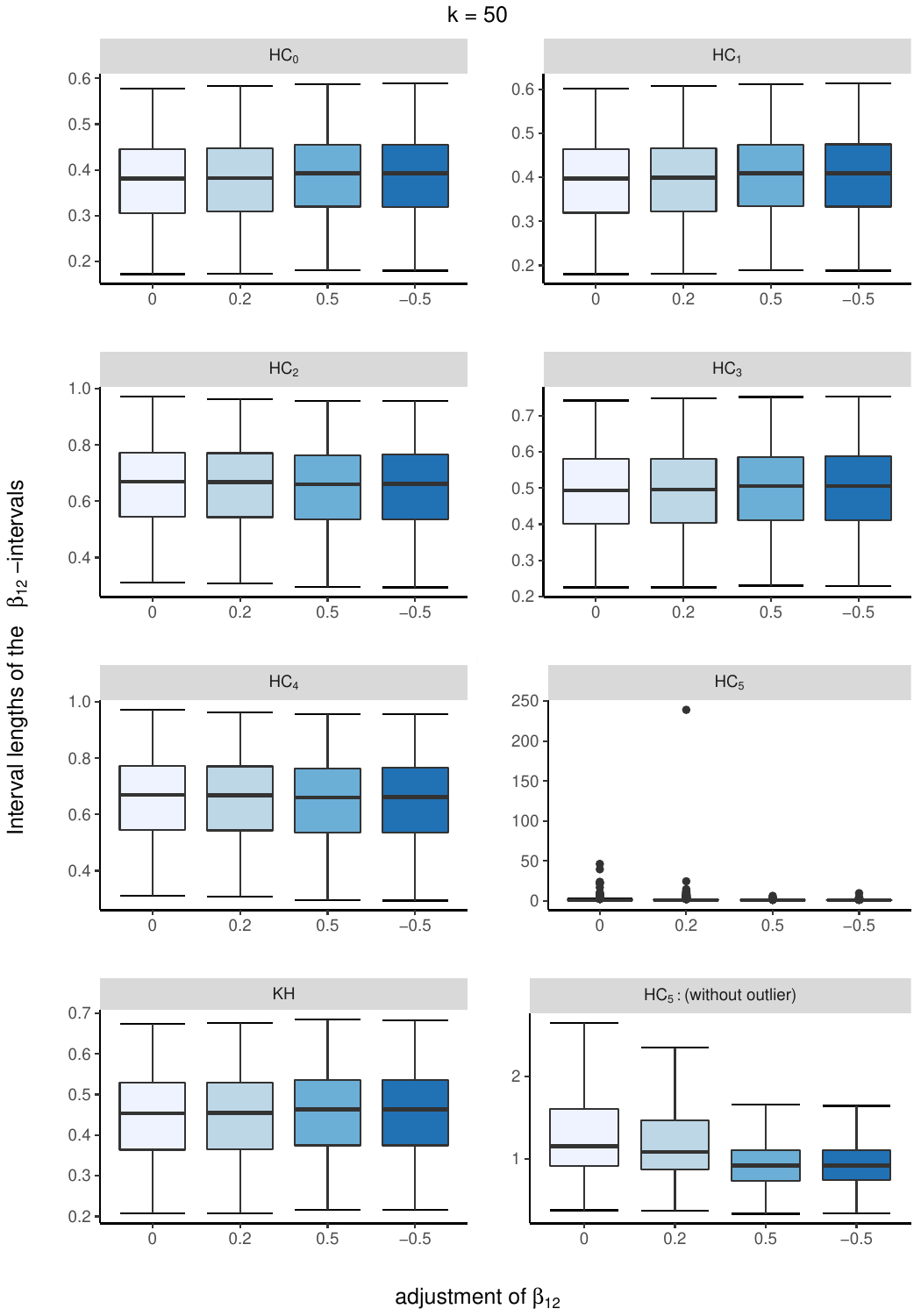}
\caption{Lengths of the \bi -intervals compared regarding the adjustments of $\beta_{12}$ for the estimators $HC_0-HC_5$ and $KH$ with $k=50$.}
\label{bi_b12_l_4}
\end{figure}

\textbf{Adjustments of $\boldsymbol{\beta_{12}}$} Besides the adjustments 0, 0.2 and 0.5 for $\beta_{12}$ the adjustment -0.5 was simulated as well, to check whether it differs from the 0.5 adjustment. This is neither the case for the interval lengths nor for the coverages of the CIs for $\beta_1$ and the CIs for $\beta_{12}$. However, the $\textbf{HC}_4$- and $\textbf{HC}_5$-CIs for $\beta_1$ have slightly lower coverage for a high absolute value of $\beta_{12}$. The effect is increasing in the number of studies $k$ (Figures \ref{b1_b12_1}-\ref{b1_b12_4}). For the CIs for $\beta_{12}$, this effect is seen not only for the $\textbf{HC}_4$- and $\textbf{HC}_5$-CIs but also for the $\textbf{HC}_2$- and $\textbf{HC}_3$-CIs. Aditionally, for $k=50$ the effect is observable for $\textbf{HC}_0$, $\textbf{HC}_1$ and $\textbf{KH}$ based CIs (Figures \ref{bi_b12_1}-\ref{bi_b12_4}).\\
The adjustment of $\beta_{12}$ has a marginal effect on the interval lengths of the CIs for $\beta_1$ and $\beta_{12}$ (Figures \ref{b1_b12_l_1}-\ref{b1_b12_l_4} and \ref{bi_b12_l_1}-\ref{bi_b12_l_4}). Only for $k=20$ the $\textbf{HC}_4$- and $\textbf{HC}_5$-CIs for $\beta_1$ have a slightly shorter length for $\beta_{12}\in\{-0.5, 0.5\}$ than for $\beta_{12}\in\{0, 0.2\}$ (Figure \ref{b1_b12_l_3}). Similarly the CIs for $\beta_{12}$ based on $\textbf{HC}_2$, $\textbf{HC}_4$ and $\textbf{HC}_5$ tend to be slightly shorter for higher absolute values of $\beta_{12}$ and $k=20$ (Figure \ref{bi_b12_l_3})\\It is also noticeable that most of the extreme outliers of $\textbf{HC}_5$-CIs occur for $\beta_{12}\in\{0,0.2\}$ (Figures \ref{b1_b12_l_4} and \ref{bi_b12_l_4}).\\

Altogether the true values of the considered parameters do not have a strong impact on the intervals of any estimator. Therefore, there is no coefficient for which an estimator performs better or worse compared to the other estimators than in the overall results.\\

\begin{figure}
\includegraphics[scale=0.7, page=1]{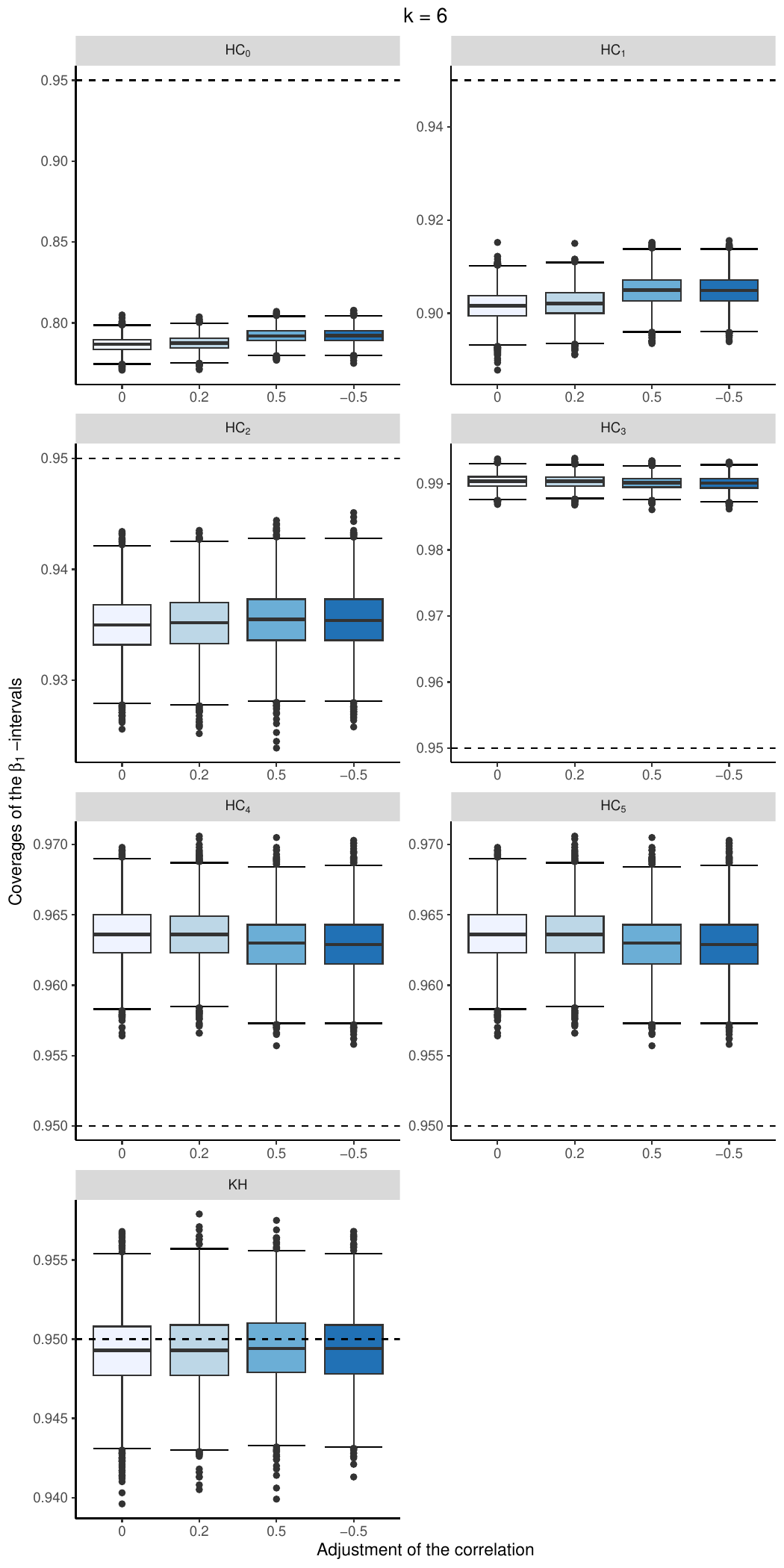}
\caption{Coverages of the \bo -intervals compared regarding the adjustments of the correlation $\rho$ for the estimators $HC_0-HC_5$ and $KH$ with $k=6$.}
\label{b1_cor_1}
\end{figure}

\begin{figure}
\includegraphics[scale=0.7, page=2]{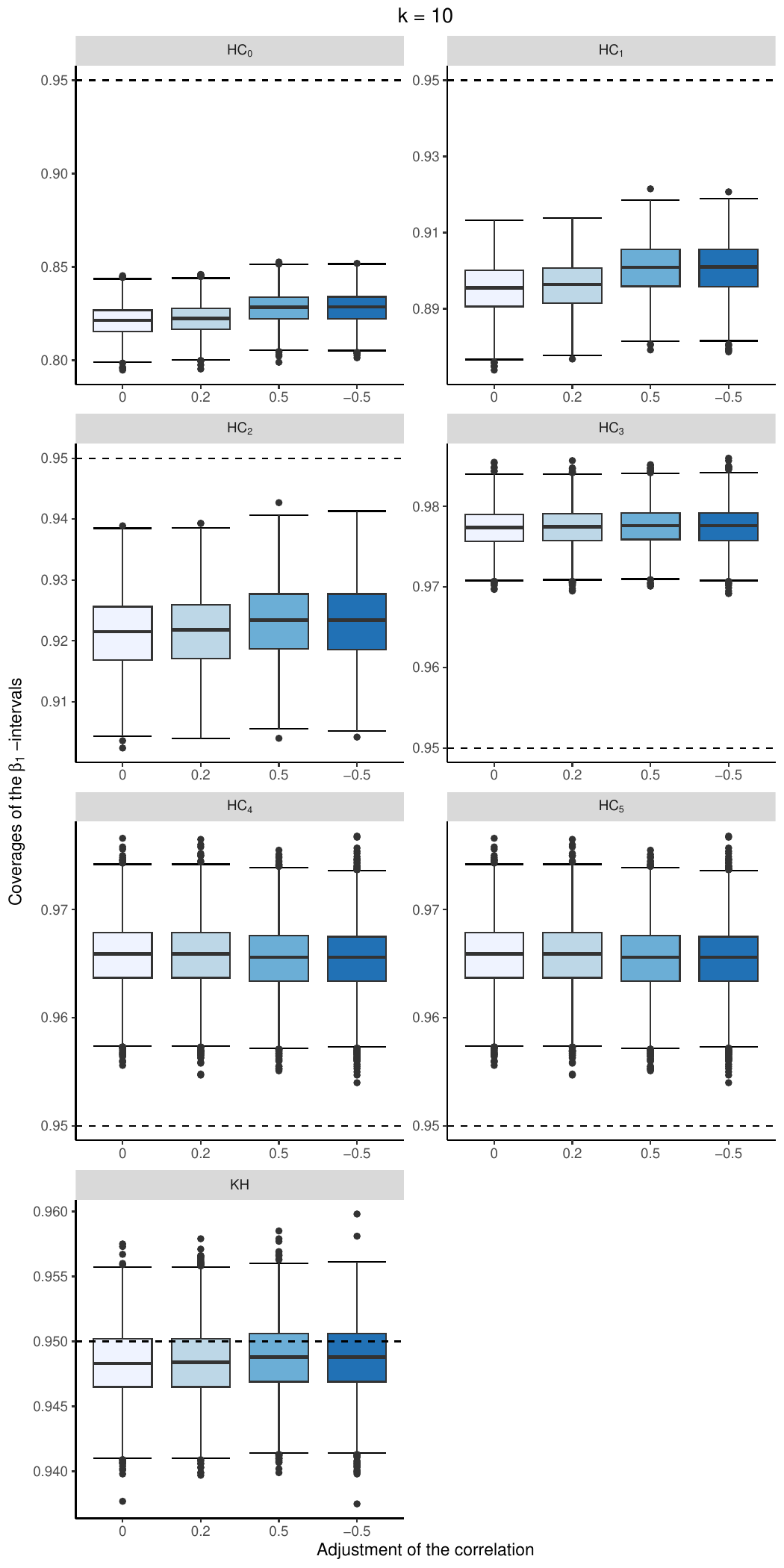}
\caption{Coverages of the \bo -intervals compared regarding the adjustments of the correlation $\rho$ for the estimators $HC_0-HC_5$ and $KH$ with $k=10$.}
\label{b1_cor_2}
\end{figure}

\begin{figure}
\includegraphics[scale=0.7, page=3]{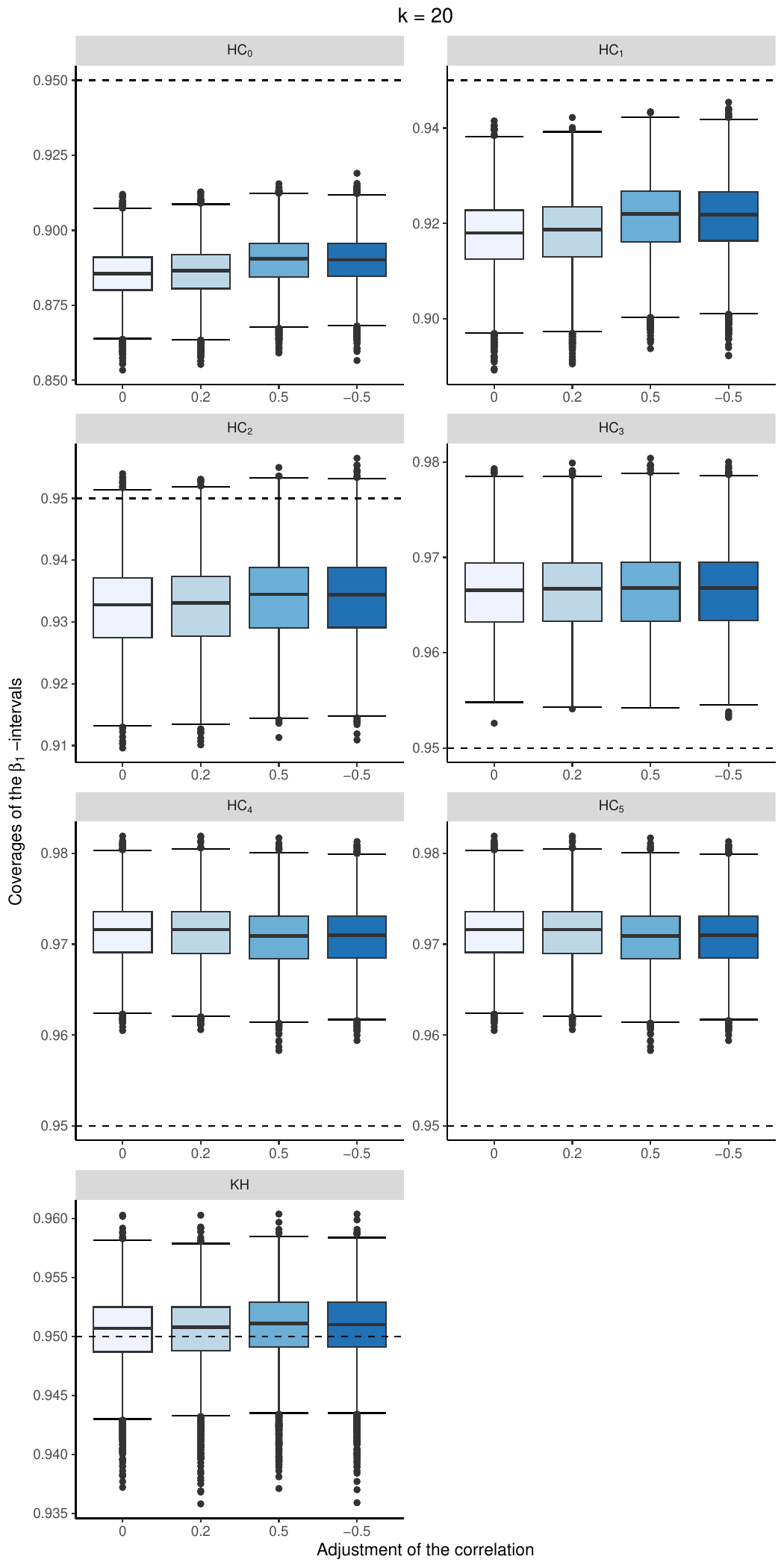}
\caption{Coverages of the \bo -intervals compared regarding the adjustments of the correlation $\rho$ for the estimators $HC_0-HC_5$ and $KH$ with $k=20$.}
\label{b1_cor_3}
\end{figure}

\begin{figure}
\includegraphics[scale=0.7, page=4]{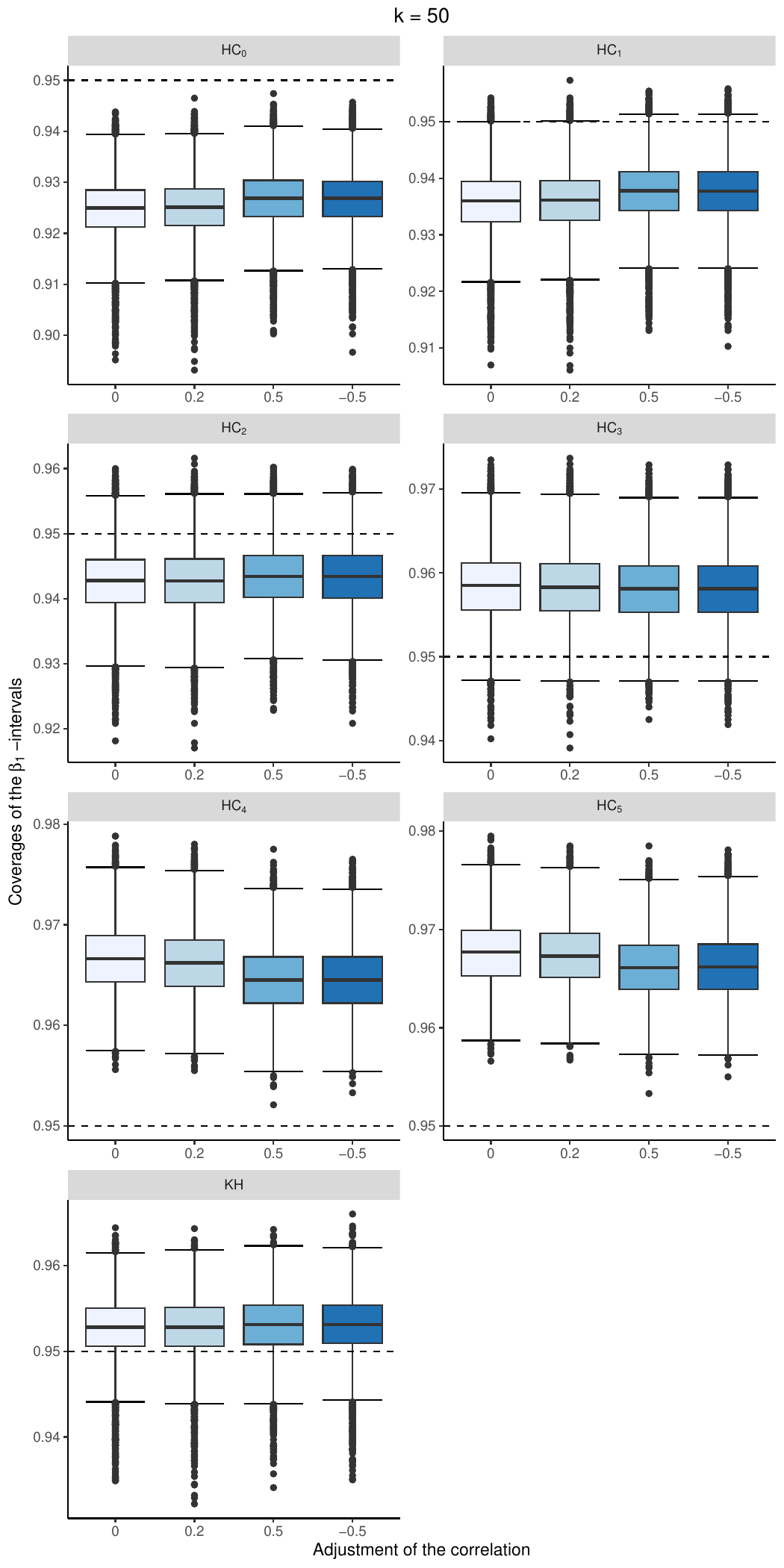}
\caption{Coverages of the \bo -intervals compared regarding the adjustments of the correlation $\rho$ for the estimators $HC_0-HC_5$ and $KH$ with $k=50$.}
\label{b1_cor_4}
\end{figure}

\begin{figure}
\includegraphics[scale=0.7, page=1]{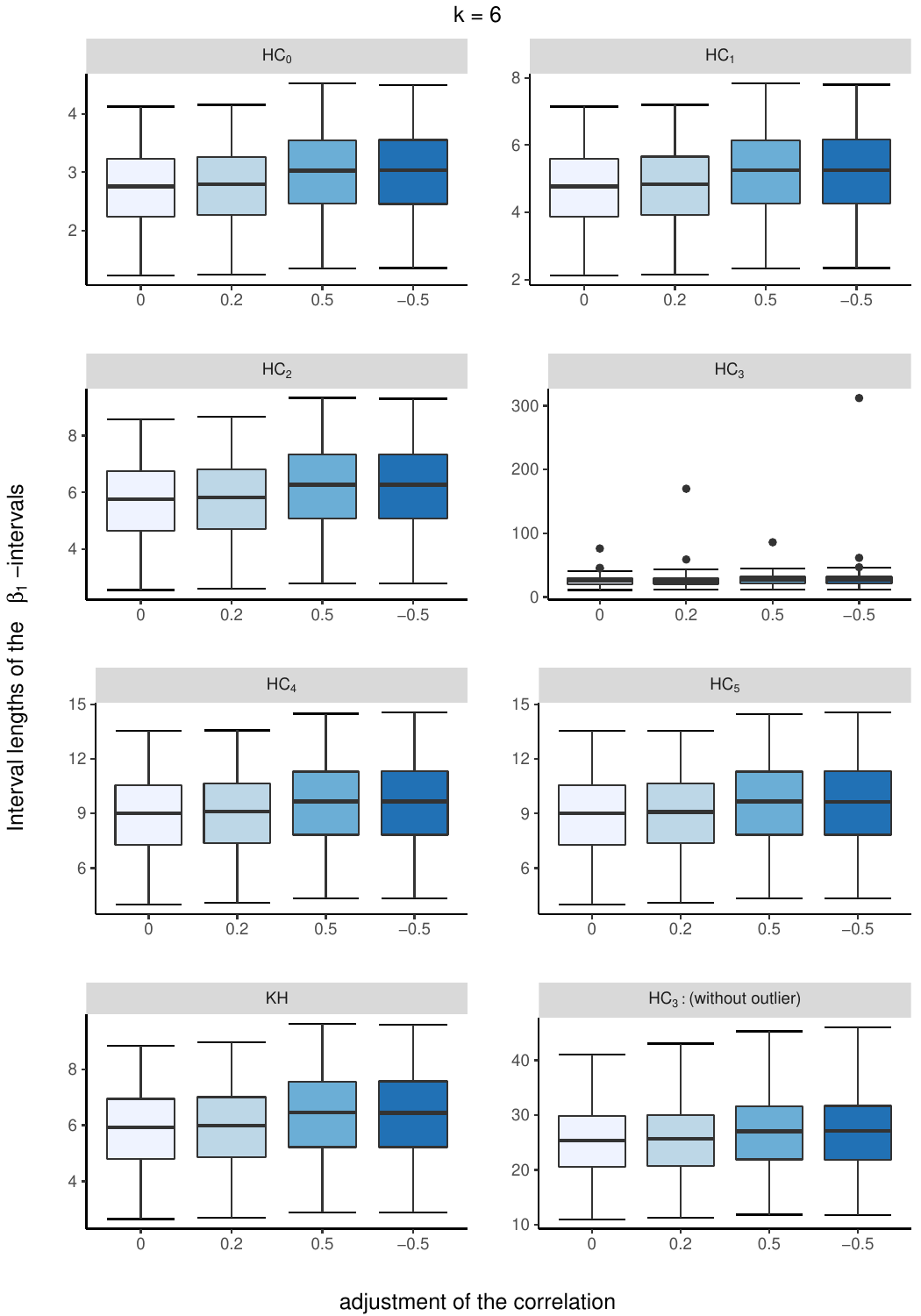}
\caption{Lengths of the \bo -intervals compared regarding the adjustments of the correlation $\rho$ for the estimators $HC_0-HC_5$ and $KH$ with $k=6$.}
\label{b1_cor_l_1}
\end{figure}

\begin{figure}
\includegraphics[scale=0.7, page=3]{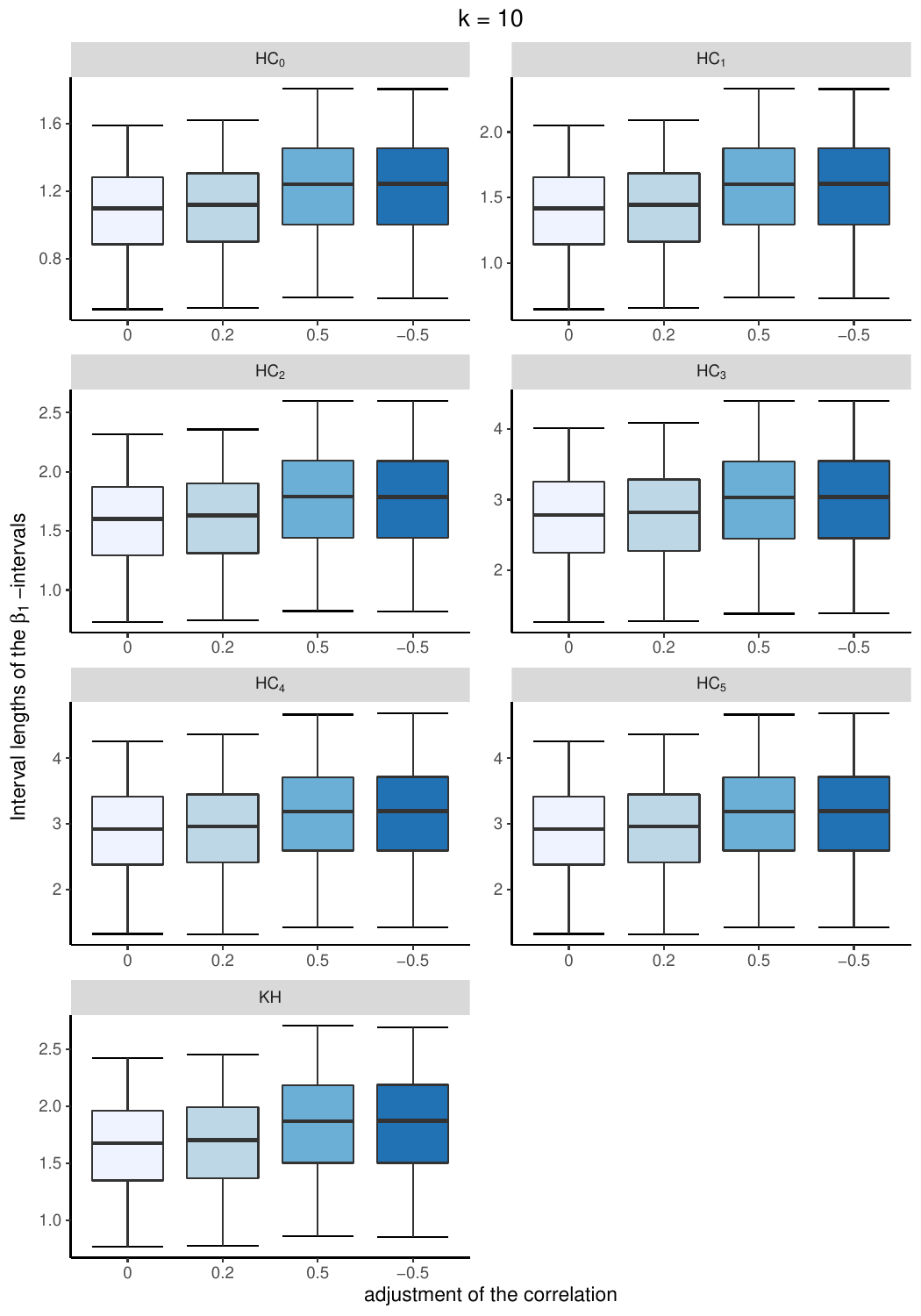}
\caption{Lengths of the \bo -intervals compared regarding the adjustments of the correlation $\rho$ for the estimators $HC_0-HC_5$ and $KH$ with $k=10$.}
\label{b1_cor_l_2}
\end{figure}

\begin{figure}
\includegraphics[scale=0.7, page=4]{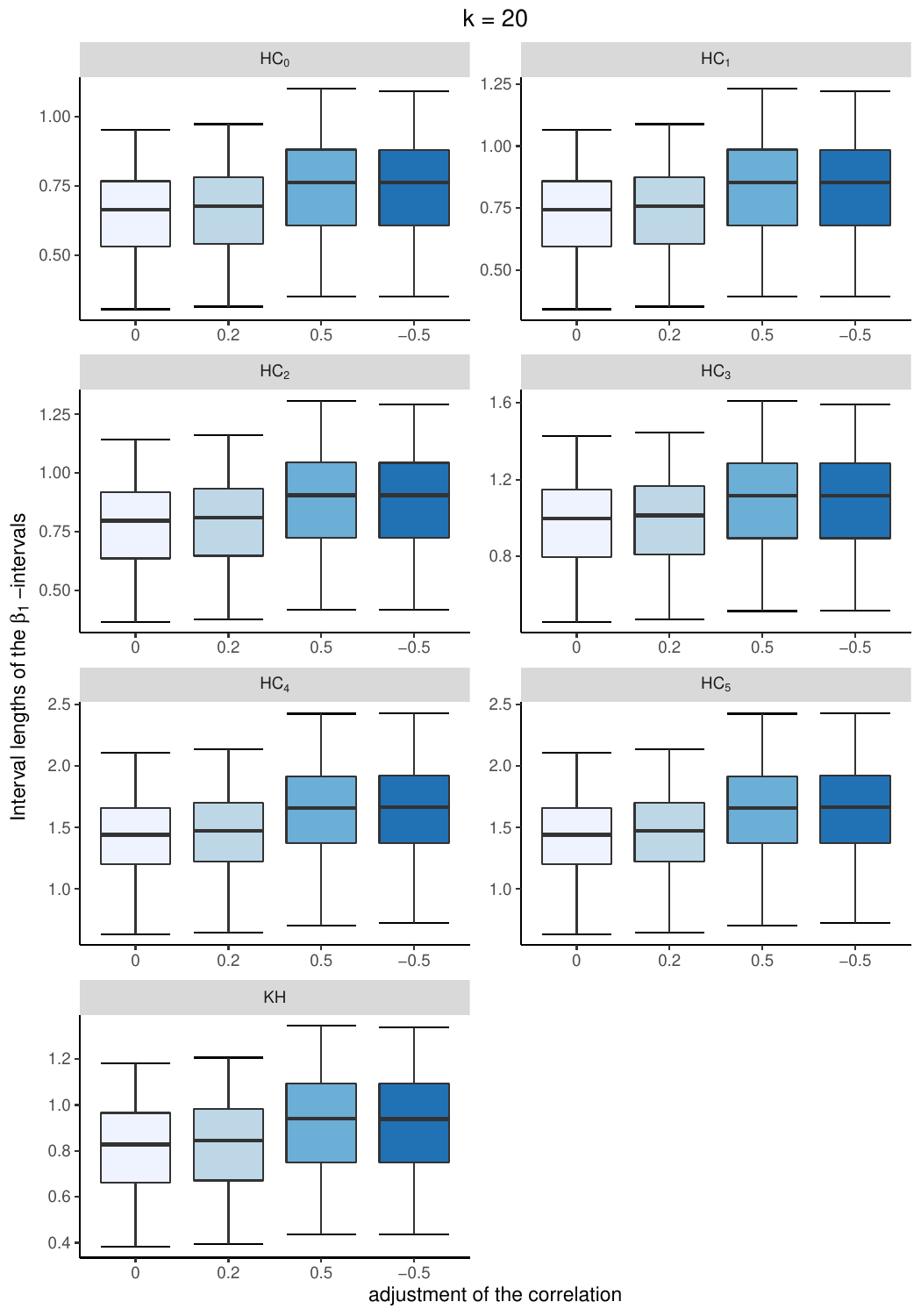}
\caption{Lengths of the \bo -intervals compared regarding the adjustments of the correlation $\rho$ for the estimators $HC_0-HC_5$ and $KH$ with $k=20$.}
\label{b1_cor_l_3}
\end{figure}

\begin{figure}
\includegraphics[scale=0.7, page=2]{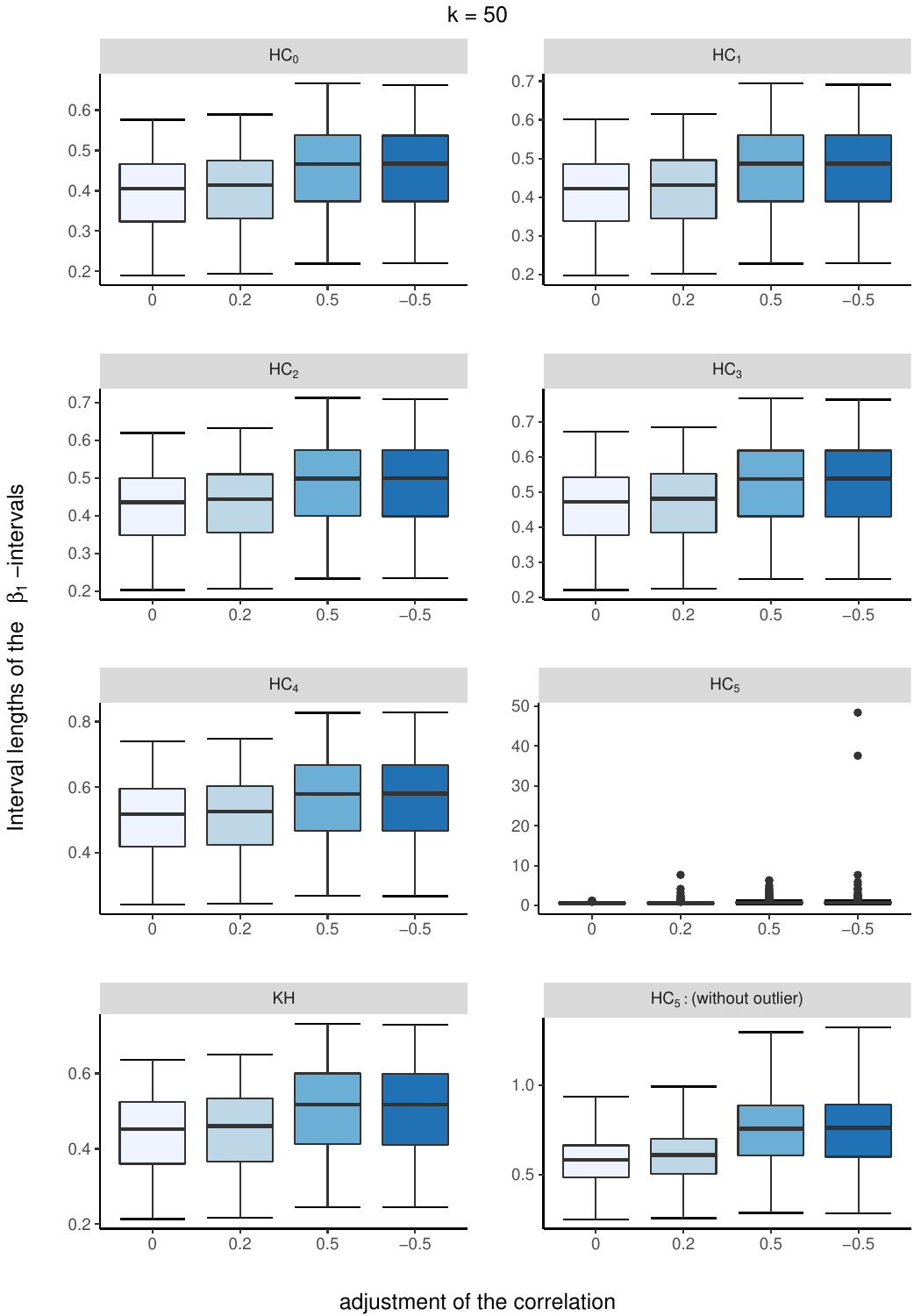}
\caption{Lengths of the \bo -intervals compared regarding the adjustments of the correlation $\rho$ for the estimators $HC_0-HC_5$ and $KH$ with $k=50$.}
\label{b1_cor_l_4}
\end{figure}

\begin{figure}
\includegraphics[scale=0.7, page=1]{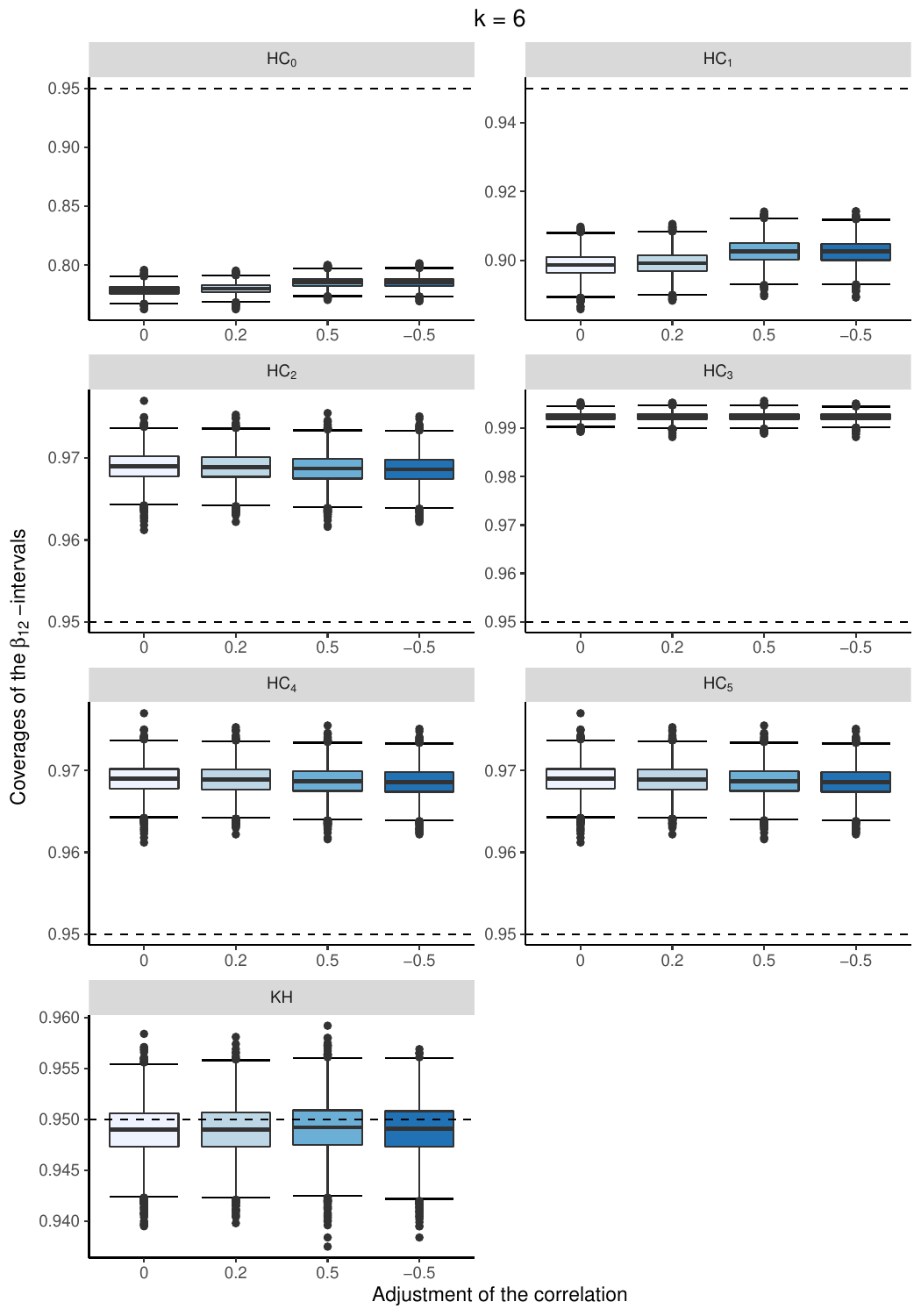}
\caption{Coverages of the \bi -intervals compared regarding the adjustments of the correlation $\rho$ for the estimators $HC_0-HC_5$ and $KH$ with $k=6$.}
\label{bi_cor_1}
\end{figure}

\begin{figure}
\includegraphics[scale=0.7, page=2]{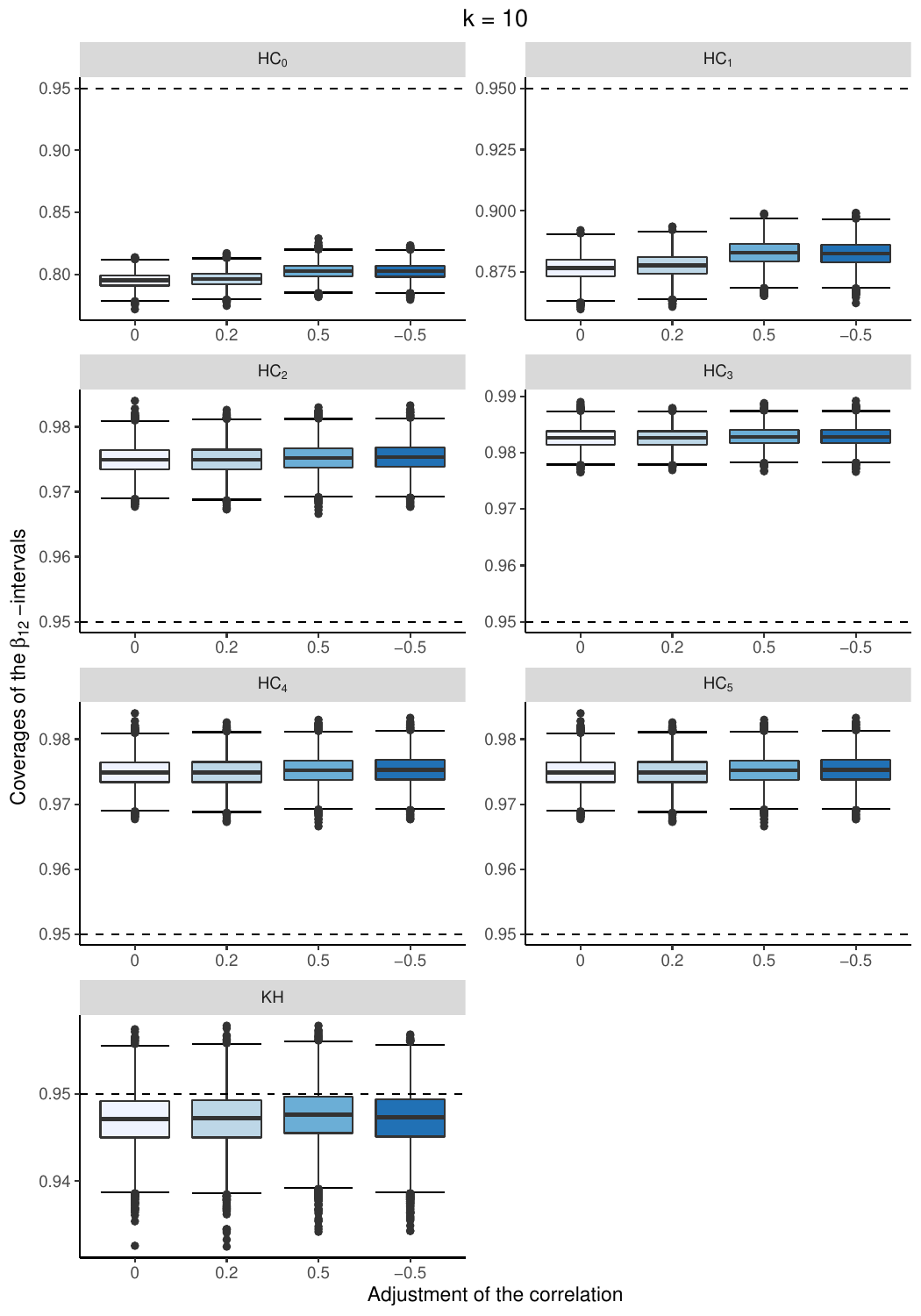}
\caption{Coverages of the \bi -intervals compared regarding the adjustments of the correlation $\rho$ for the estimators $HC_0-HC_5$ and $KH$ with $k=10$.}
\label{bi_cor_2}
\end{figure}

\begin{figure}
\includegraphics[scale=0.7, page=3]{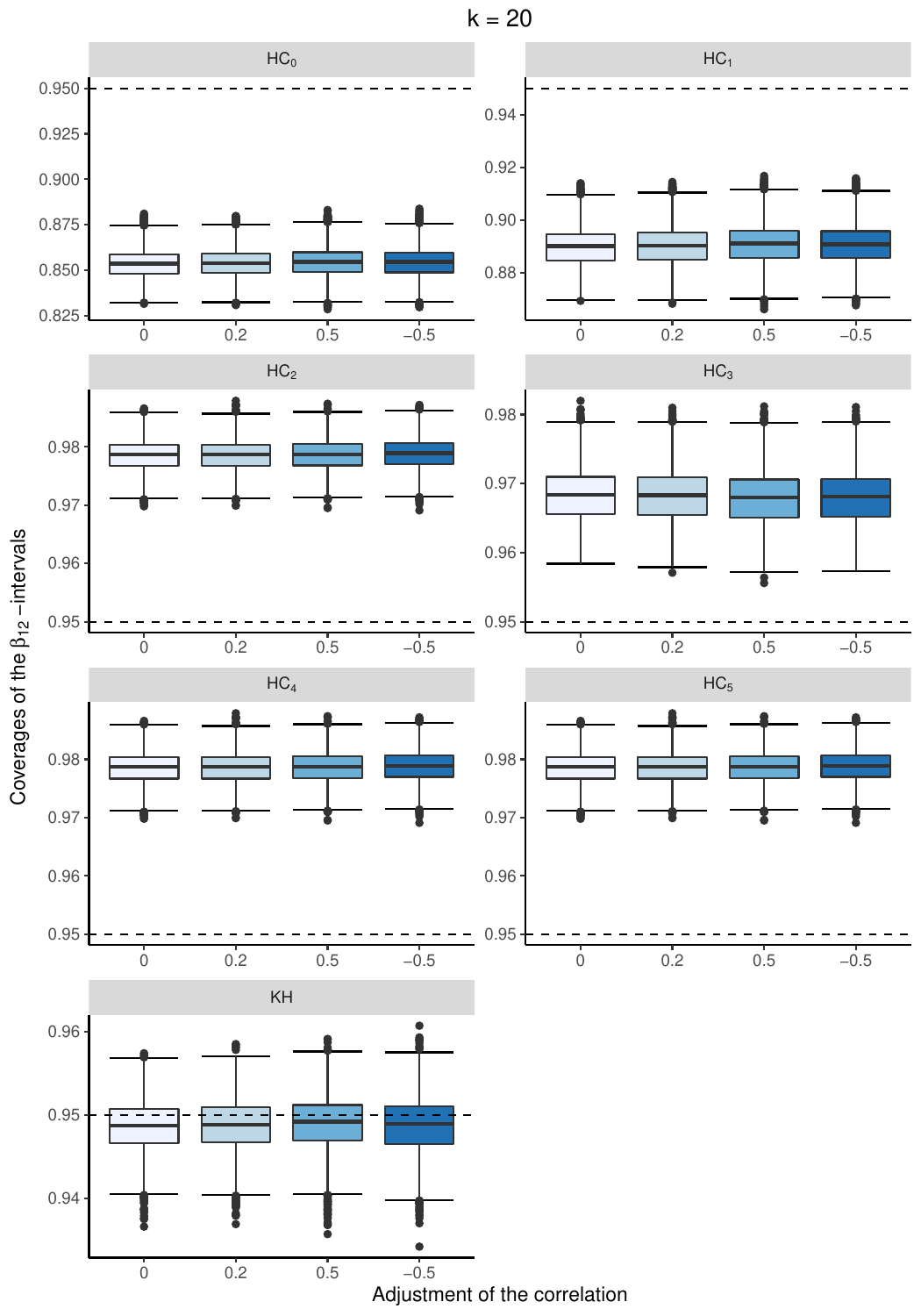}
\caption{Coverages of the \bi -intervals compared regarding the adjustments of the correlation $\rho$ for the estimators $HC_0-HC_5$ and $KH$ with $k=20$.}
\label{bi_cor_3}
\end{figure}

\begin{figure}
\includegraphics[scale=0.7, page=4]{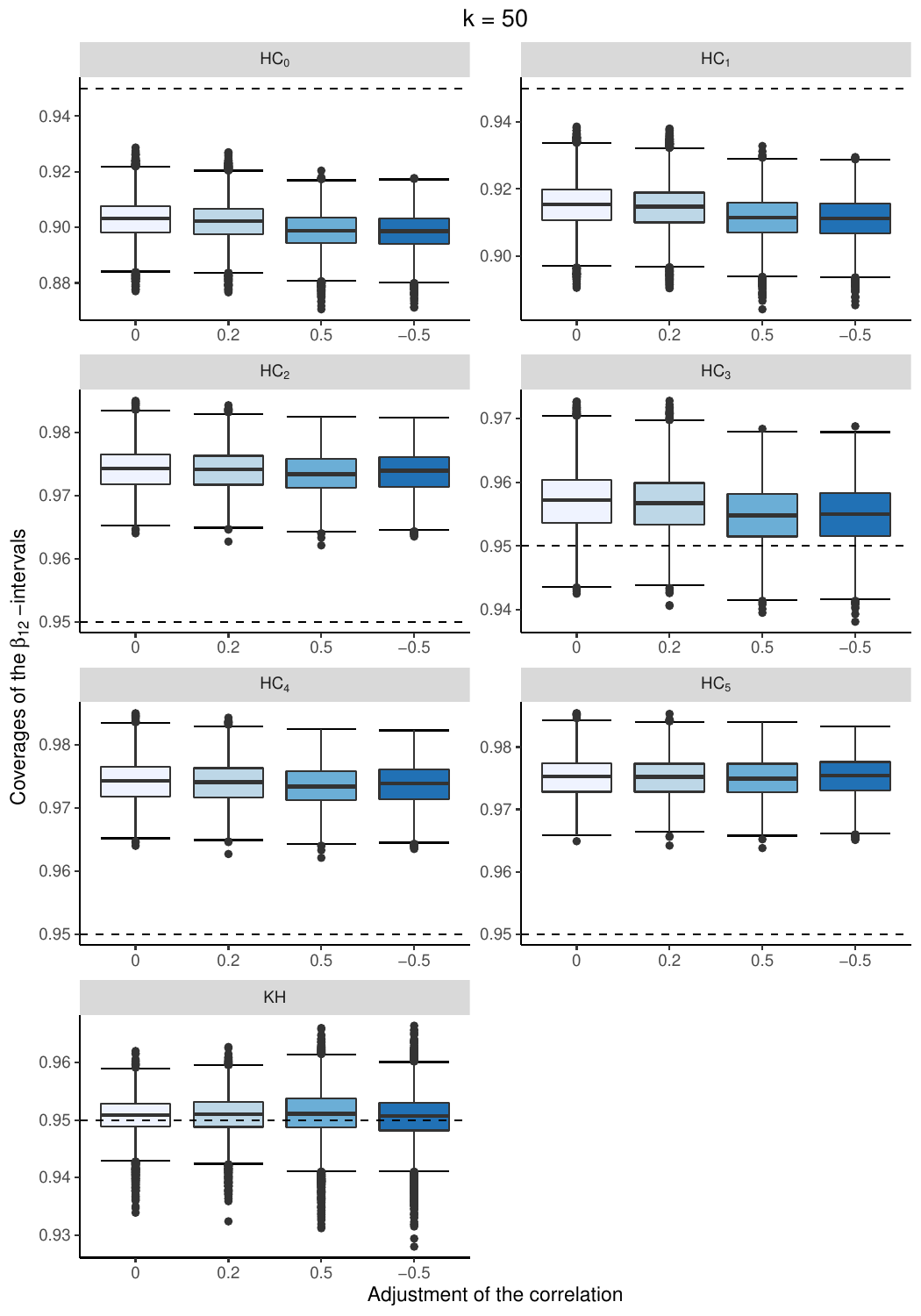}
\caption{Coverages of the \bi -intervals compared regarding the adjustments of the correlation $\rho$ for the estimators $HC_0-HC_5$ and $KH$ with $k=50$.}
\label{bi_cor_4}
\end{figure}

\begin{figure}
\includegraphics[scale=0.7, page=1]{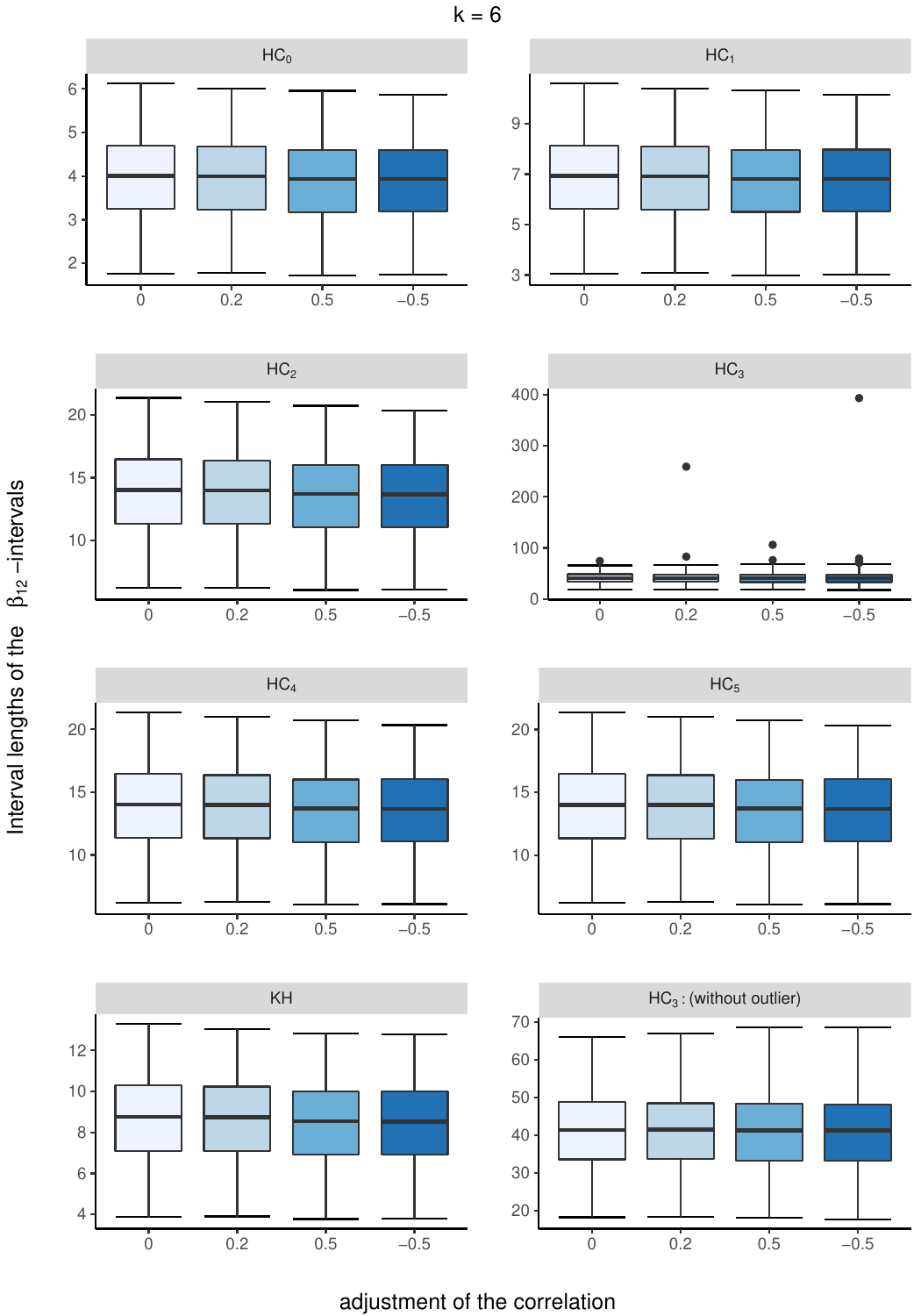}
\caption{Lengths of the \bi -intervals compared regarding the adjustments of the correlation $\rho$ for the estimators $HC_0-HC_5$ and $KH$ with $k=6$.}
\label{bi_cor_l_1}
\end{figure}

\begin{figure}
\includegraphics[scale=0.7, page=3]{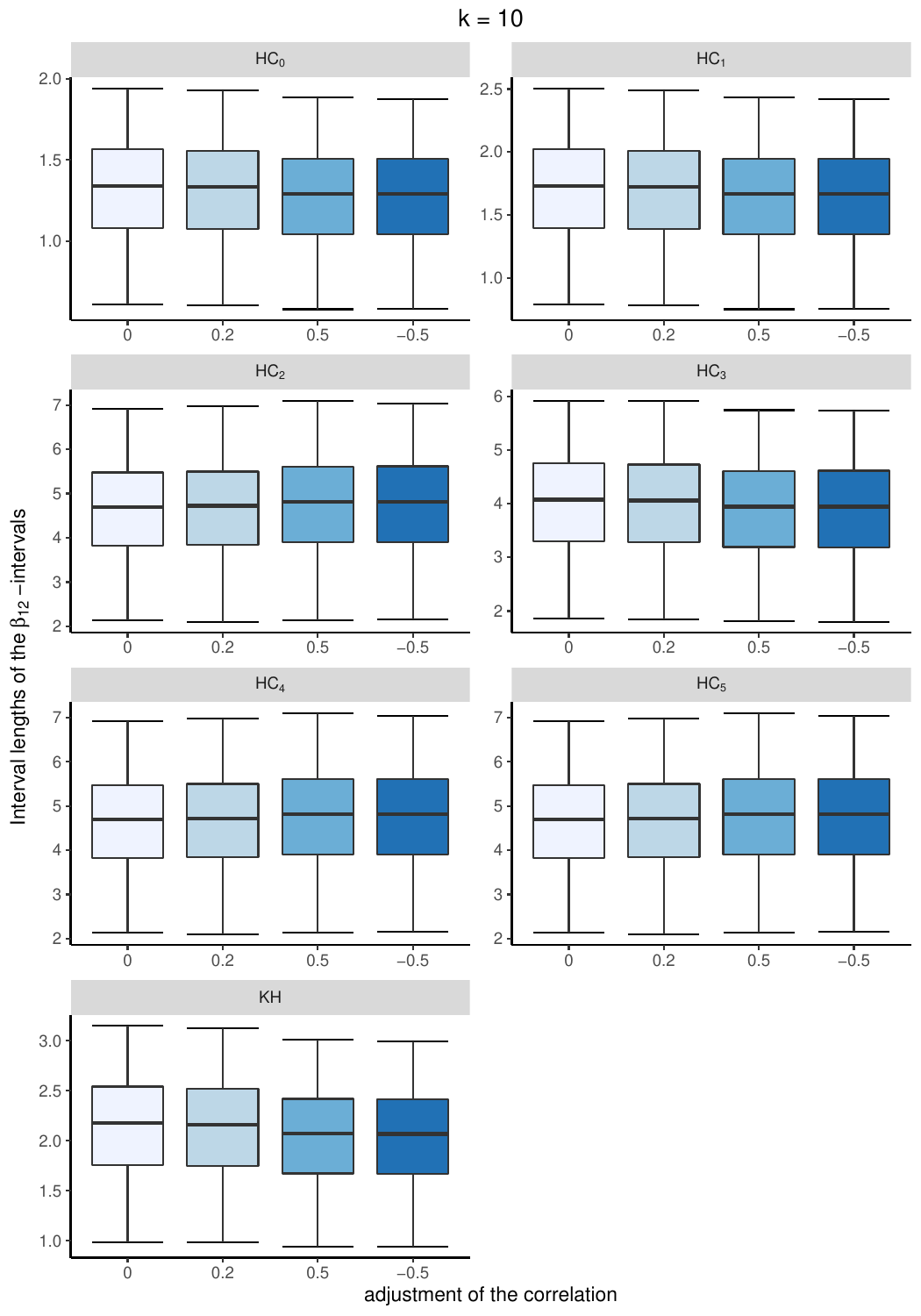}
\caption{Lengths of the \bi -intervals compared regarding the adjustments of the correlation $\rho$ for the estimators $HC_0-HC_5$ and $KH$ with $k=10$.}
\label{bi_cor_l_2}
\end{figure}

\begin{figure}
\includegraphics[scale=0.7, page=4]{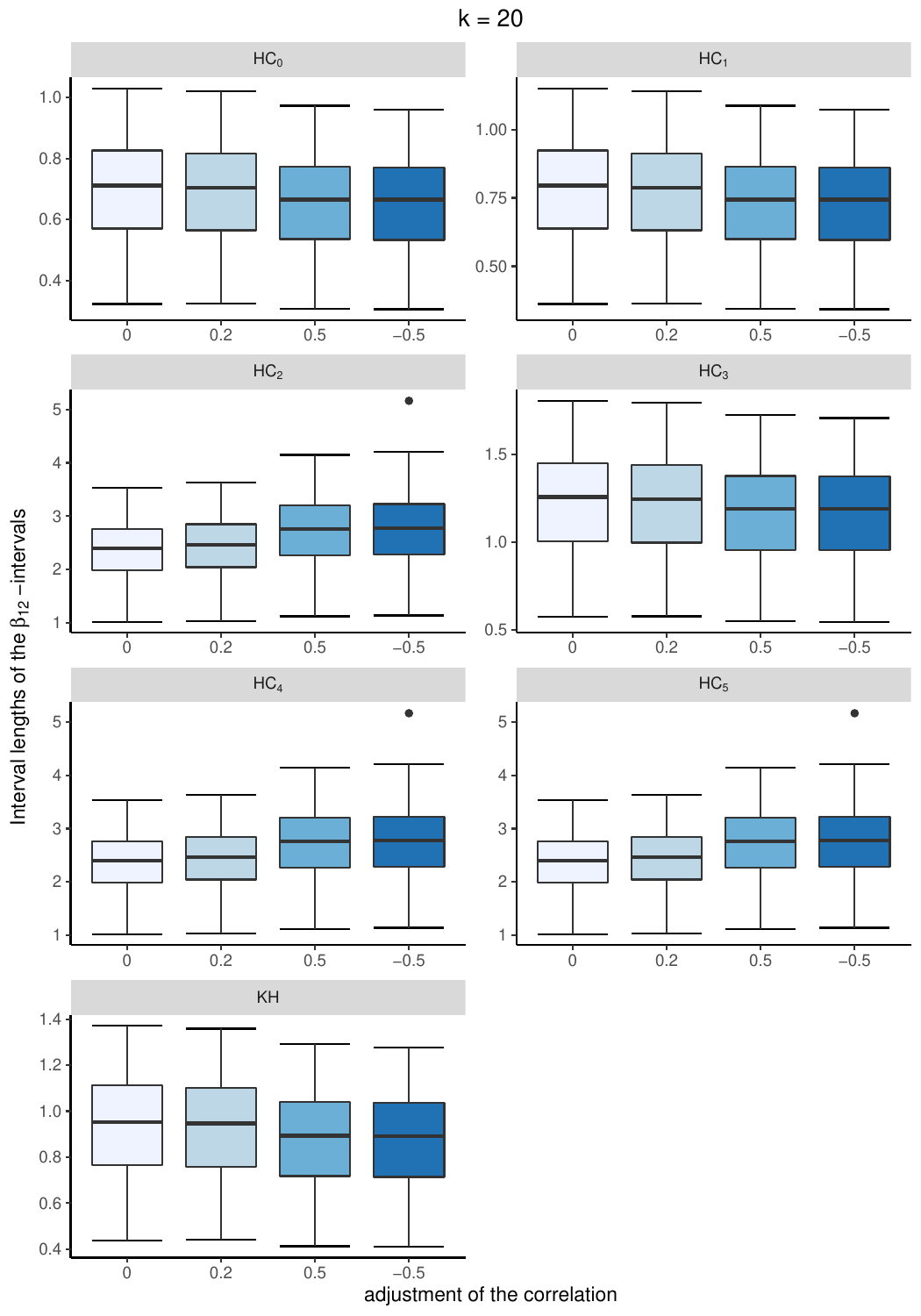}
\caption{Lengths of the \bi -intervals compared regarding the adjustments of the correlation $\rho$ for the estimators $HC_0-HC_5$ and $KH$ with $k=20$.}
\label{bi_cor_l_3}
\end{figure}

\begin{figure}
\includegraphics[scale=0.7, page=2]{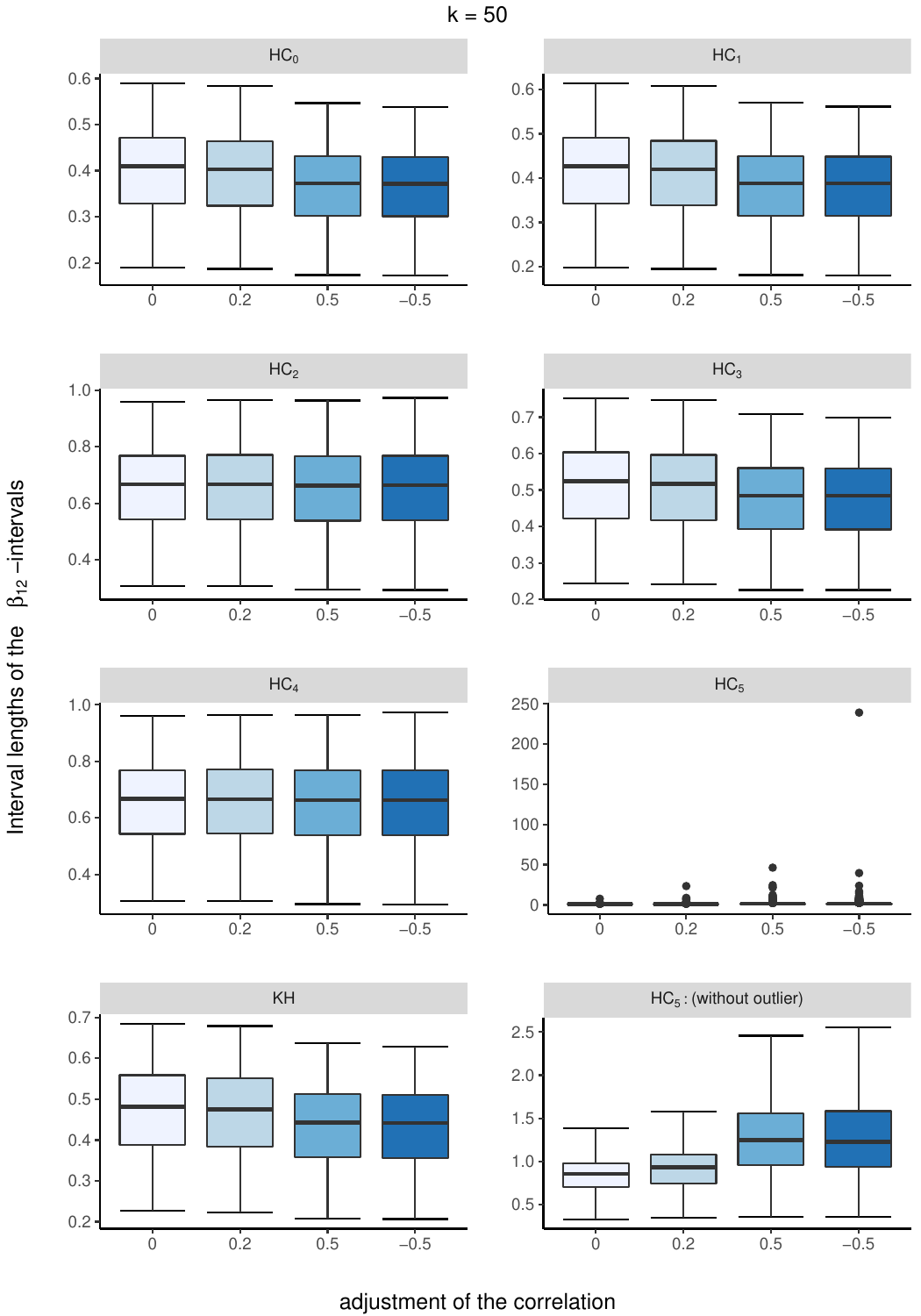}
\caption{Lengths of the \bi -intervals compared regarding the adjustments of the correlation $\rho$ for the estimators $HC_0-HC_5$ and $KH$ with $k=50$.}
\label{bi_cor_l_4}
\end{figure}

\textbf{Adjustments of the correlation $\boldsymbol{\rho}$} Examined adjustments of $\rho$ are 0, 0.2, 0.5 and -0.5. Intervals for $\beta_1$ that are based on $\textbf{HC}_3-\textbf{HC}_5$ tend to have a lower coverages for higher correlations, whereas CIs based on  $\textbf{HC}_0-\textbf{HC}_2$ tend to have a higher coverages for $\lvert\rho\rvert=0.5$. For $\textbf{HC}_2$ and $\textbf{HC}_3$ the respective effect is only marginal. The coverages do not differ regarding the sign of the correlation for all estimators (Figures \ref{b1_cor_1}-\ref{b1_cor_4}). Large correlations induce longer CIs for $\beta_1$ for all number of studies and estimators (Figures \ref{b1_cor_l_1}-\ref{b1_cor_l_4}).\\ The impact of the correlation on the CIs for $\beta_{12}$ is a little different. For $k=6$ the coverages of the $\textbf{HC}_0$- and $\textbf{HC}_1$-CIs are higher for larger values of $\lvert\rho\rvert$ (Figure \ref{bi_cor_1}). Concerning the other estimators no trend or rather a marginal negative trend in $\lvert\rho\rvert$ for $\textbf{HC}_2$, $\textbf{HC}_4$ and $\textbf{HC}_5$ is observable. The same holds for $k=10$, but in this situation the trend of $\textbf{HC}_2$-$\textbf{HC}_5$ is increasing (Figure \ref{bi_cor_2}). When observing $k=20$ there is no effect of the correlation on any estimators' confidence interval (Figure \ref{bi_cor_3}). However, for $k=50$ the $\textbf{HC}_0-\textbf{HC}_5$ based CIs have a lower coverage for the higher non negative values of $\rho$. The coverage of the CIs with $\rho=-0.5$ based on $\textbf{HC}_2-\textbf{HC}_5$ is marginally higher than with $\rho=0.5$. $\textbf{KH}$-CIs do not differ in their median coverage regarding the positive adjustments of $\rho$, but the coverage of the CIs with $\rho=-0.5$ is marginally lower (Figure \ref{bi_cor_4}).\\  The sign of $\rho$ does not effect the lengths. $\textbf{HC}_0$, $\textbf{HC}_1$, $\textbf{HC}_3$ and $\textbf{KH}$ based CIs have shorter lengths for larger values of $\lvert\rho\rvert$ and all $k$. Intervals based on $\textbf{HC}_2$, $\textbf{HC}_4$ and $\textbf{HC}_5$ have marginally decreasing lengths in $\lvert\rho\rvert$ for $k=6$, slightly increasing lengths for $k\in\{10,20\}$ and again marginally decreasing lengths for $k=50$ (Figures \ref{bi_cor_l_1}-\ref{bi_cor_l_4}). It is also interesting to notice, that most of the extreme outliers of $\textbf{HC}_5$ occur for high correlations (Figures \ref{b1_cor_l_4} and \ref{bi_cor_l_4}).\\

\begin{figure}
\includegraphics[scale=0.7, page=1]{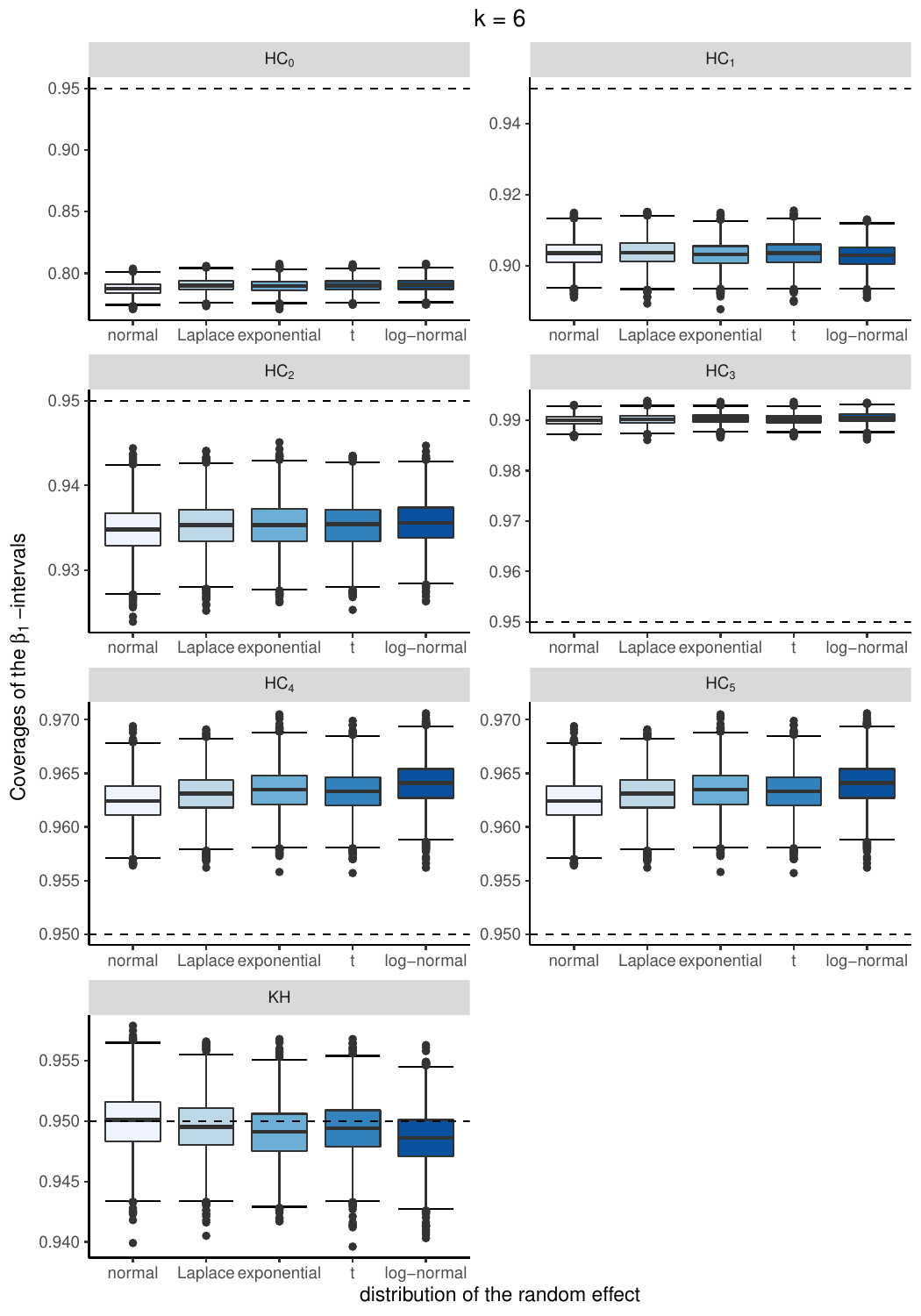}
\caption{Coverages of the \bo -intervals compared regarding the adjustments of $u_i$ for the estimators $HC_0-HC_5$ and $KH$ with $k=6$.}
\label{b1_err_1}
\end{figure}

\begin{figure}
\includegraphics[scale=0.7, page=2]{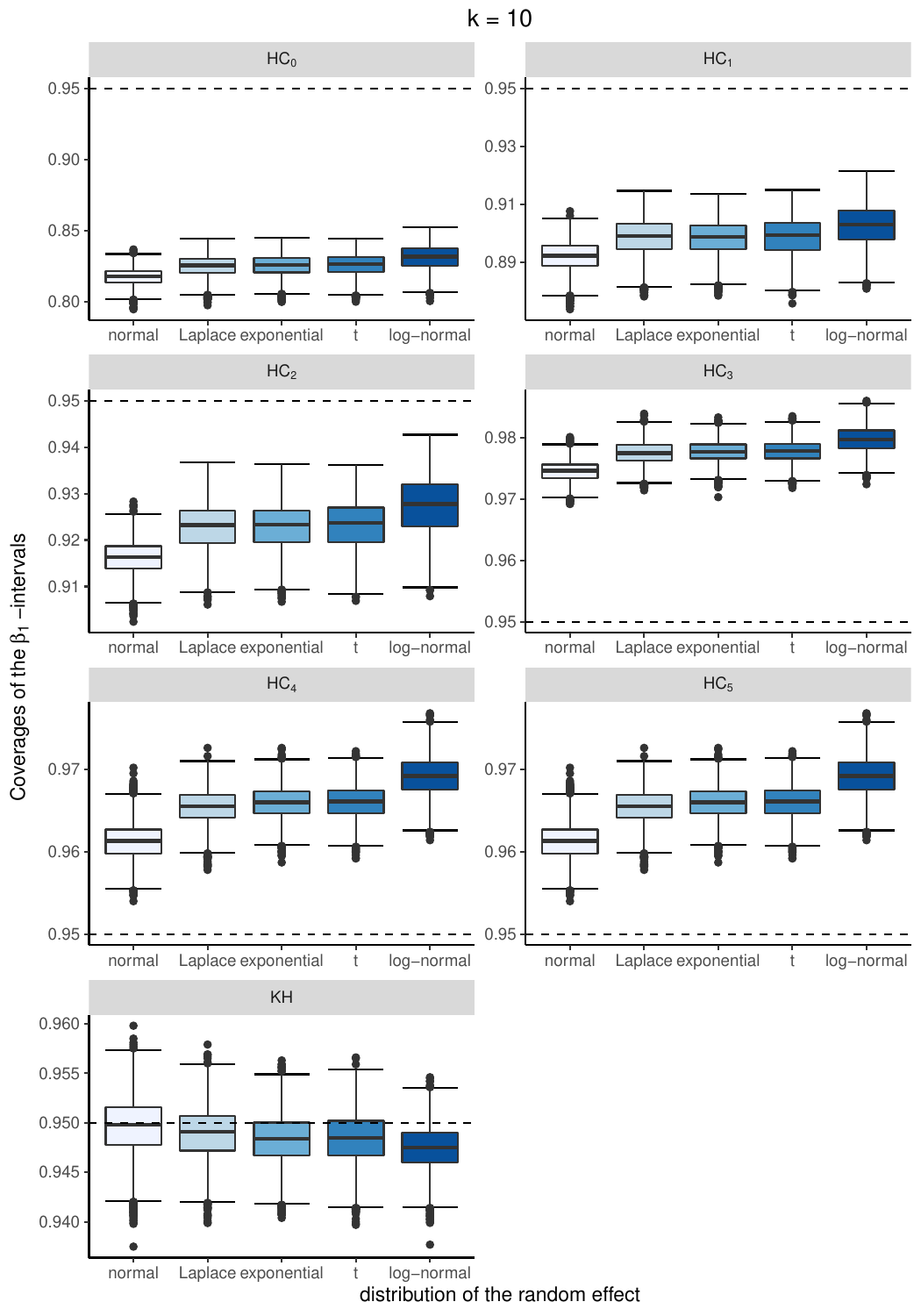}
\caption{Coverages of the \bo -intervals compared regarding the adjustments of $u_i$ for the estimators $HC_0-HC_5$ and $KH$ with $k=10$.}
\label{b1_err_2}
\end{figure}

\begin{figure}
\includegraphics[scale=0.7, page=3]{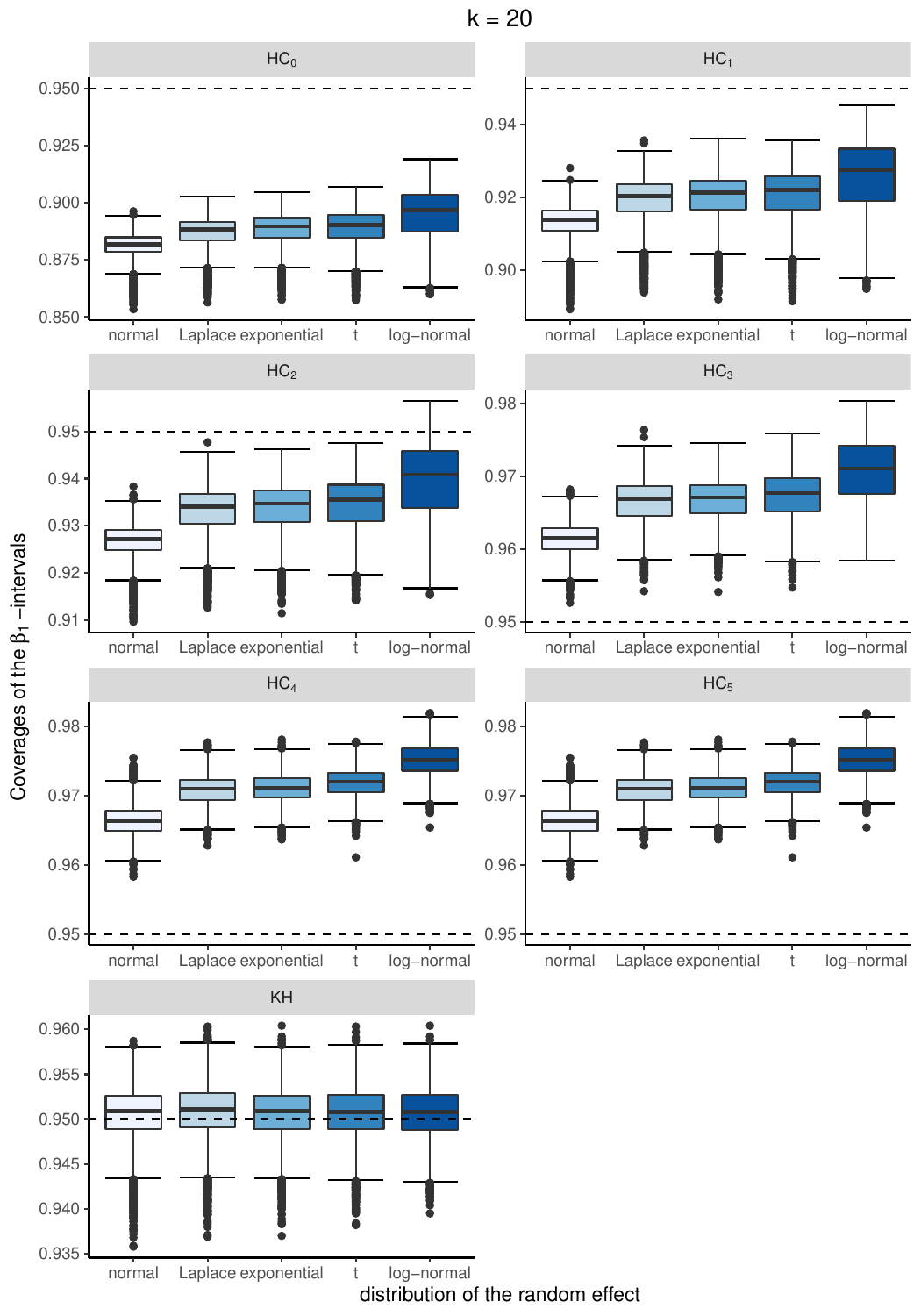}
\caption{Coverages of the \bo -intervals compared regarding the adjustments of $u_i$ for the estimators $HC_0-HC_5$ and $KH$ with $k=20$.}
\label{b1_err_3}
\end{figure}

\begin{figure}
\includegraphics[scale=0.7, page=4]{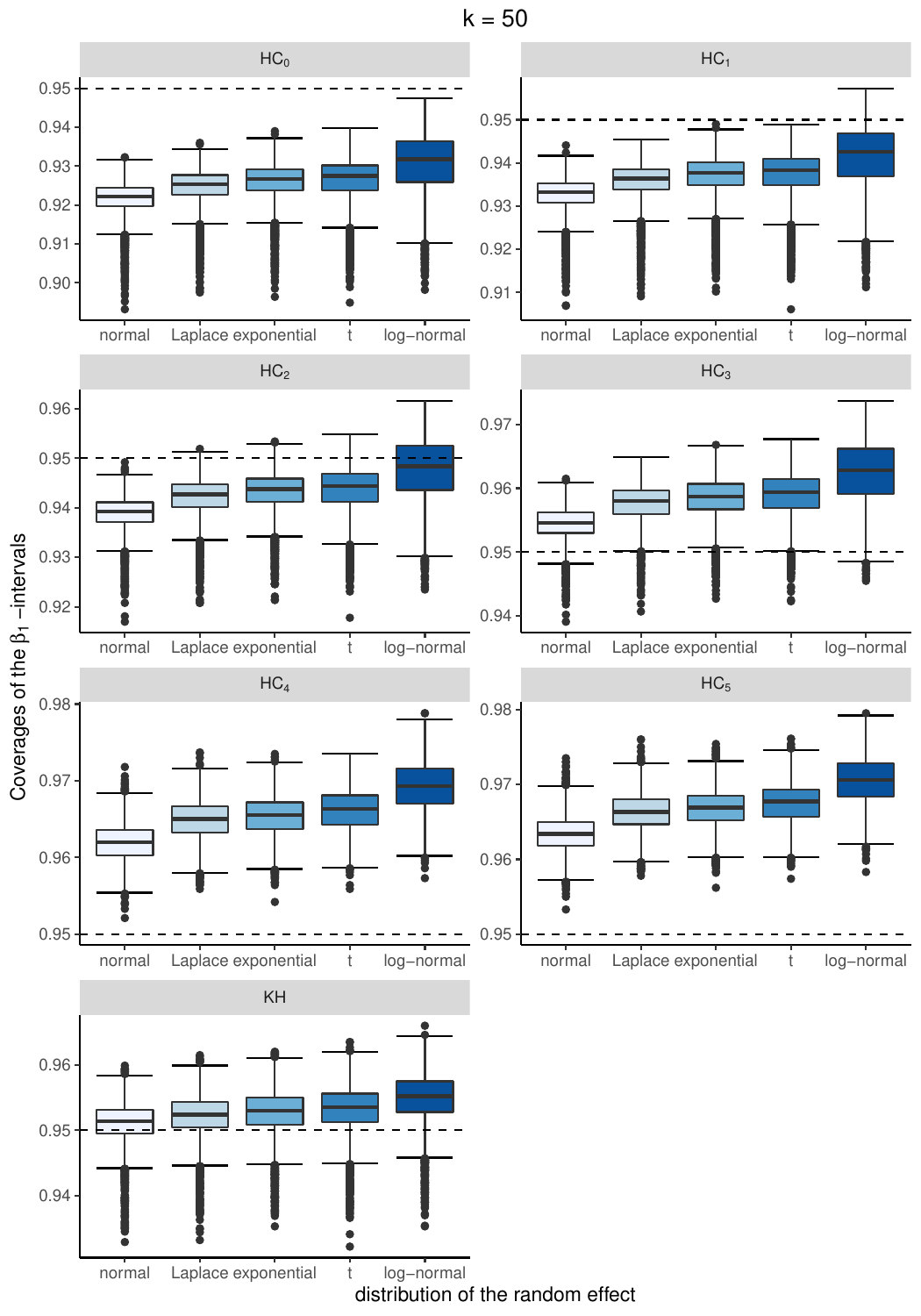}
\caption{Coverages of the \bo -intervals compared regarding the adjustments of $u_i$ for the estimators $HC_0-HC_5$ and $KH$ with $k=50$.}
\label{b1_err_4}
\end{figure}

\begin{figure}
\includegraphics[scale=0.7, page=1]{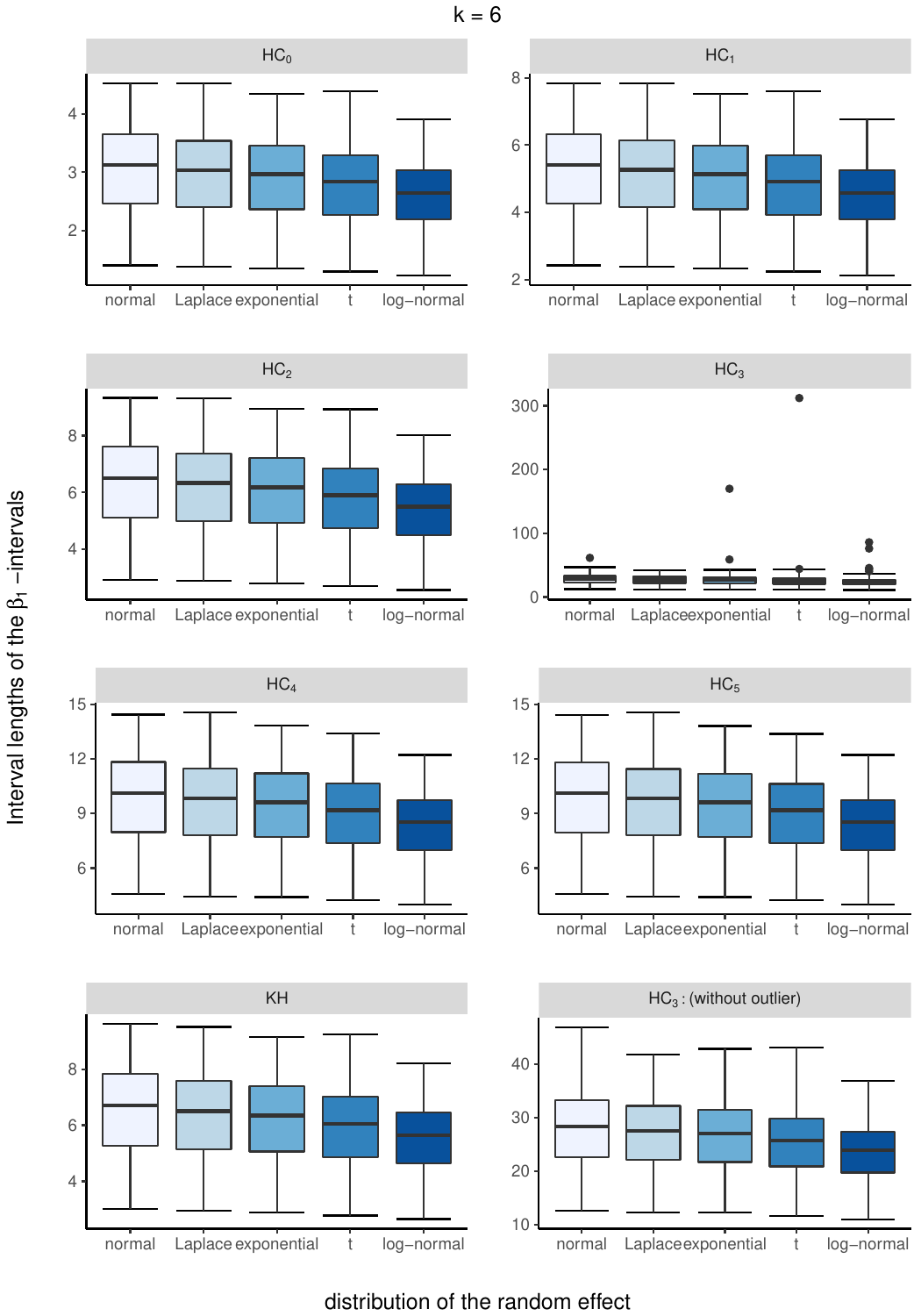}
\caption{Lengths of the \bo -intervals compared regarding the adjustments of $u_i$ for the estimators $HC_0-HC_5$ and $KH$ with $k=6$.}
\label{b1_err_l_1}
\end{figure}

\begin{figure}
\includegraphics[scale=0.7, page=3]{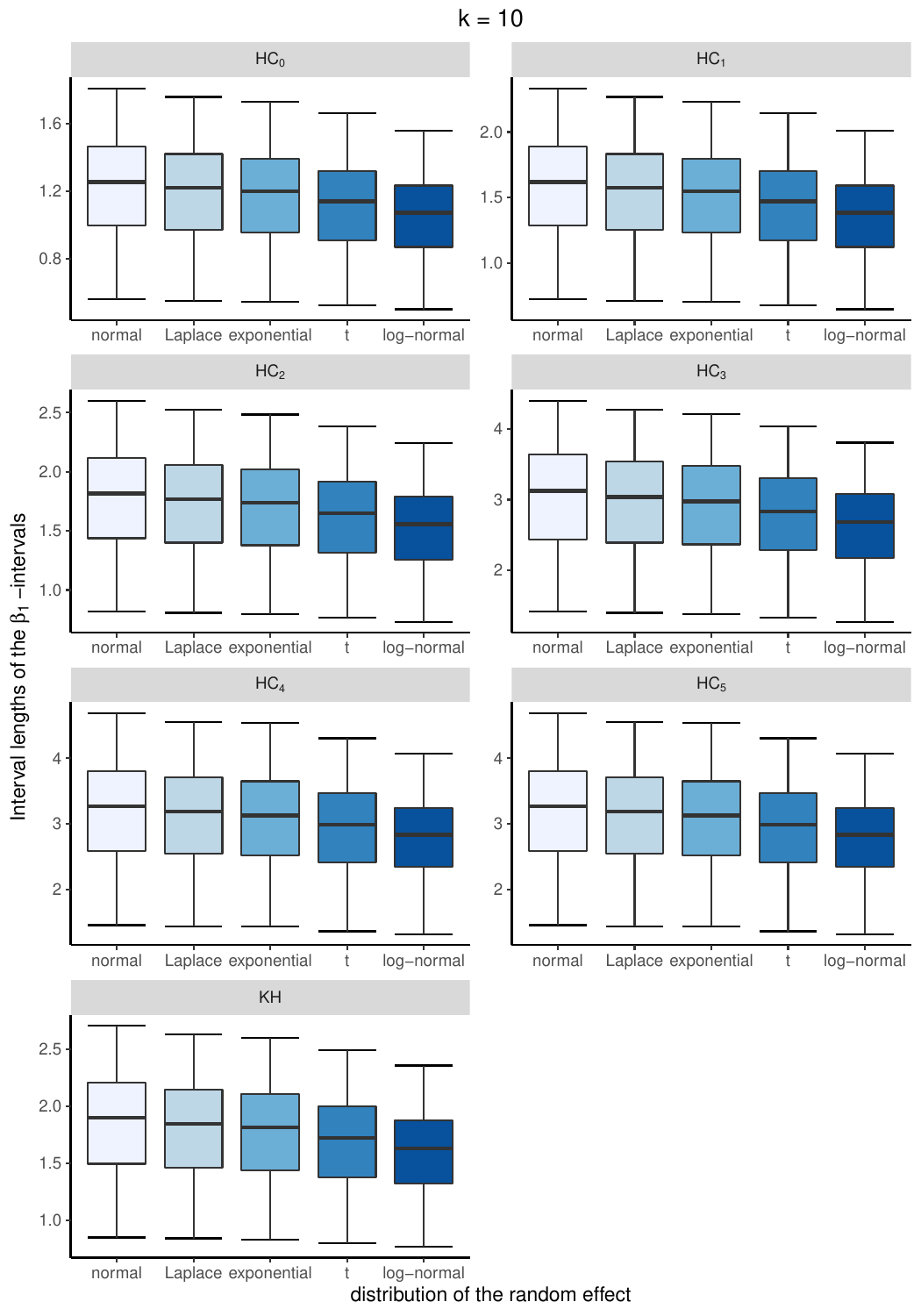}
\caption{Lengths of the \bo -intervals compared regarding the adjustments of $u_i$ for the estimators $HC_0-HC_5$ and $KH$ with $k=10$.}
\label{b1_err_l_2}
\end{figure}

\begin{figure}
\includegraphics[scale=0.7, page=4]{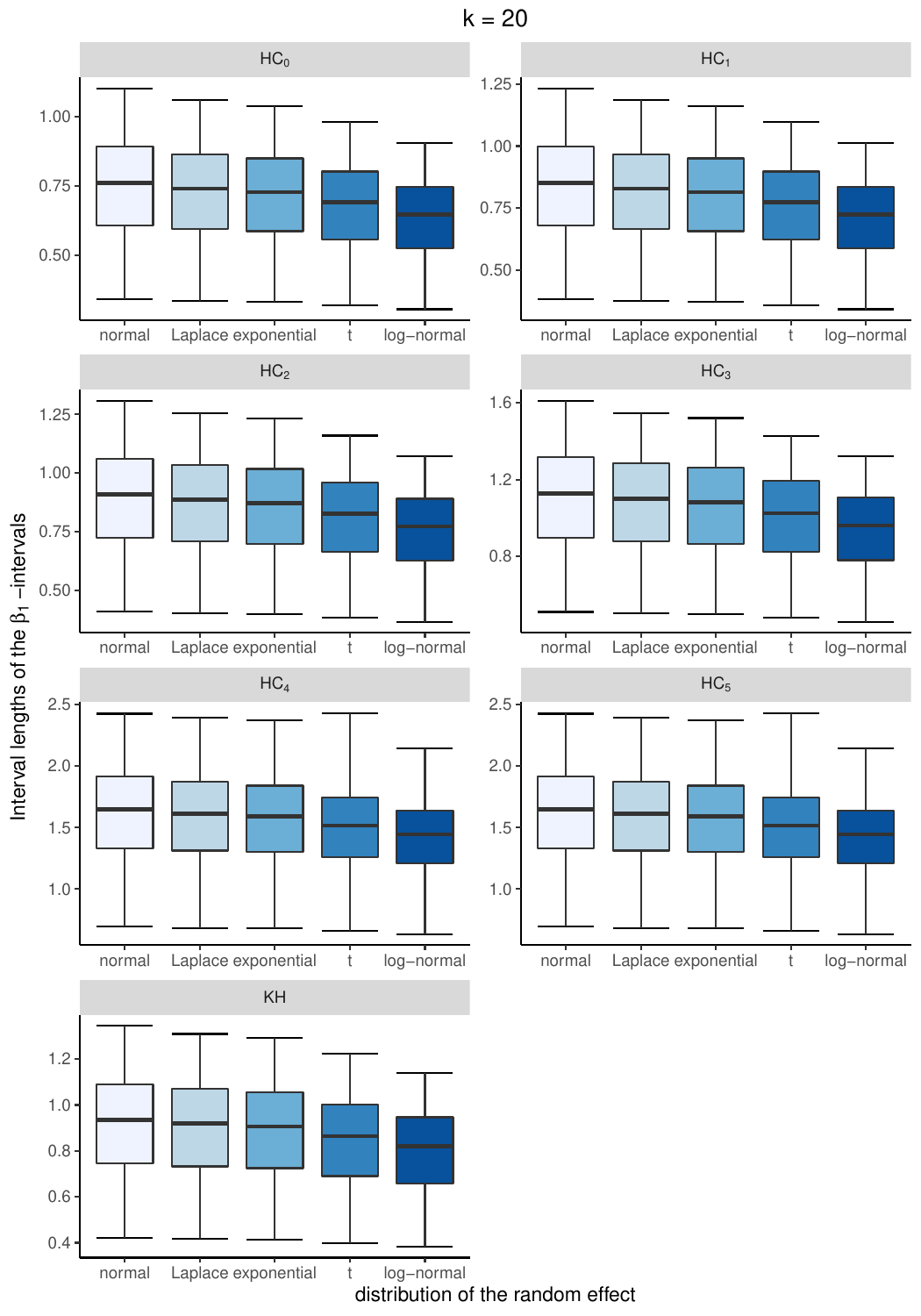}
\caption{Lengths of the \bo -intervals compared regarding the adjustments of $u_i$ for the estimators $HC_0-HC_5$ and $KH$ with $k=20$.}
\label{b1_err_l_3}
\end{figure}

\begin{figure}
\includegraphics[scale=0.7, page=2]{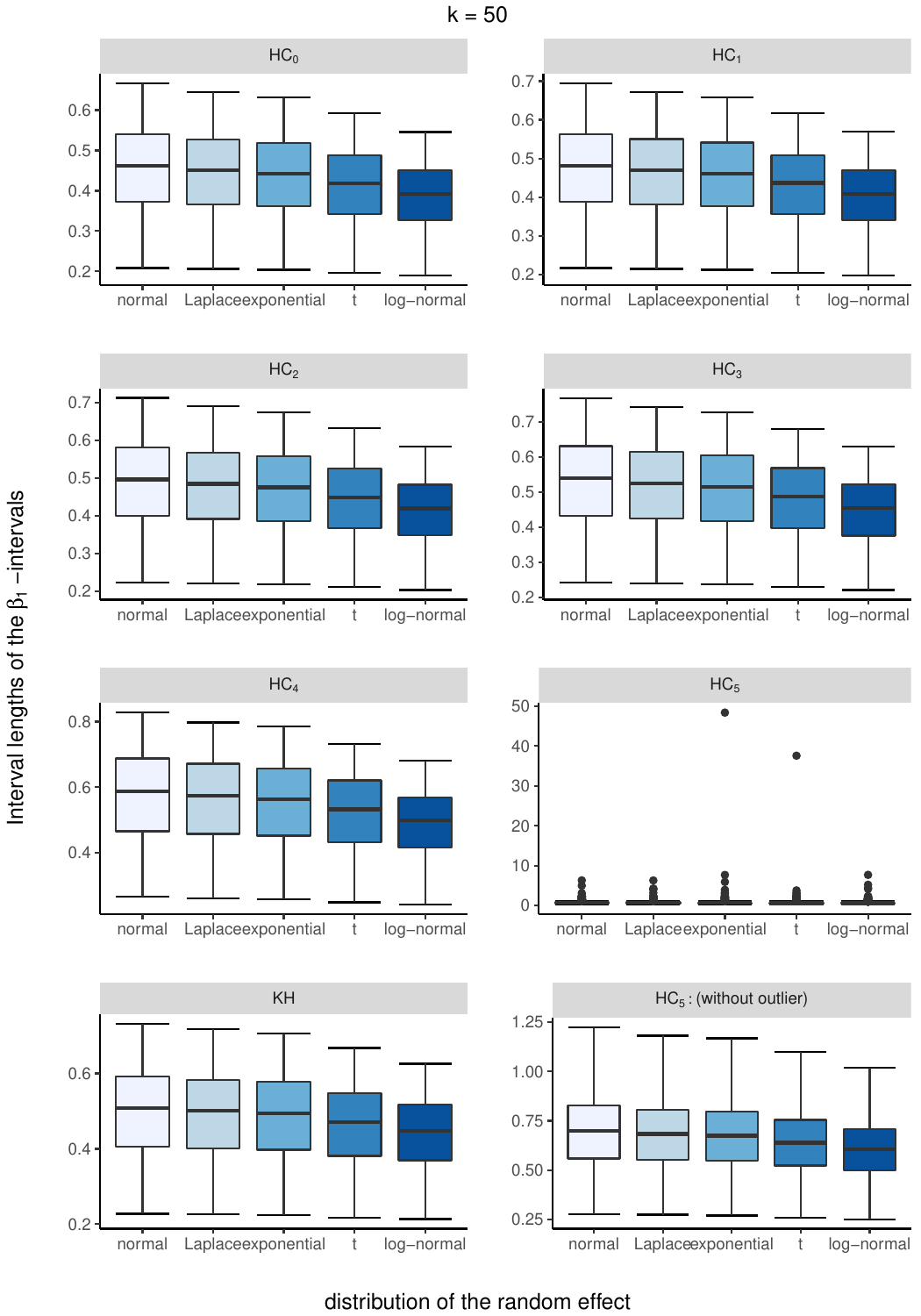}
\caption{Lengths of the \bo -intervals compared regarding the adjustments of $u_i$ for the estimators $HC_0-HC_5$ and $KH$ with $k=50$.}
\label{b1_err_l_4}
\end{figure}

\begin{figure}
\includegraphics[scale=0.7, page=1]{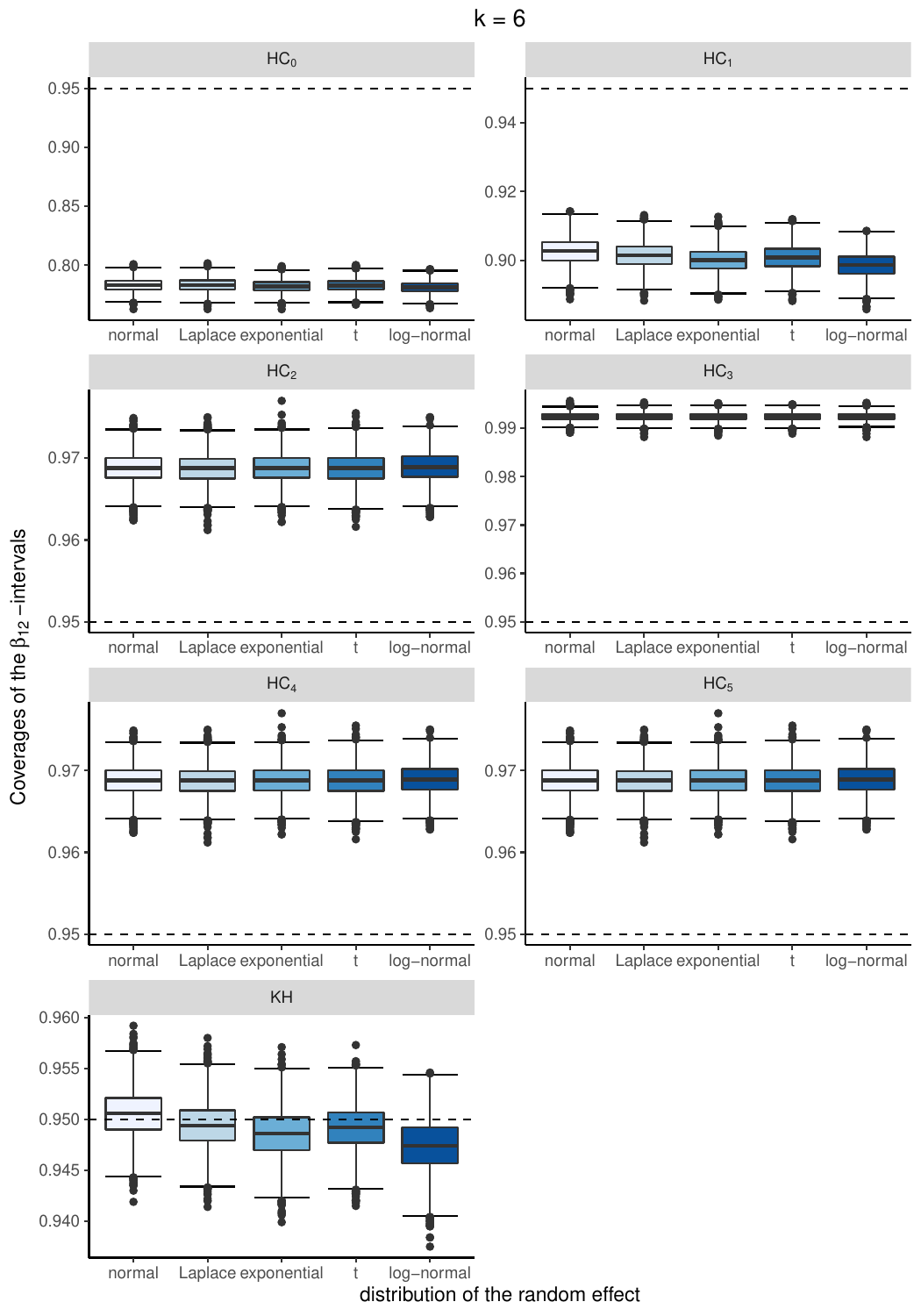}
\caption{Coverages of the \bi -intervals compared regarding the adjustments of $u_i$ for the estimators $HC_0-HC_5$ and $KH$ with $k=6$.}
\label{bi_err_1}
\end{figure}

\textbf{Adjustments of the random effect distribution}
Simulated random effect distributions are the standard normal distribution and standardized Laplace-, exponential, $t_3$- and log-normal-distributions. In comparison with the other simulated distributions, the coverages of CIs for $\beta_1$ based on $\textbf{HC}_0-\textbf{HC}_5$ are on average the smallest with normal distributed $u_i$ and highest with log-normal distributed random effects. The coverages do not differ much in respect of the other random effect distributions but in most situations the coverages for Laplace random effects are slightly lower than coverages for exponential random effects, which again are lower than coverages with $t_3$ random effects. The trend is slightly stronger for a larger number of studies $k$. The $\textbf{KH}$-CIs for $k\in\{6, 10\}$ have the highest coverage with normal distributed random effects and the lowest with log-normal random effects. Especially for $k=10$ the coverages of the $\textbf{KH}$-CIs with non-normal distributed random effects tend to be lower. In $73.71\%$ of the adjustments with non-normal distributed random effects the coverages of the $\textbf{KH}$-CIs are below 0.95. Considering only the adjustments with log-normal random effects this holds for $85.16\%$ of the adjustments. Consequently, for non-normal random effects it is questionable whether the coverages of the $\textbf{KH}$-CIs can still be assessed as acceptable.\\ For $k=20$ the $\textbf{KH}$-CIs show no observable differences between the random effect distributions, whereas for $k=50$ the order of the median coverages is the same as for the other estimators (Figures \ref{b1_err_1}-\ref{b1_err_4}). \\
Coverages of the CIs for $\beta_{12}$ are affected similarly by the random effect distribution for $k\in\{10,20,50\}$. In accordance with the CIs for $\beta_1$, coverages of the $\textbf{HC}_0-\textbf{HC}_5$ based CIs for $\beta_{12}$ are the lowest with normal distributed $u_i$ and the highest with log-normal $u_i$. The coverages of the CIs with exponential distributed $u_i$ decrease compared to the other intervals in $k$. For $k=10$ they have the second highest median coverage, but for $k=50$ the second lowest. In the situation of $k=6$, $\textbf{HC}_2-\textbf{HC}_5$ based CIs coverages are not affected by the random effects distribution. The $\textbf{HC}_0$- and $\textbf{HC}_1$-CIs have the largest median coverage with normal $u_i$ and the smallest with log-normal. This also holds for $\textbf{KH}$-CIs with $k\in\{6, 10, 20\}$. For $k=10$ the coverages of the $\textbf{KH}$-CIs for $\beta_{12}$ are below 0.95 in $88.96\%$ of the adjustments with non-normal random effects and in $96.63\%$ of the adjustments with log-normal random effects. Thus, in these situations the coverages of the $\textbf{KH}$-CIs for $\beta_{12}$ are even less adequate than for the intervals for $\beta_1$. If $k=50$, the $\textbf{KH}$-CIs are not affected by the random effects' distribution (Figures \ref{bi_err_1}-\ref{bi_err_4}).\\The median lengths of both coefficients' CIs can be ordered in the following way for all $k$ and estimators: normal > Laplace > exponential > $t_3$ > log-normal (Figures \ref{b1_err_l_1}-\ref{b1_err_l_4} and \ref{bi_err_l_1}-\ref{bi_err_l_4}). Thus, for $\textbf{HC}_0-\textbf{HC}_5$ the confidence intervals have better properties, when the random effect distribution is different from a normal distribution. Therefore the quantile used  as critical value is suitable, even if the distribution of the $u_i$ is not normal. In contrast the $\textbf{KH}$-CIs depend more on the normality assumption  for smaller numbers of studies ($k\in\{6,10\}$), especially for $k=10$. Due to the high share of coverages of the $\textbf{KH}$-CIs below the nominal confidence level, for non-normal and particularly log-normal random effects it is arguable whether $\textbf{KH}$ is the best estimator in this situation. If a precise control of the nominal confidence level is required $\textbf{HC}_3$ may be preferable, since for $k=10$ its CIs have higher coverages than 0.95 in every adjustment and are the shorter compared to the $\textbf{HC}_4$- and $\textbf{HC}_5$-CIs. For $k\in\{20,50\}$ the performance of the $\textbf{KH}$-CIs is still the best for all distributions of $u_i$.

In summary the estimators are affected by most parameter adjustments in the same way or only slightly different. Only the number of studies $k$ shows a strong varying effect on the coverages of some estimators. Besides the number of studies, the study size and the heterogeneity parameter $\tau^2$ have impact on the interval lengths. However, the trend is the same for all estimators and reducible to the direct impact of these parameters on components of the confidence interval in equation (6) 
of the main paper. The results of the different random effect distributions indicate that the $\textbf{HC}$ estimators are more robust against deviations from the normal distribution. For small numbers of studies it is questionable whether the coverages of the $\textbf{KH}$-CIs for non-normal random effects are still adequate. In this situation $\textbf{HC}_3-\textbf{HC}_5$ might be more suitable compared to $\textbf{KH}$. In all other simulated adjustments, there is no situation where a certain estimator performs superior compared to its overall performance.

\clearpage

\clearpage

\clearpage

\clearpage

 \clearpage

\clearpage

\clearpage

\clearpage

 \clearpage

\clearpage

\clearpage

\clearpage

\clearpage

\clearpage

\clearpage

\clearpage

\clearpage

\clearpage

\clearpage

\clearpage

 \clearpage

\clearpage

\clearpage

\clearpage

 \clearpage

\clearpage

\clearpage

\clearpage

\clearpage

\clearpage

\clearpage

\clearpage

\clearpage

\clearpage

\clearpage

\clearpage

\clearpage

\clearpage

\clearpage

\clearpage

\clearpage

\clearpage

\clearpage

\clearpage

\clearpage

\clearpage

\clearpage

\clearpage

\clearpage

\clearpage

\clearpage

\clearpage

\clearpage

\clearpage

\clearpage

\clearpage

 \clearpage

\clearpage

\clearpage

\clearpage

\clearpage

\clearpage

\clearpage

\clearpage

\clearpage

\clearpage

\clearpage

\clearpage

\clearpage

\clearpage

\clearpage

\clearpage

\clearpage

\clearpage

\clearpage

\clearpage

\clearpage

\clearpage

\clearpage

\clearpage

\clearpage

\clearpage

\clearpage

\clearpage

 \clearpage

\clearpage

\clearpage

\clearpage

 \clearpage

\clearpage

\clearpage

\clearpage

\clearpage

\clearpage

\clearpage

\clearpage

\clearpage

\clearpage

\clearpage

\clearpage

 \clearpage

\clearpage

\clearpage

\clearpage

\clearpage

\clearpage

\clearpage

\clearpage

\clearpage

\clearpage

\begin{figure}
\includegraphics[scale=0.7, page=2]{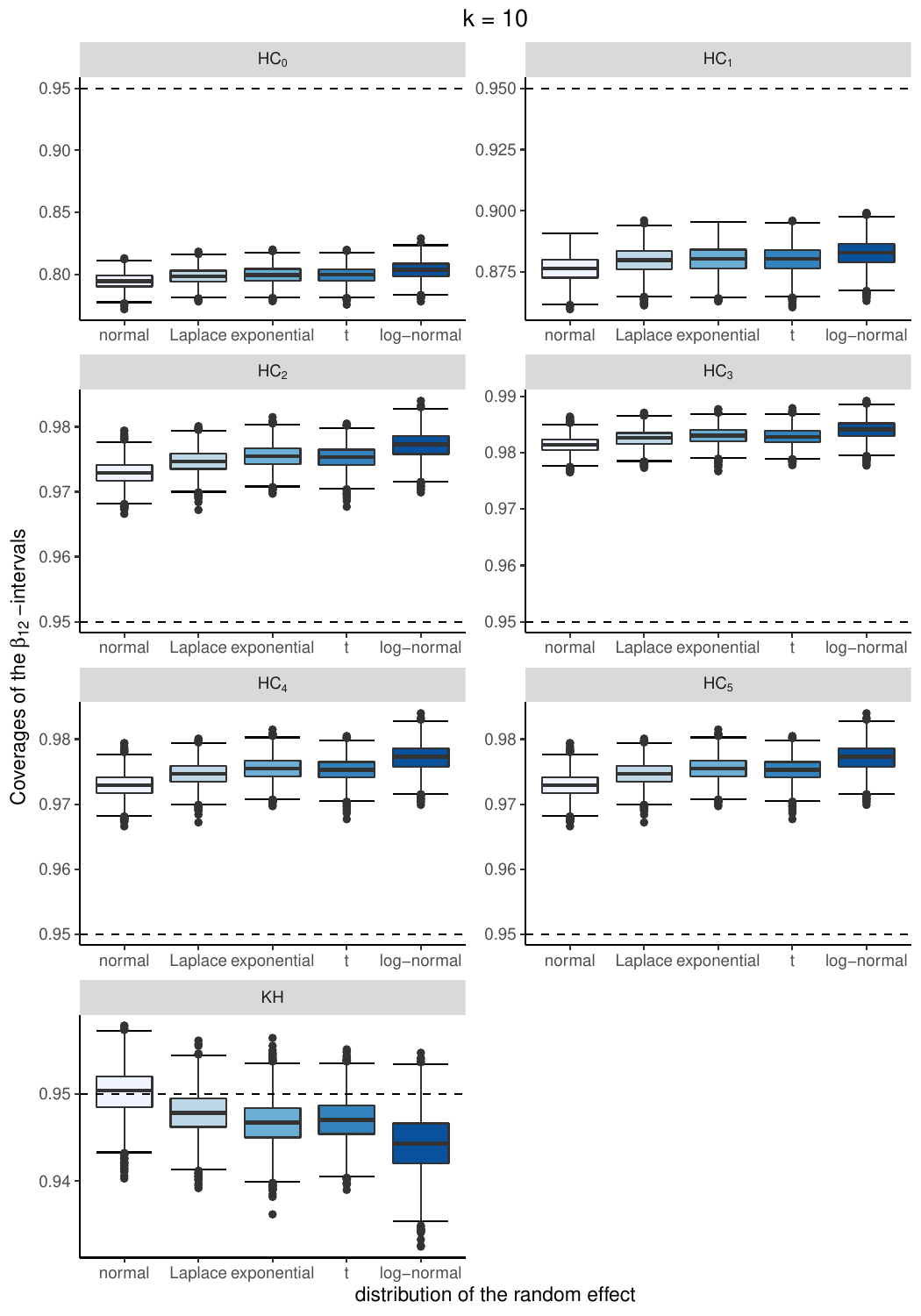}
\caption{Coverages of the \bi -intervals compared regarding the adjustments of $u_i$ for the estimators $HC_0-HC_5$ and $KH$ with $k=10$.}
\label{bi_err_2}
\end{figure}

\clearpage

\begin{figure}
\includegraphics[scale=0.7, page=3]{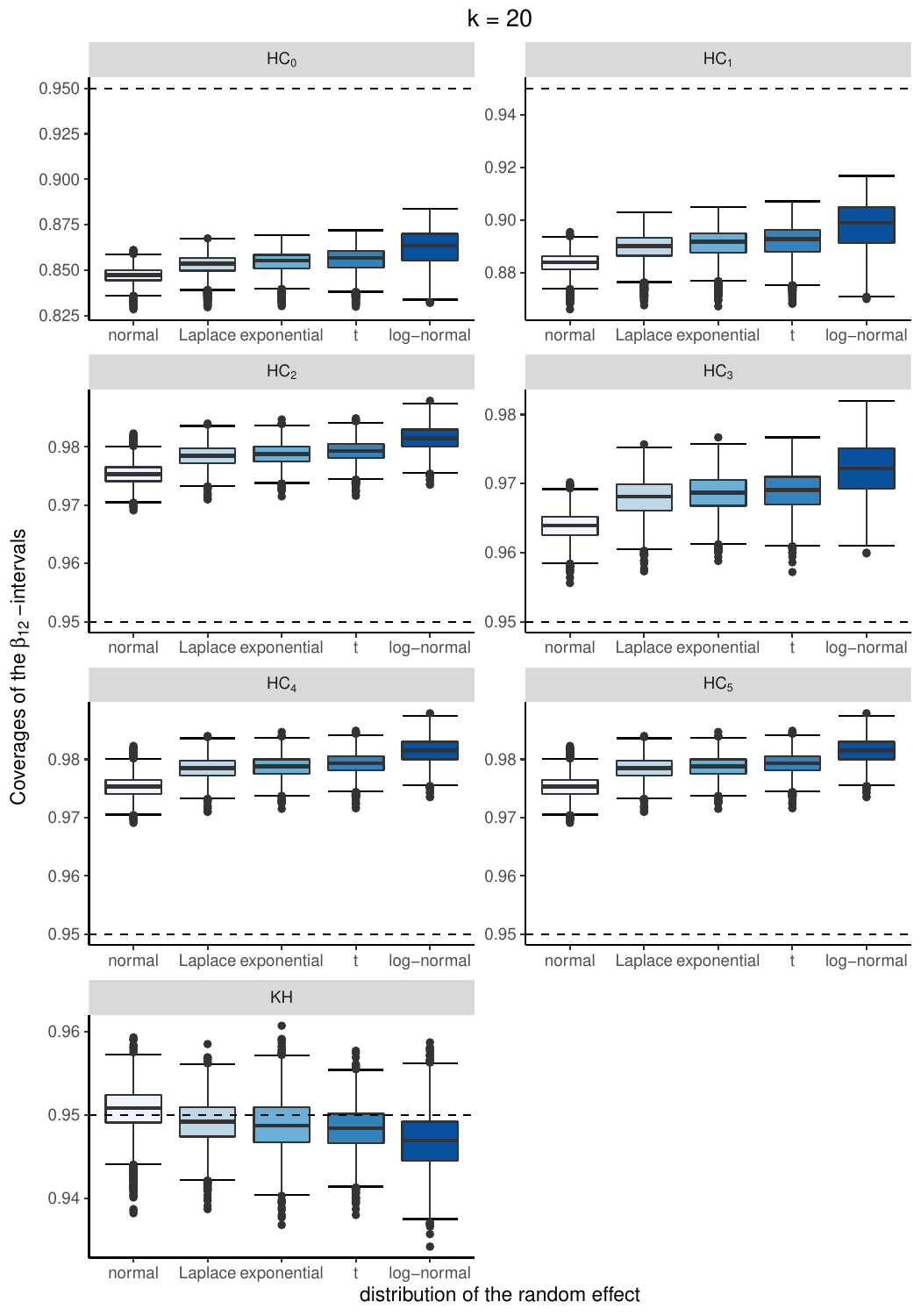}
\caption{Coverages of the \bi -intervals compared regarding the adjustments of $u_i$ for the estimators $HC_0-HC_5$ and $KH$ with $k=20$.}
\label{bi_err_3}
\end{figure}

\clearpage

\begin{figure}
\includegraphics[scale=0.7, page=4]{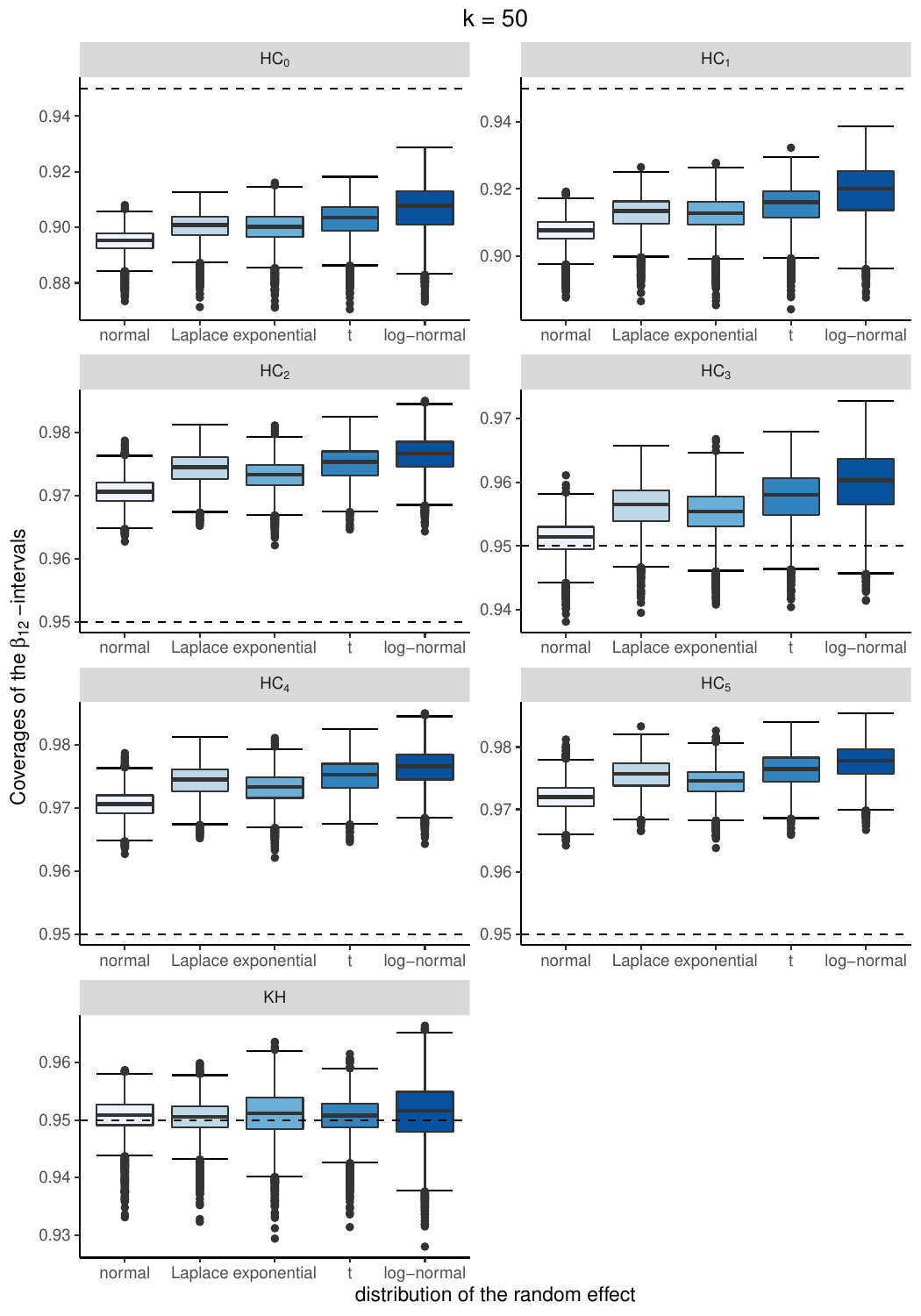}
\caption{Coverages of the \bi -intervals compared regarding the adjustments of $u_i$ for the estimators $HC_0-HC_5$ and $KH$ with $k=50$.}
\label{bi_err_4}
\end{figure}


\clearpage

\begin{figure}
\includegraphics[scale=0.7, page=1]{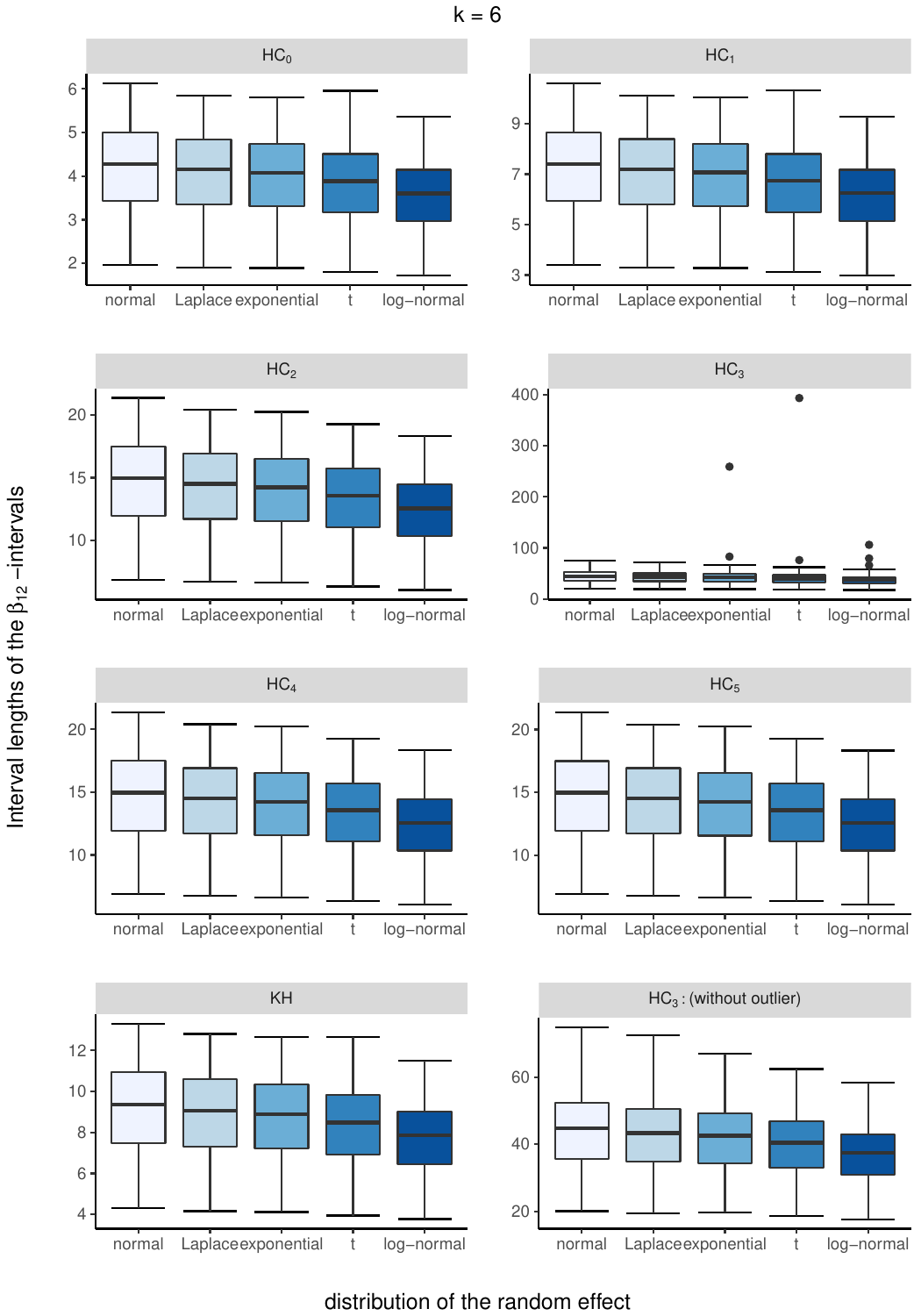}
\caption{Lengths of the \bi -intervals compared regarding the adjustments of $u_i$ for the estimators $HC_0-HC_5$ and $KH$ with $k=6$.}
\label{bi_err_l_1}
\end{figure}

\clearpage

\begin{figure}
\includegraphics[scale=0.7, page=3]{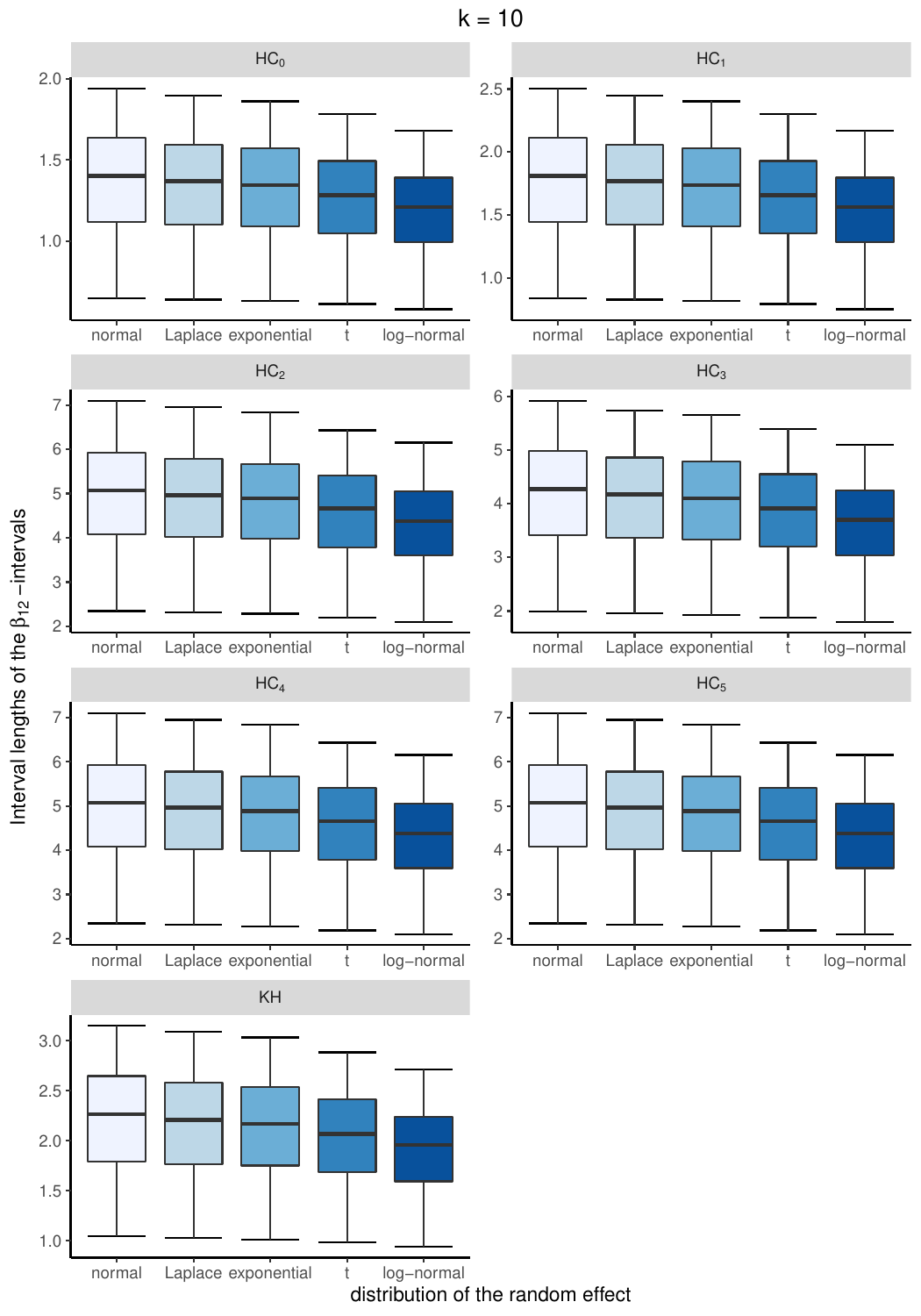}
\caption{Lengths of the \bi -intervals compared regarding the adjustments of $u_i$ for the estimators $HC_0-HC_5$ and $KH$ with $k=10$.}
\label{bi_err_l_2}
\end{figure}

\clearpage

\begin{figure}
\includegraphics[scale=0.7, page=4]{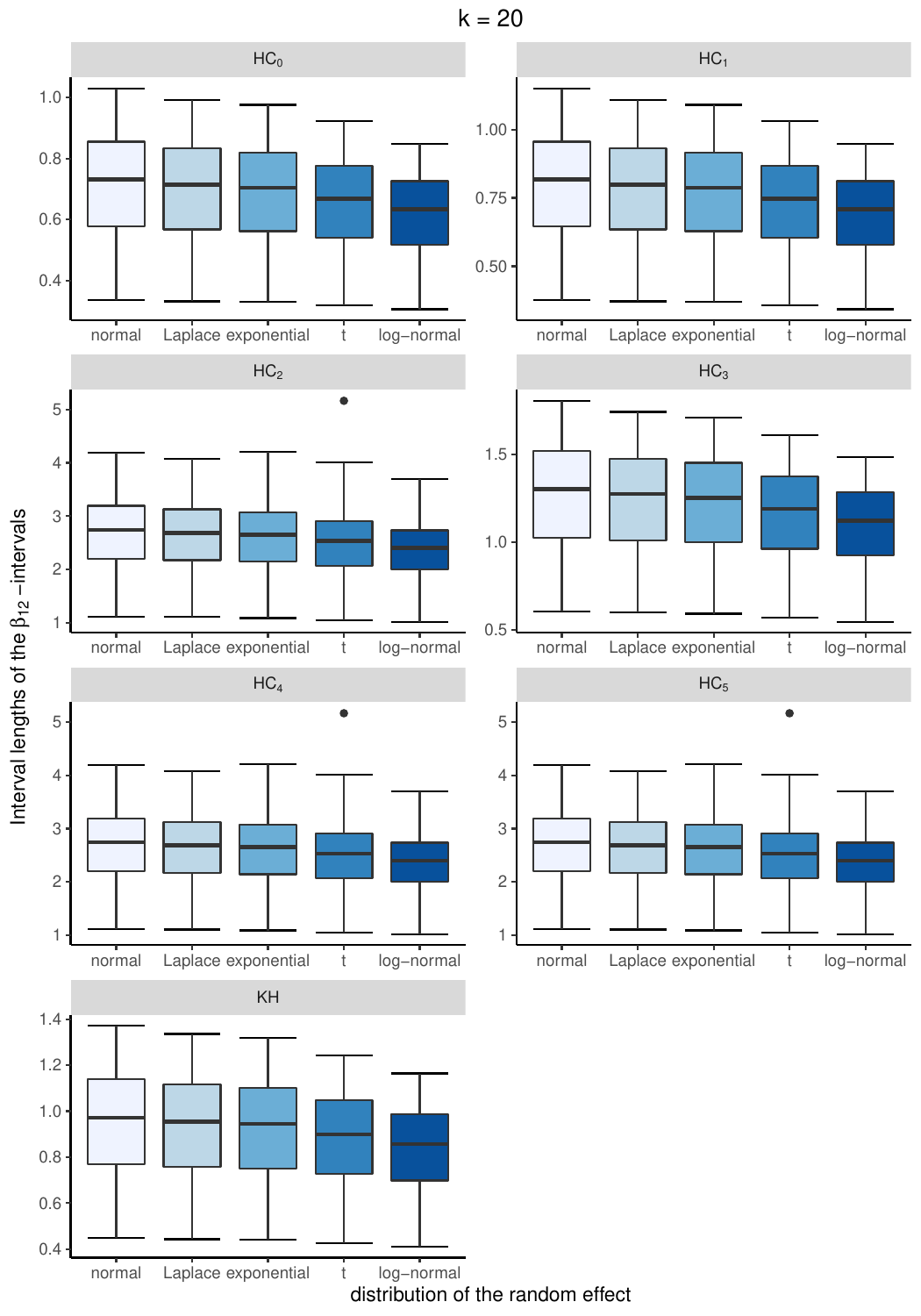}
\caption{Lengths of the \bi -intervals compared regarding the adjustments of $u_i$ for the estimators $HC_0-HC_5$ and $KH$ with $k=20$.}
\label{bi_err_l_3}
\end{figure}

\clearpage

\begin{figure}
\includegraphics[scale=0.7, page=2]{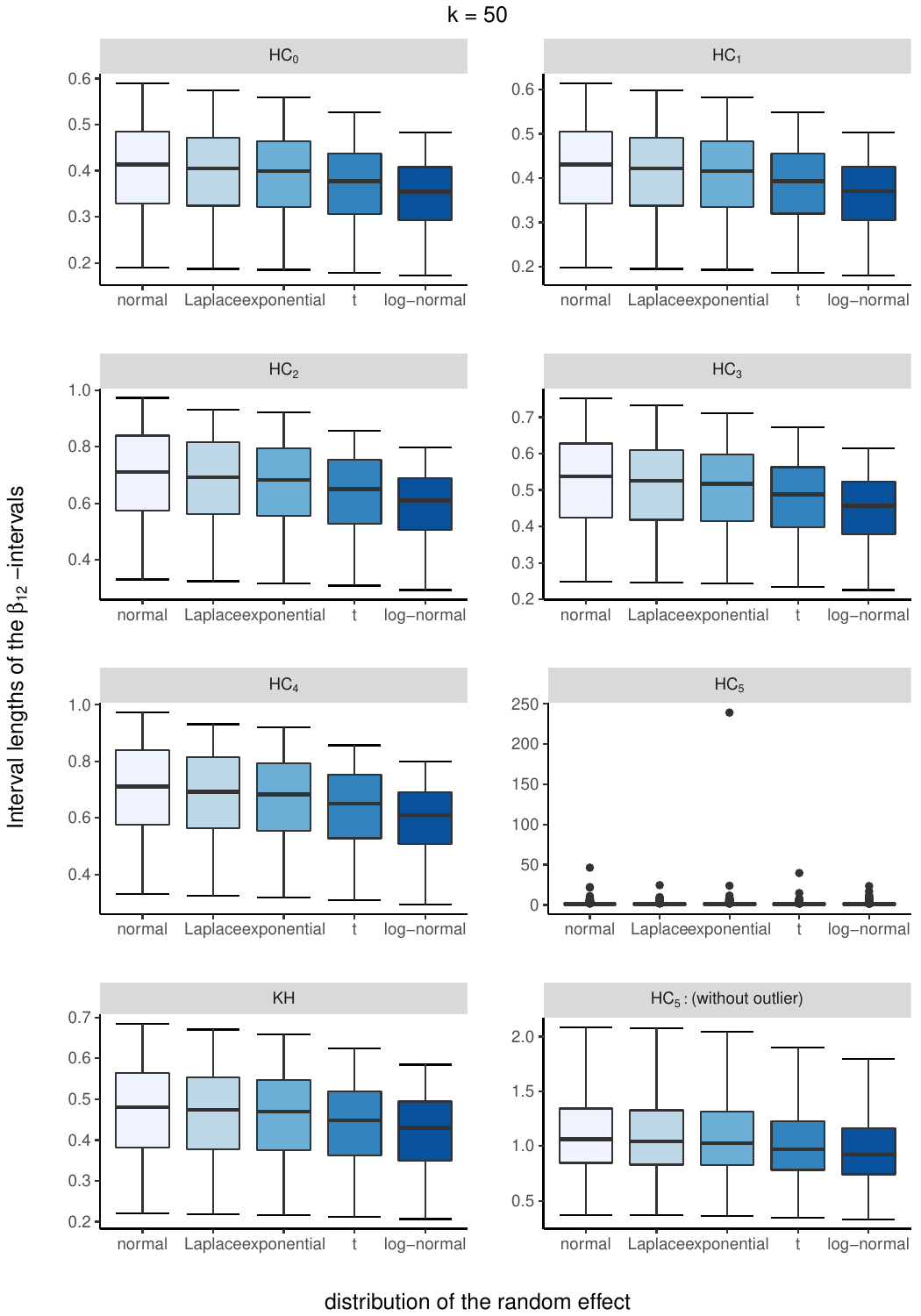}
\caption{Lengths of the \bi -intervals compared regarding the adjustments of $u_i$ for the estimators $HC_0-HC_5$ and $KH$ with $k=50$.}
\label{bi_err_l_4}
\end{figure}

\clearpage

\section{Additional Simulations}
In additional simulations we considered the effects of fitting the wrong model and of very high correlations. The setup of the simulation was the same as in the main simulation, except for the adjustments mentioned below:

To assess the effect of fitting a wrong model, models with only two moderators (short model) and models with two covariates and their interaction (long model) were fitted. The true interaction was simulated as either $\beta_{12}=0$ or $\beta_{12}=0.5$. Hence there were two types of a wrong model simulated: (i) when $\beta_{12}=0$ but the long model was fitted and (ii) when $\beta_{12}=0.5$ but the short model was fitted. For both the respective right model was simulated as well. In order to analyse the effect of very high correlations, correlations of $\rho=0$ and $\rho=0.9$ were simulated. 

Since most of the flexible parameters did not alter the results of the main simulation much, the simulation was conducted for less parameter combinations than the main simulation. Like $\beta_{12}$ also $\beta_1$ and $\beta_2$ were chosen from $\{0, 0.5\}$. The heterogeneity parameter $\tau^2$ was set to either 0.1, 0.5 or 0.9. Simulated random effects distributions were a normal distribution and the log normal distribution. The vectors of study sizes $\boldsymbol{n}_{15}, \boldsymbol{n}_{25}, \boldsymbol{n}_{50}$ and the number of studies $k$ were chosen like in the main simulation. Every parameter combination was simulated $N=1,000$ times. Besides the mentioned modifications the simulation was conducted like the main simulation.

To assess the fit of the models, the short and the long model were compared by parallel boxplots of the coverages and interval lengths of $\beta_0$ and $\beta_1$ for each estimator $\textbf{HC}_0-\textbf{HC}_5$ and $\textbf{KH}$. 

\begin{figure}[t]
\includegraphics[trim = 0cm 1cm 0cm 1.8cm, clip, scale = 0.6, page = 1]{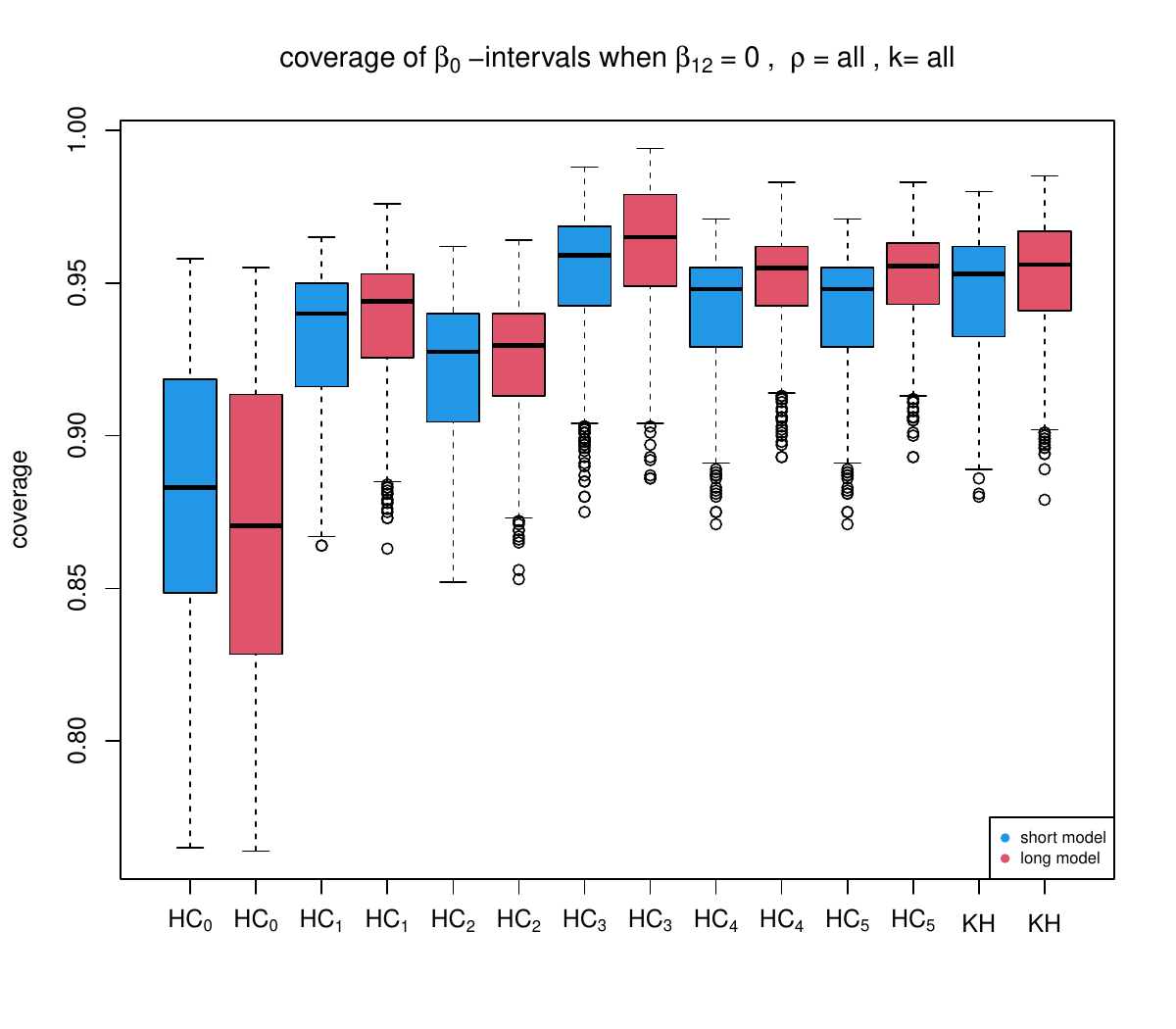}
\includegraphics[trim = 0cm 1cm 0cm 1.8cm, clip, scale = 0.6, page = 2]{figures/comp_boxplot_plots.pdf}
\caption{Coverages (upper plot) and interval lengths (lower plot) of $\beta_0$ compared for a model without interaction (short) and a model with interaction (long) when $\beta_{12}=0$.}
\label{extra_b0_noint}
\end{figure}

\begin{figure}[t]
\includegraphics[trim = 0cm 1cm 0cm 1.8cm, clip, scale = 0.6, page = 3]{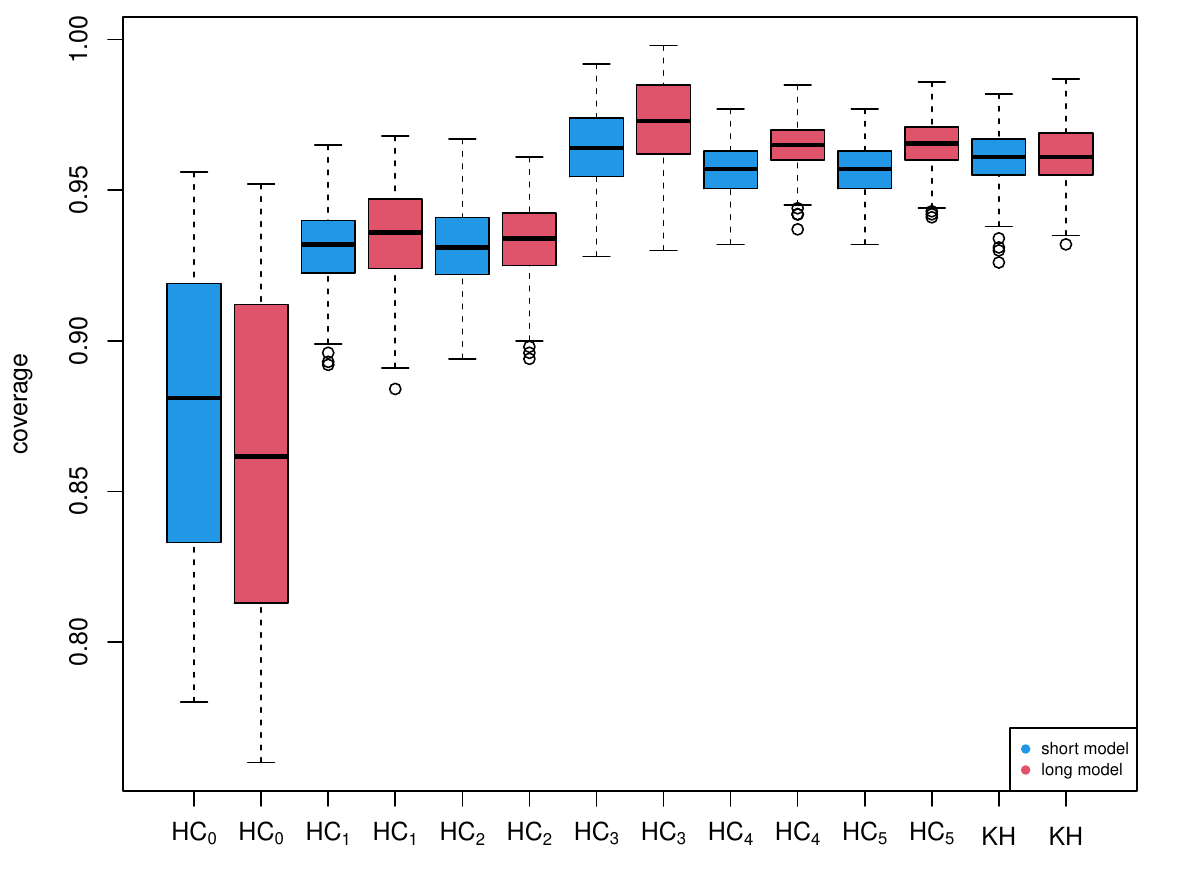}
\includegraphics[trim = 0cm 1cm 0cm 1.8cm, clip, scale = 0.6, page = 4]{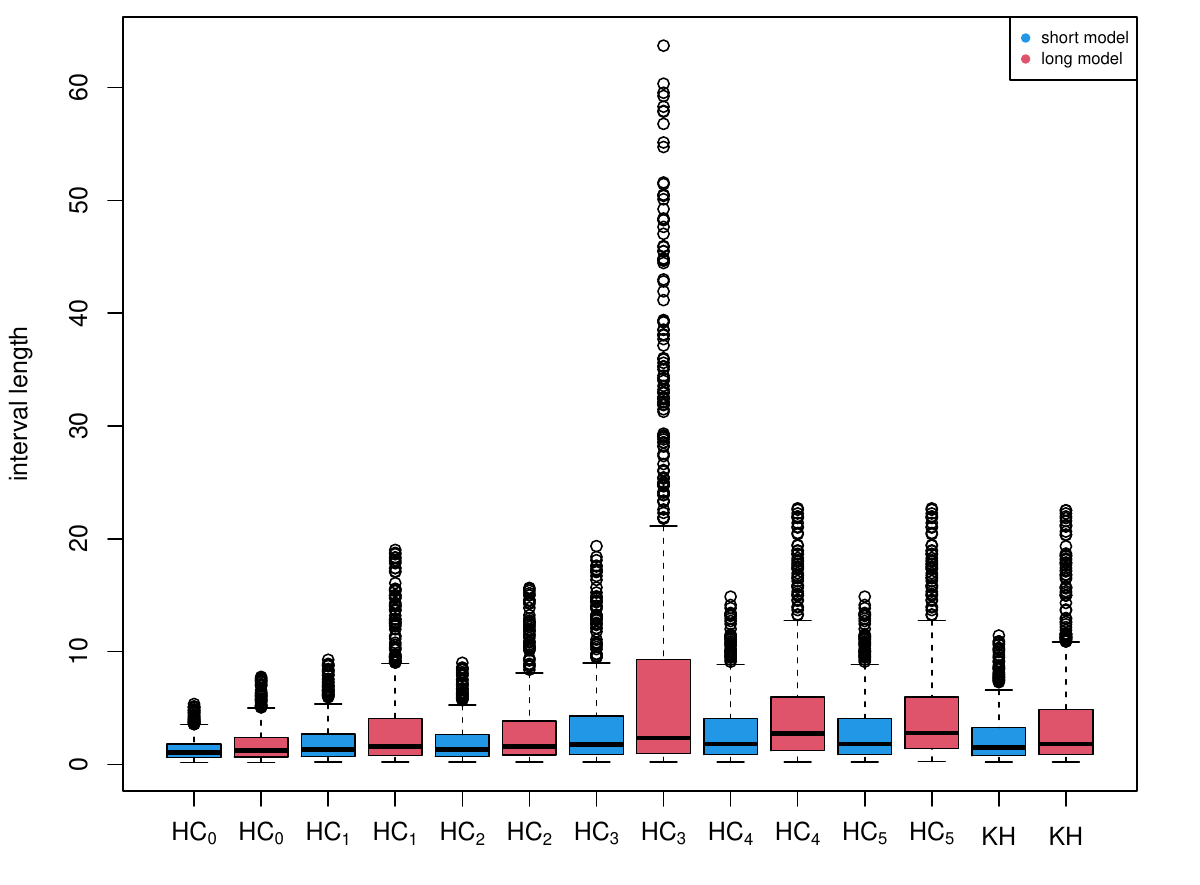}
\caption{Coverages (upper plot) and interval lengths (lower plot) of $\beta_1$ compared for a model without interaction (short) and a model with interaction (long) when $\beta_{12}=0$.}
\label{extra_b1_noint}
\end{figure}

Figure \ref{extra_b0_noint} shows the coverages and interval lengths of $\beta_0$ when the true model contains no interaction term. For all estimators but $\textbf{HC}_0$ the coverages of $\beta_0$ are slightly higher when an interaction term is fitted. On the other hand the intervals tend to be longer. The same holds for the intervals of $\beta_1$ (see Figure \ref{extra_b1_noint}). 

\begin{figure}[t]
\includegraphics[trim = 0cm 1cm 0cm 1.8cm, clip, scale = 0.6, page = 5]{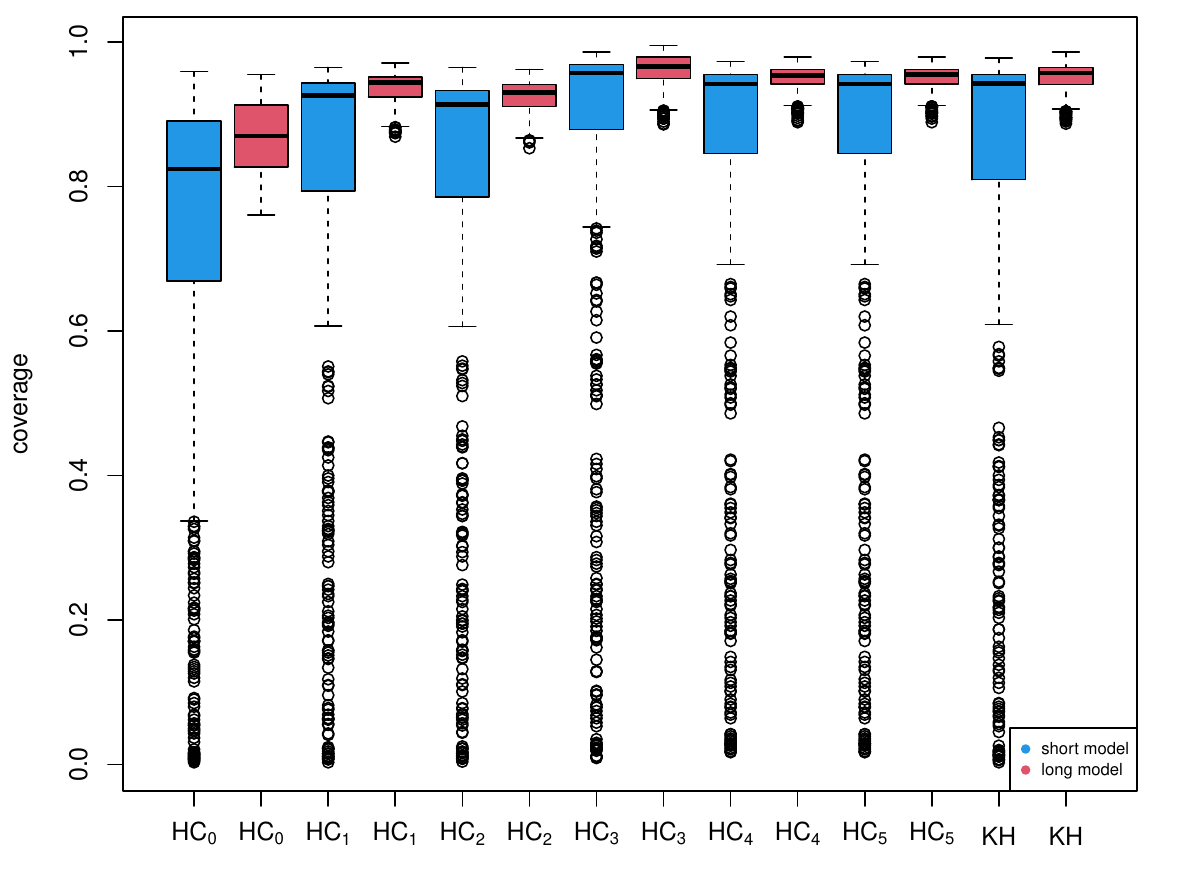}
\includegraphics[trim = 0cm 1cm 0cm 1.8cm, clip, scale = 0.6, page = 6]{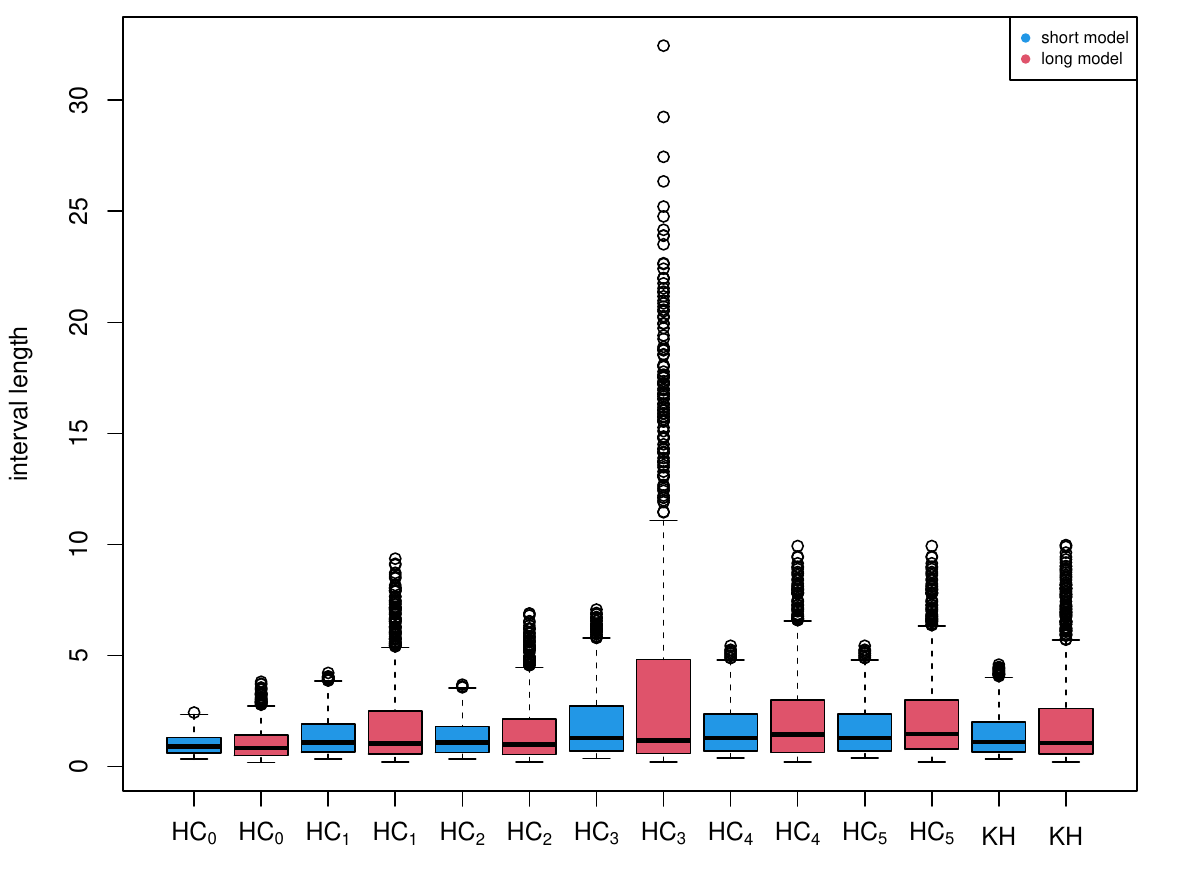}
\caption{Coverages (upper plot) and interval lengths (lower plot) of $\beta_0$ compared for a model without interaction (short) and a model with interaction (long) when $\beta_{12}=0.5$.}
\label{extra_b0_int}
\end{figure}

\begin{figure}[t]
\includegraphics[trim = 0cm 1cm 0cm 1.8cm, clip, scale = 0.6, page = 7]{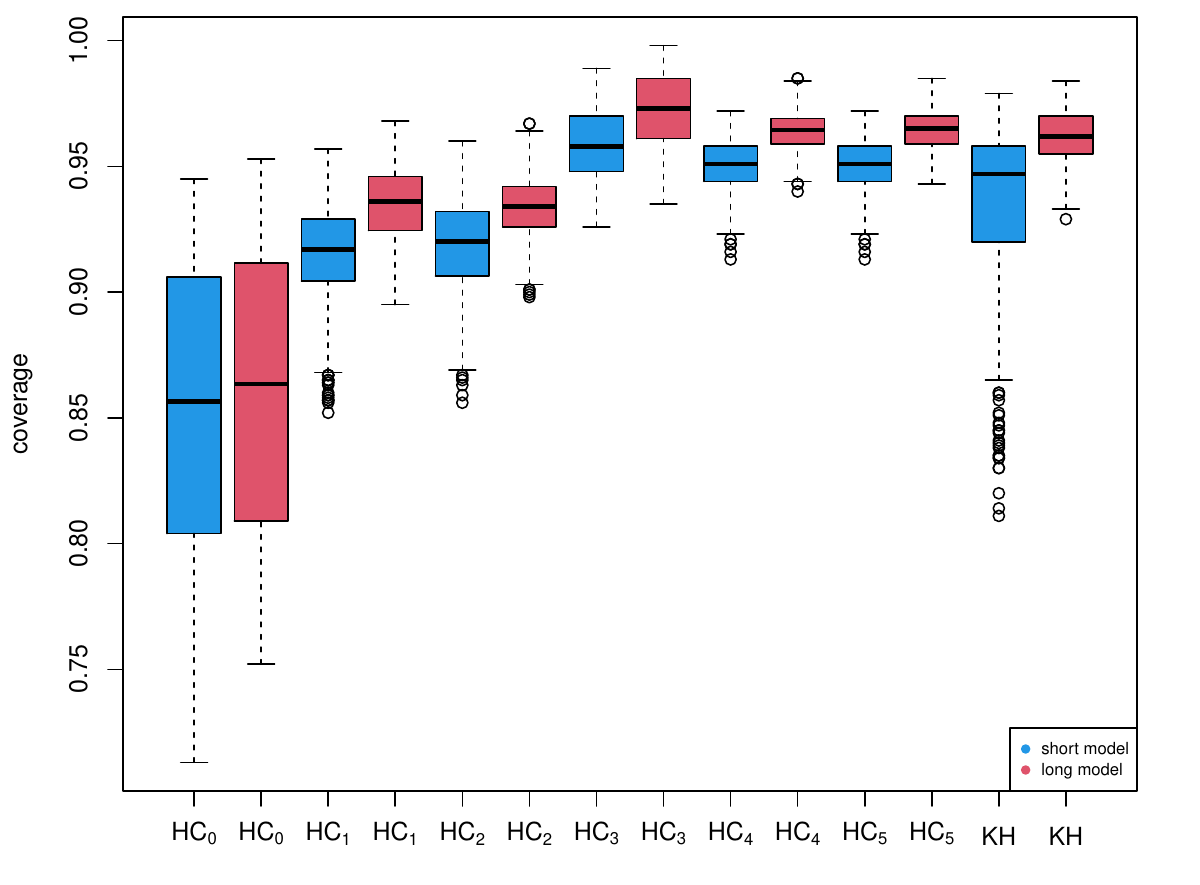}
\includegraphics[trim = 0cm 1cm 0cm 1.8cm, clip, scale = 0.6, page = 8]{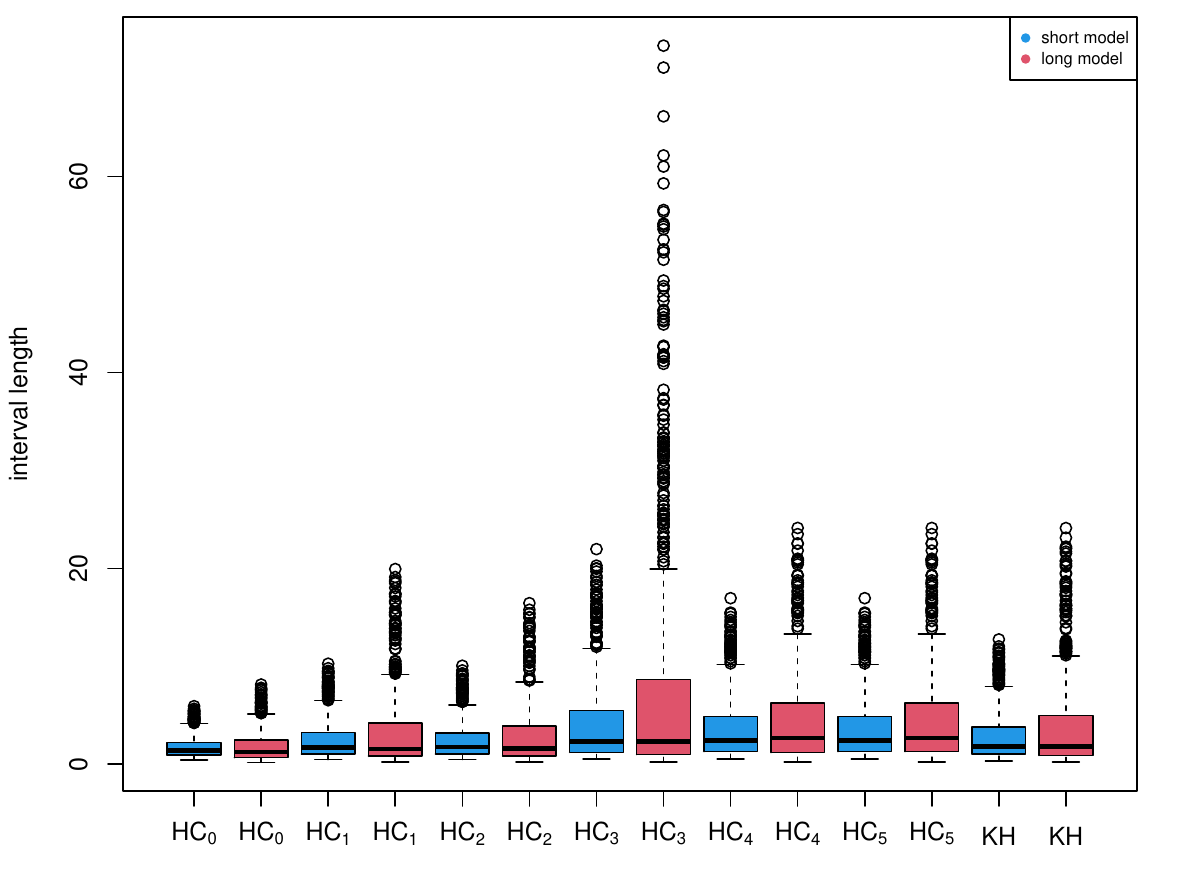}
\caption{Coverages (upper plot) and interval lengths (lower plot) of $\beta_1$ compared for a model without interaction (short) and a model with interaction (long) when $\beta_{12}=0.5$.}
\label{extra_b1_int}
\end{figure}

When the true model contains an interaction ($\beta_{12}=0.5$), the coverage of $\beta_0$ tends to be lower if a short model is fitted, compared to the long model. As Figure \ref{extra_b0_int} shows there are many huge outliers towards a coverage of 0. Also the CIs tend to be longer. The coverage of $\beta_1$ is higher when the long (correct) model is fitted for all estimators. Again the CIs tend to be longer when the model with interaction is fitted (see Figure \ref{extra_b1_int}). 

\begin{figure}[t]
\includegraphics[trim = 0cm 1cm 0cm 1.8cm, clip, scale = 0.6, page = 9]{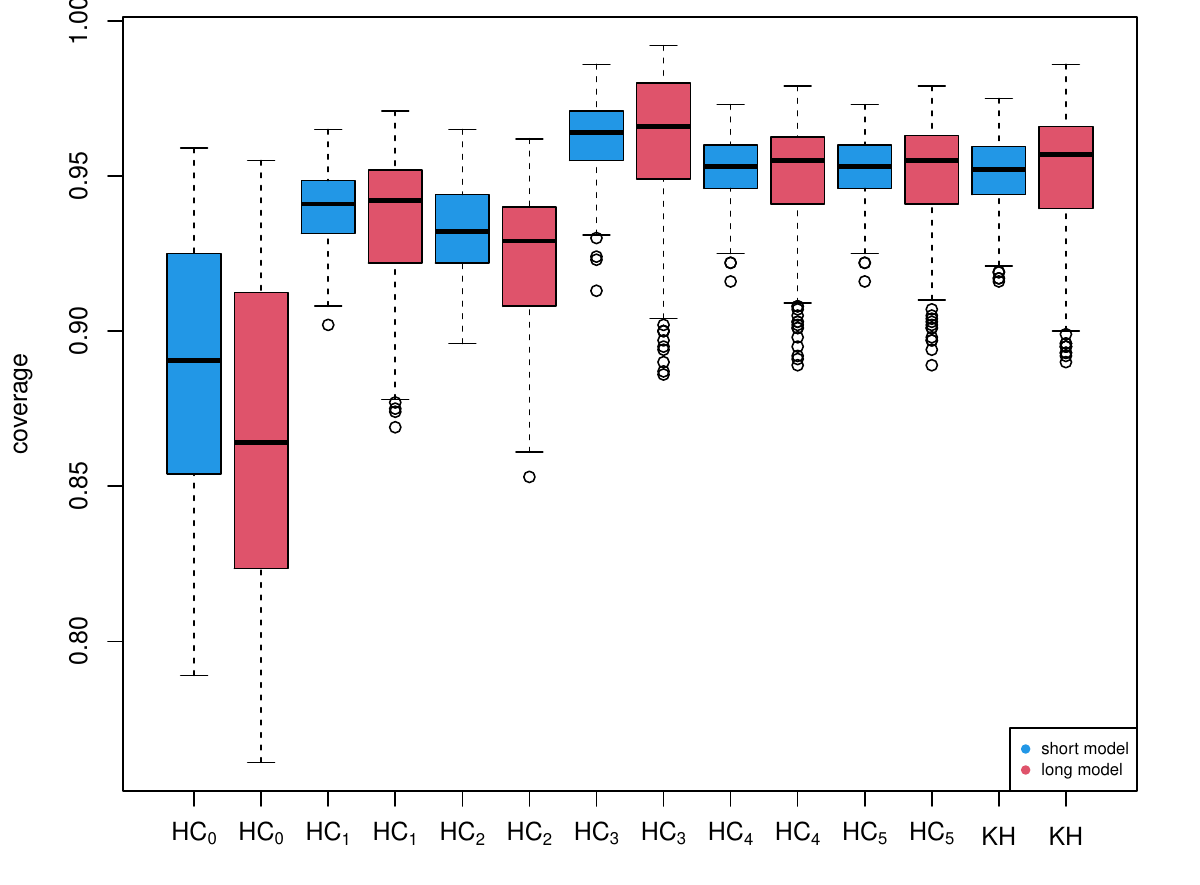}
\includegraphics[trim = 0cm 1cm 0cm 1.8cm, clip, scale = 0.6, page = 10]{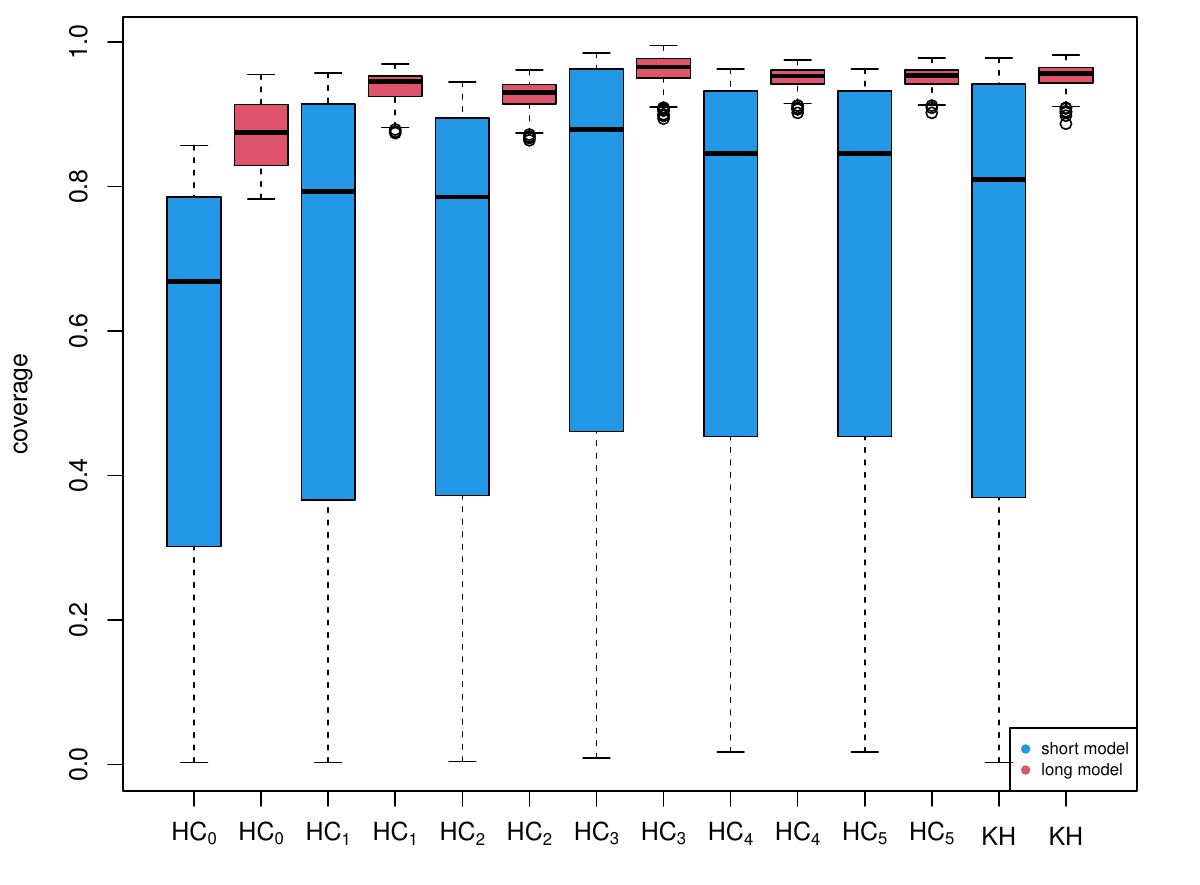}
\caption{Coverages  of $\beta_0$ compared for a model without interaction (short) and a model with interaction (long) and correlations of $\rho=0$ (upper plot) and $\rho=0.9$ (lower plot) when $\beta_{12}=0.5$.}
\label{extra_b0_corr}
\end{figure}

\begin{figure}[t]
\includegraphics[trim = 0cm 1cm 0cm 1.8cm, clip, scale = 0.6, page = 11]{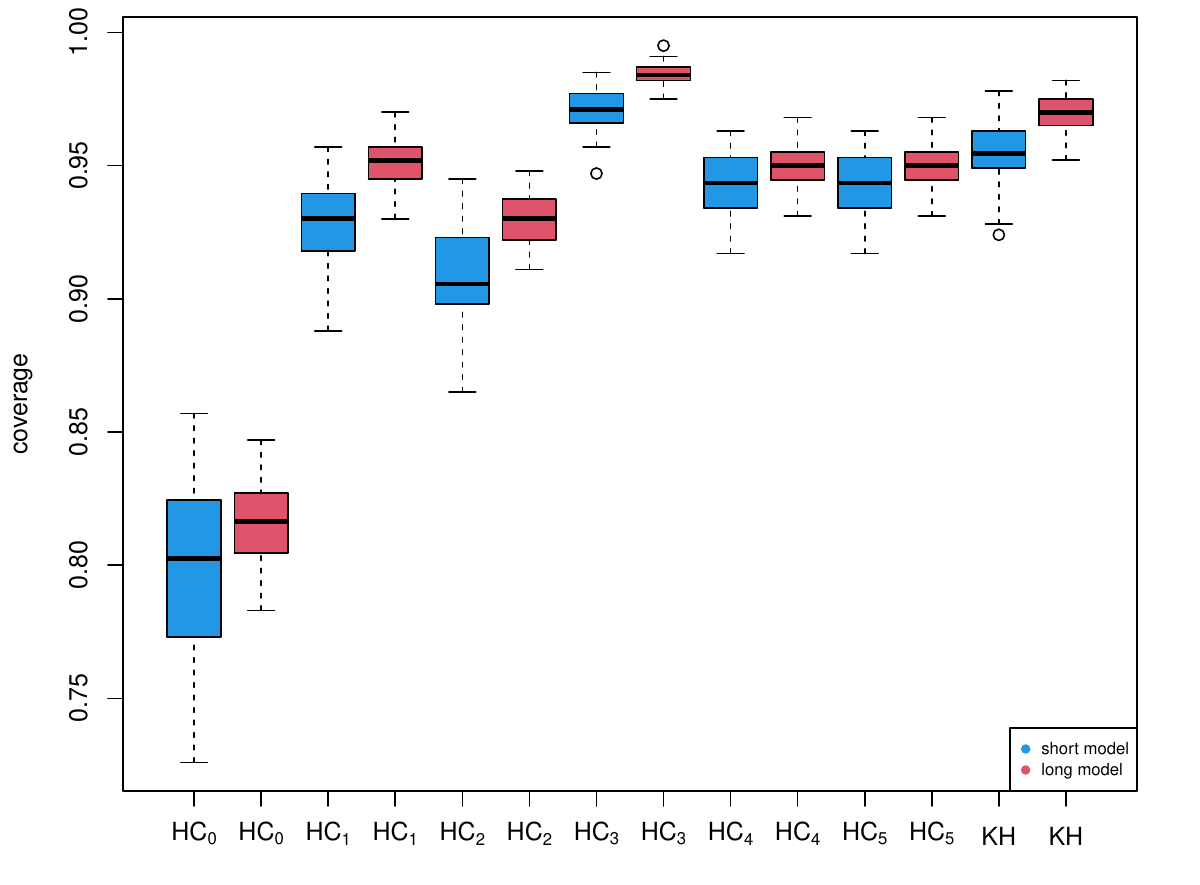}
\includegraphics[trim = 0cm 1cm 0cm 1.8cm, clip, scale = 0.6, page = 12]{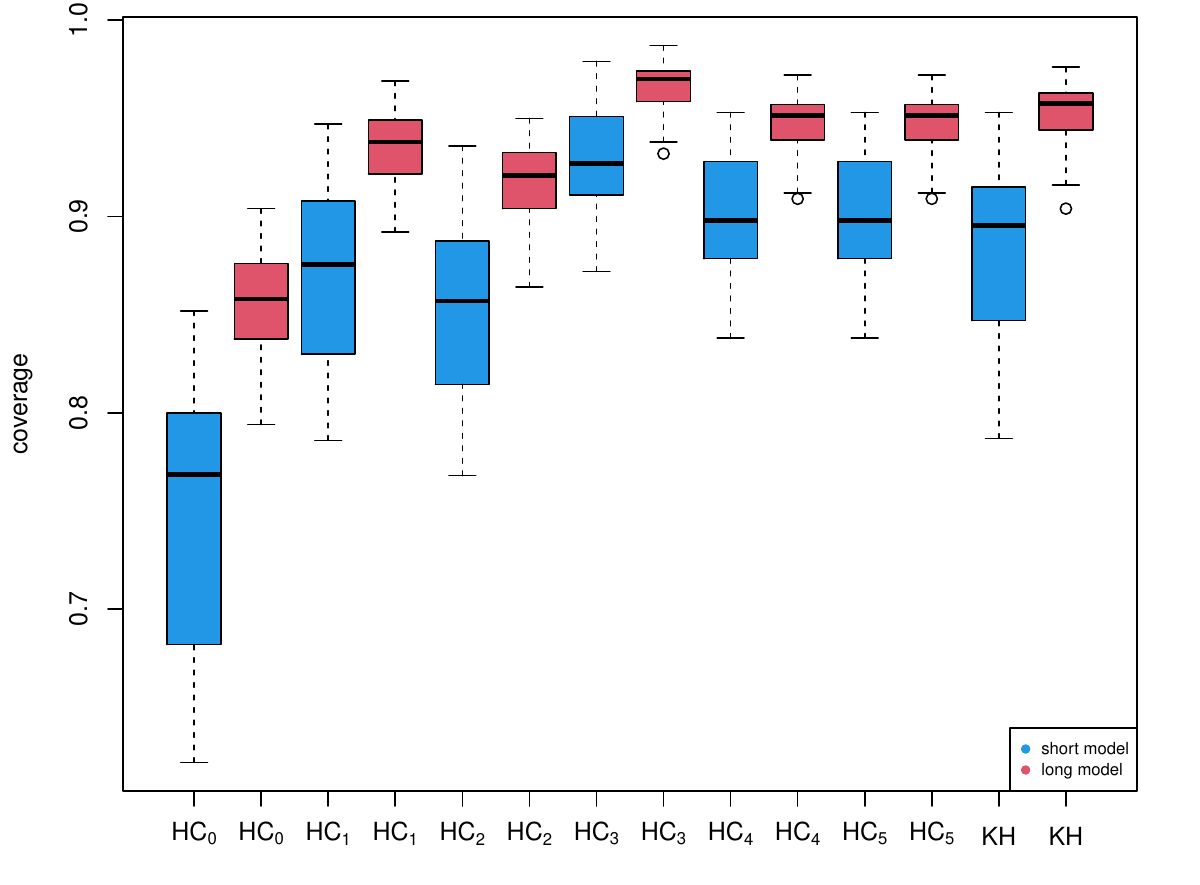}
\caption{Coverages of $\beta_0$ compared for a model without interaction (short) and a model with interaction (long) when $\rho=0.9$, $\beta_{12}=0.5$ and and the number of studies $k=6$ (upper plot) or $k=10$ (lower plot).}
\label{extra_cov_lcorr_sk}
\end{figure}

\begin{figure}[t]
\includegraphics[trim = 0cm 1cm 0cm 1.8cm, clip, scale = 0.6, page = 13]{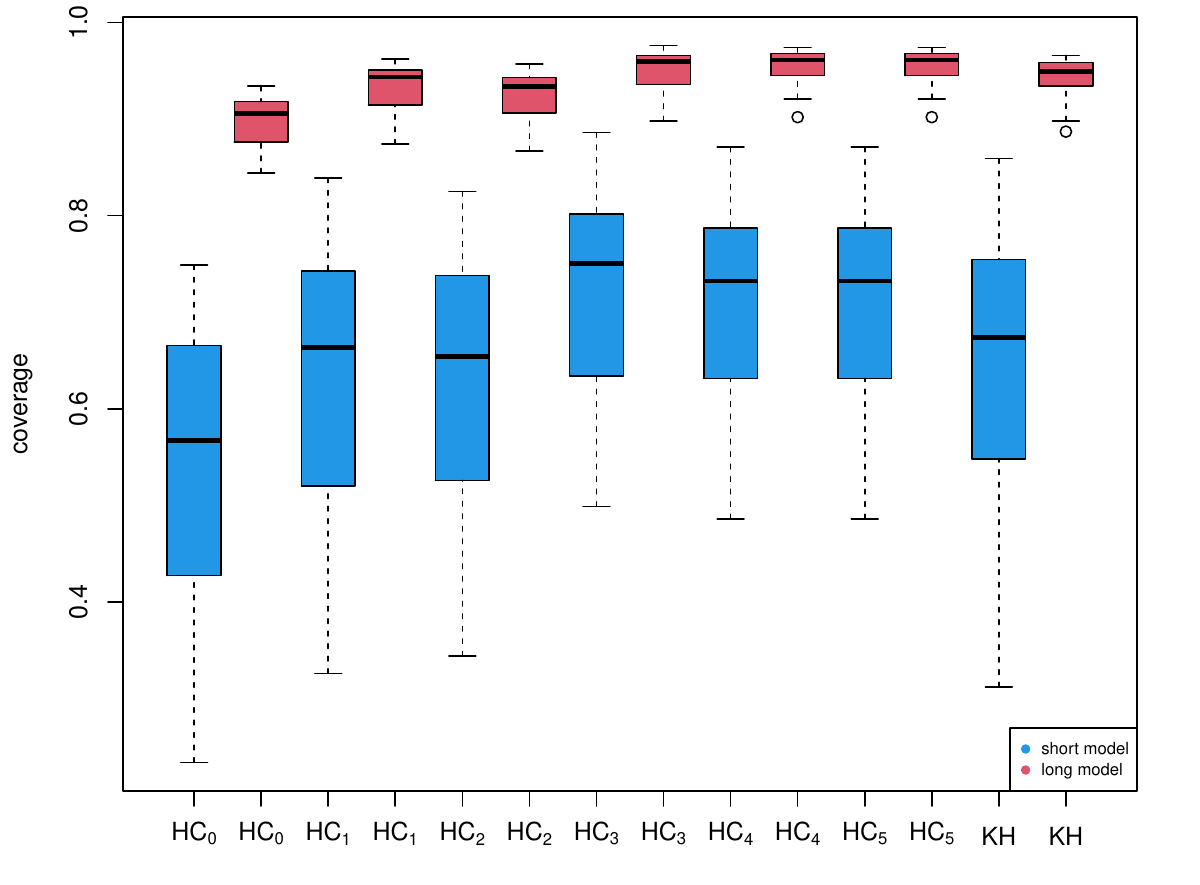}
\includegraphics[trim = 0cm 1cm 0cm 1.8cm, clip, scale = 0.6, page = 14]{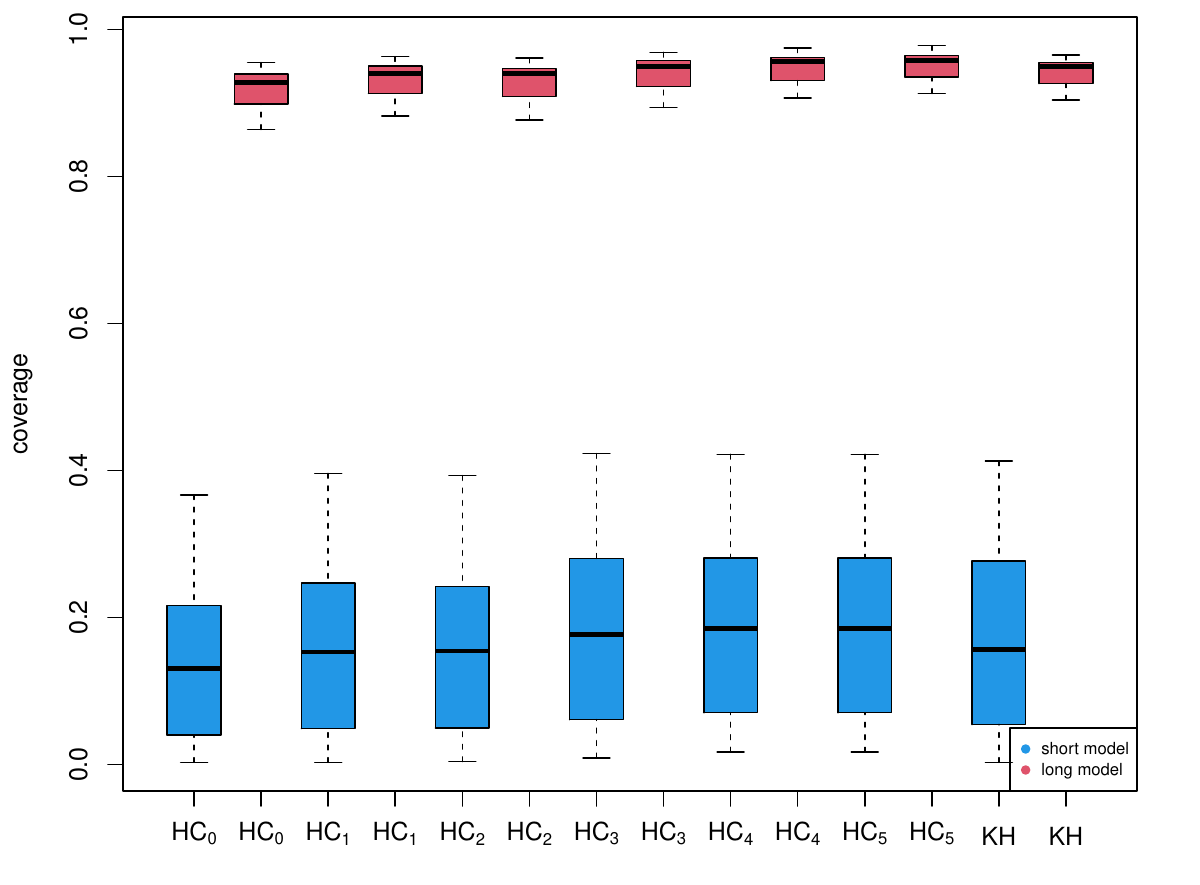}
\caption{Coverages of $\beta_0$ compared for a model without interaction (short) and a model with interaction (long) when $\rho=0.9$, $\beta_{12}=0.5$ and and the number of studies $k=20$ (upper plot) or $k=50$ (lower plot).}
\label{extra_cov_lcorr_lk}
\end{figure}

\begin{figure}[t]
\includegraphics[trim = 0cm 1cm 0cm 1.8cm, clip, scale = 0.6, page = 15]{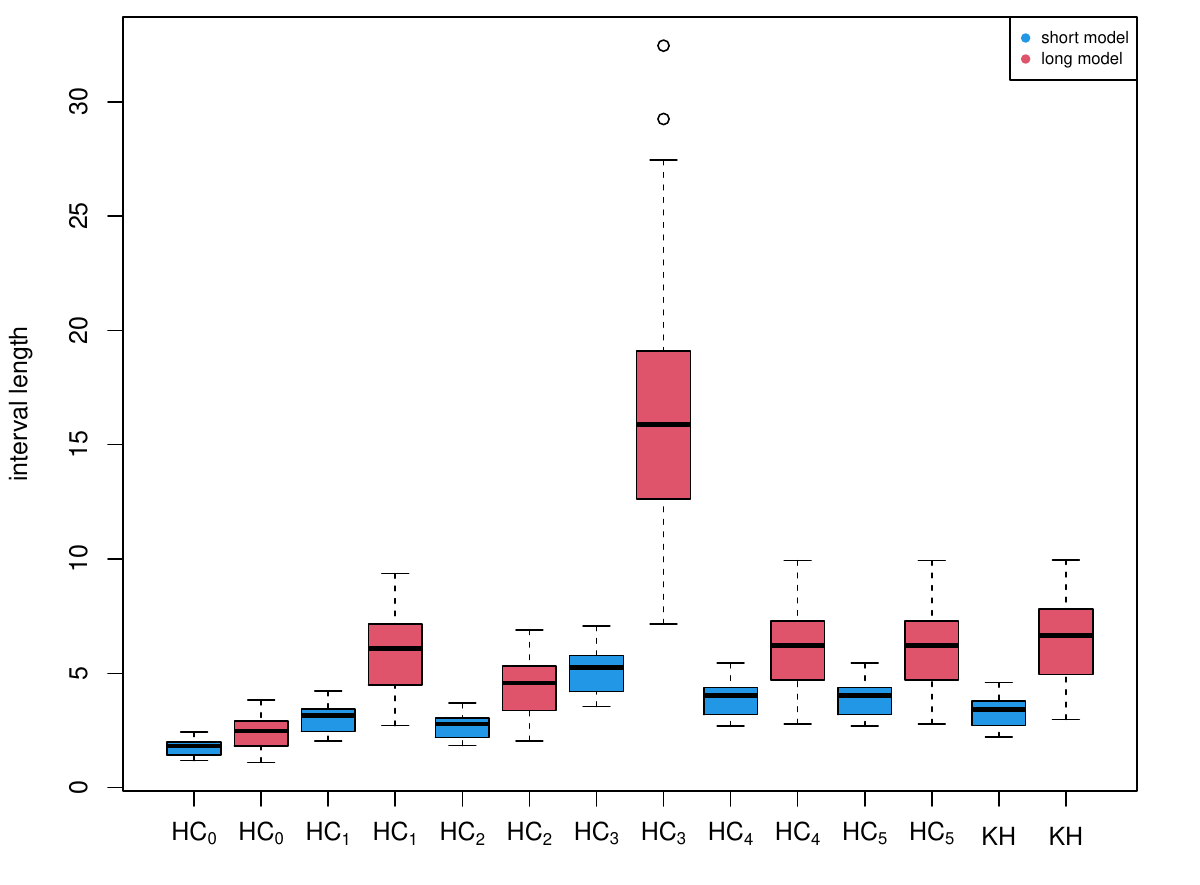}
\includegraphics[trim = 0cm 1cm 0cm 1.8cm, clip, scale = 0.6, page = 16]{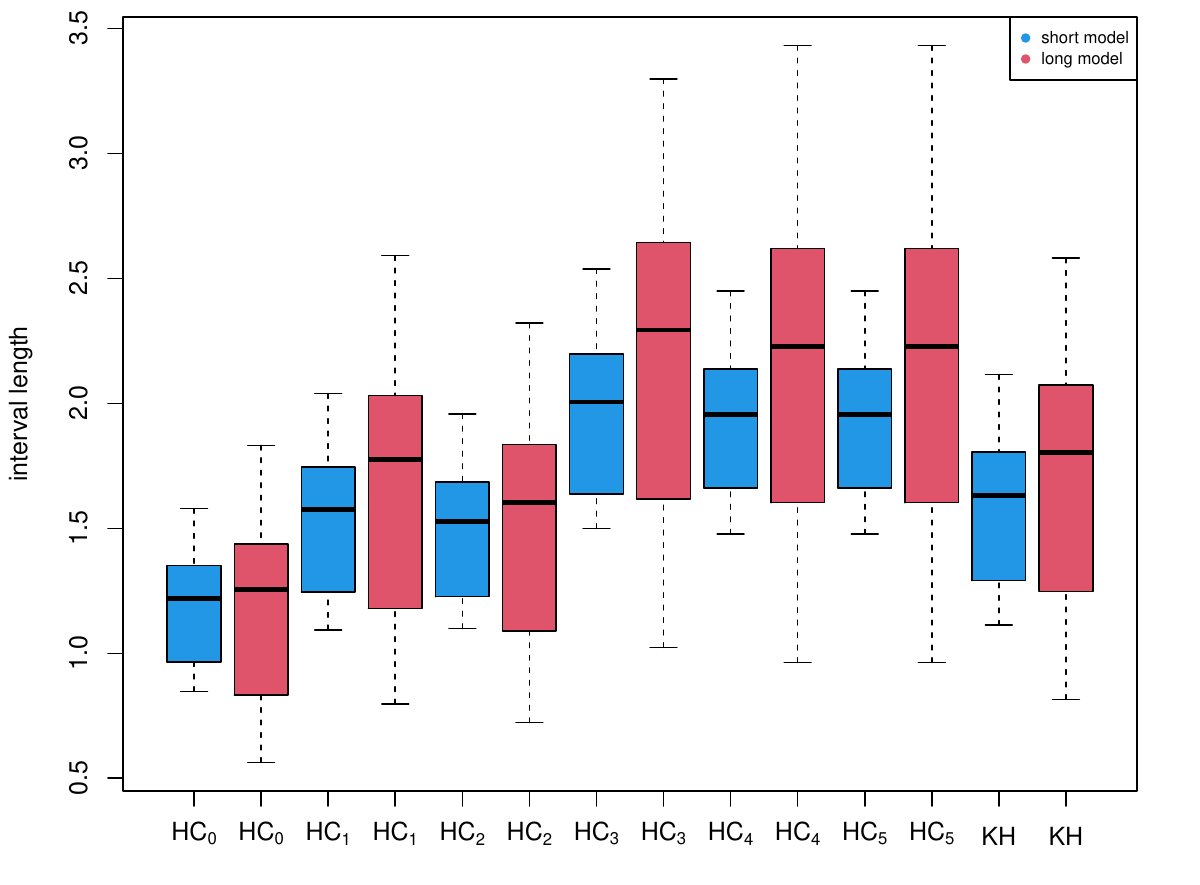}
\caption{Interval lengths  of $\beta_0$ compared for a model without interaction (short) and a model with interaction (long) when $\rho=0.9$, $\beta_{12}=0.5$ and the number of studies $k=6$ (upper plot) or $k=10$ (lower plot).}
\label{extra_lgt_lcorr_sk}
\end{figure}

\begin{figure}[t]
\includegraphics[trim = 0cm 1cm 0cm 1.8cm, clip, scale = 0.6, page = 17]{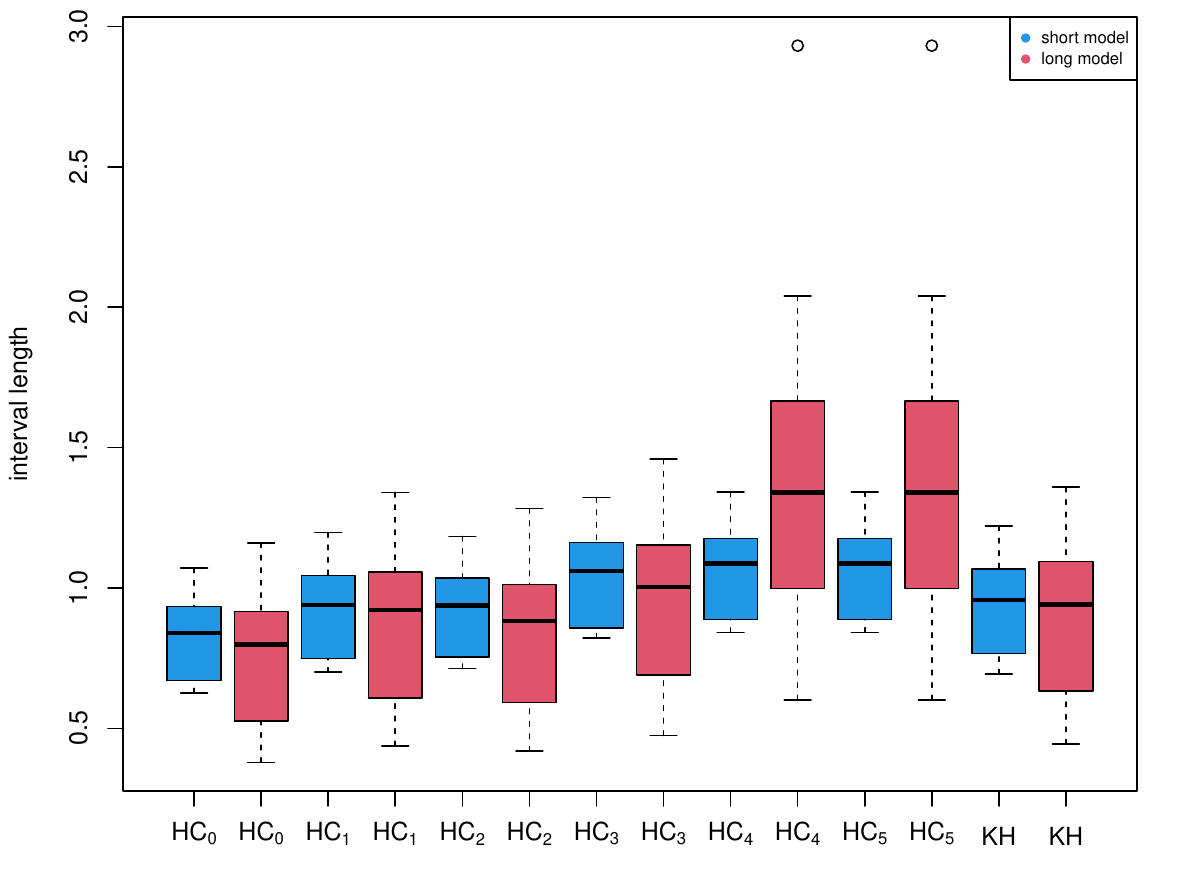}
\includegraphics[trim = 0cm 1cm 0cm 1.8cm, clip, scale = 0.6, page = 18]{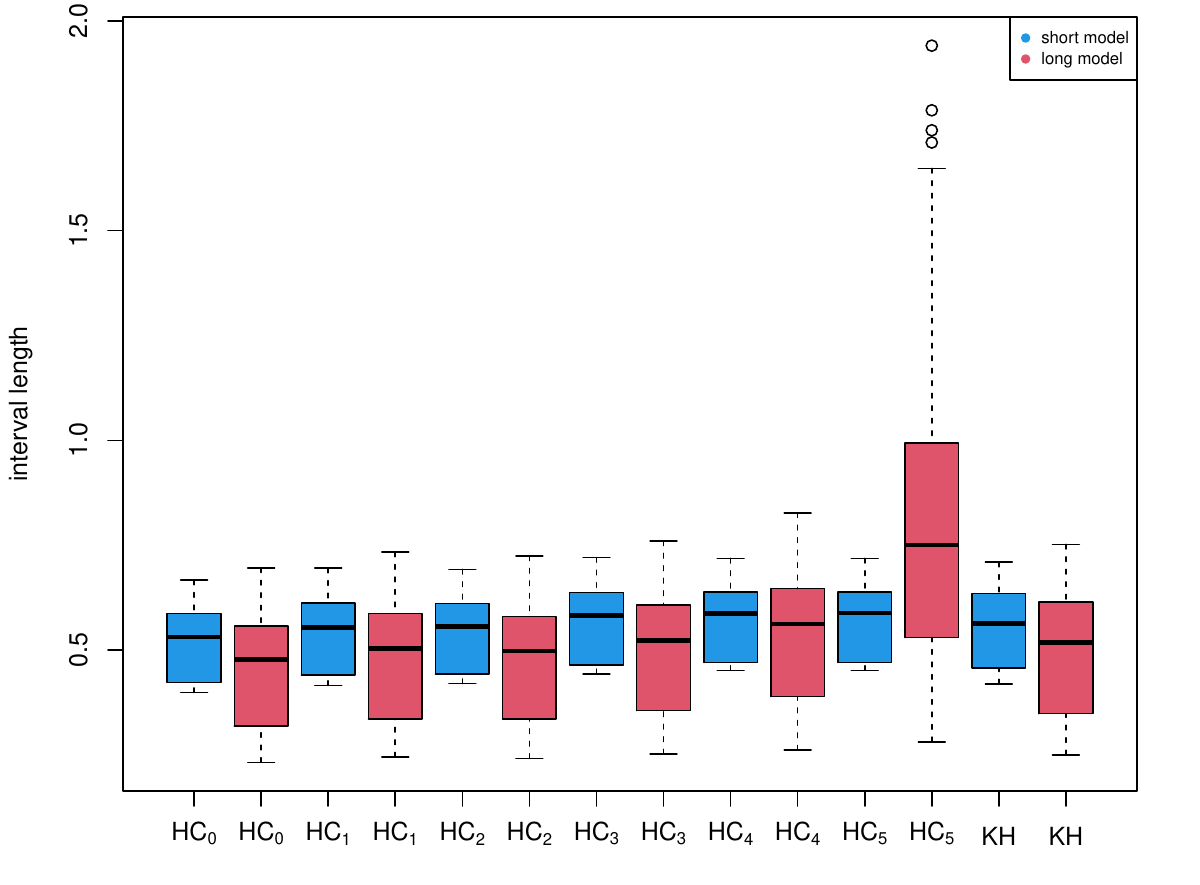}
\caption{Interval lengths of $\beta_0$ compared for a model without interaction (short) and a model with interaction (long) when $\rho=0.9$, $\beta_{12}=0.5$ and the number of studies $k=20$ (upper plot) or $k=50$ (lower plot).}
\label{extra_lgt_lcorr_lk}
\end{figure}

In Figure \ref{extra_b0_corr} the coverages of $\beta_0$ in the short model are shown for low and high correlations separately. It reveals that the low coverages for $\beta_0$ only occur when a large correlation between the moderators is present. To analyse this issue in more detail, this situation is considered again in Figures \ref{extra_cov_lcorr_sk}-\ref{extra_lgt_lcorr_lk} for each number of studies $k$ separately. As Figures \ref{extra_cov_lcorr_sk} and  \ref{extra_cov_lcorr_lk} show, the coverages are decreasing in the number of studies. The coverages close to zero occur only for $k=50$. This is probably caused by the bias in the estimation of $\beta_0$ when the moderator $\beta_{12}$ is omitted. Since the interval lengths are also decreasing in the number of studies (see Figures \ref{extra_lgt_lcorr_sk} and \ref{extra_lgt_lcorr_lk}) the true parameter is covered more often for small $k$ because the estimation is assumed to be uncertain. For large $k$ the estimation is assumed to be more precise and therefore the bias causes a lower coverage.

Finally, the coverages and lengths of $\beta_0$ and $\beta_1$ in a model with interaction are compared for $\rho=0$ and $\rho=0.9$. The coverages of the $\beta_0$-CIs and $\beta_1$-CIs based on $\textbf{HC}_0-\textbf{HC}_2$ tend to be higher for large correlations, whereas for the other estimators they are slightly lower or there is almost no difference. The lengths tend to be higher when large correlations are present for both coefficients. This result is in accordance with the results from the main simulation (see Figures \ref{b1_cor_1}-\ref{b1_cor_l_4}).

Concluding, fitting a longer model is always related to longer confidence intervals. However, neglecting an interaction may cause a large bias in the estimation and therefore lower coverages, especially when the moderators are correlated. Hence, when in doubt an interaction should always be included in the model.  Only when the correct model is fitted do very high correlations have little impact on the performance of the confidence intervals.

\clearpage

\clearpage

\clearpage

\clearpage

\clearpage

\clearpage

\clearpage

\clearpage

\clearpage

\clearpage

\begin{figure}[t]
\includegraphics[trim = 0cm 1cm 0cm 1.8cm, clip, scale = 0.6, page = 1]{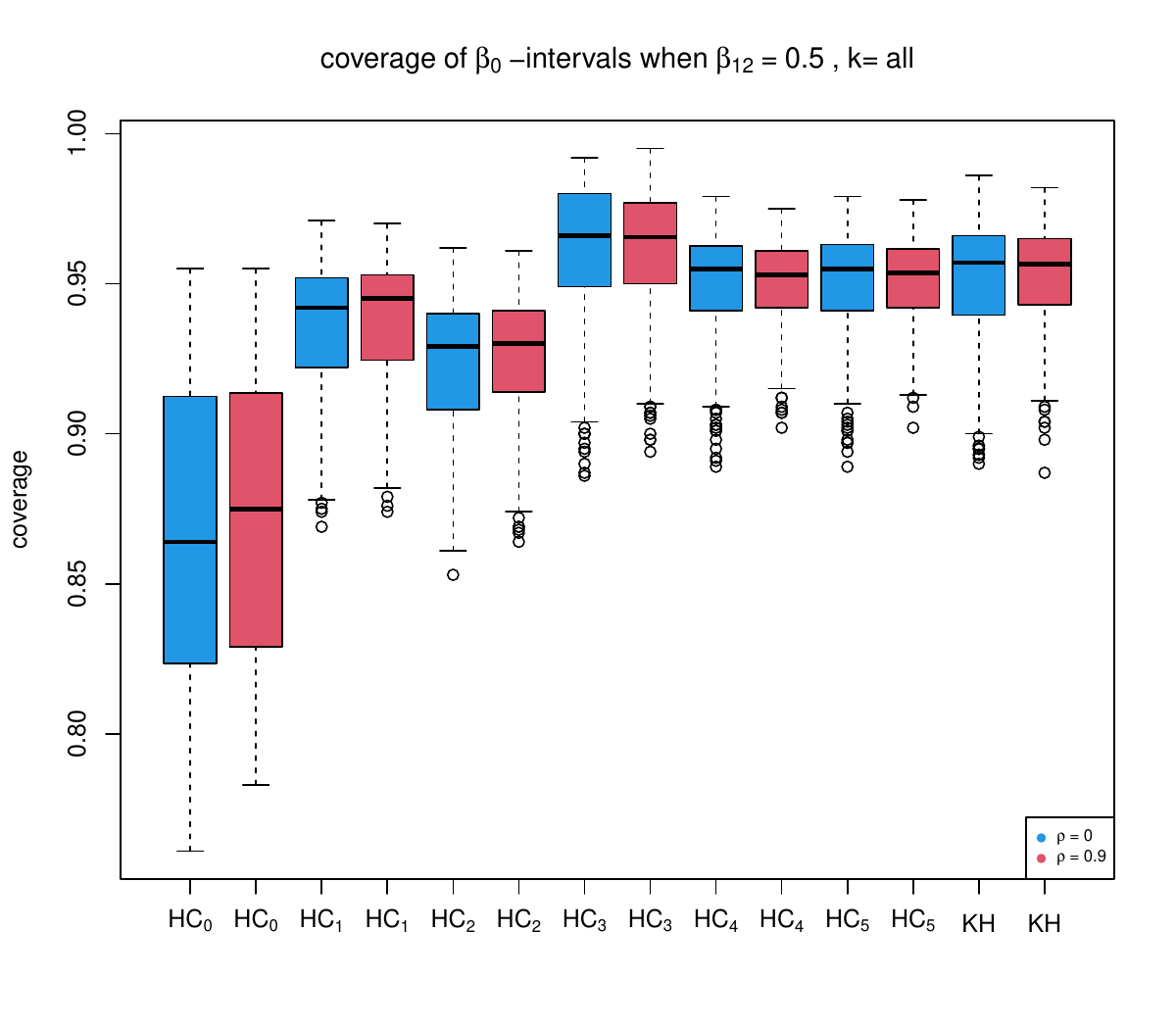}
\includegraphics[trim = 0cm 1cm 0cm 1.8cm, clip, scale = 0.6, page = 2]{figures/comp_corr_plots.pdf}
\caption{Coverages (upper plot) and interval lengths (lower plot) of $\beta_0$ in a model with interaction compared for correlations of $\rho=0$ and $\rho=0.9$ when $\beta_{12}=0.5$.}
\label{extra_corr_b0}
\end{figure}

\clearpage

\begin{figure}[t]
\includegraphics[trim = 0cm 1cm 0cm 1.8cm, clip, scale = 0.6, page = 3]{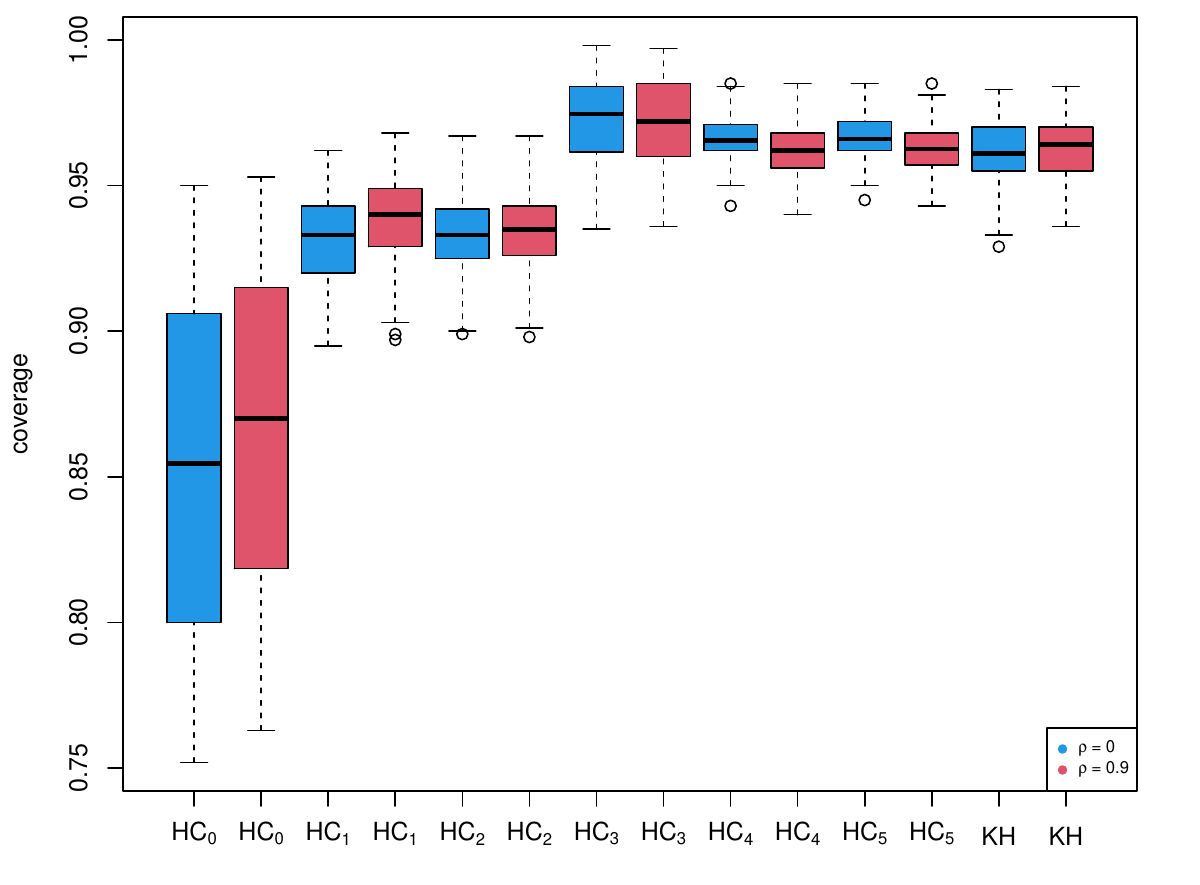}
\includegraphics[trim = 0cm 1cm 0cm 1.8cm, clip, scale = 0.6, page = 4]{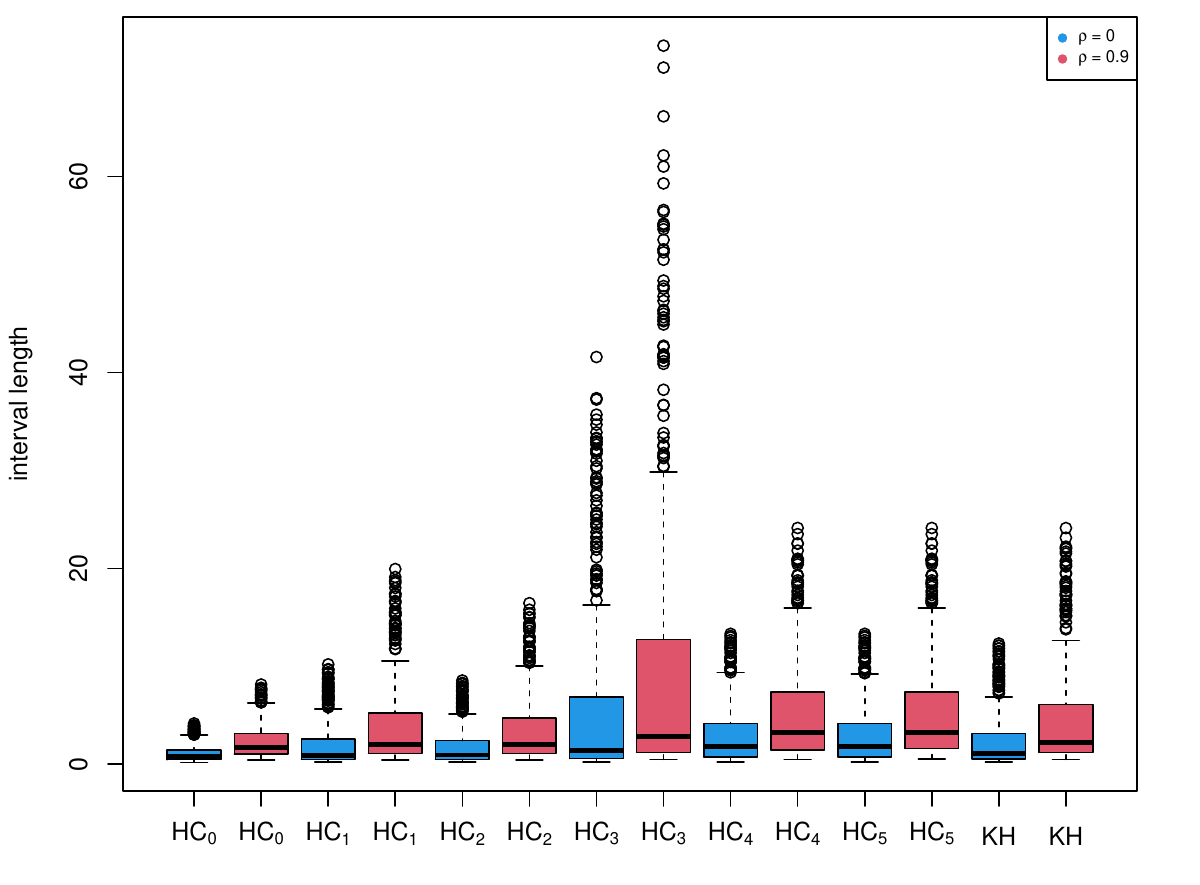}
\caption{Coverages (upper plot) and interval lengths (lower plot) of $\beta_1$ in a model with interaction compared for correlations of $\rho=0$ and $\rho=0.9$ when $\beta_{12}=0.5$.}
\label{extra_corr_b1}
\end{figure}

\end{document}